\documentclass{aastex62}
\usepackage{hyperref}
\usepackage{color}
\usepackage{natbib}
\usepackage{amssymb}

\received{January 1, 0000}
\revised{January 1, 0000}
\accepted{January 1, 0000}
\submitjournal{ApJ}

\shorttitle{DL TARDIS}
\shortauthors{Chen et al.}
\begin{document}

\title{Artificial Intelligence Assisted Inversion (AIAI) of Synthetic Type Ia Supernova Spectra}
\correspondingauthor{Lifan Wang}
\email{lifan@tamu.edu}

\author{Xingzhuo Chen}
\affiliation{George P. and Cynthia Woods Mitchell Institute for Fundamental Physics \& Astronomy, \\
Texas A. \& M. University, Department of Physics and Astronomy, 4242 TAMU, College Station, TX 77843, USA}
\affiliation{College of Physical Science and Technology,\\
Sichuan University, Chengdu, 610064, People’s Republic of China}
\affiliation{Purple Mountain Observatory, Nanjing 210008, China}

\author{Lei Hu}
\affiliation{Purple Mountain Observatory, Nanjing 210008, China}

\author{Lifan Wang}
\affiliation{George P. and Cynthia Woods Mitchell Institute for Fundamental Physics \& Astronomy, \\
Texas A. \& M. University, Department of Physics and Astronomy, 4242 TAMU, College Station, TX 77843, USA}
%\affiliation{Purple Mountain Observatory, Nanjing 210008, China}

\begin{abstract}
    We generate $\sim$ 100,000 model spectra of Type Ia Supernovae (SNIa) to form a spectral library for the purpose of building an Artificial Intelligence Assisted Inversion (AIAI) algorithm for theoretical models. As a first attempt, we restrict our studies to time around $B$-band maximum and compute theoretical spectra with a broad spectral wavelength coverage from 2000 $-$ 10000 ${\rm \AA}$ using the code TARDIS. Based on the library of theoretically calculated spectra, we construct the AIAI algorithm with a Multi-Residual Convolutional Neural Network (MRNN) to retrieve the contributions of different ionic species to the heavily blended spectral profiles of the theoretical spectra. 
    The AIAI is found to be very powerful in distinguishing spectral patterns due to coupled atomic transitions and has the capacity of quantitatively measuring the contributions from different ionic species. By applying the AIAI algorithm to a set of well observed SNIa spectra, we demonstrate that the model can yield powerful constraints on the chemical structures of these SNIa.
    Using the chemical structures deduced from AIAI, we successfully reconstructed the observed data, thus confirming the validity of the method.
    We show that the light curve decline rate of SNIa is correlated with the amount of $^{56}$Ni above the photosphere in the ejecta. We detect a clear decrease of $^{56}$Ni mass with time that can be attributed to its radioactive decay. 
  %  \textcolor{red}
    Our Code and model spectra are made available on the website \href{https://github.com/GeronimoChen/AIAI-Supernova}{https://github.com/GeronimoChen/AIAI-Supernova}.
     %We are also able to calculate theoretical luminosities of SNIa using the chemical structures determined from spectral profiles of SNIa. The theoretical luminosities are found to be correlated with the luminosities derived from supernova light curves, but with an overall offset of around 0.7 mag. This offset is likely to be due to physical approximations made in TARDIS. This study suggests that a first principle theoretical prediction of SNIa luminosity based on spectral profiles around optical maximum is possible,  provided that this magnitude discrepancy can be eliminated by theoretical models involving more realistic treatment of the radiative processes.
    
\end{abstract}


\section{Introduction}

%\textcolor{red}{Big changes are marked with red color, small changes are listed here: 1. Correct the citation "Bulla, Sim \& Kroer 2015". 2. Remove a space at section 2.2 after the TARDIS damping parameter configuration descriptions. 3. modify "the following (the) equation" 4. Corret the word "Further more" to "Furthermore". 5. Correct "Figure C". }

%Change MDD model to Initial-Guess model
Type Ia supernovae (SNIa) have been used as standard candles for cosmological probes \citep{riess1998observational,perlmutter1999measurements,WangCMAGIC2003,WangStrovink,Rubin:2013}. The tremendous success is built upon empirical methods \citep{Pskovskii:1977,Phillips:1993ApJ...413L.105P,riess1996precise,perlmutter1999measurements}. In contrast to such success, theoretical models of SNIa have not yet produced satisfactorily fits with precision matching that of observational data. In particular, it has been extremely difficult in establishing a quantitative approach to reliably model the spectral features of SNIa and determine their luminosities based on physical models. 

The root of these difficulties resides in the complexity of radiation transfer through SNIa ejecta. The process involves quantitative models of the chemical and kinematic structures of SN ejecta and detailed radiative transfer calculations. The spectral features of SNIa are normally broadened to about 10,000 km/s by ejecta motion. The atomic lines are heavily blended such that it is hard or impossible to separate spectral features arising from different atomic transitions based on conventional spectral feature measurement. A large number of input parameters and physical uncertainties affect the spectral feature formation of theoretical models. It is difficult to explore the relevant parameter space in great detail to search for the  models that provide the best fits to observations. This is true even for the simple code SYNOW \cite{branch2009comparative}. For example, if we want to optimize over 10 free parameters with each parameter modeled in 10 different parameter values, a grid search for the optimal model would require 10 billion models to be calculated. This is formidably difficult. 

Further complications arise from the unclear explosion physics of SNIa. The origin of SNIa is believed to involve one or two white dwarfs in a binary system, either the merge of two white dwarf stars (double degenerate scenario) or the accretion of a white dwarf from a non-degenerate companion star (single degenerate scenario). For some SNIa, the double-detnotation scenario seems to provide excellent fits to observations \citep{Shen:2018}. It is possible that the progenitors of SNIa form a diverse class of objects, and we do not yet know how to relate different physical systems to observations. 

The SNIa explosion occurs when the temperature and density inside the progenitor white dwarf become appropriate for carbon ignition. At least for some spectroscopically normal SNIa \cite{branch2009comparative}, the explosion is likely to involve an early deflagration phase followed by a phase of detonation \cite[the delayed-detonation model, hereafter DDT,][]{KhokhlovDDT}. The physics of the transition from deflagration to detonation is still not clear. Recently, a general theory of turbulence-induced deflagration-to-detonation transition (tDDT) in unconfined systems is published by \cite[][]{Poludnenko:2011,Poludnenkoeaau7365}. The tDDT has the potential of generating a class of models based on first principle explosion mechanisms, but its application to SNIa modeling has not yet been explored. A variety of chemical elements are synthesized during the explosions. The decay of radioactive material serves as the energy source to power the radiation of the SNe. 

Nonetheless, first principle calculations based on parameterized explosion models and detailed radiative transfer may prove to be useful \citep{Khokhlov:1991A&A...245..114K,Hoeflich:1996ApJ...472L..81H,Hoeflich:2017ApJ...846...58H}, although the density at which the transition to detonation occurs is set as a free parameter in these models. Such models usually cost days to weeks to calculate even with today's fastest computer. This makes a thorough exploration of model parameter space impossible. For example, the progenitor metallicity especially the C/O ratio plays an important role in the production of radioactive material \citep{Timmes:2003}, as well as the density at the time of detonation and the mass of the progenitor. Departure from spherical symmetry may also make the models viewing angle dependent \citep{Wang:1996ApJ...467..435W,Wang:2008ARAA, Yang:2019arXiv190310820Y, Cikota:2019MNRAS.tmp.2014C}. Due to these reasons, it is difficult to perform thorough parameter space searches to optimize model fits to observational data. Currently, the best fits to observations are usually derived from only a small number of model trials, and there is ample room for further improvement.

As an example, W7 \citet{W7} is an early model of SNIa that is still widely used in SNIa spectral syntheses. In W7 and other deflagration models (e.g., WS15, WS30, \citep{NucleosynthesisReview}), the flame speed of the deflagration is 0.01 to 0.3 of the acoustic speed in the exploding white dwarf  \citep{NucleosynthesisReview}. 
The DDT models were introduced to produce enough intermediate mass elements (IME) and radioactive material, and have been validated by radiative transfer modeling \citep{DDSequenceSpectra,Hoeflich:1996ApJ...472L..81H} and comparisons to observational data. 
These models have now been extended to 2-D and 3-D hydrodynamic simulations  \citep[e.g.,][]{3dimDDTScience}. Several variations of the DDT model, such as Gravitationally Confined Detonation (GCD) model \citep{GCD}, are proposed to address the diversity and asymmetry of spectral behavior of SNIa \citep{GCD2}. The complexities intrinsic to these physical processes explain the difficulties in identifying exactly identical spectral twins despite the fact that the broadband photometries of the majority of all SNIa can be precisely modelled empirically with a one or two parameter light curve family. 

In this study, we want to focus on spectral fitting around optical maximum. Around this phase, the ejecta of SNIa can be assumed to be expanding homologously \citep[e.g.,][]{Hoeflich:1996ApJ...472L..81H,DanielKasonLineAnalysis}. 
For $1-$dimensional ejecta structure, the SN is spherically symmetric and the radiative process can be modeled by Monte Carlo algorithms \citep{Lucy:1971ApJ...163...95L,MazzaliMCMC,Mazzali:1993A&A...279..447M,MazzilMCMC3}.
Other codes, with varying levels of physical details and complexities that have been applied to SNIa include as examples,  Hydra \citep{Hoeflich:1996ApJ...472L..81H}, SYNAPPS \citep{SYNAPPS}, PHOENIX \citep{PHOENIX} and CMFGEN \citep{CMFGEN}. Many spectral models have been computed for some well-observed SNIa, {\it e.g.}, SN1990N \citep{MazzilMCMC2}, SN1992A \citep{MazzilMCMC4}, SN1991bg \citep{MazzilMCMC2}, SN2005bl \citep{SN2005bl}, SN1984A \citep{SN1984A}, SN1999by \citep{SN1999by}, and more recently SN~2011fe \citep{rho11fe}.  Sometime both deflagration and detonation models are explored. 
Within the context of these models, the abundance stratification can be studied as the photosphere recedes in mass coordinate while the ejecta expand and become progressively optically thin \citep{AbundanceTomography}. 
This ``Abundance Tomography" was applied to some well-observed SNIa, such as SN2011fe \citep{rho11fe}, SN2002bo \citep{AbundanceTomography}, and SN2011ay \citep{AbunTomoSN2011ayTardis}. 
Also, 3-dimensional time-dependent radiation transfer programs such as SEDONA \citep{SEDONA} and ARTIS \citep{ARTIS} have been constructed, which also enable calculations of polarization spectra \citep{HoeflichPol1996ApJ...459..307H,PolarSynthesis,Wang:2008ARAA}.

Recent advancement in computer sciences, especially in Artificial Intelligence (AI), opens a new possibility for theoretical modeling of SN spectra. In general, a large number of parameters are needed to properly describe an SNIa. These parameters needs to cover critical ejecta properties such as the density structure, the chemical abundances at different layers, the expansion velocity, the radioactive heating, the geometric symmetry, and the temperature structure of the ejecta, etc. We present in this paper an attempt to construct a Artificial Intelligence Assisted Inversion (AIAI) model to study a library of theoretical spectra of SNIa around optical maximum. The AIAI trains a deep learning neural network to inverse the modeling procedure and deduce the correlations between the resulting spectra and input model parameters.

The spectral lines of SNIa are usually blended and form ``pseudo continua" that make it very difficult to isolate contributions from individual ions for even the most isolated spectral features. Most spectral lines have contributions from multiple ionic species and atomic transitions; their profiles are strongly affected by the density and kinematic structures of the ejecta. The variation of spectral features are governed by fundamental physics although it is difficult to establish a one-to-one match between a given atomic transition and the associated spectral line. The spectral features and their variation with time can be studied by voluminous realization of the atomic processes in numerous theoretical models under different physical conditions. A statistical study of the theoretical model realizations may very well unveil the hidden correlations between the atomic processes and the heavily blended observable spectral features. A neural network is by design remarkably suitable for such a study. 
 
In the study, we find that a Multi-Residual Neural Network (MRNN) \citep{MResNet} can be trained to provide the best correlation between spectral features and input physical parameters. This MRNN is further tested with a different set of theoretical models to verify model stability and reliability. We then apply the MRNN to a set of observed spectra of SNIa to derive the model parameters for the observational data. 

By calculating a large number of simulated spectra and using them as input to train the deep learning neural network, we demonstrate that AIAI can indeed reveal the underlying chemical structures of SNIa ejecta. Many of the radiative transfer codes for supernova atmosphere are technically expensive that prohibits large amount of models to be calculated. For our purpose, and as a first attempt of AIAI, we choose the code Temperature And Radiation Diffuse In Supernovae, also known as TARDIS \citep{TARDIS}, to generate the theoretical models. TARDIS is a 1-dimensional radiation transfer code using Monte Carlo algorithm. In previous studies, TARDIS has been applied to the modelling of normal SNIa \citep{TardisNormal}, type Iax SNe \citep{AbunTomoSN2011ayTardis,IaxFewParameter}, type II SNe \citep{TardisII} and kilonovae \citep{TardisKilonova}. The agreements to observations in these models have been moderately successful considering  simplicity of TARDIS and the limited coverage of the model parameter space. 

We restrict our studies to SNIa with UV coverage. The UV is important as it is very sensitive to the density and chemical structure of the SNe. Optical data alone may provide tight constraints on some intermediate mass elements but are less efficient in constraining iron group elements. 

For further test and application of AIAI, we apply the MRNN trained models to a set of observational data.
We generate neural network matched spectral models to 6 SNIa (SN2011fe, SN2011iv, SN2015F, SN2011by, SN2013dy, ASASSN-14lp) with well observed UV and optical (2,000 ${\rm \AA}$ - 10,000 ${\rm \AA}$) spectra, and 15 SNIa with wavelength coverage of $3,000 - 5,200\ {\rm \AA}$ around their $B-$band maximum luminosity. Moreover, we applied MRNN to predict the $B-$band absolute luminosities of these SNIa, based on the ejecta structure deduced from their spectral profiles. Such predictions are still rough due largely to approximations inherit to TARDIS, but future refinement may prove to be useful if these approximations are systematic and can be calibrated by other models with more complete treatment of radiative transfer physics. 

The paper is structured as the following: 
The TARDIS configuration and the model SN spectral library are introduced in Section~\ref{sec:DDmodel}. 
The neural network structure and its performance on synthetic spectra are discussed in Section~\ref{sec:DeepLearning}. 
In Section~\ref{sec:fitresults}, we apply the results of the neural network and present the resulting abundance and ejecta structure of a sample of SNIa near $B-$band maximum luminosity. 
In Section~\ref{sec:Results} we present further applications of AIAI and show the correlations between the ejecta structure and the luminosity of the SNIa with UV/optical coverage, the spectral evolution of SN~2011fe and SN~2013dy near maximum, and the predicted luminosities of all selected SNe in comparison with the absolute luminosities derived from well calibrated optical light curves. 
Section~\ref{sec:Summary} gives the conclusions and discussions. 

\section{The Generation of Model Spectral Library}\label{sec:DDmodel}

\subsection{TARDIS Spectral Syntheses}

Our study needs a library of SN spectra. This library spectra should capture the radiative processes involved in SNIa as much as possible, and cover a broad range of the physical properties of the ejecta. The parameter space is large and the number of models to be calculated can be huge.
Codes that are very CPU demanding are apparently inappropriate for this study, at least for the current attempt which is still an exploratory first step. For the determination of chemical structures, the primary requirement is that the code should be approximately correct in producing physical models to the spectral profiles of the most important observable ionic species. Luckily, there are several codes that fits this criteria. For example, the code SYNOW \citep{branch2009comparative,Parrent:2010,Thomas:2013} can run at very high speed and generate spectral libraries of a broad range of ejecta parameters. However, the code only allows a very crude description of the ejecta geometry. The input parameters are given in terms of optical depth of some reference lines of certain ionic species. Quantitative constraints to parameters related to ejecta structure and supernova luminosity is difficult using the available versions of SYNOW.

With the computing power available to us, the Monte-Carlo code TARDIS \citep{TARDIS} is a good compromise that matches the requirement of our study. The input parameters for TARDIS include the elemental abundances and density structures of the ejecta which can be flexibly modified. A spectrum can be calculated for any given day during the photospheric phase, once the location of the photosphere and the luminosity at that day are provided. TARDIS calculates the electron density, level population of ions and atoms, and the temperature of different velocity 
layers. The code offers several options for photon transfer through the ejecta. In this study, we use dilute local thermodynamic equilibrium {\tt\string dilute-lte} to calculate the atomic level population and nebular local thermodynamic equilibrium {\tt\string nebular} to calculate the ionization fraction, both equilibria are part of TARDIS \citet{TARDIS}. 
The program then evaluates the transition probabilities of atomic energy-levels. TARDIS generates energy packets (an ensemble of photons with same energy) at an inner boundary which propagate through the SN ejecta, calculates the electron optical depth by path-integrating the electron density on the trajectory, and simulates the photon-atom interaction optical depth by randomly sampling the atomic transition probabilities. When the optical depth of an energy packet reaches 1 during electron scattering process, its energy and direction is reassigned following Compton scattering process. When the optical depth of an energy packet reaches 1 during line-interaction, its energy and direction is reassigned following the transition probabilities of atomic lines. 
By collecting the emitted energy packets, the code then compares the resulting luminosity to the input luminosity, and update the photospheric temperature and the temperature throughout the ejecta. 
There are two line-interaction strategies, {\tt\string downbranch} and {\tt\string macroatom}, available in TARDIS. 
In {\tt\string downbranch} \citep{MazzilMCMC3}, photo-excited atoms are allowed to re-emit photon with same excited energy or other de-excitation transition energies, and the de-excitation channel is selected according to transition probabilities. 
Based on {\tt\string downbranch}, {\tt\string macroatom} \citep{macroatom} is a more sophisticated line-interaction strategy which allows upward and downward internal transitions with different probabilities for a photo-excited particle. 
Considering {\tt\string macroatom} is closer to the physical reality while the computational time difference between two line-interaction strategies are not much, we choose {\tt\string macroatom} for all of our calculations. 

\subsection{The Initial Guess Model} \label{IGM}

Our goal is to generate a large number of models that cover the parameter space of observed SNe as much as possible. For this purpose we need a fiducial ejecta model of which we will perturb the parameters of that model to account for ejecta diversity.

\begin{figure}
    \plottwo{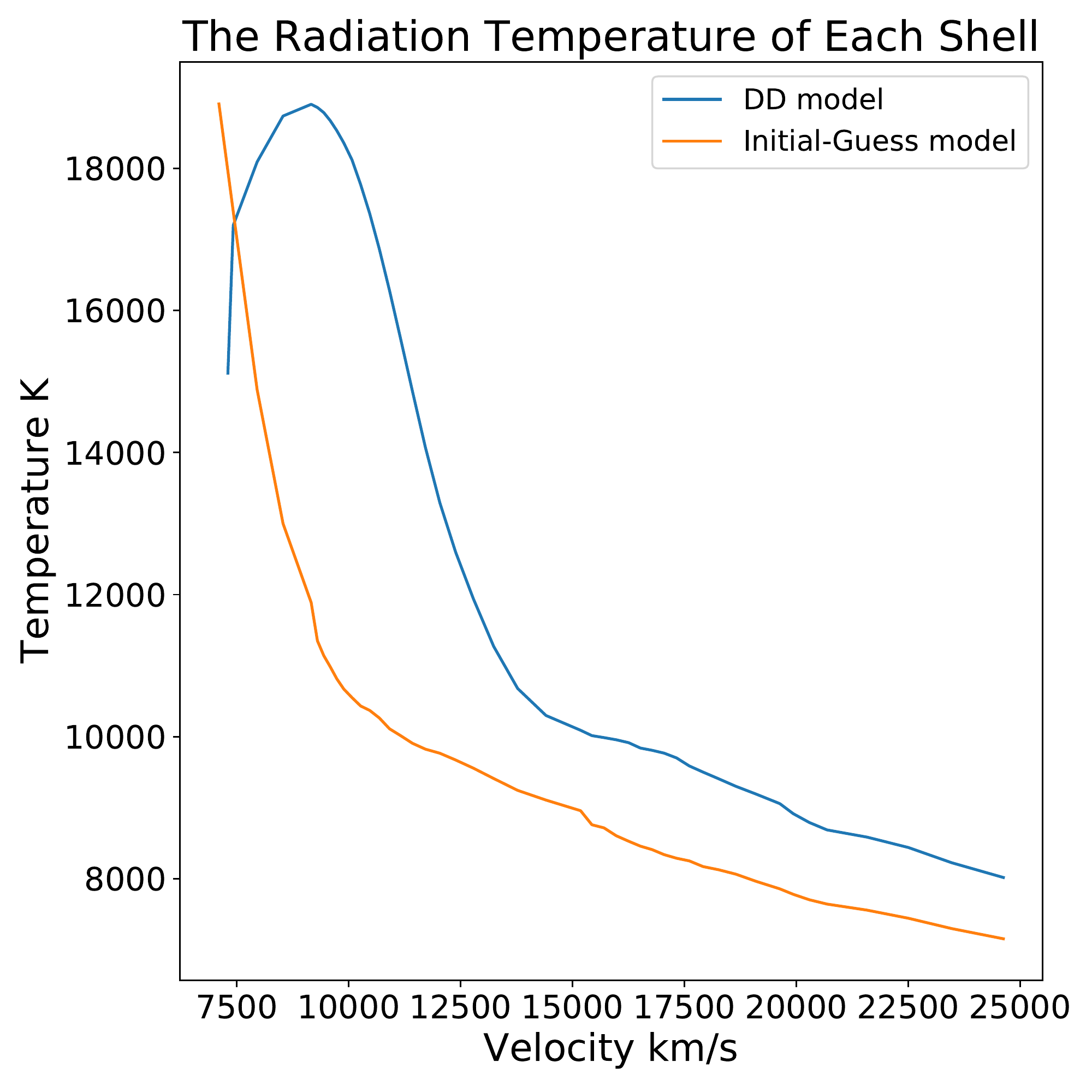}{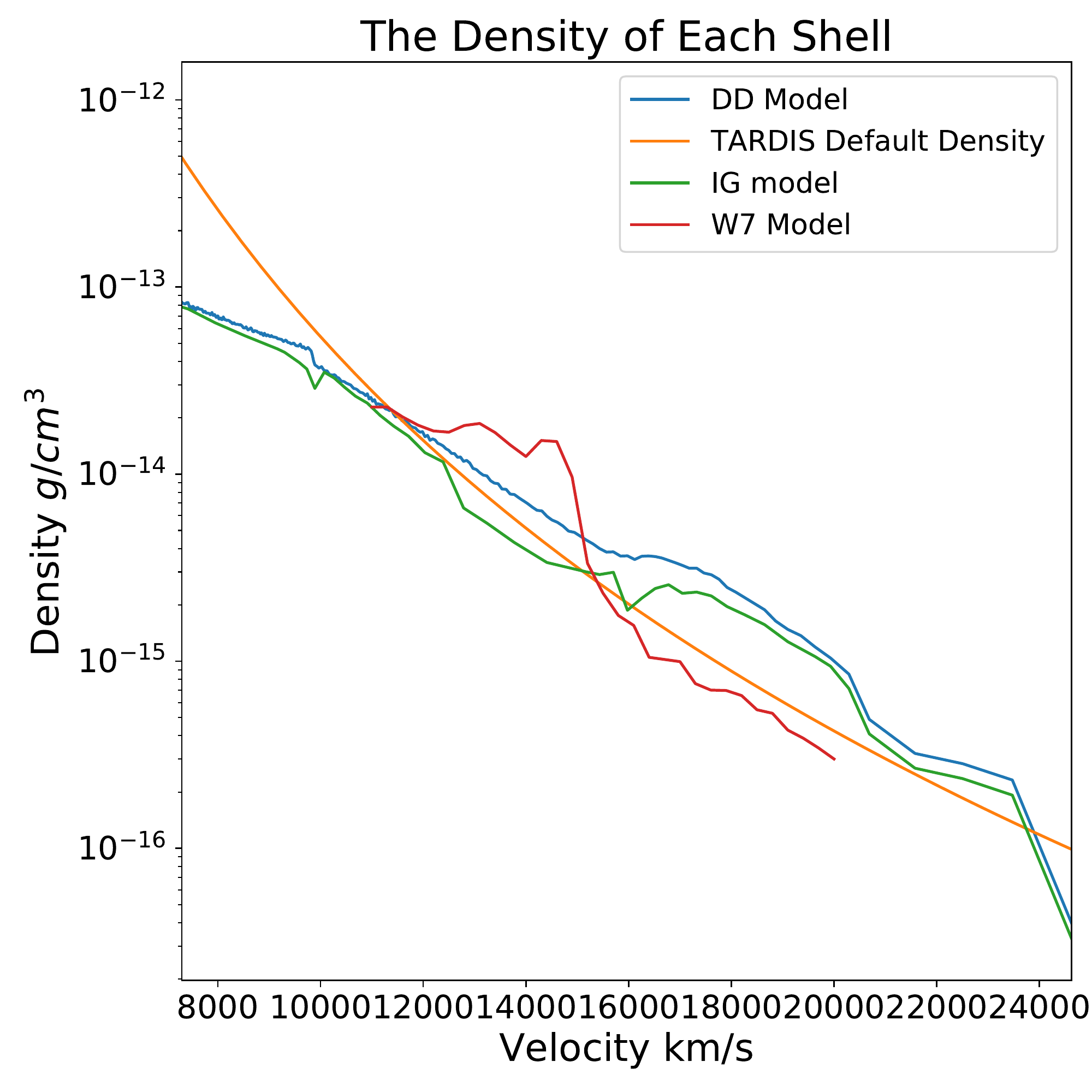}
    \plottwo{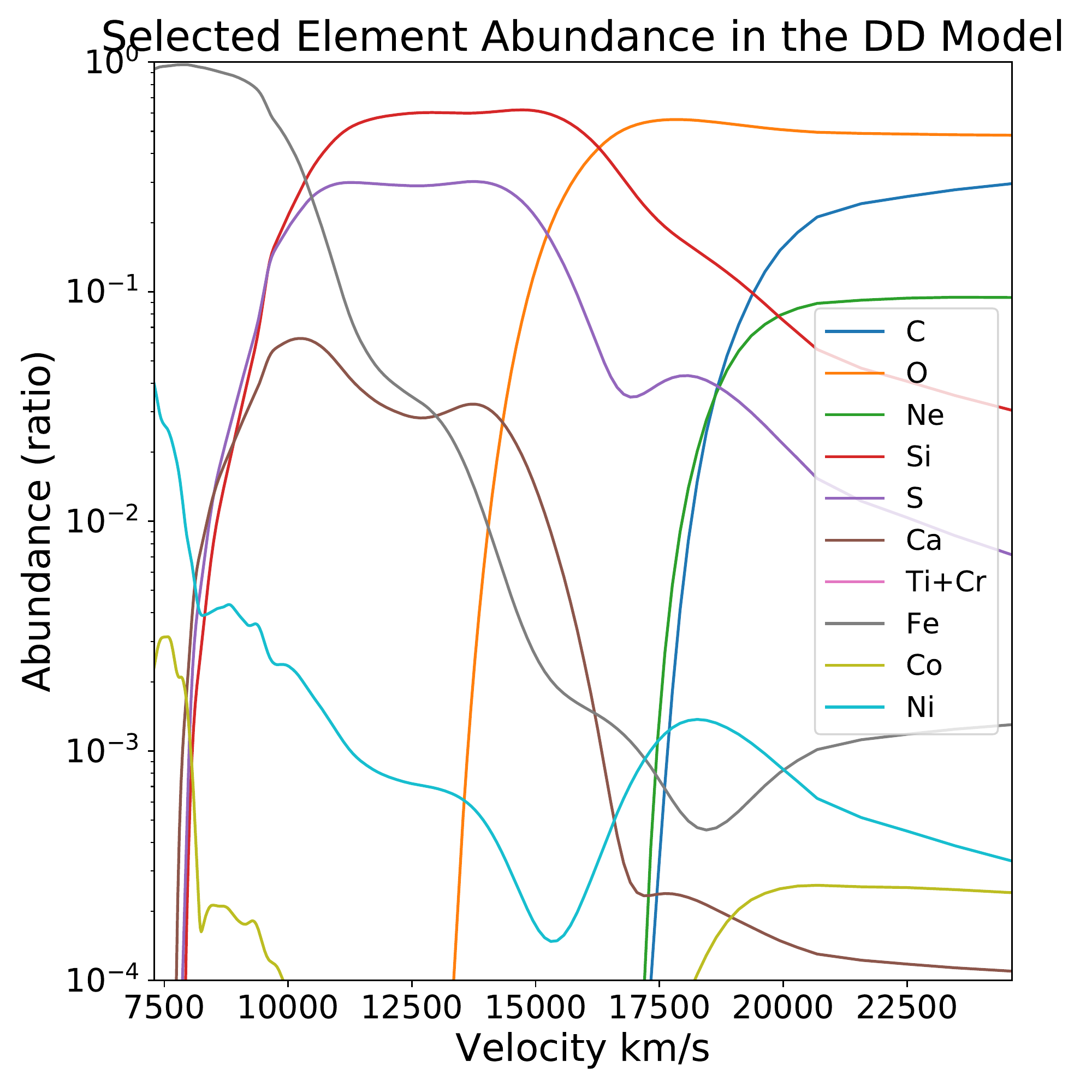}{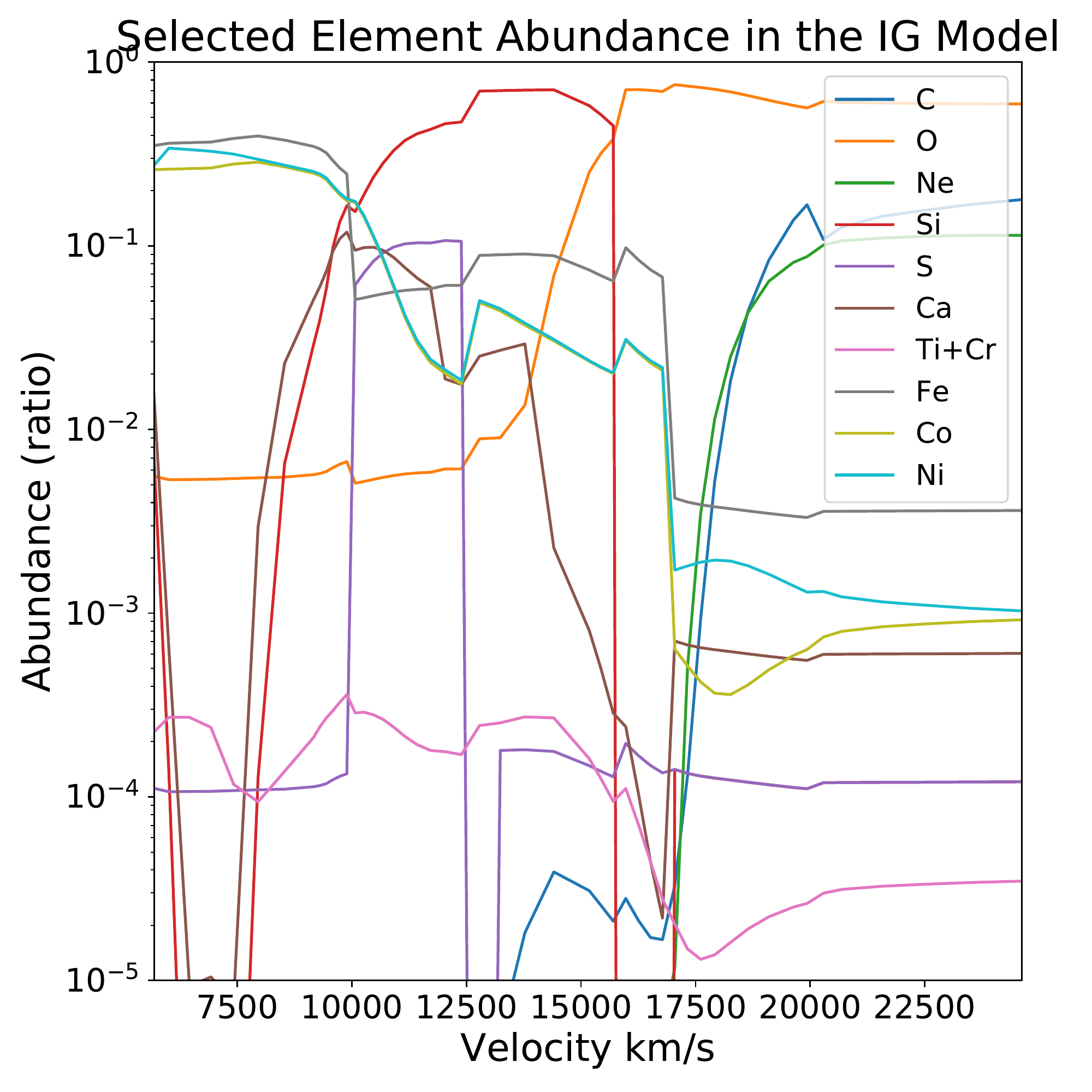}
    \plotone{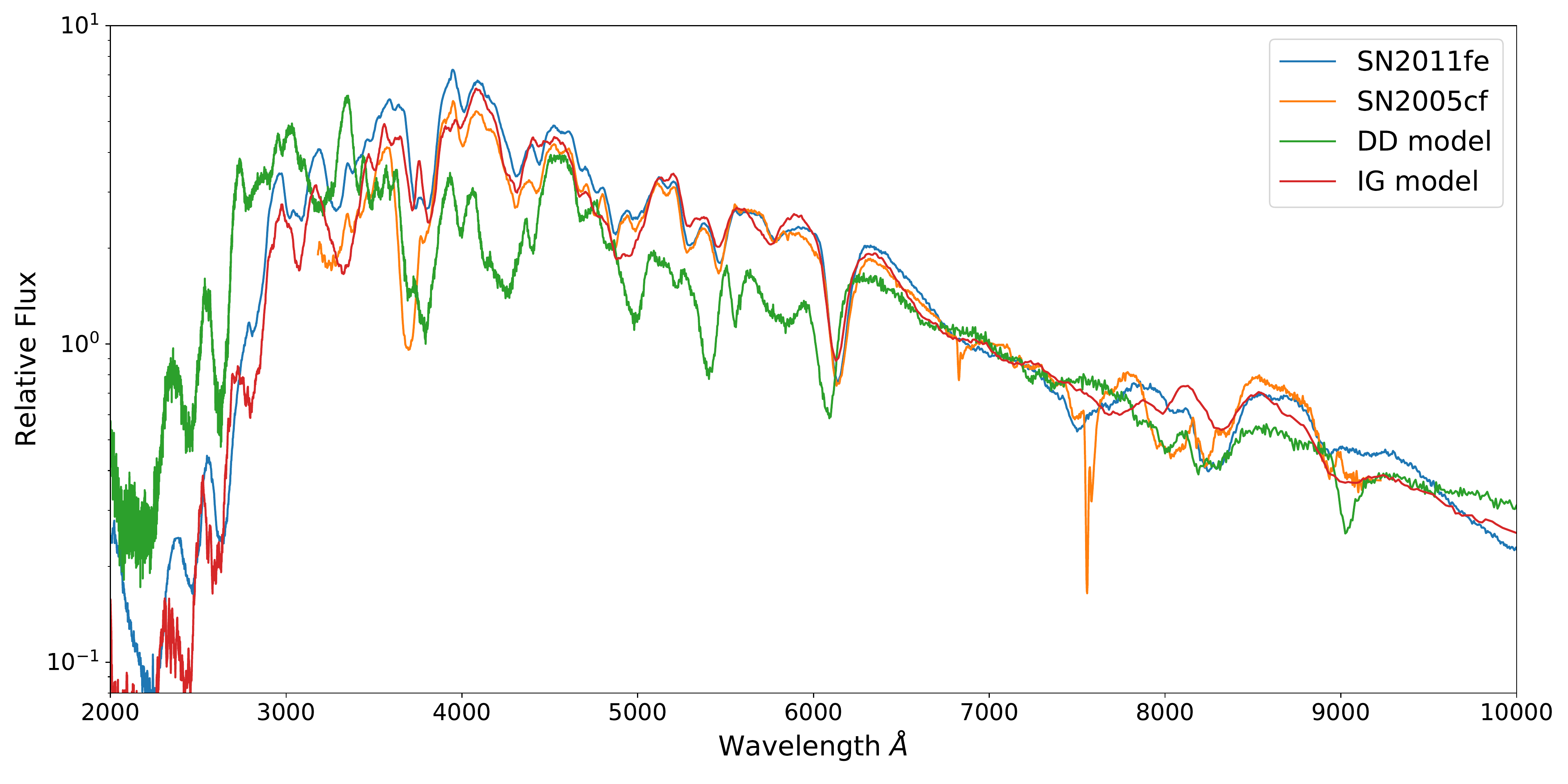}
    \caption{\textbf{(a) Upper Left:} The temperature in each shell of the DDT model {\tt\string 5p0z22d20\_20\_27g} and the IGM calculated with TARDIS. 
    \textbf{(b) Upper Right:} The density profile of the DD model, IGM, W7 model and TARDIS' default 7-order approximated W7 model ($\rho = 3\times 10^{29}\times v^{-7} \times (t / 0.000231481)^{-3} g\cdot cm^{-3}$, with $t$ in days and $v$ in km/s). 
    All of the densities are normal type Ia SN at 19 days after explosion. 
    \textbf{(c) Middle Left:} The elemental abundances of the DD model. 
    \textbf{(d) Middle Right:} The elemental abundances of the IGM, the elemental abundance ratios are normalized to 1 in each shell. 
    \textbf{(e) Lower Panel:} The TARDIS synthesized spectra of the DD model and the IGM, and the observed SN~2011fe / SN~2005cf spectra. }\label{fig:MDDmodel}
\end{figure}

We restricted our studies to models at \textcolor{black}{16 - 23} days after explosion, which correspond to the time around optical maximum. 
We used the Delayed Detonation (DD) model {\tt\string 5p0z22d20\_20\_27g} \citep{Khokhlov:1991A&A...245..114K,Hoeflich:1996ApJ...472L..81H}, and the W7 model \citep{W7} as the ejecta density profiles to derive an Initial Guess Model (IGM) for the ejecta structure (see Figure~\ref{fig:MDDmodel}).  These initial models were chosen for their success in providing reasonable fits to a number of supernovae in previous studies \citep[e.g.,][]{Hoeflich:1996ApJ...472L..81H}. The exact details of these models are not important as they only serve as a starting point for the ejecta structures and will be heavily modified later to generate the spectral library for further analyses. 

The IGM was derived by comparing model spectral shapes with the data of two well-observed SNe: SN~2011fe and SN~2005cf. The spectrum of SN~2011fe was acquired at UT $2011-9-10\ 09:22:00$ (0.39 days after $B-$band maximum luminosity) by HST/STIS \citep{rho11fe}. The SN~2005cf spectrum was acquired at UT $2005-6-12\ 00:00:00$ (the $B-$band maximum luminosity) from SSO-2.3m/DBS \citep{SN2005cf}.
The original chemical profiles of (DD) model {\tt\string 5p0z22d20\_20\_27g} is not optimized for SN~2011fe and SN~2005cf. The density and chemical structure of model {\tt\string 5p0z22d20\_20\_27g} was adjusted manually to match the strongest observed spectral features of SN~2011fe and SN~2005cf. Major changes to the elemental abundances were introduced when the profiles of strong lines such as \ion{Si}{2} 6,355 \AA\ were examined closely. Increasing iron group elements of DD {\tt\string 5p0z22d20\_20\_27g} improves significantly the spectral match in the UV of these two SNe. 
 
The observed spectra were found to be well fitted by a model with the photospheric velocity being $v_{ph}\ =\ \ 7,300$ km/s and the integrated luminosity between $6,500\ {\rm \AA} \sim 7,500\ {\rm \AA}$ being $10^{8.52}$ times solar luminosity. 
Note that to ensure the convergence of the temperature structure, the temperature profiles were calculated with 40 iterations using the default temperature convergence parameters ({\tt\string type:damped. damping\_constant:1. threshold:0.05. fraction:0.8. hold\_iterations:3. t\_inner\_damping\_constant:1.}). The default iteration in TARDIS is 20 which is usually sufficient to reach temperature convergence. 

The temperature and density profiles of the IGM model that fits major spectral features of SN~2011ef and SN~2005cf are shown in Figure~\ref{fig:MDDmodel} (a) and (b), respectively. 
The chemical profiles of DD {\tt\string 5p0z22d20\_20\_27g} and the IGM are shown in Figure~\ref{fig:MDDmodel}~(c) and (d), respectively. The simulated spectra of the IGM are shown in Figure~\ref{fig:MDDmodel}~(d) where we also show the spectra computed using DD {\tt\string 5p0z22d20\_20\_27g} for comparison. Note that in Figure~\ref{fig:MDDmodel}~(d), the flux levels of the models and the data were arbitrarily scaled to match the spectral features. The fits to the observed spectra of SN~2011fe and SN~2005cf across most spectral lines and UV continua are apparently better for the IGM model than for the original DD {\tt\string 5p0z22d20\_20\_27g} model. Note that we did not attempt to construct a quantitative model to the elemental structure at this stage. This manual step only modified the masses of a limited number of elements (listed in Section~\ref{tab:Residual2}) with  prominent spectral features to achieve a crude fits to the data. 

\subsection{The Model Spectral Library} \label{Sec:SpecLib}

For our deep learning neural network, the IGM was used as a baseline model that was perturbed to generate the library of model spectra. It is impossible to build a complete model grid with varying elemental abundances at all velocity layers considering the large number of chemical elements involved, so we simplified the model into limited number of velocity zones and use random sampling of parameter space to cover a broad range of physical possibilities. For the structure of the ejecta, we divided the ejecta into four distinct zones defined by different velocity boundaries; the velocity ranges of Zones 1, 2, 3, and 4 are 5690-10000 km/s, 10000-13200 km/s, 13200-17000, and 17000-24000 km/s, respectively. The ejecta structure includes all 23 elements with atomic number from 6 to 28. With four velocity zones, this leads to 92 free parameters on the masses of the chemical elements in each zone. The other parameters to run TARDIS but are not included in MRNN training are the luminosity, date of explosion, and photospheric velocity. The total number of parameter space dimensions is 95. It is impossible to construct a grid of models in such a vast dimension. The total number of models would reach a staggering value of 2$^{95}\ \sim\ 4\times10^{28}$ even if only two grid values are sampled for each dimension.

The chemical compositions and densities were allowed to fluctuate independently within each velocity zone but inside each zone the velocity dependence of the density of each element is scaled to that of the  same element of the IGM. For elements other than iron group, we used the following equation
\begin{equation}
    \label{eq:zones}
\rho_{ik}\ = \ \rho_{ik}^{IGM} \ \times\ 3U_k, 
\end{equation}
where $i$ denotes each different chemical elements, $k$ denotes the four different zones with $k\ = \ 1,\ 2,\ 3,\ 4$ stands for the velocity shells between 5690-10000 km/s, 10000-13200 km/s, 13200-17000 km/s, and 17000-24000 km/s, respectively, and $U_k$ is a random number drawn from a [0,1) uniform distribution. The locations of the velocity boundaries were chosen to approximately match major chemical layers in the IGM, and $\rho_i^{IGM}$ is the density profile of element $i$ in the IGM of shell $i$. The total density including all the elements is calculated from the sum over all elements. With Equation~(\ref{eq:zones}), the masses of the elements in the 4-zones were artificially scaled from 0 to 300\%\ relative to their respective values in the IGM. 

Because Zone 1 contains mostly Fe, Co and Ni and is partially inside the location of the photosphere, we choose the variation of Fe, Co and Ni in Zone 1 to obey 
\begin{equation}
\label{eq:zonesIron}
\rho_{ik}\ = \ \rho_i^{IGM}\ \times\ 3^{(1-2U_k)}, 
\end{equation}
where $U_k$ is again a random number drawn from a uniform distribution between 0 and 1.
Equation~(\ref{eq:zonesIron}) samples the chemical structures around $\rho_i^{IGM}$  more frequently than large deviations from it. Hereafter, the elemental mass ratios between the library models and the IGM are delineated as "multiplication factor". 

Another input parameter of TARDIS is the time after explosion. For the current study we restricted the models to time around optical maximum. This parameter was drawn from a uniform distribution between 16 to 23. The location of the photosphere was defined by interpolating the values given in Table~\ref{tab:photosphere}, with an additional random number from a uniform distribution between -120 km/s and 120 km/s to sample the range of possible variations of the location of the photosphere. The ranges of the location of the photosphere in Table~\ref{tab:photosphere} were derived using the IGM as input to TARDIS for which the corresponding photospheric velocities provide reasonable matches to the observed spectra of SN~2011fe between day~16 and day~23 after explosion.
In this process, we found the photosphere recesses approximately 300 km/s per day, which is consistent with the results of previous model fits by others \citep[e.g.,][]{rho11fe}. 

TARDIS also requires the luminosity of the SN as an input parameter. For the spectral library, the luminosities in the range of 6500\ -\ 7500 {\rm \AA}\ of the different models observe a uniform distribution in log-space between $10^{8.64} $ to $ 10^{8.7}$ solar luminosity. 

TARDIS assumes a predefined sharp inner boundary and does not calculate the gamma-ray transport in the ejecta. The radial temperature profile is only approximately correct, and in some regions may be severely incorrect. This caveat makes it difficult to constrain physical quantities derived from TARDIS based on first principle physics. The models are calculated independently at each epoch. This may introduce error when comparing the models with observed luminosities. However, we expect the spectral profiles to be governed by fundamental physics and the model spectra bear the imprints of the atomic processes. We rely on AIAI to extract the  finger prints of these atomic processes through analyses of a large set of spectral models. Furthermore, some of the problems indigenous to TARDIS may be coped with by comparing a subset of TARDIS generated models with more sophisticated radiative transfer codes such as PHOENIX \citep{PHOENIX}, CMFGEN \citep{CMFGEN} and HYDRA \citep{Hoeflich:1996ApJ...472L..81H} to identify systematic differences among these models. These systematic comparisons may lead to a large library of supernova models with more physical processes being taken into account, although 
the calculations of a large library of PHOENIX, CMFGEN or HYDRA models are too computationally expensive to be practical.
We leave a comparative study of different models to future studies.

We used TARDIS's default temperature convergence strategy in calculating these models but with 15 temperature iterations ({\it i.e.}, {\tt\string type:damped. damping\_constant:1. threshold:0.05. fraction:0.8. hold\_iterations:3. t\_inner\_damping\_constant:1. }). Additionally, in order to generate spectra with different signal-to-noise ratios, we set the number of Monte-Carlo packets to vary from $1.5\ \times\ 10^5$ to $2\ \times\ 10^6$. The wavelength coverage of the model spectra was set to 2000 - 10000 \AA.

\begin{deluxetable}{ccccccccc}
    \tablecaption{Photospheric Velocity and Explosion Date}\label{tab:photosphere}
    \tablehead{\colhead{Days After Explosion} & \colhead {16} & \colhead {17} & \colhead {18} & \colhead {19} & \colhead {20} & \colhead {21} & \colhead {22} & \colhead {23}}
    \startdata
    Photosphere Velocity (km/s) & 8090 & 7850 & 7430 & 7100 & 6650 & 6290 & 6050 & 5690 \\
    \enddata 
\end{deluxetable}

\subsection{Model Spectra Computation}

The calculation of a single spectrum of the spectral library  with about 10$^7$ energy packets takes approximately 0.5 to 2 CPU-hours on our workstation which is mounted with two Intel Xeon E5-2650 v4 CPUs. After about one month's calculation utilizing 80 cores (2 chips of Intel Xeon E5-2650 v4, and one chip of Intel Xeon E5-2650 v1), a total of 99510 spectra were generated with varying elemental abundances, photospheric velocities, explosion dates, and luminosities, as prescribed in the previous Section. 

\subsection{Response of Spectral Profiles to the Variations of Input Abundances}
\begin{figure}
    \plottwo{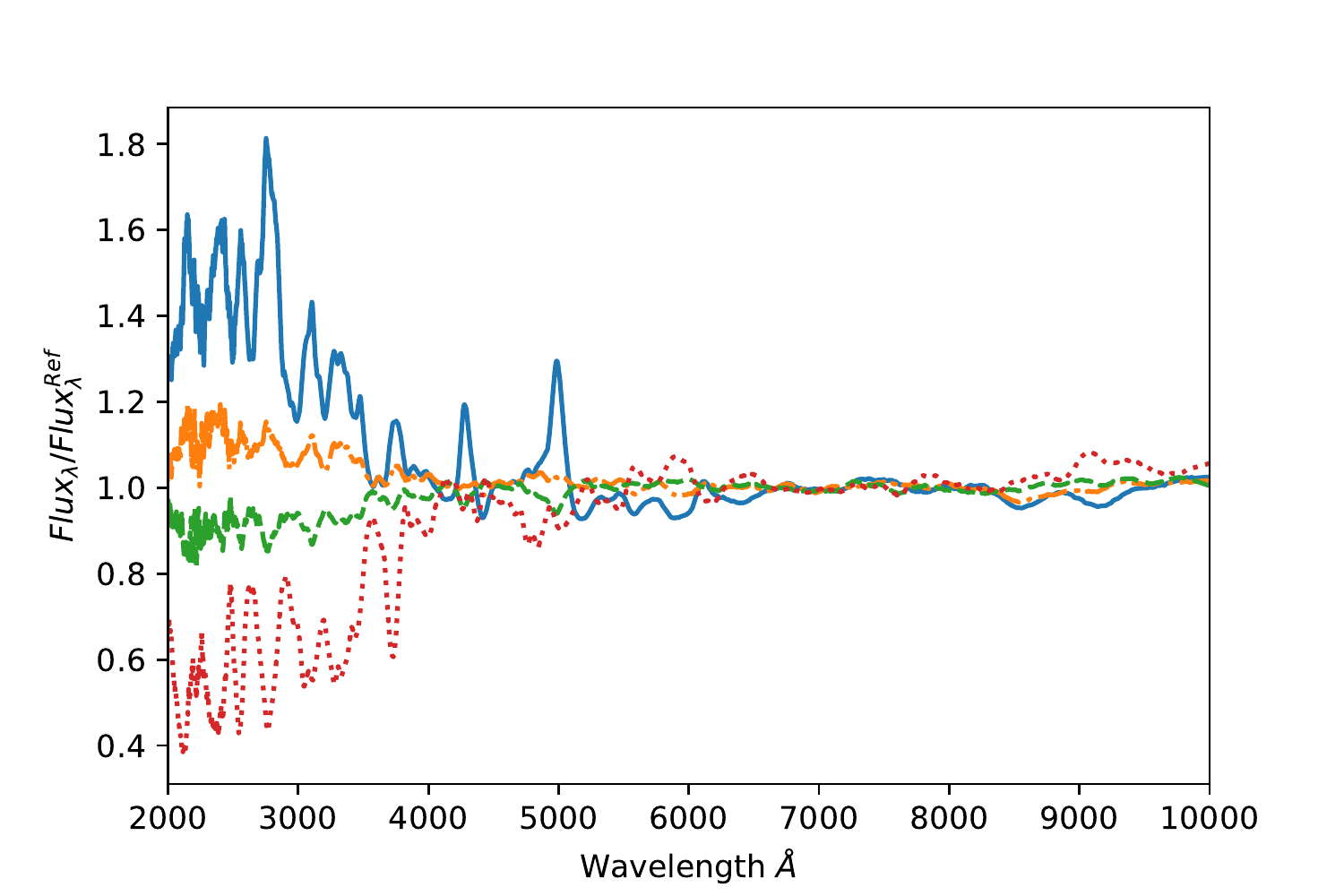}{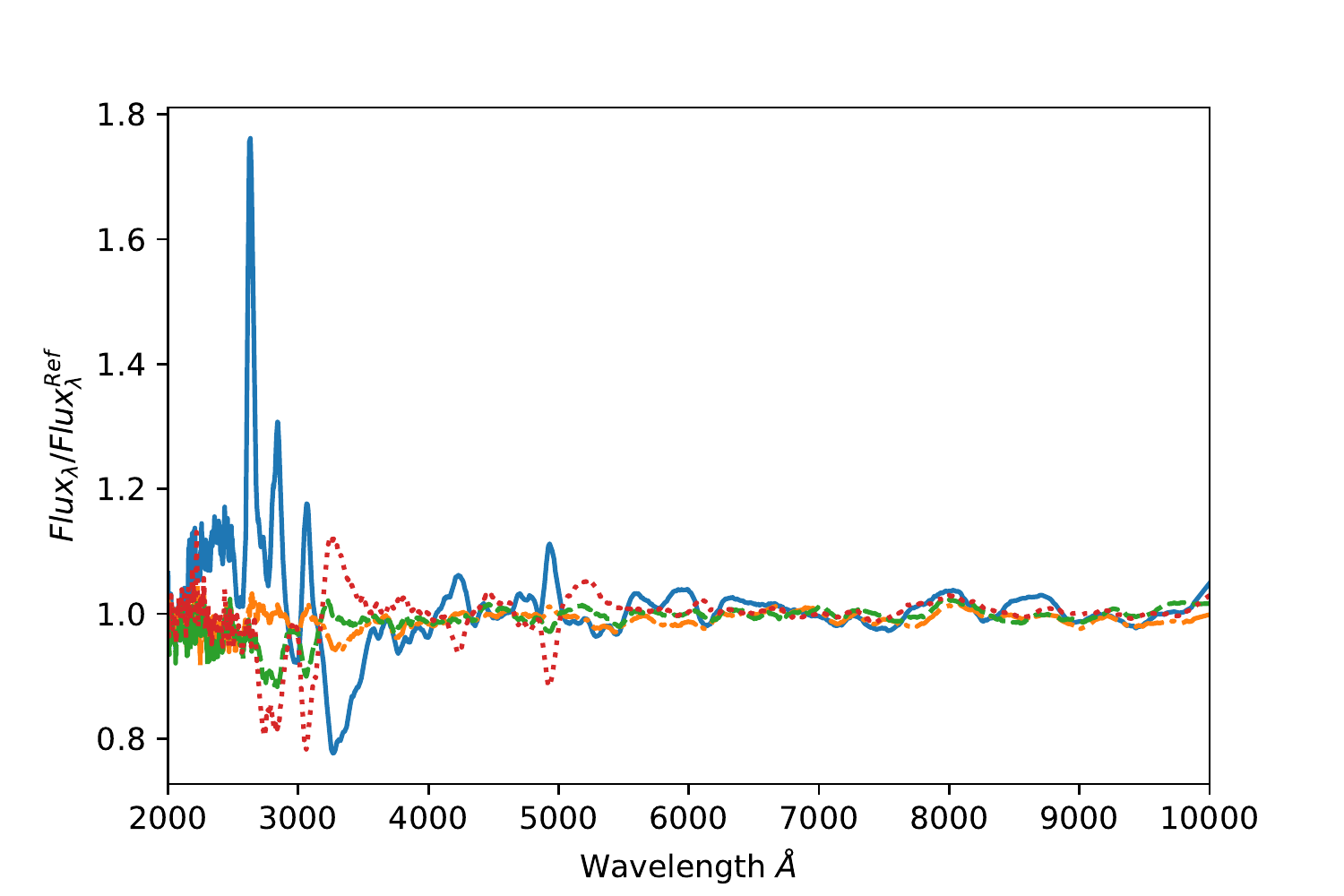}
    \plottwo{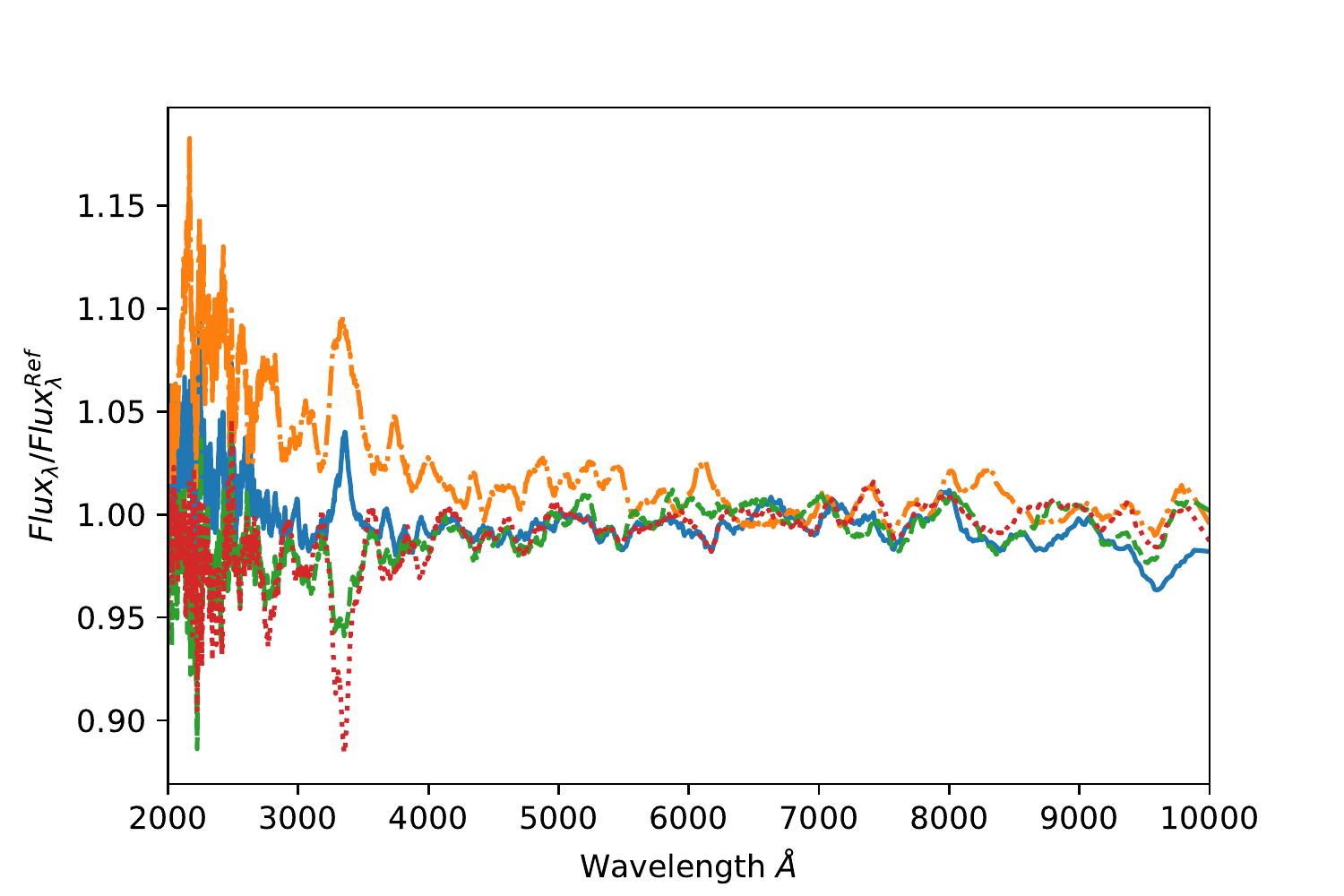}{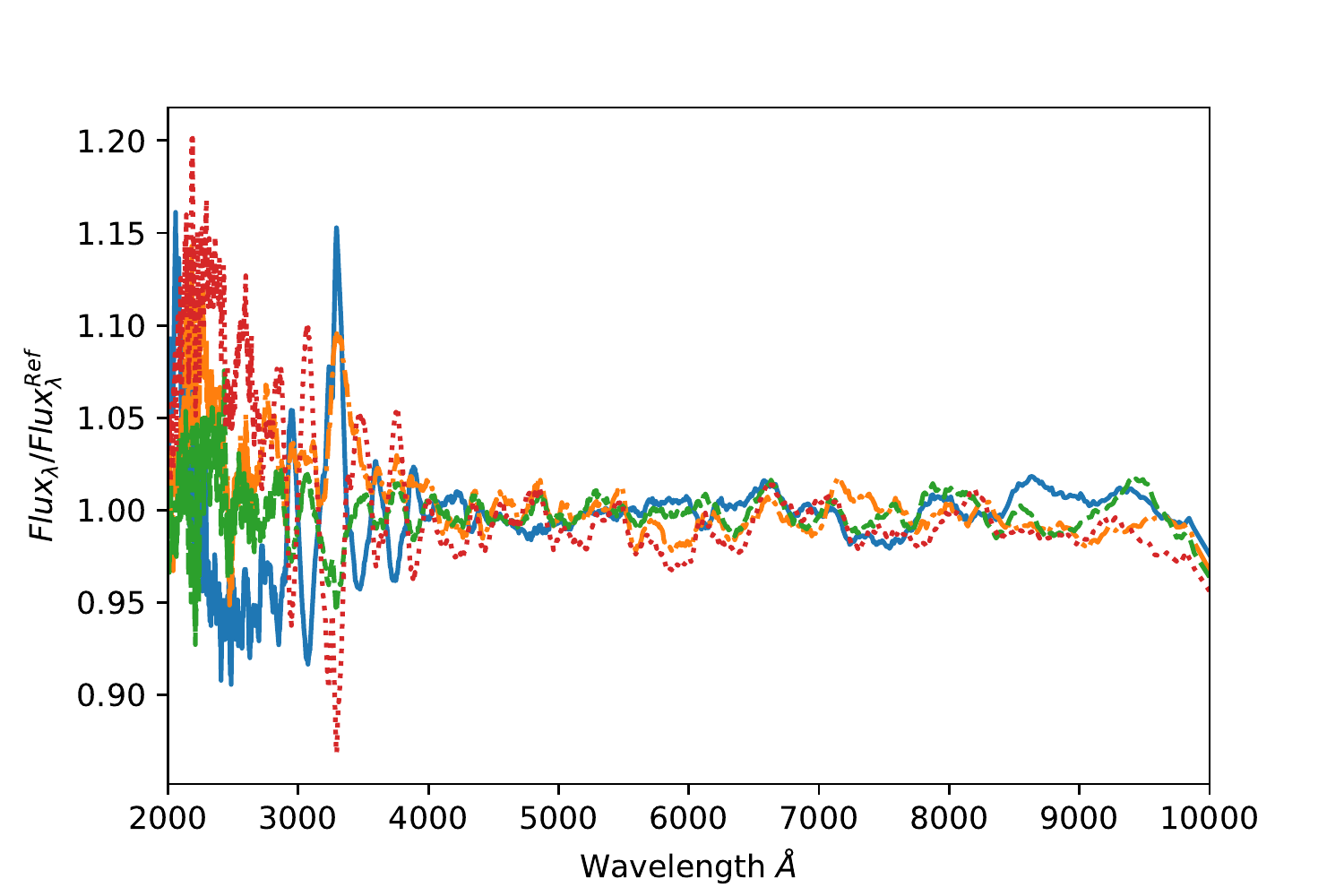}
    \caption{The effect on spectral profiles by only varying Fe in Zone 2 ({\bf Upper Left}), Fe in Zone 3 ({\bf Upper Right}), Ni in Zone 2 ({\bf Lower Left}), and Ni in Zone 3 ({\bf Lower Right}). The spectra were normalized by a reference spectrum which provides a good fit to the observed spectrum of SN~2011fe at day 0.4. Note that solid blue lines with a scaling factor of 0 are models equivalent of removing that component from the ejecta. The dash-dotted orange lines, dash green lines, and dotted red lines are for models with that component enhanced by a factor of 0.8, 1.5, and 3, respectively.}\label{fig:SingleElem}
\end{figure}

The working hypothesis of this study is that the spectral profiles are sensitive to the input parameters and there is predictive power of the TARDIS models, albeit the spectral profiles may respond to the variations of chemical elements in a very complicated way. We demonstrate in Figure~\ref{fig:SingleElem} that this is indeed the case. To show the effect of varying chemical elements, a single element at a given layer was artificially altered while all other parameters remained the same as in the reference model. Here the reference model was chosen from the spectral library that best matches SN~2011fe at day 0.4 (see Section \ref{sec:fitresults}). Figure~\ref{fig:SingleElem} shows the results of artificially altering $Fe$ and $Ni$ by 0, 0.8, 1.5, and 3 times with respect to a reference spectrum. 
  
We see that the spectral patterns of Fe and Co are very different. The spectra vary strongly the UV wavelength and creates spectral ``wiggles" that are characteristic to the input chemical elements. The features generated by elements in different velocity zones are also distinctively different. 
  
%\begin{figure}
%    \includegraphics[width=5in,angle=90]{CNNsingle_element.eps} 
%    \caption{From Upper Left clockwise, the panels show the effect on spectral profiles by only varying $Fe$ in Zone 2 {\bf (a)}, $Fe$ in Zone 3 {\bf (b)}, $Co$ in Zone 2 {\bf (c)}, and $Co$ in Zone 3 {\bf (d)}. The spectra were normalized by a reference spectrum which provides a good fit to the observed spectrum of SN~2011fe at day 0.4. Note that solid black lines with a scaling factor of 0 are models equivalent of removing that component from the ejecta. The dotted blue lines, dash green lines, and dash-dotted red lines are for models with that component enhanced by a factor of 2, 4, and 8, respectively. 
%     }\label{fig:SingleElem}
%\end{figure}

\section{The Multi-Residual Connected CNN Model}\label{sec:DeepLearning}

In this Section, we apply the Multi-Residual Convolutional Neural Network (MRNN) \citep{MResNet} to the spectral library synthesized through the procedures outlined in Section~\ref{sec:DDmodel} and train deep-learning models to infer the ejecta structure from the synthesized spectral library. 

\subsection{Model Data Pre-processing}

The spectra generated by TARDIS are in units of $erg/s/ {\rm \AA}$, which represents the luminosities of the SNe. They have a typical value of $\sim 10^{38}$ around optical maximum. This study explores the models in relative flux scale as TARDIS is more reliable in modeling spectral features than absolute luminosity. The flux of each spectrum is normalized by dividing its average flux between 6500 and 7500 ${\rm \AA}$. Only the overall spectral shapes are of importance here. The absolute level of the flux is ignored throughout the neural network analyses. A discussion on the correlation between spectral features and luminosity will be given after the establishment of the neural network (see Section~\ref{sec:AbsoluteLumi}).

Moreover, to account for the problems related to TARDIS  Monte-Carlo noise when comparing to observational data, we further employed two methods for data augmentation: Gaussian noisification and Savitzky-Golay filtering \citep{SGfilter}. 
Gaussian noisification is done by applying a Gaussian noise for each wavelength bin following the formula
$F_{n}(\lambda)\ = \ F_o(\lambda)\ \times\ (1+\mathcal{N}(0, {F_o(\lambda)})/F_o^{1/2}(5500)/S)$, where $\mathcal{N}(0, {F_o})$ is a normal distribution with $\mu\ = 0$ and $\sigma^2\ = \ F_o$, $S$ is a measure of the signal-to-noise ratio at the reference wavelength 5500 \AA\ for the spectra with noise added, and
%\textcolor{red}{LW: ($F_{n}=F_{o}\times 1.1^{\phi}$: this formula is approximately $Nf = No (1 + 0.1 \mathcal{N})$, which is correct for S = 10 above.}
$F_{n}$ and $F_{o}$ are the flux of the noise added flux and the original TARDIS model flux, respectively. 
Savitzky-Golay filtering is achieved with smoothing windows randomly selected from (7, 9, 11, 13, 15, 17, 19, 21, 23, 25) and the order randomly chosen from (2, 3, 4, 5, 6). 
With this data augmentation strategy, we generated a new dataset that is 8 times larger than the original spectral library.

\subsection{The Neural Network Structure}\label{subsec:NNstructure}

Deep learning techniques have been developed for stellar spectroscopy \citep{DLstellar,DLstellar2}. For stellar spectroscopy, most of the spectral features are narrow and well separated, and the required network depth is minimal. 
\citet{DLstellar} applied a CNN with 2 convolution layers and 2 fully-connected layers to the infrared spectra data of APOGEE \citep{Nidever:2015} sky survey program and later Gaia-ESO database \citep{DLstellar2}. 
\citet{Dynamic} applied a CNN with 8 convolution layers and 2 fully-connected layers to mineral Raman spectrum classifications. 
\citet{LAMOST} utilized 5 layered CNN (combined with Support Vector Machine, SVM) to detect hot subdwarf stars from LAMOST DR4 spectra. 
In all of the above studies, the spectral features are much less blended than in the case of SNe. The SN spectra we aim to model are characterized by broad spectral lines due to the high velocity of the ejecta and the blending of atomic multiplets; models of SNIa require a more complicated neural network architecture to reach sufficient accuracy and sensitivity. 

Between the input and output, a typical Convolutional Neural Network (CNN) contains stacked convolutional layers above one or two fully-connected layers, incorporated with pooling layers and activation layers. Starting from the input layer, the data propagate through the convolutional layers via convolution with convolution cores, and pass through the activation layer with non-linear functions (i.e., Recitified Linear Activation $f(x)=0\ when\ x\leqslant0; x\ when\ x>0$). In the pooling layer, the dimensions of the data are reduced by binning adjacent data while preserving the maximum value (MaxPooling method) or the average value (AveragePooling method). The fully-connected layer shares the same structure as normal neural networks, which calculates linear combinations using the weights and biases between every neurons in the adjacent layers.

The trainable parameters (mainly the weights and biases in the fully connected layer, and the convolution cores in the convolution layers) are randomly assigned at the outset, and is updated in the training process through forward and backward propagation. 
Forward propagation calculates the output using the current neural network parameters and the input, then compares the network outputs (predictions) with the target values (the real values) under a loss function (i.e., Mean Squared Error, MSE: $Loss\ =\ Mean(\hat{M}_{scaled}-\hat{M}_{predict})^2$). As all the calculations in neural network are analytical, the gradients of all trainable parameters can be deduced. In the back propagation process, the trainable parameters are updated by multiplying a pre-defined learning rate to their gradients (for a review, see \citet{DeepLearning}). Limited by computational efficiency, every forward-backward-propagation iteration only avail a subset of the training data set (batch), so a full review of the training dataset consists of several batches. 

It is stated \citep{NetworkLayers} and experimentally demonstrated (i.e. AlexNet \citep{AlexNet}, VGG16 \citep{VGG16}) that additional layers endow the neural network with exponential increase in agility. Nonetheless, a deeper network (a network with more layers) suffers from "gradient explosion" or "gradient diminish" problems: during the forward propagation calculations, an input signal may be consecutively multiplied with $<$ 1 or $>$ 1 numbers, which result in the signal turn to zero or infinity in machine accuracy and affect the calculation on gradients. As a consequence, \citet{BatchNorm} introduced batch normalization layers, which normalize the input batch in every dimensions to be average zero and standard derivative one. Additionally, \citet{ResNet} suggested to add the output from the previous convolution layers onto the current layer output, and use this 2 convolutional layered structure as the building block for CNN structure. Such Residual Neural Networks (ResNet) structure shows great accuracy and easy to optimize even in 152-layered scenario \citep{ResNet}. Based on ResNet, \citet{DNN} proposed a Densely Connected Neural Networks (DNN) with another signal shortcut method: directly concatenate the output of all the previous layers as the input of the current layer, and insert low-dimension layers served as "bottle-necks" to break the cumulative dimension increase caused by the consecutive concatenations. 

Consequently, we adopt the Multi-Residual Neural Network (MRNN) structure \citep{MResNet} for our studies. The structure of MRNN is relatively simple. Comparing to ResNet which allows signal shortcut in 2 layers, MRNN introduces signal shortcut in multiple layers by adding all the previous layers' (or blocks' ) outputs together to be the next layer's input. Such a architecture is much simpler than DNN, and requires less computation time to find a suitable network structure. Secondly, MRNN balances the training demands and the accuracy. According to the test on CIFAR-10 and CIFAR-100 image classification dataset \citep{cifardata}, MRNN shows an equivalent performance compared to a much deeper ResNet structure\citep{MResNet}, and better than all plain CNN structure available at that time. However, we notice our CNN and MRNN with similar depth consume comparable amount of CPU time to finish one epoch of training. After these preliminary assessments of the performance of CNN, ResNet, MRNN, and DNN, we chose MRNN to probe the best-performance element prediction.

All the models were trained using {\tt\string keras}\citep{keras} with {\tt\string tensorflow} \citep{tensorflow} as backend. In the training step to buildup the initial neural network structure, we selected the iron abundance in zone 3 and the TARDIS synthesized spectra between wavelength range 2000 and 10000 \AA. We chose mean squared error (MSE) as the loss function, and "adam" \citep{adam} as the optimizer. We adopted a two-step training scheme in order to achieve convergence while avoiding over-fitting (the model performs exceptionally well on the training data set but fail on the testing dataset). In the first step, the learning rate is 3 $\times\ 10^{-5}$ and decays $10^{-8}$ per time step, the batch size is 4,000, the training session jumps to the second step when there is no progress in the loss function of the testing dataset after 10 epochs. In the second step, the learning rate is 3 $\times\ 10^{-7}$ and decays $10^{-8}$ per time step, the batch size is 10,000, the training session stops when there is no progress in the loss function of the testing dataset after 5 epochs. 

\begin{figure}
    \centering
    \includegraphics[width=0.8\linewidth]{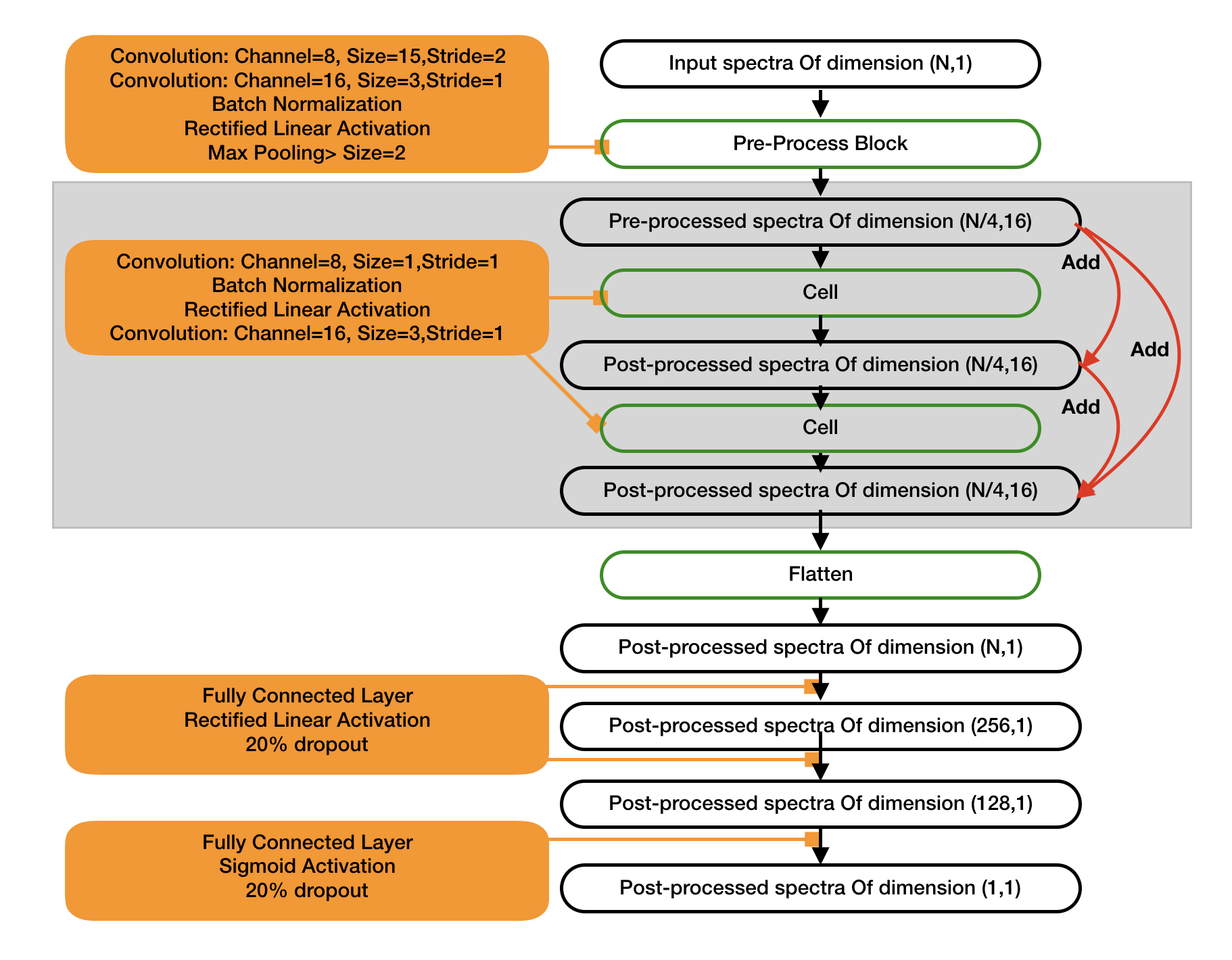}
    \caption{The MRNN structure. The black boxes are data tensors transferring in the network, green boxes are the "block" structures which contain multiple neural network layers whose details are shown in the yellow boxes. Note that the input and output of the "cell" structure have the same array size, and can be repeated multiple times in MRNN together with the boxes in the shaded area. The outputs of each cell are added with all the inputs of previous cells to form the input of the next layer, as is marked by the red arrows. For the Densely Connected Neural Network, the "adding" processes are replaced with "Concatenate" process, so the data size of the Intermediate, the Processed and, the first Fully-Connected layers are doubled. }\label{fig:MRNNfigure}

\end{figure}

The network architecture is shown in Figure \ref{fig:MRNNfigure}. 
The neural-network structure for the absolute luminosities is similar, but with a different training schedule, details are discussed in Section~\ref{sec:AbsoluteLumi}. 
We also tried to train the neural network to predict the photospheric velocities and the date of maximum, however we failed to use this two predictions to re-fit the observed spectra. For the spectral fits in Section~\ref{sec:Results}, we adopted the explosion date from light curves, and use grid search to find the best photospheric velocities instead. 

We trained the neural network using spectral data with two different wavelength ranges. One set used the entire spectral wavelength range from 2,000 \AA\ to 10,000 \AA\ (WR-Full, hereafter), the other only used the wavelength range from 3,000 \AA\ to 5,200 \AA\ (WR-Blue, hereafter). These two sets of data are useful for applications to observational data with different spectral coverage. 

\begin{figure}
    \includegraphics[width=0.5\linewidth]{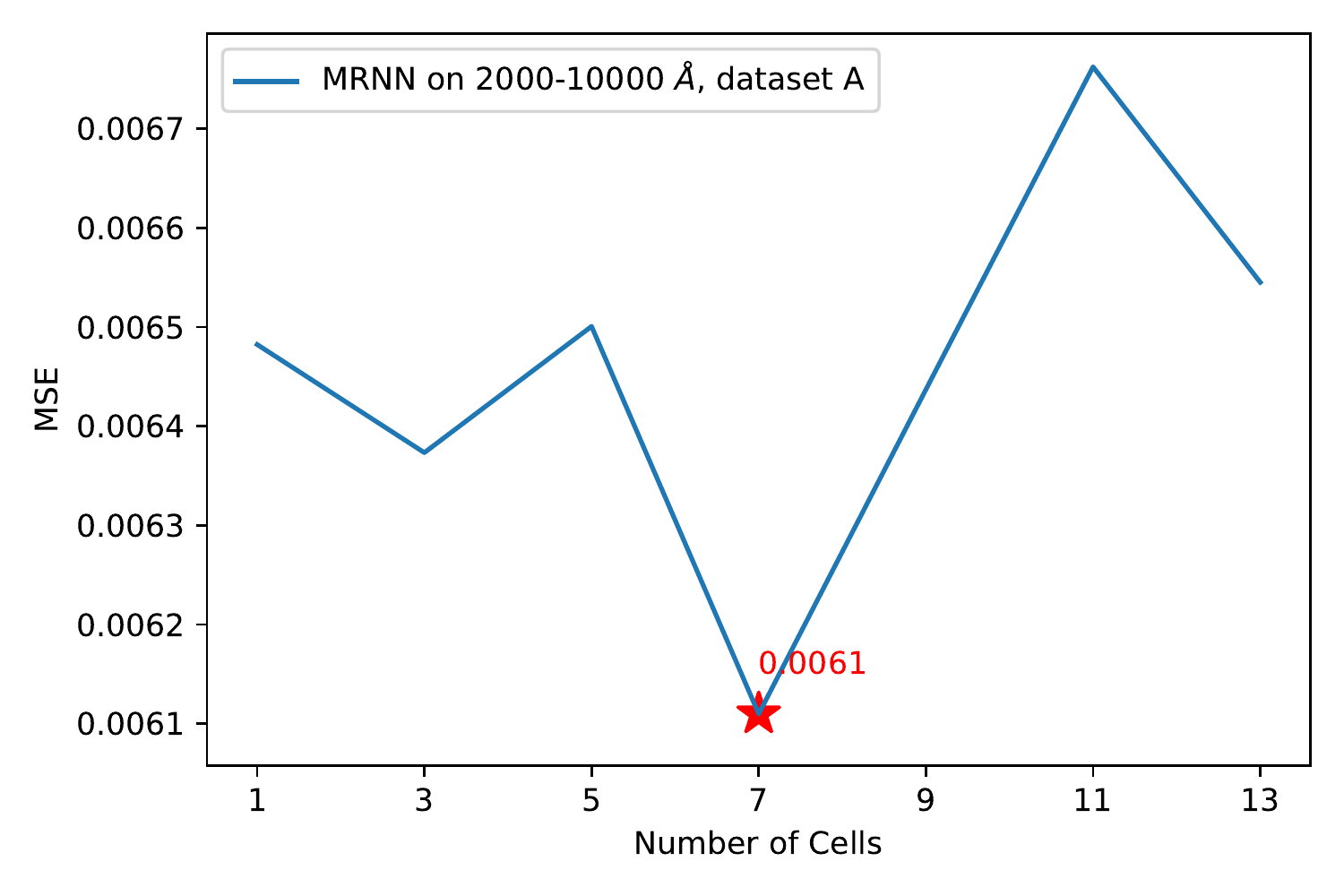}
    \includegraphics[width=0.5\linewidth]{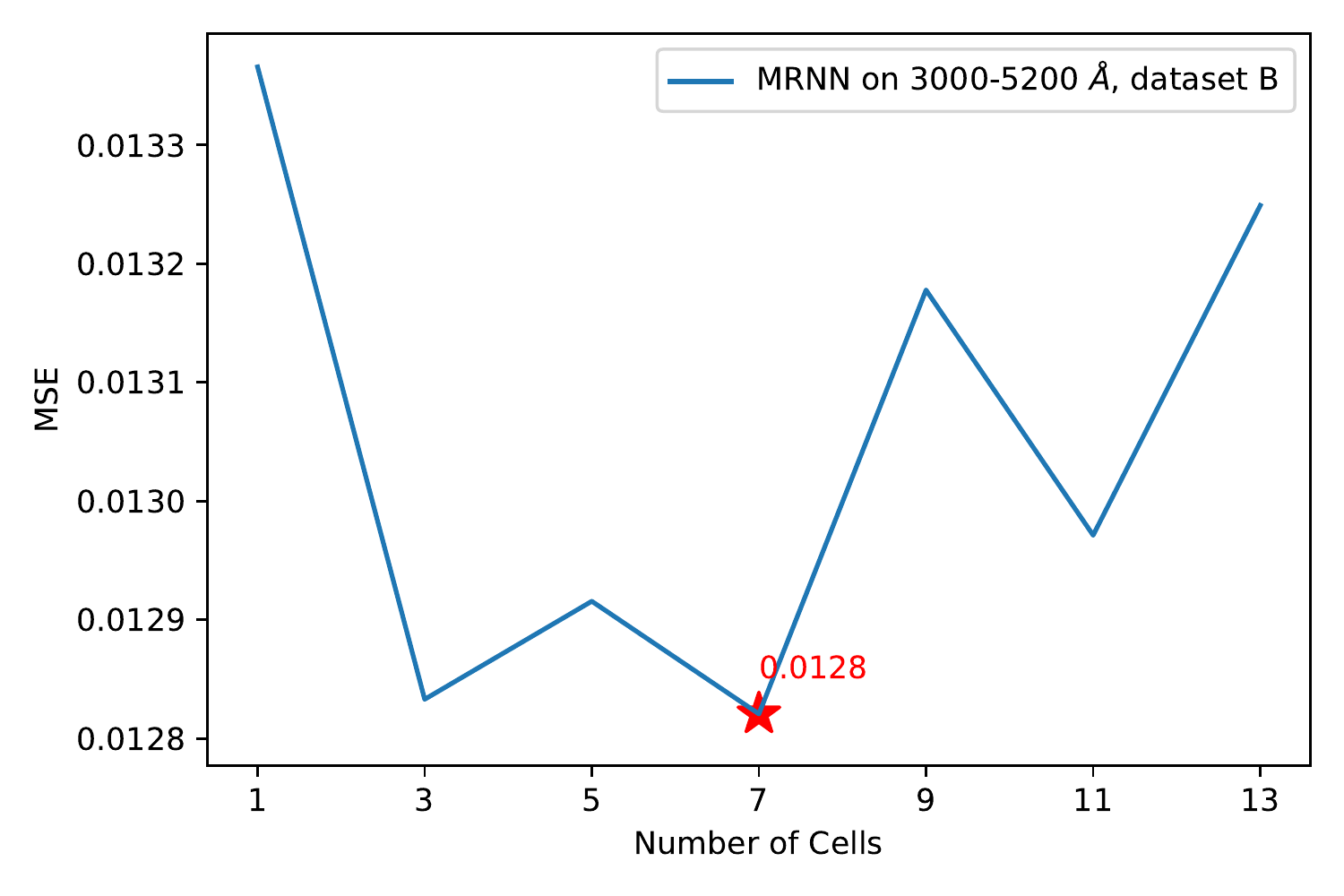}
    \caption{ The mean squared error (MSE) of Fe in Zone 3 on the testing dataset of the MRNN model. In this trial, 20,000 spectra were used in the training set and 2666 spectra were used for testing. \textbf{Left Panel:} The MSE from the MRNN trained on 2000-10000 \AA spectra (WR-Full), the MRNN with 7 cells performs best, which shows a 0.0061 MSE. \textbf{Right Panel:} The MSE from the MRNN trained on 3000-5200 \AA spectra (WR-Blue). The MRNN with 7 cells performs best, with a 0.0128 MSE. }\label{fig:DNNperf}
\end{figure}

In the next step, we chose the number of "cells" to be 1, 3, 5, 7, 9, 11, and 13 and explore the performances of MRNN. We randomly selected 20,000 spectra for training and reserved 2,666 spectra for testing. We found that MRNN with 7 cells performs the best among all models, as is shown in Figure~\ref{fig:DNNperf}. Increasing the number of cells actually deteriorates the fits. 

We thus adopted the MRNN for the training on the chemical elements from 6 to 28 and the $B$-band absolute magnitudes. The chemical elements in the four different zones (plus the luminosity) were trained separately. In principle, the parameters can be trained jointly. However, the size of such a neural-network could be too large for our computational capability, so we opted to choose a fast and robust neural-network structure and applied it to all prediction tasks. 

We have also tried the Densely Connected Neural Network (DNN)\citep{DNN} on the 2000-10000 \AA\ models, by simply changing the "adding" process in the MRNN by "concatenation". 
The MSE when using 3 cells for calculation is 0.0080, which is higher than MRNN with similar depth. 
However, due to our limited RAM capacity and the complexity in modifying DNN's cell structures, "bottle neck"\citep{DNN} positions and other hyper-parameters, we didn't explore more of this network structure. 

\subsection{Target Parameters for Machine Learning}

Our goal is to find the ejecta structures that best fit the observed spectral features of an SNIa at around optical maximum. We choose the "multiplication factor" (as is mentioned in Section~\ref{sec:DDmodel}) in the zone of our concern as the neural network output. The "multiplication factor" was restricted to have a range from 0 to 3. Our experiments with the models and data indicate that this range is sufficient to ensure coverage of a broad range of SNIa. Then, we applied the following normalization strategy in order to constrain the output of the neural network into (0,1):
%attain a relatively unbiased distribution
%\textcolor{red}{Why is this relatively unbiased?}
\begin{equation}\label{eq:scaled}
    \hat{M}_{scaled}=\tanh\left(\frac{(\hat{m}-\mu(\hat{m}))}{\sigma(\hat{m})}\right)/2+0.5,
\end{equation}
where $\hat{M}_{scaled}$ is the scaled elemental abundances of the neural network output, $\mu$ and $\sigma$ are the average and the standard derivative of the multiplication factor $\hat{m}$, respectively. This non-linear normalization strategy allows the trained model to predict values outside the parameter space (extrapolation) within a small parameter range, but suppresses erratic values derived by the neural network; elemental abundances less than zero or approaching infinity are remapped to values close to 0 and 1, respectively. 
%\textcolor{blue}{Although normalizing the output data with as-mentioned non-linear function causes biased distribution, it remaps the outliers 0 or infinity into [0,1], which makes the neural network more robust for further improvements on the dataset. }

\subsection{Training Results}\label{sec:trainresult}

\begin{figure}[htb!]
\includegraphics[width=0.5\textwidth]{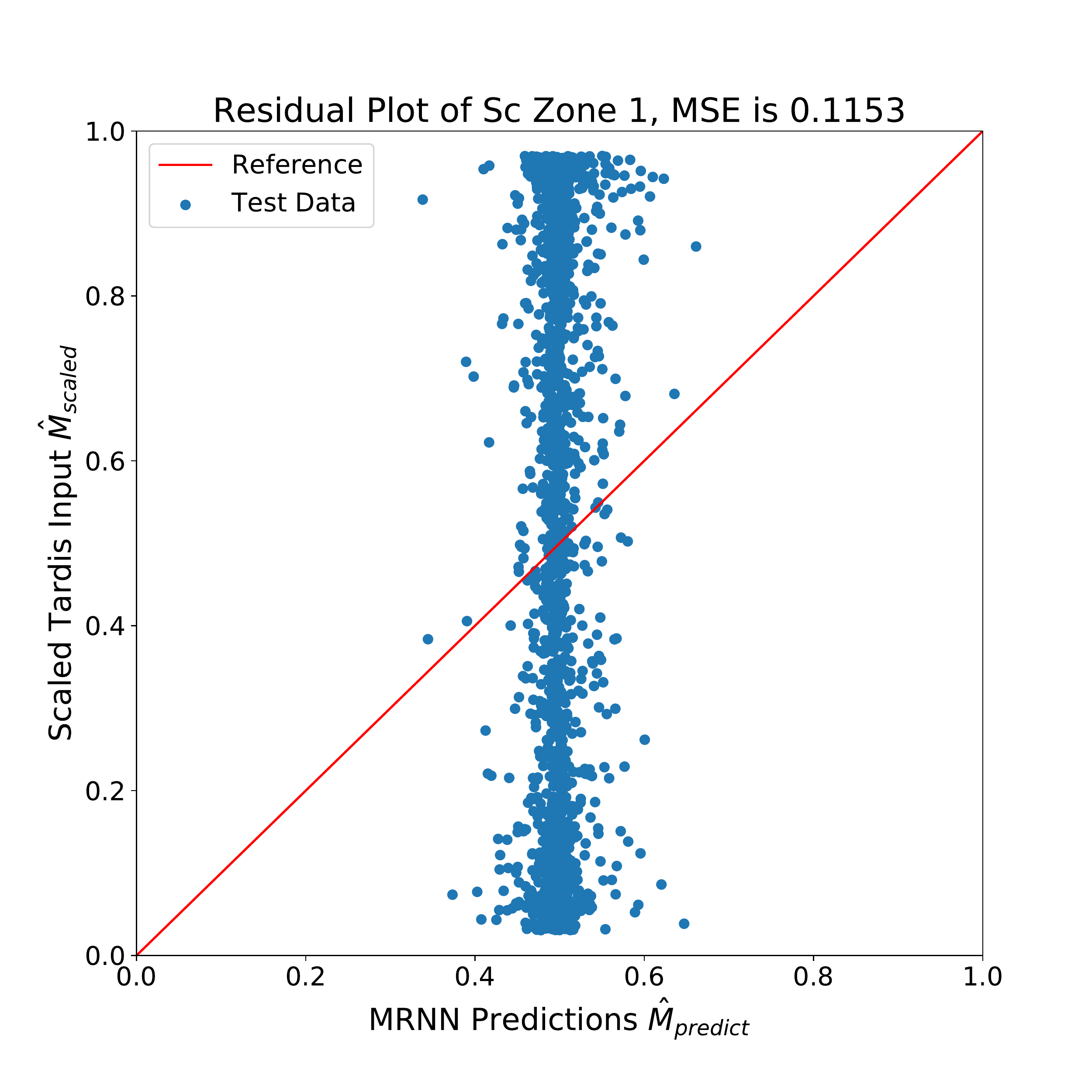}
\includegraphics[width=0.5\textwidth]{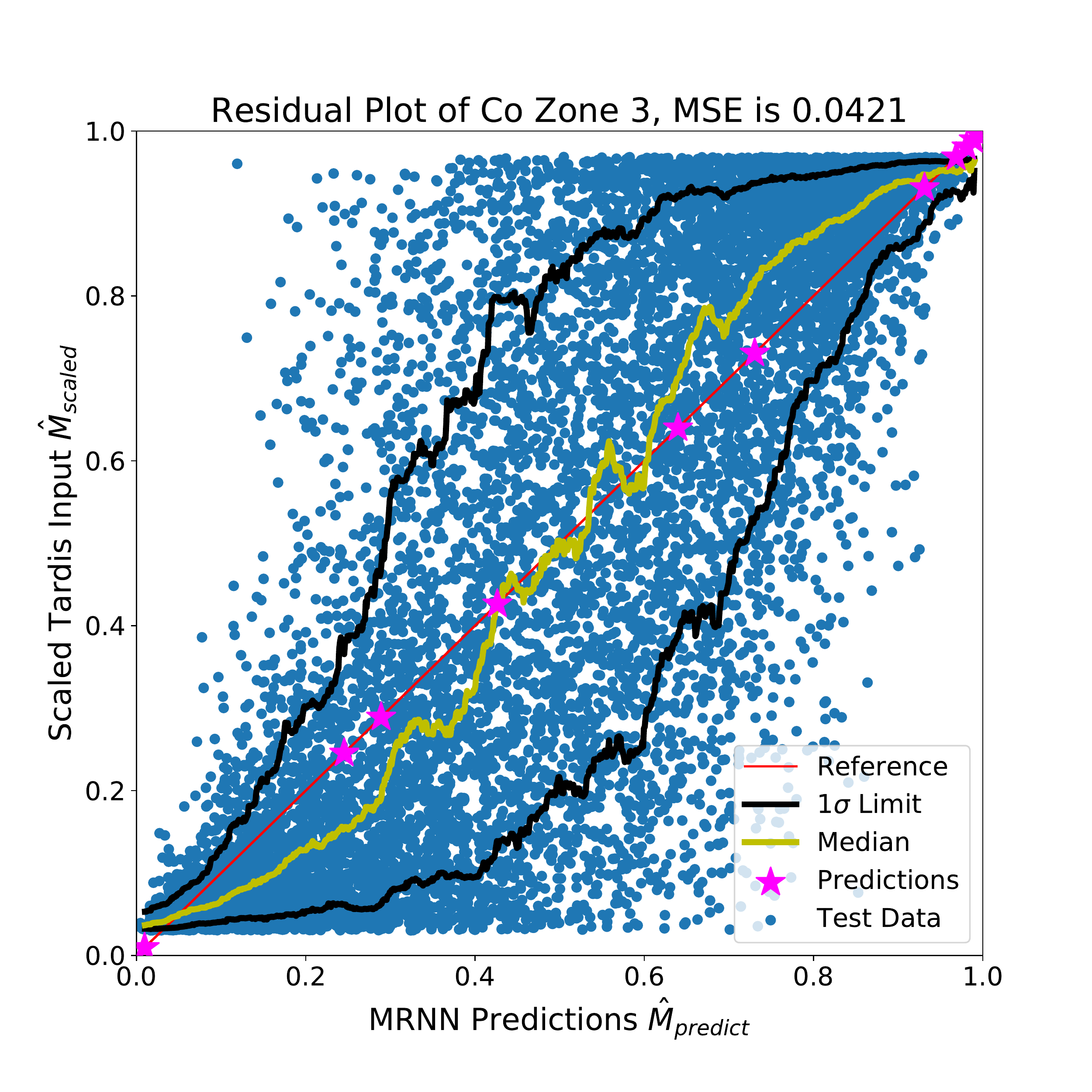}
\caption{The correlation between the scaled elemental abundances of MRNN predictions and TARDIS inputs. X-axis is the prediction from the MRNN model, Y-axis is the input into TARDIS for spectral calculations. In both panels, the diagonal red-lines indicate the ideal model predictions, and the blue dots were derived from the test dataset. Yellow and black lines show the median and $1-\sigma$ error limit of the given prediction in the testing set. Predictions from the selected 11 SNIa spectra with wavelength coverage from 2000 - 10000 \AA\ (which are analyzed in \S~\ref{sec:fitresults}) are shown as magenta stars. \textbf{(a) Left Panel:} The correlation plot of Sc in Zone 1, MSE is 0.1153. This model was trained on 10000 spectra and tested on 1,829 spectra. The predicted values are close to 0.5, which violates the "real" value for spectral synthesis, indicating the poor performance of the neural-network on this element-zone. \textbf{(b) Right Panel:} The plot showing the correlation of scaled elemental abundances of Co in Zone 3, with a MSE value of 0.042. This model is trained on 89559 spectra and tested on 9951 spectra. }\label{fig:ResidualSelecter}
\end{figure}

\begin{figure}[htb!]
\includegraphics[width=0.5\textwidth]{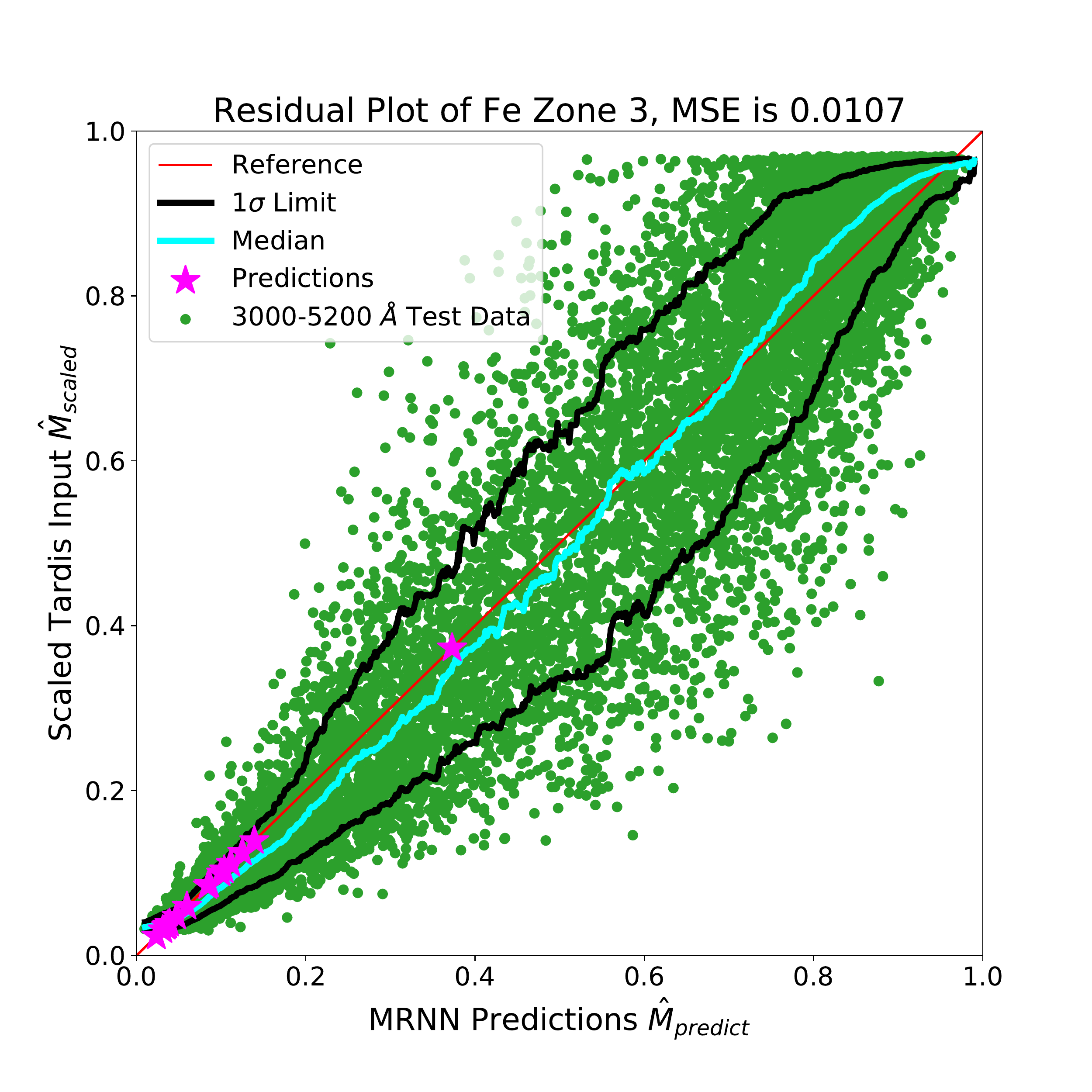}
\includegraphics[width=0.5\textwidth]{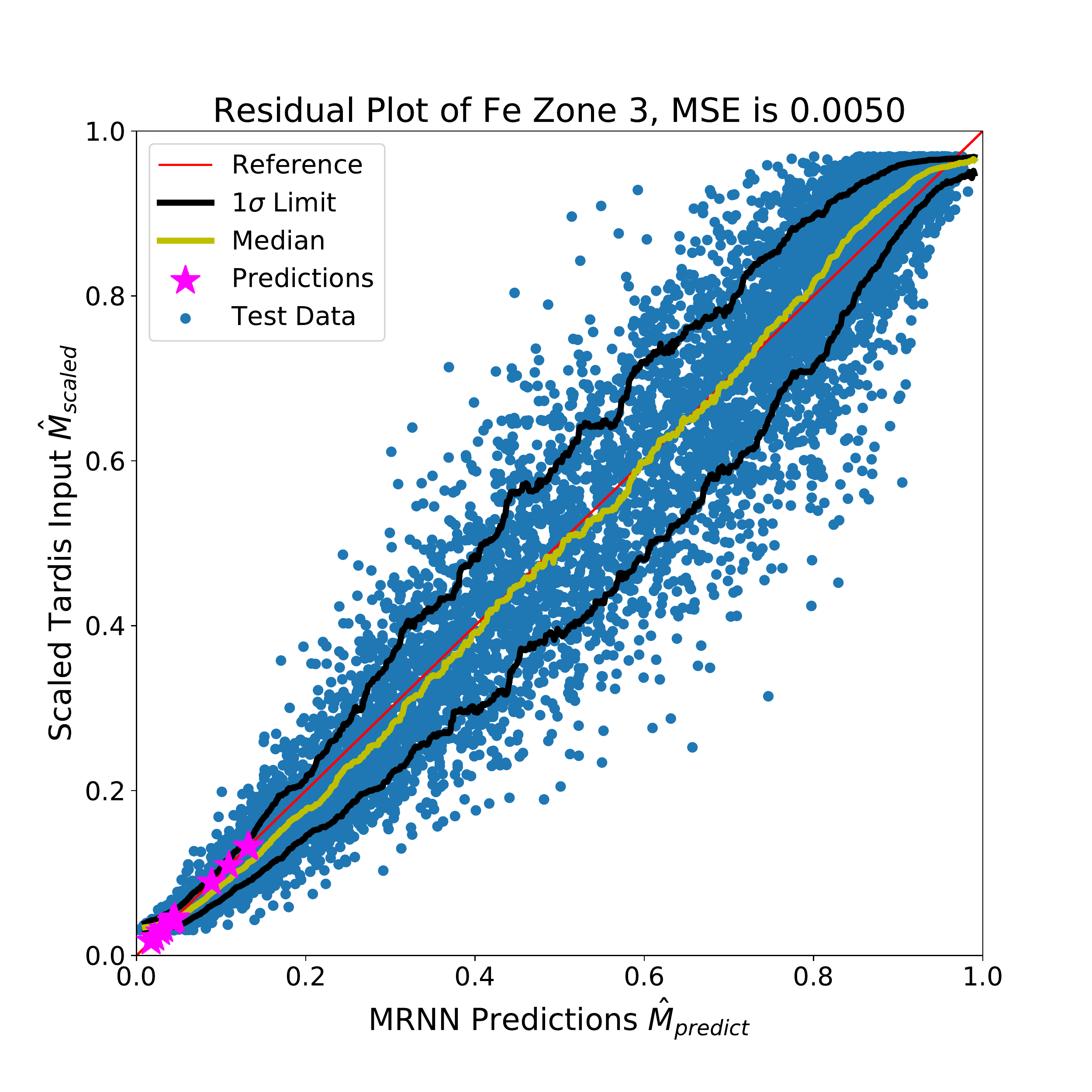}
\caption{Same as \ref{fig:ResidualSelecter}, but for Fe in Zone 3. \textbf{(a) Left Panel:} The correlation between scaled elemental abundances of MRNN predictions and TARDIS inputs for neural network trained for wavelength range  3000-5200 \AA. The MSE is 0.0107. \textbf{(b) Right Panel:} The same plot for wavelength range 2000-10000 \AA. The MSE is 0.0050. }\label{fig:Residual}
\end{figure}

Not all the elements in the IGM are significantly influencing the spectra. As a first trial, we adopted a subset of model spectra and a simplistic neural-network structure to probe the effect of various chemical elements. In this trial, 10000 spectra were selected as the training set and 1829 spectra as testing set, and we chose the MRNN with one cell structure. All the elements from atomic number 6 (Carbon) to 28 (Nickel) in the 4 velocity Zones are trained on the training set and verified on the testing set. 
By comparing the MSEs from the testing set, as well as the correlation between the neural network-predicted scaled elemental abundance (Equation~\ref{eq:scaled}) and the corresponding value of the TARDIS input (Figure~\ref{fig:ResidualSelecter} and \ref{fig:Residual}), we found models with MSE larger than 0.1 to be poorly trained with little predictive power on elemental abundances. Consequently, we chose only the elements in zones with MSEs less than 0.1 for further training. There are 34 and 31 trainable chemical constituents located in the 4 velocity zones for train sets WR-Full and WR-Blue, respectively.  The MSEs of these chemical constituents are shown in Table~\ref{tab:Residual} and \ref{tab:Residual2} for train sets WR-Full and WR-Blue, respectively.  In the tables, we show also the correlation coefficients for each trainable elements which can be used to gauge the reliability of the results. 

\begin{deluxetable}{|cccc|cccc|}[htb!]
    \tablecaption{MSE and Correlation for Selected Elements and Zones of WR-Full (Wavelength Range 2,000-10,000 \AA)}
    \tablehead{\colhead{Element} & \colhead{Zone Number} & \colhead{MSE} & \colhead{Corr} 
    & \colhead{Element} & \colhead{Zone Number} & \colhead{MSE}  & \colhead{Corr} }
    \startdata
    O  & 4 & 0.071 & 0.610 & Mg & 2 & 0.010 &  0.957 \\
    Mg & 3 & 0.075 & 0.585 & Mg & 4 & 0.040 &  0.809 \\
    Si & 1 & 0.051 & 0.747 & Si & 2 & 0.013 &  0.942 \\
    Si & 3 & 0.019 & 0.915 & S  & 2 & 0.067 &  0.641 \\
    Ar & 2 & 0.092 & 0.425 & Ca & 1 & 0.054 &  0.726 \\
    Ca & 2 & 0.018 & 0.920 & Ca & 3 & 0.048 &  0.762 \\
    Ca & 4 & 0.019 & 0.913 & Sc & 4 & 0.058 &  0.698 \\
    Ti & 1 & 0.086 & 0.498 & Ti & 2 & 0.055 &  0.718 \\
    Ti & 3 & 0.046 & 0.772 & Ti & 4 & 0.044 &  0.787 \\
    V  & 1 & 0.050 & 0.748 & V  & 2 & 0.019 &  0.909 \\
    Mn & 1 & 0.026 & 0.875 & Mn & 2 & 0.017 &  0.926 \\
    Fe & 1 & 0.004 & 0.984 & Fe & 2 & 0.005 &  0.981 \\
    Fe & 3 & 0.050 & 0.978 & Fe & 4 & 0.025 &  0.885 \\
    Co & 1 & 0.088 & 0.958 & Co & 2 & 0.029 &  0.864 \\
    Co & 3 & 0.042 & 0.796 & Co & 4 & 0.078 &  0.567 \\
    Ni & 1 & 0.009 & 0.960 & Ni & 2 & 0.028 &  0.869 \\
    Ni & 3 & 0.050 & 0.754 & Ni & 4 & 0.075 &  0.586 \\
    \enddata
    \tablecomments{Neural Networks are trained on the 89,559 training set and all the MSE are calculated on the 9,951 testing data set. }\label{tab:Residual}
\end{deluxetable}

\begin{deluxetable}{|cccc|cccc|}[htb!]
    \tablecaption{MSE and Correlation for Selected Elements and Zones of WR-Blue (Wavelength range 3,000-5,200 \AA)}
    \tablehead{\colhead{Element} & \colhead{Zone Number} & \colhead{MSE} & \colhead{Corr} 
    & \colhead{Element} & \colhead{Zone Number} & \colhead{MSE} & \colhead{Corr} }
    \startdata
    Mg & 2 & 0.070 & 0.627 & Mg & 3 & 0.086 & 0.487 \\
    Mg & 4 & 0.051 & 0.747 & Si & 1 & 0.085 & 0.511 \\
    Si & 2 & 0.058 & 0.701 & Si & 3 & 0.053 & 0.728 \\
    S  & 2 & 0.071 & 0.607 & Ar & 2 & 0.093 & 0.420 \\
    Ca & 1 & 0.062 & 0.480 & Ca & 2 & 0.072 & 0.605 \\
    Ca & 3 & 0.073 & 0.605 & Ca & 4 & 0.052 & 0.739 \\
    Sc & 4 & 0.062 & 0.680 & Ti & 2 & 0.066 & 0.653 \\
    Ti & 3 & 0.061 & 0.685 & Ti & 4 & 0.046 & 0.773 \\
    V  & 1 & 0.087 & 0.499 & V  & 2 & 0.046 & 0.773 \\
    Mn & 1 & 0.069 & 0.640 & Mn & 2 & 0.058 & 0.702 \\
    Fe & 1 & 0.037 & 0.808 & Fe & 2 & 0.024 & 0.893 \\
    Fe & 3 & 0.011 & 0.952 & Fe & 4 & 0.036 & 0.829 \\
    Co & 1 & 0.030 & 0.845 & Co & 2 & 0.045 & 0.783 \\
    Co & 3 & 0.046 & 0.776 & Ni & 1 & 0.039 & 0.793 \\
    Ni & 2 & 0.056 & 0.714 & Ni & 3 & 0.051 & 0.747 \\
    Ni & 4 & 0.080 & 0.543 &    &   &       &       \\
    \enddata
    \tablecomments{Neural Networks are trained on the 89,559 training set and all the MSE are calculated on the 9,951 testing data set. }\label{tab:Residual2}
\end{deluxetable}

As discussed in Section~\ref{subsec:NNstructure}, we adopted the MRNN with 7 cells for elemental abundance estimation. 
The training dataset contains 89,559 spectra (90\% of the total), and the testing data set contains 9,951 spectra (10\% of the total). For a typical neural network, it takes approximately 1 hour to finish 200 training epochs on 2 Tesla P100 GPUs.

In Table~\ref{tab:Residual} and Table~\ref{tab:Residual2}, we list the MSE on the testing dataset of the 34 and 31 selected elements and zones for two sets of neural networks built for WR-Full and WR-Blue, respectively. As an example, the scaled elemental abundances of the MRNN prediction and those of the TARDIS input of Fe in Zone~3 are shown in Figure~\ref{fig:Residual}. 
The neural network predicted chemical abundances clearly correlates with the values used to generate the model spectra. The neural network using the full 2000-10000\AA\ wavelength coverage outperforms the ones with partial coverage from 3000-5200\AA. 
%\textcolor{red}{The Pearson Correlation Coefficients are xxx and xxx for the models with wavelength coverages from 2,000-10,000\AA\ and 3,000-5,200\AA, respectively. Linear fits to the data points in Figure~\ref{fig:Residual} yields slopes of yyy and yyy for the two models, respectively. }
The fact that relation between MRNN prediction and the original TARDIS input elemental mass is approximately given by $\hat{M}\ \approx\ \hat{m}$ in all cases suggests that the training dataset yields results that are consistent with the testing dataset. This ensures that overfitting is not severely affecting the MRNN we have constructed.
However, overfitting does appear to be an issue for elements that have weak spectral lines and where the correlation between the MRNN predicted and the original TARDIS input values are weak. This can be seen from Figure~\ref{fig:ResidualSelecter} where the predicted mass of Sc is biased toward the mean value of the true values of TARDIS input ($\sim\ 0.5$). For Co in Zone 3 the correlation can be detected but the predicted values are biased towards values higher and lower than the true TARDIS inputs for input values close to 0 and 1, respectively. Such bias is much weaker when the correlations are strong, as shown in Figure~\ref{fig:Residual}) for Fe in Zone 3. This bias can in principle be corrected by using the median values of the scaled TARDIS input $\hat{M}$ to estimate the original model input. 

In the current study, we did not correct this bias. Instead, we used the original MRNN predictions directly as the estimates of the mass of the input chemical elements but set the confidence levels of the estimates according to the correlations that can be derived from the test dataset. We did not attempt to estimate the confidence intervals using Bayesian statistics based on Markov Chain Monte-Carlo algorithm (for example, see \citet{emcee}), due primarily to limited processing power of our computers. 
Instead, we estimated the $1-\sigma$ error by using the testing set itself. For each input model abundances such as shown as examples in the vertical axes of Figure~\ref{fig:Residual}), 
we derive the $1-\sigma$ confidence levels based on the dispersion of the predicted values (the horizontal axis). 
To be more specific, we collected all the data in the test dataset that agree with the MRNN predictions to within [-0.02, +0.02] range to build a histogram of the scaled TARDIS input mass, and adopted the position of the 15.8\% , 50\% (median) and 84.1\% of the histogram as $1-\sigma$ error, median value, and 84.1\% estimates.
%\textcolor{blue}{After using observed spectra for predictions, we firstly count the predicted values of our 9951-spectra testing set, then collect the testing spectra of which the prediction is within [-0.02,+0.02] of the observed spectra prediction. }
%\textcolor{blue}{With these collected predictions, we trace back their relating "real" value and use the "real" value to plot the histogram. }
%\textcolor{blue}{Finally, we adopt the 15.8\% , 50\% (median) and 84.1\% of the histogram as $1-\sigma$ error. }
The $1-\sigma$ errors of Fe in zone 3 are overplotted in Figure~\ref{fig:Residual} as an example. The errors for other elements in various zones are available in the online material. 

Assuming that the TARDIS model spectra capture major spectral features of observed SNIa, we may apply the neural network trained by theoretical models straightforwardly to observational data. 
When an observed spectra is inserted into the well-trained MRNN, a set of chemical abundances and their associated uncertainties can be derived. Unfortunately we do not have a large enough library of observed spectra to train the link between observational data and theoretical models. The consistency and reliability of the derived abundances can only be evaluated by comparing the results for different supernovae and their spectral time sequences. 

Alternatively, we may consider the chemical elements determined by feeding the observed spectra to the MRNN as {\it empirical parameters that yields optimal TARDIS fits to the observed spectra}. These parameters do not necessarily serve as true estimates of the elemental abundances of the supernovae, but nonetheless, they can be used as model based empirical parameters to analyze the properties of SNIa. 

Note that the neural network approach is different from a simple observation-to-theory spectral match. The neural network model aims to match spectral features that are associated with differential chemical structure variations. Each chemical element in a certain zone is trained separately and the neural network thus trained are most sensitive to the changes of the spectral features related to this particular chemical element throughout the spectral range under study. Even though the spectral features of various elements are highly blended, the differential changes of spectral features due to varying chemical abundances are still detectable by the neural network we have constructed. Furthermore, the theoretical models may have intrinsic shortcomings and do not include all the essential physics. The assumption of a sharp photosphere, for example, cannot be correct in a more strict sense. The lack of time dependence of the radiative transfer may also limit the precision of the theoretical models. However, our approach to elemental abundance may be less sensitive to these problems by construction although the current study can not establish a quantitative assessment of the uncertainties caused by approximations intrinsic to TARDIS models. 

Moreover the neural network allows for theoretical luminosity to be derived based on theoretical model once the chemical structure of the ejecta is fully constrained. From the MRNN, the latter can be determined by spectral features without the need of knowing the absolute level of the spectral fluxes. 
When the global spectral profiles between observations and models are matched, the luminosity (or the temperature of the photosphere) is then completely constrained theoretically. This provides a theoretical luminosity of the supernova understudy that is independent of the flux calibration of the spectra.

\section{Applications of the Neural Network to Observational Data}
\label{sec:fitresults}

Assuming the AIAI method constructed using MRNN in Section~\ref{sec:DeepLearning} to be correct,  we may apply it to observed spectra as a sanity check of the method. We stress again that this step is detached from the deep learning neural network and is not a validation of the MRNN. 

The parameter space describing the ejecta structure is so large that our spectral library covers it only sparsely. The chance of having a perfect fit to any specific observation with spectra already in the library is low. The model spectra are thus  recalculated using TARDIS with the ejecta structures determined from the MRNN. To derive the optimal TARDIS models, we also allowed the photospheric positions and the luminosities of the supernovae to vary.

\subsection{Applications to SNe with Wavelength Coverage $2000\ -\ 10000$\AA}

There are six well observed SNe with a wavelength coverage from 2000-10000\AA. They are shown in Table~\ref{tab:Stretch}. The UV data are all acquired by the HST. These data are rebinned to 1 \AA/pixel and are normalized by dividing their respective average flux between 6,500 and 7,500 \AA, similar to what was done for TARDIS model spectra during neural network training.  

\begin{deluxetable}{ccccccc}[htb!]
    \tablecaption{Extinction and Stretch of Ia SNe with 2000-10000 \AA spectra\label{tab:Stretch}}
    \tablehead{\colhead{SN name} & \colhead{Redshift} & \colhead{MW E(B-V)} 
    & \colhead{Host E(B-V)}  & \colhead{Stretch} & \colhead{Reference} }
    \startdata
    SN2011fe    & 0.000804  & 0     & 0     & $ 1.062 \pm 0.005  $ & \citep{SN2015F,2011feLC} \\
    SN2011by    & 0.002843  & 0.013 & 0.039 & $ 0.93  \pm 0.02   $ & \citep{HSTspec,SN2011by,SN2011by2} \\
    SN2013dy    & 0.00389   & 0.135 & 0.206 & $ 1.098 \pm 0.008  $ & \citep{SN2013dy,SN2013dy500} \\
    SN2015F     & 0.0049    & 0.175 & 0.035 & $ 0.912 \pm 0.005  $ & \citep{SN2015F} \\
    SN2011iv    & 0.006494  & 0     & 0     & $ 0.830 \pm 0.007  $ & \citep{SN2011ivHST} \\
    ASASSN-14lp & 0.0051    & 0.33  & 0.021 & $ 1.101 \pm 0.004  $ & \citep{asassn14lp,asassn14lplate} \\
    \enddata
\end{deluxetable}

\subsubsection{SN~2011fe}

SN~2011fe was detected at Aug. 24, 2011 at M101 galaxy at a distance of approximately 6.4 Mpc \citep{SN2011feNature}. 
Its luminosity decline in B-band within 15 days after the $B$-band maximum is $\Delta M_{B,15}=1.12\pm0.05$ mag \citep{2011feLC}. 
Based on optical and radio observations, SN~2011fe is not heavily affected by any interstellar (ISM)/circumstellar material (CSM) or Galactic dust extinction \citep{SN2011feEVLA,SN2011fePatat}. 
Consequently, we didn't introduce any host galaxy extinction correction for SN~2011fe spectra. 

\begin{figure}[htb!]
    \includegraphics[width=\linewidth]{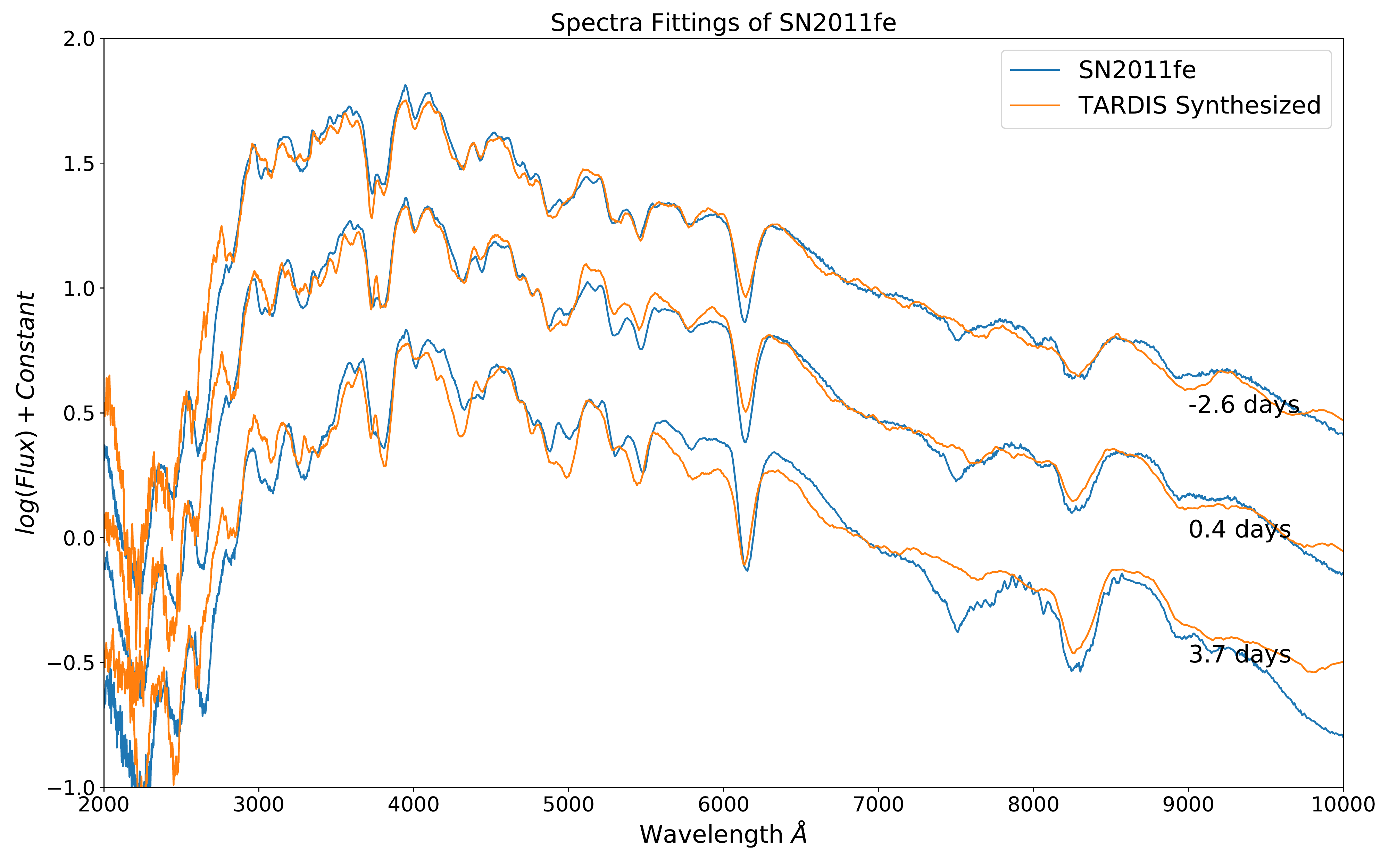}
    \caption{The observed spectra (blue line) and the TARDIS synthesized spectra (orange line) of SN~2011fe at $-$2.6 days (Upper), 0.4 day (Middle) and 3.1 days (Lower). The TARDIS-synthetic spectra were filtered by a Savitzky-Golay filter with window=9, order=1 to remove Monte Carlo noise. The spectra are in logarithmic scale and are arbitrarily shifted in the vertical axis for the different dates. }\label{fig:SN2011feSpec}
\end{figure}

We chose the HST spectra at $-2.6$ days, 0.4 days and 3.7 days relative to the $B$-band maximum date for the elemental abundance calculations and spectral fittings. 
The results are shown in Figure~\ref{fig:SN2011feSpec}. 

\clearpage

\begin{figure}[htb!]
    \includegraphics[width=0.33\linewidth]{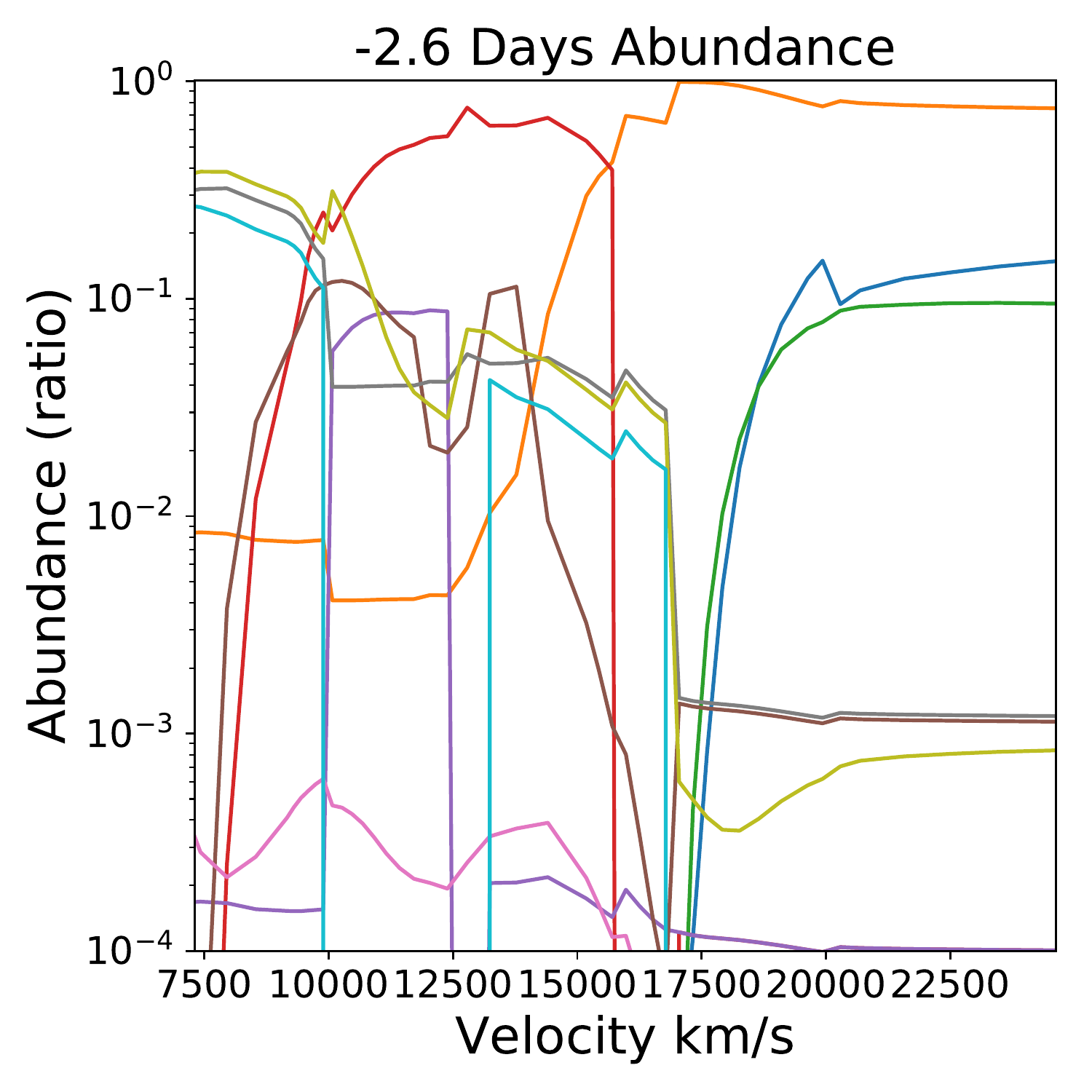}
    \includegraphics[width=0.33\linewidth]{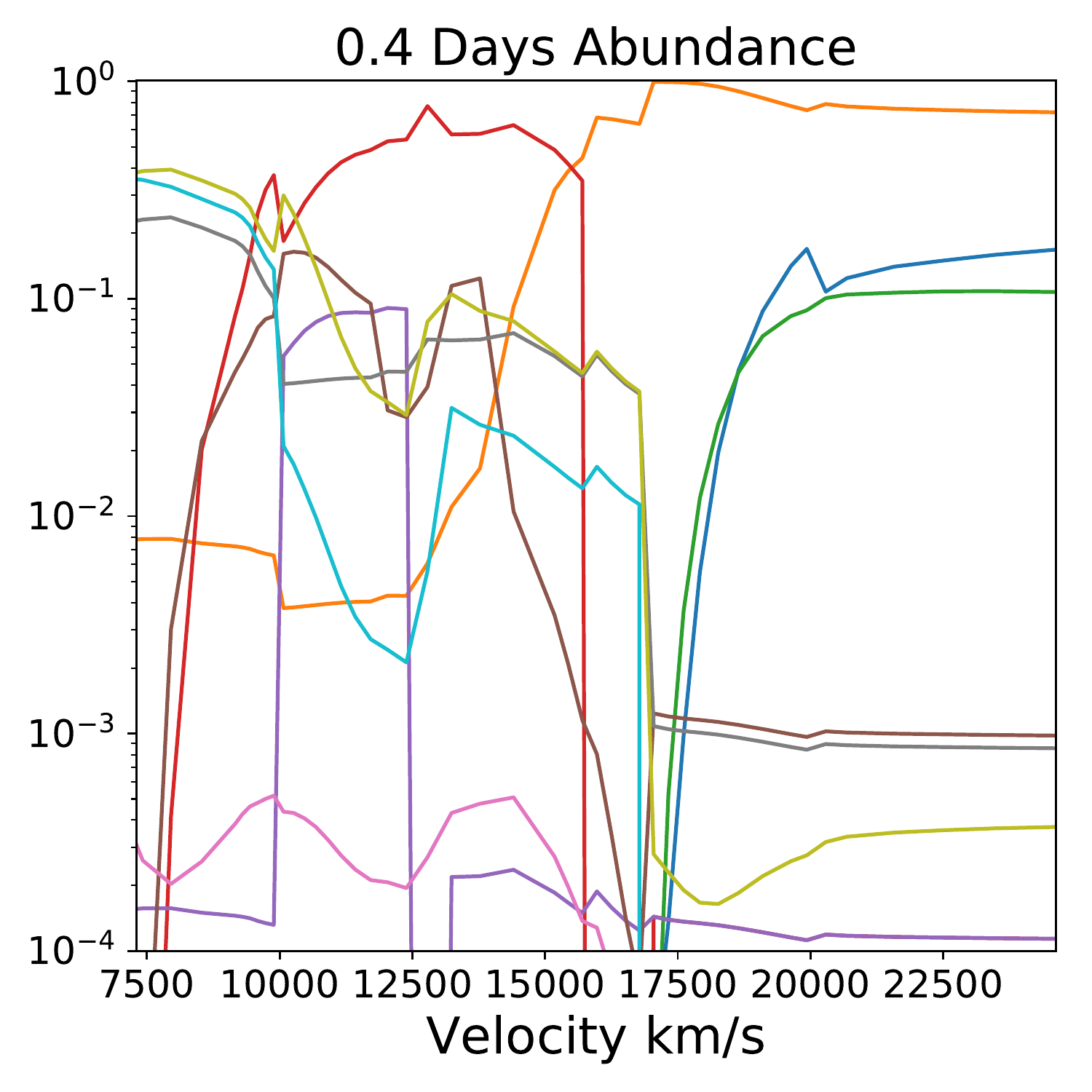}
    \includegraphics[width=0.33\linewidth]{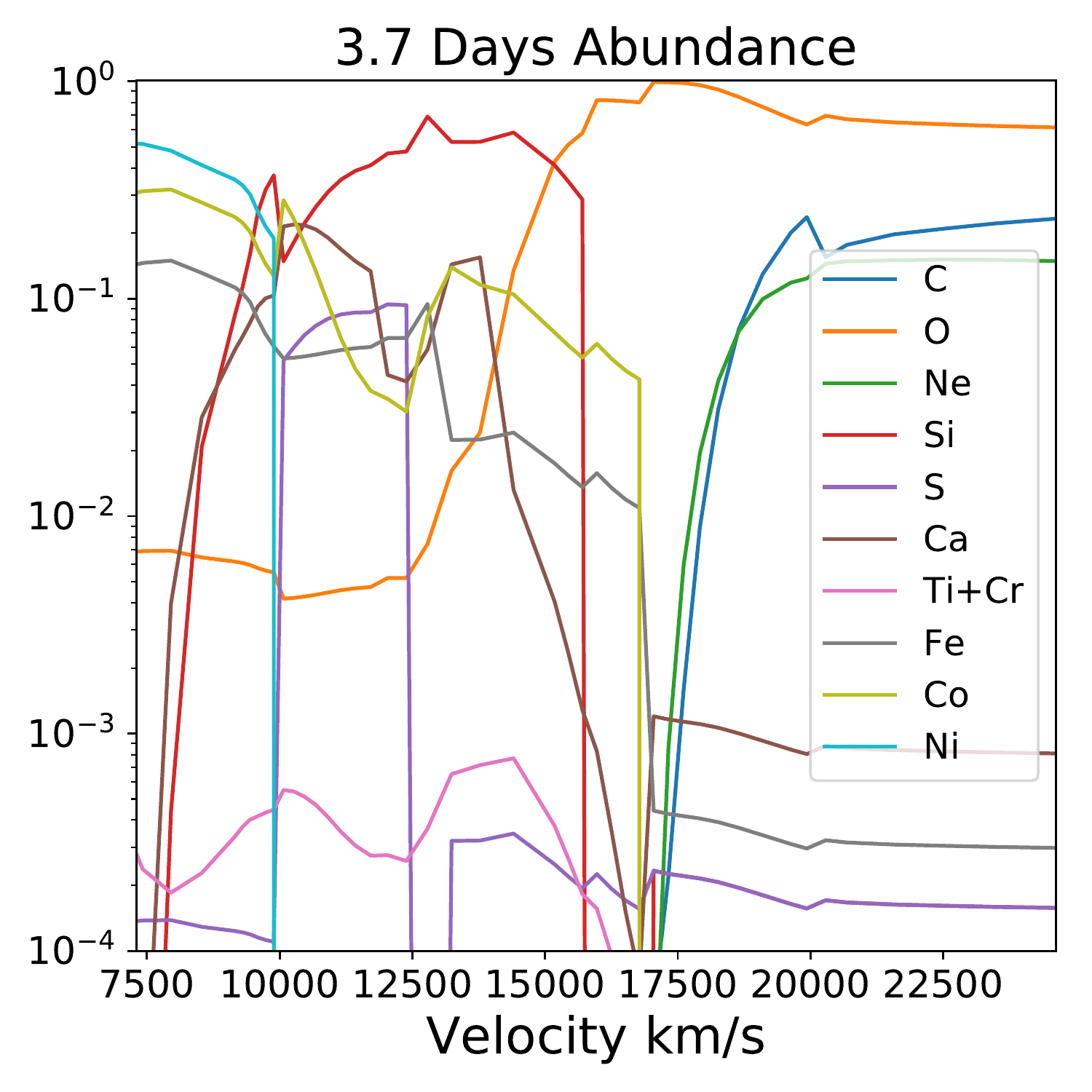}
    \caption{The elemental abundances predicted by the neural networks from SN~2011fe at $-$2.6 days (a, Left Panel), 0.4 days (b, Middle Panel), and 3.7 days (c, Right Panel). These elemental abundances and the densities in Figure~\ref{fig:SN2011feTemp} were used for the synthetic spectra shown in Figure~\ref{fig:SN2011feSpec}. }\label{fig:SN2011feElem}
\end{figure}

\begin{figure}[htb!]
    \includegraphics[width=0.244\linewidth]{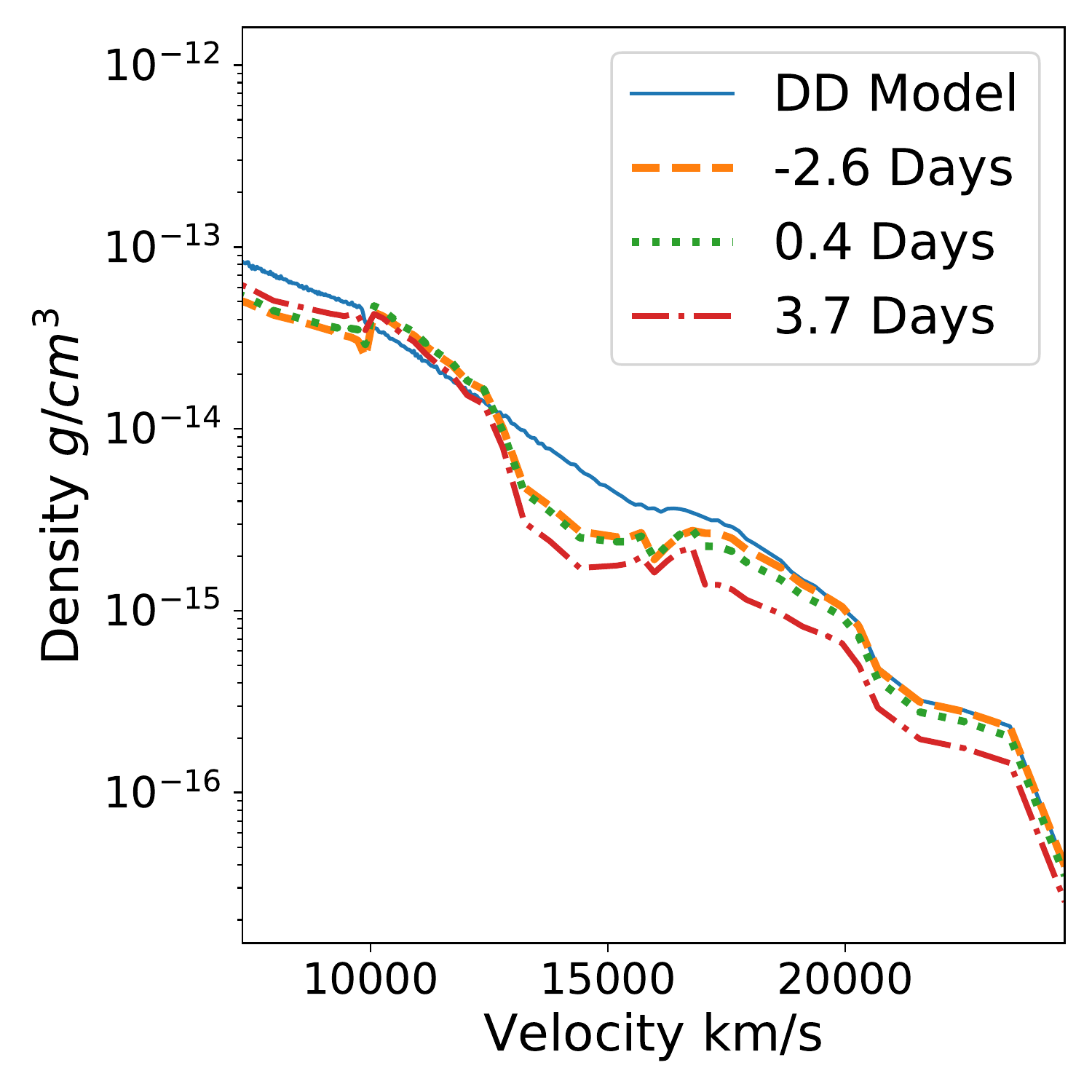}
    \includegraphics[width=0.244\linewidth]{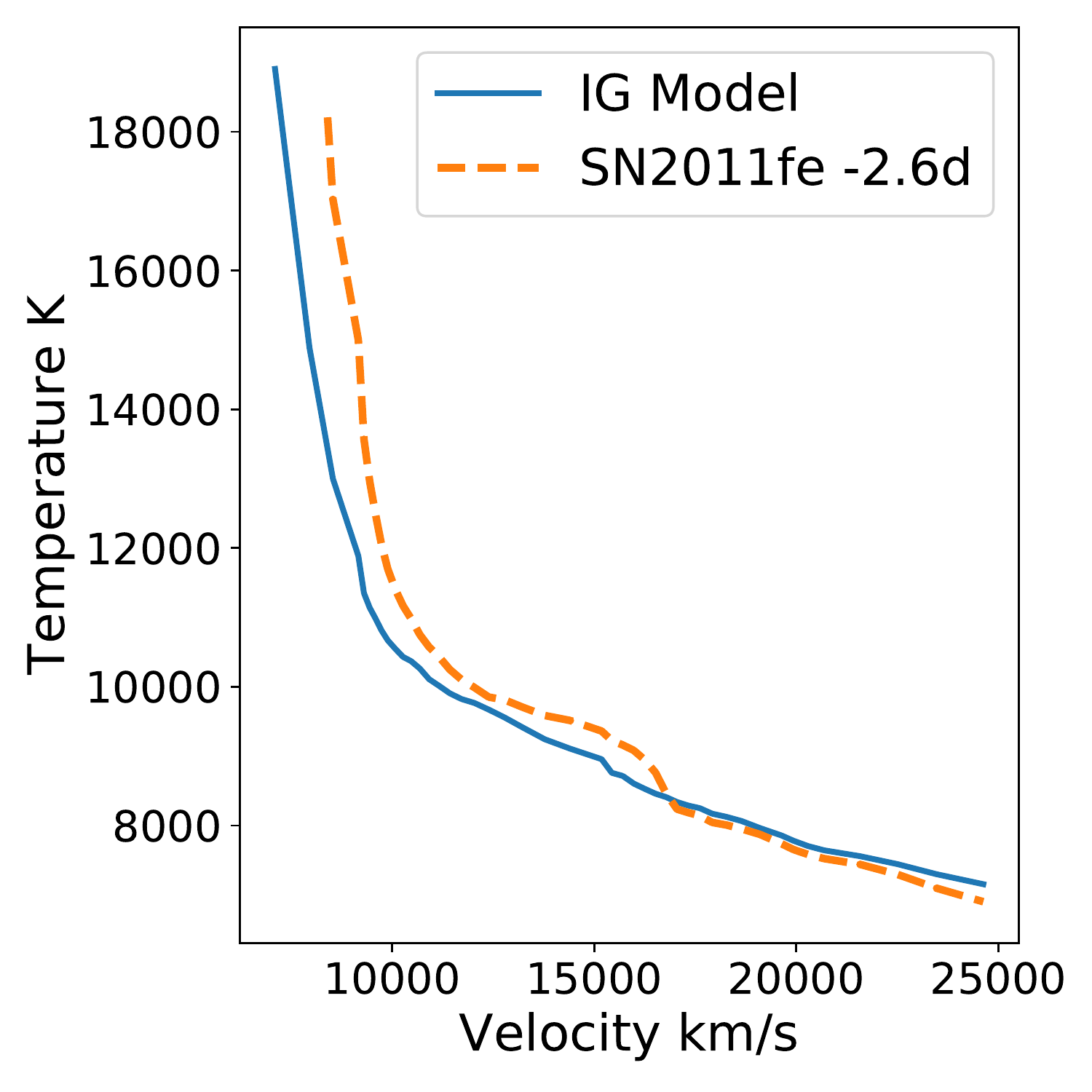}
    \includegraphics[width=0.244\linewidth]{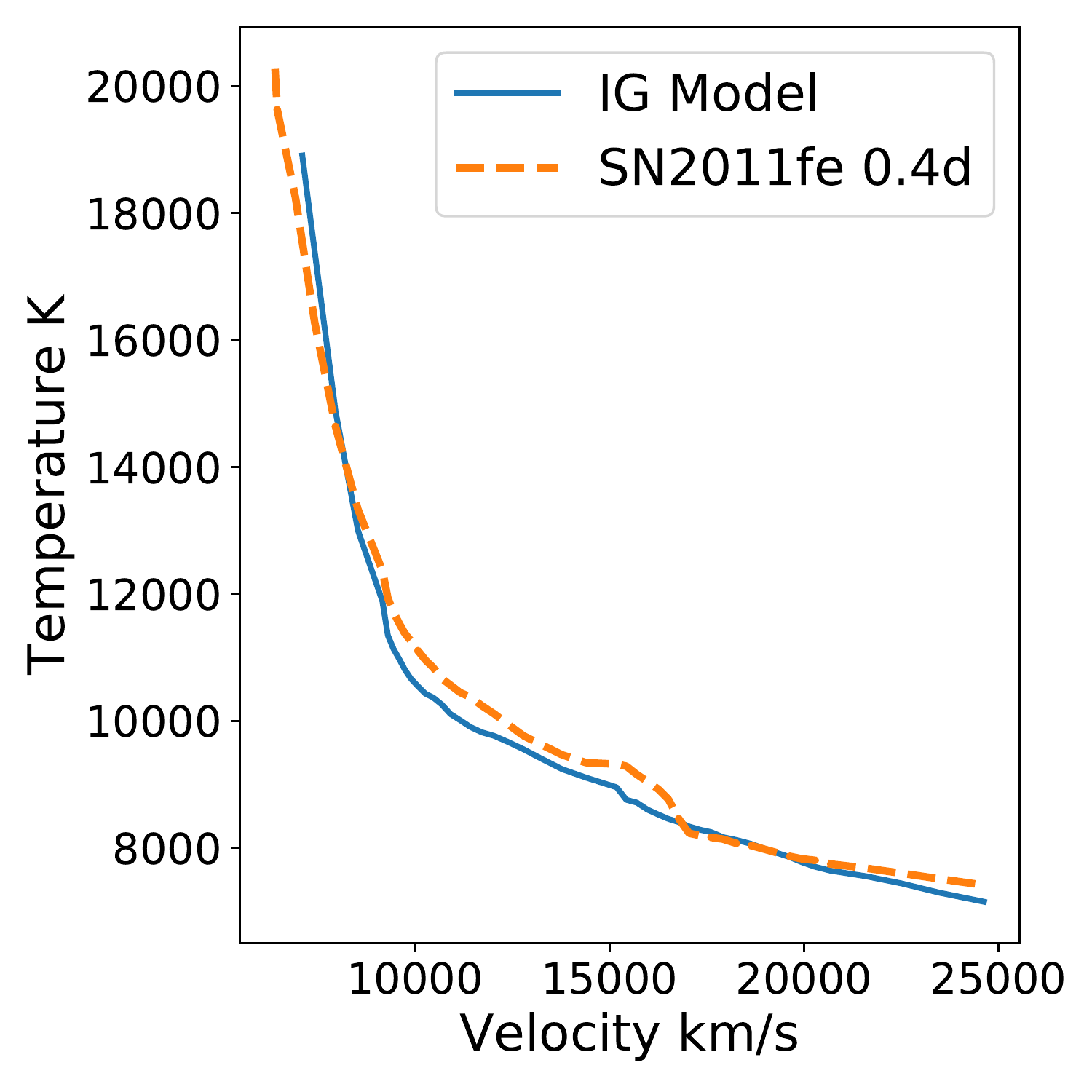}
    \includegraphics[width=0.244\linewidth]{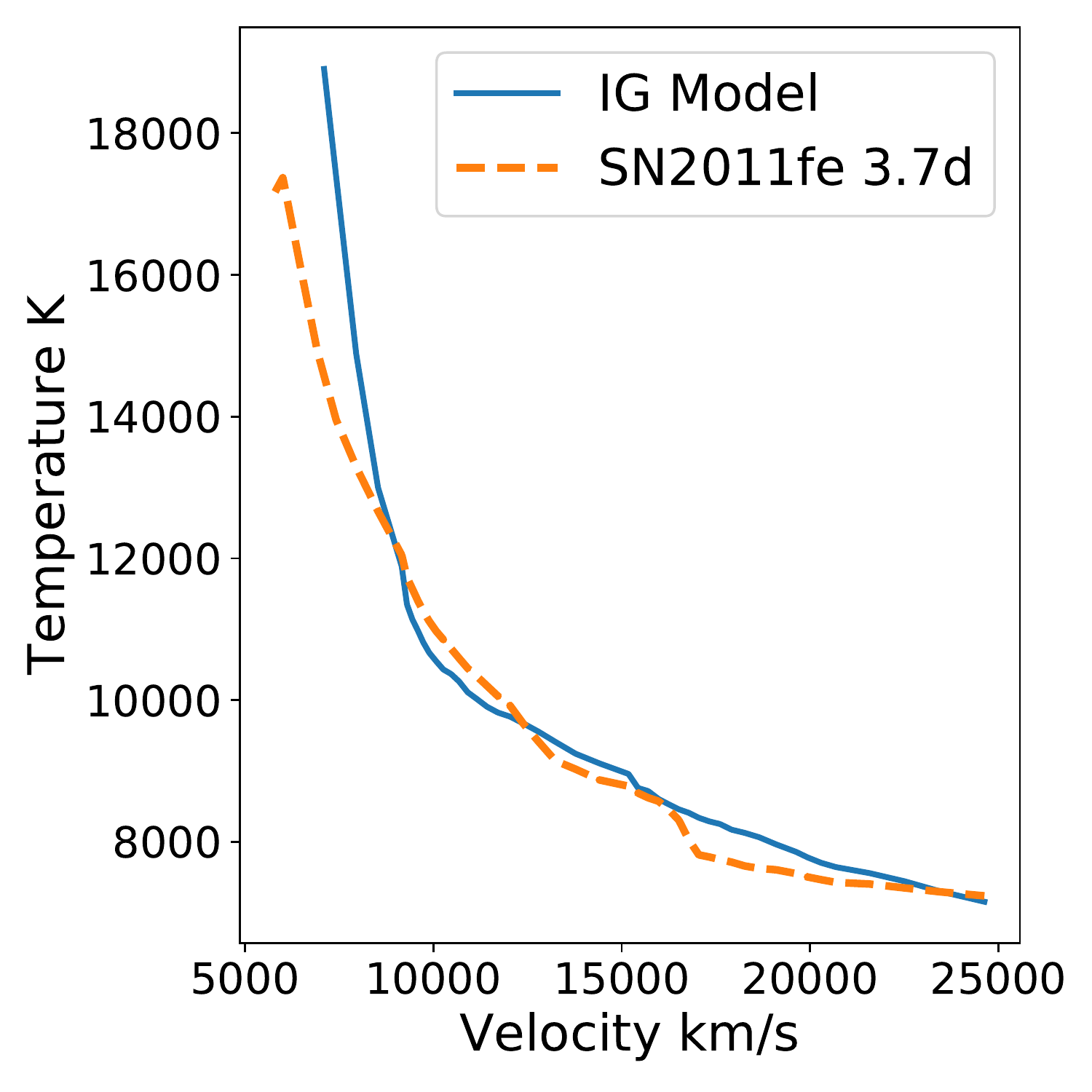}
    \caption{\textbf{From Left To Right:} \textbf{(a)} The density structures of SN~2011fe at $-$2.6 days (orange line), 0.4 days (cyan line) and 3.7 days (red line) as predicted by the neural networks. The DD model is shown in blue line for comparison. \textbf{(b)} The temperature structure for SN~2011fe at $-$2.6 days.  \textbf{(c)} The temperature structure for SN~2011fe at 0.4 days. \textbf{(d)} The temperature structure for SN2011fe at 3.7 days. Note that for convenience of comparisons we have converted all density profiles to 19 days after explosion assuming homogeneous expansion, using the relation $\rho \propto t^{-3}$.
    The temperature profiles for the IGM are shown in blue lines for comparison, and MRNN fits are shown as red lines.}\label{fig:SN2011feTemp}
\end{figure}

The temperature profiles of the TARDIS models of SN~2011fe in the three spectroscopically observed phases are shown in Figure~\ref{fig:SN2011feTemp}. The three supernova structures shown in Figure~\ref{fig:SN2011feElem} are predicted by the neural-network individually for each epoch, the density structures may not strictly observe homologous expansion. However, it is encouraging to see that the density structures in Figure~\ref{fig:SN2011feTemp}~(a) shows adequate similarity with each other, and the corrections compared to the DD model appears to be consistent at different velocity layers; this cross-validates the prediction of the neural networks. 

%The flux levels in the UV are systematically higher in the model predictions. This is consistent with the presence of a high velocity tail at above 25,000 km/sec \citep{rho11fe} which may suppress the UV flux. 

\subsubsection{SN~2011iv}

The transitional supernova (between type Ia normal and type Ia-91bg like) SN~2011iv is located at NGC~1404 with a $B$-band decline rate $\Delta M_{B,15}\ =\ 1.69\pm0.05$ \citep{SN2011iv}. 
According to \citet{SN2011ivHST}, this supernova has negligible dust extinction effect in the line of sight so we did not apply extra extinction corrections on it (see Table~\ref{tab:Stretch}). 

\begin{figure}[htb!]
    \includegraphics[width=\linewidth]{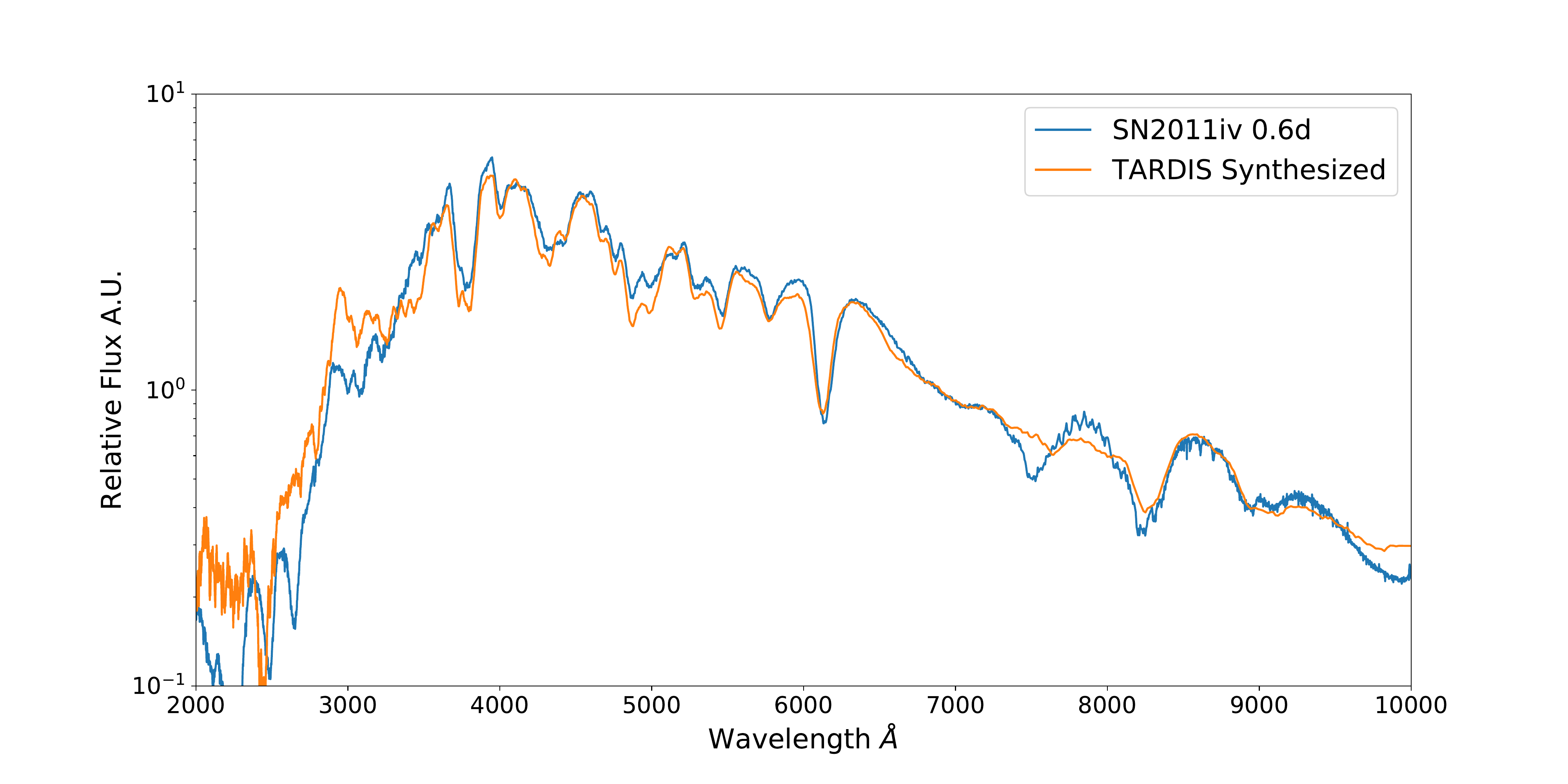}
    \includegraphics[width=0.33\linewidth]{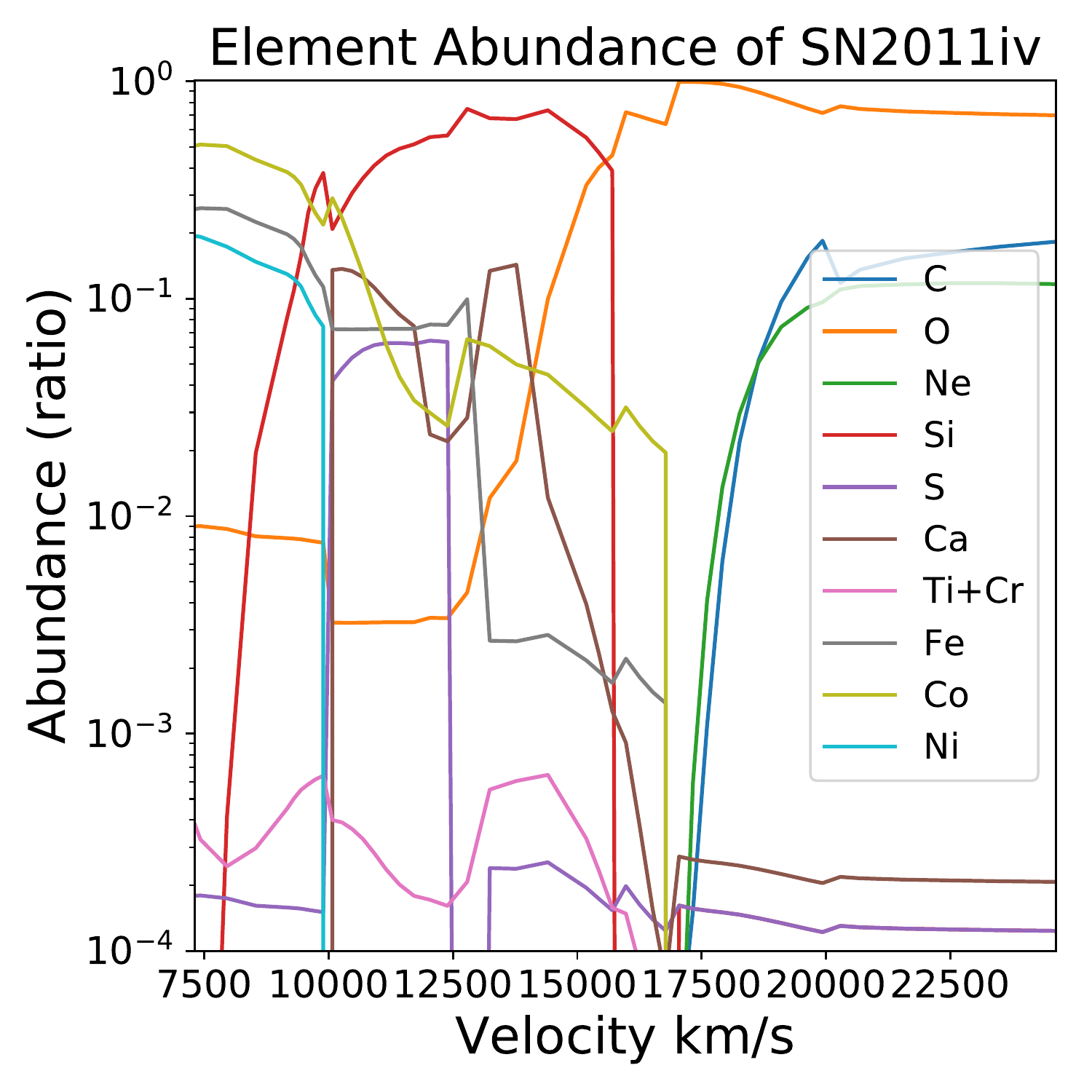}
    \includegraphics[width=0.33\linewidth]{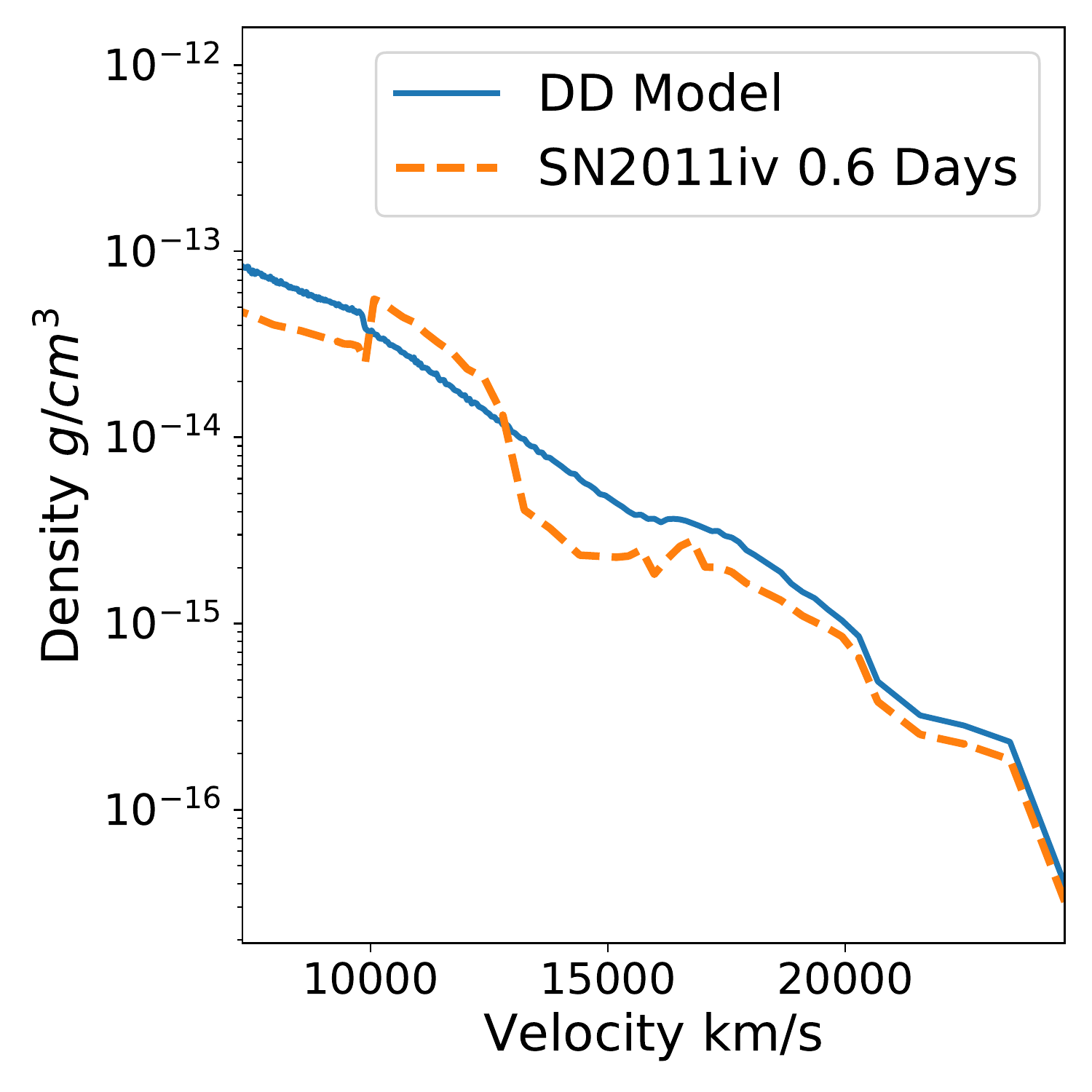}
    \includegraphics[width=0.33\linewidth]{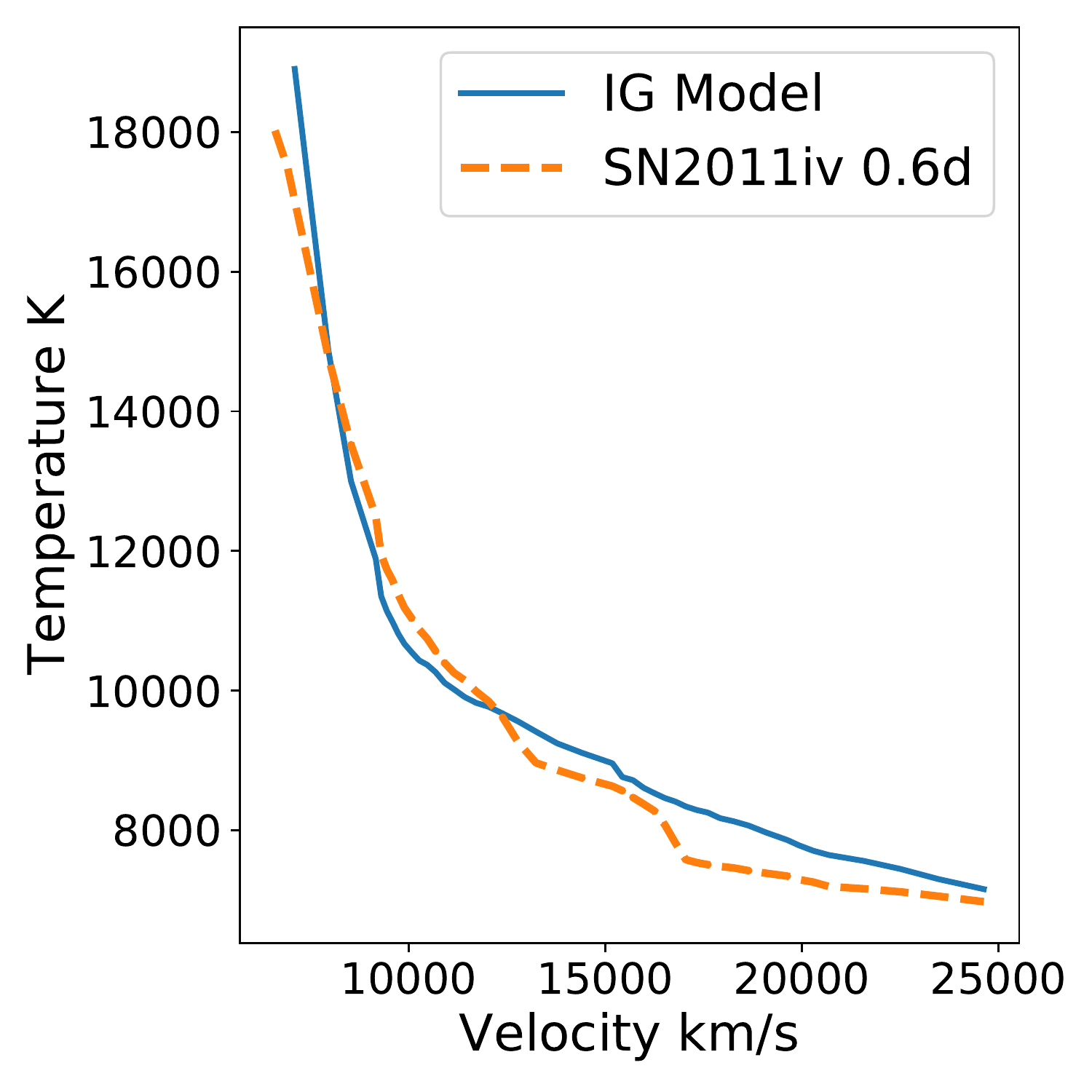}
    \caption{\textbf{a, Upper Panel:} The observed spectrum (blue line) and the TARDIS synthetic spectrum (orange line) of SN~2011iv at 0.6 days after $B$-maximum. \textbf{b, Lower Left:} The elemental abundances of SN~2011iv predicted from neural networks. \textbf{c, Lower Middle:} Density structure of SN~2011iv predicted from neural networks (orange line) and the DD model density structure for comparison (blue line). Both densities are converted to that of day 19 using $\rho \propto t^{-3}$ relation. \textbf{c, Lower Right:} The temperature structure for SN2011iv spectral fitting (orange line) and the IGM temperature structure for comparison (blue line). }\label{fig:SN2011ivSpec}
\end{figure}

In the elemental prediction and spectral fitting process, we adopted the combined spectra of HST and the Magellan telescope at 0.6 days after the $B$-band maximum date \citet{SN2011ivHST}. 
The results are shown in Figure~\ref{fig:SN2011ivSpec}

The model spectrum agrees well with the observed spectrum reasonably across major spectral features in the optical. The disagreement across OI 7300 \AA\ line is obvious. This is likely due to an insufficient amount of oxygen and the feature is not well fit even when the oxygen abundance is enhanced to three times of the IGM. A similar problem may also be seen in Figure~\ref{fig:SN2011feSpec} for SN~2011fe. We see also that the spectral features below 3,000 \AA\ are poorly fit, suggesting again the elemental structure may not have appropriately covered this transition Type SNIa. We will improve these fits in future studies with the construction of a larger spectral library. 

\subsubsection{SN~2011by}

SN~2011by in NGC~3972 has a luminosity decline rate $\Delta M_{B,15}\ =\ 1.14\pm 0.03$ \citep{BSNIPV}. 
SN~2011by has remarkably similar optical spectra and light curves to those of SN~2011fe and are identified as optical "twins" \citep{SN2011by}. 
The only prominent difference is that SN~2011fe is significantly more luminous in the UV (1600 $\ <\ \lambda\ <\ 2500\ {\rm \AA}$) than SN~2011by 
before and around peak brightness \cite{SN2011by2}. 
However, based on the distance deduced from Cepheid Variables in NGC~3972, SN~2011by is about $0.335\pm0.069$ mags dimmer than SN~2011fe \citep{SN2011by2}. This apparent magnitude difference can be a concern for supernova cosmology as its origins are unknown and are thus difficult to be corrected. 

\begin{figure}[htb!]
    \includegraphics[width=\linewidth]{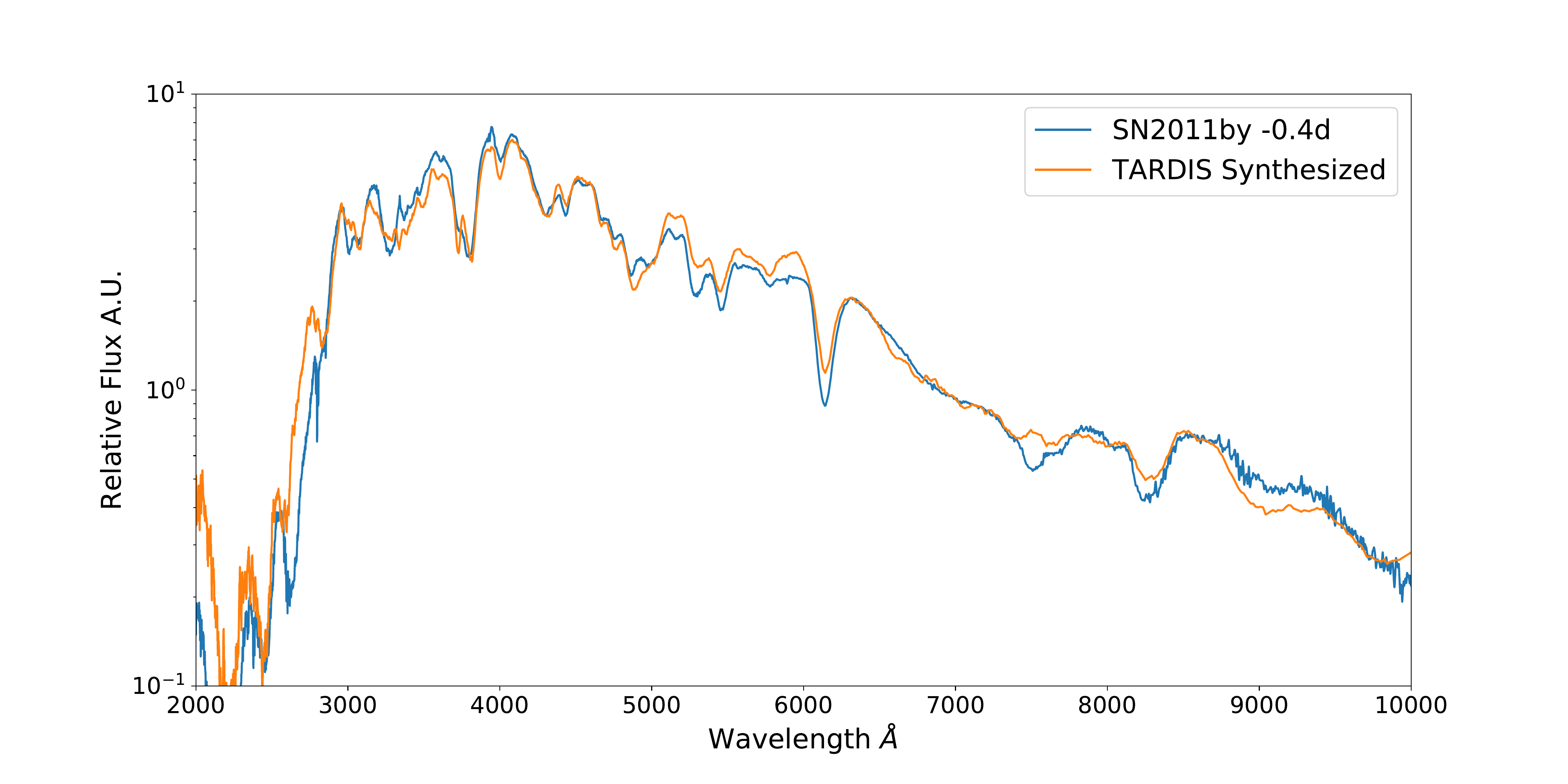}
    \includegraphics[width=0.33\linewidth]{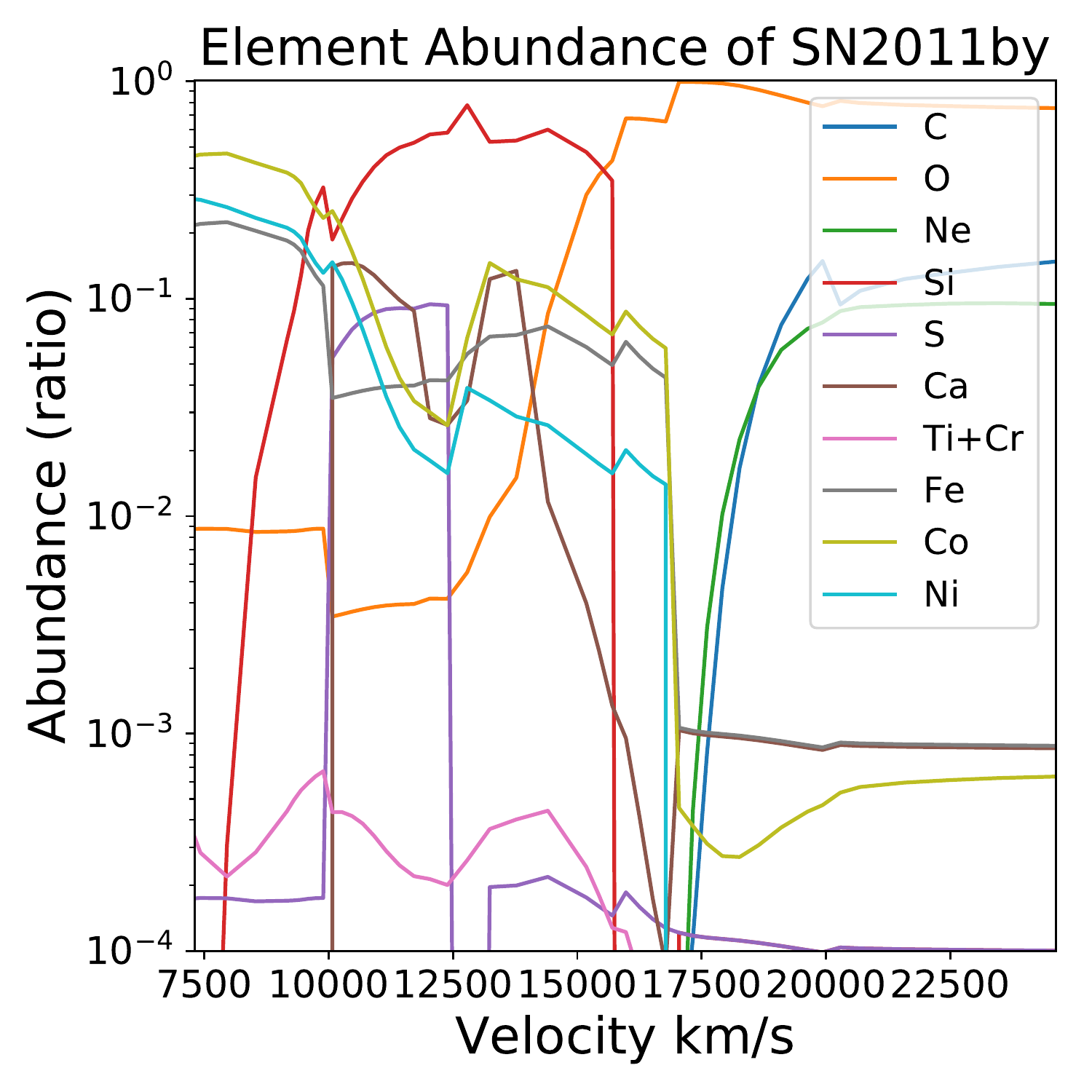}
    \includegraphics[width=0.33\linewidth]{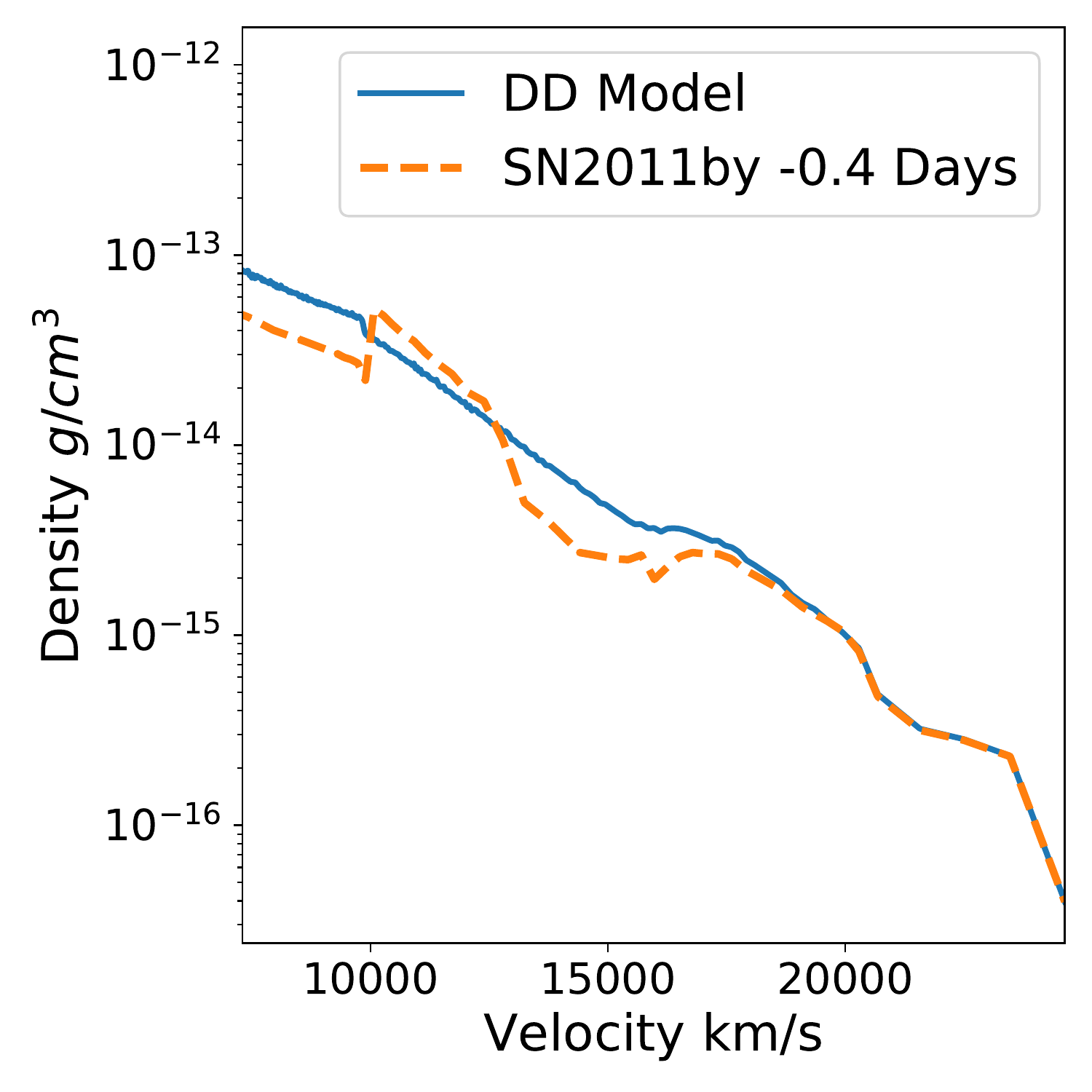}
    \includegraphics[width=0.33\linewidth]{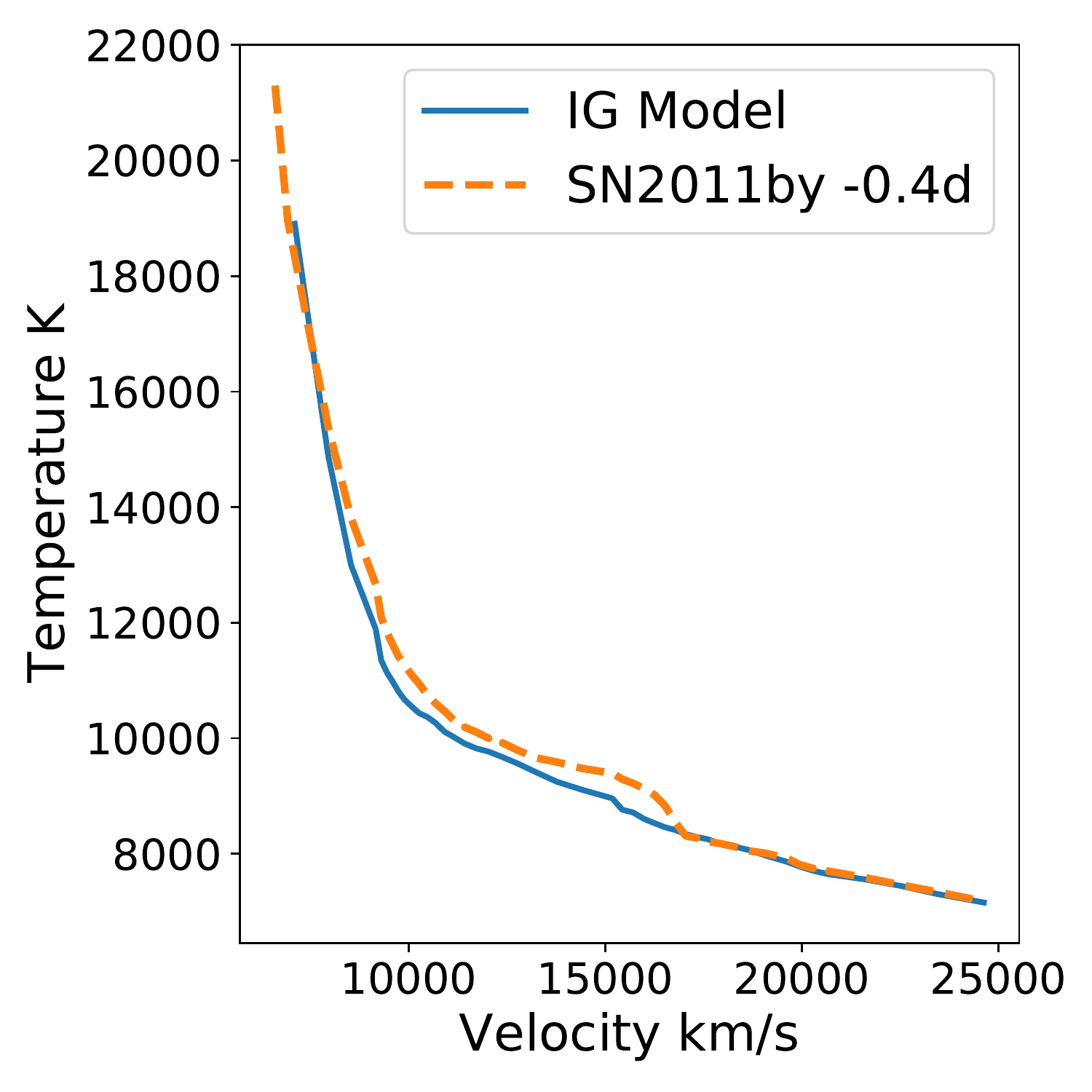}
    \caption{\textbf{a, Upper Panel:} The extinction-corrected observed spectra (blue line) and the TARDIS synthesized spectra (orange line) of SN2011by at -0.4 days. \textbf{b, Lower Left:} The element abundance of SN2011by predicted from neural networks. \textbf{c, Lower Middle:} Density structure of SN2011by predicted from neural networks (orange line) and the DD model density structure as comparison. Both  densities are converted to 19 days after explosion using $\rho \propto t^{-3}$ relation. \textbf{d, Lower Right:} Comparison of the temperature structure of SN2011by spectral model (orange line) and that of the IG model (blue line). }\label{fig:SN2011bySpec}
\end{figure}

As in \citet{SN2011by2}, we adopted the \citet{F99} extinction model with $R_{v}\ =\ 3.1$ and $E(B-V)\ =\ 0.039$ to correct the host galaxy extinction, and Milky Way extinction models of \citet{GCC09}  with $E(B-V)\ =\ 0.013$ to correct for Milky Way extinction. 
The HST spectrum of SN~2011by at 0.4 days before the $B$-band maximum is employed to derive the elemental abundances using MRNN. 
The extinction-corrected observed spectrum and the corresponding synthetic spectrum are shown in Figure~\ref{fig:SN2011bySpec}. 

It is worth noticing that the absolute luminosity of SN~2011by found by the MRNN process is slightly more luminous than that of SN~2011fe. Our MRNN spectral fits thus do not provide a theoretical explanation to the apparent luminosity difference of the two SNe. This may be caused by model uncertainties and the fact that we adopted an extinction correction without further iterations to improve the overall model fits in the UV (see Figure~\ref{fig:SN2011bySpec}). Indeed, the model spectrum is too bright in the wavelength range shorter than 2800 \AA, which may again suggest that there is room for improvement by enlarging the ranges of the elemental abundances for the spectral library to better sample the physical conditions of the observed spectrum. 

\subsubsection{SN~2015F}

\begin{figure}[htb!]
    \includegraphics[width=\linewidth]{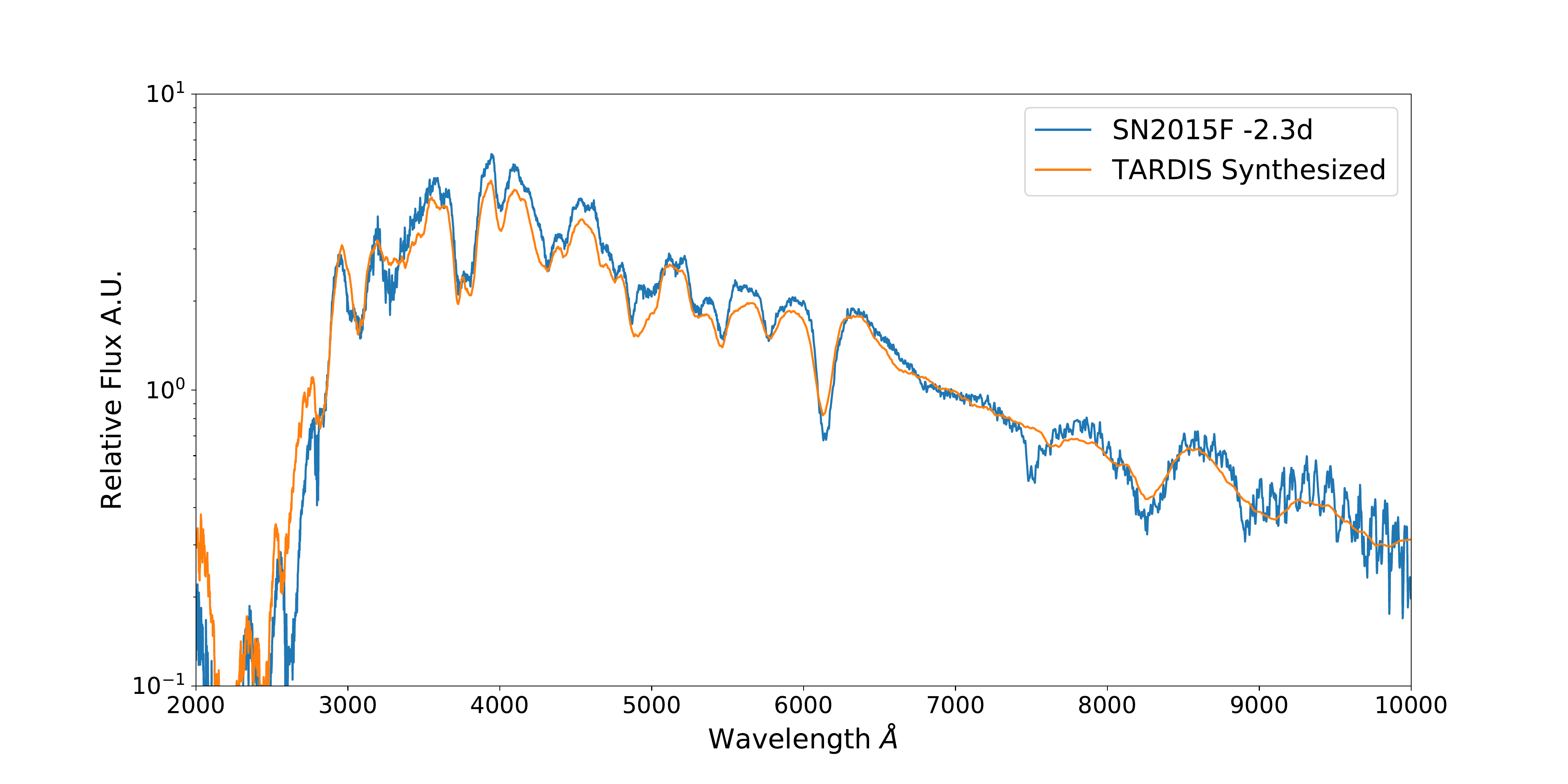}
    \includegraphics[width=0.33\linewidth]{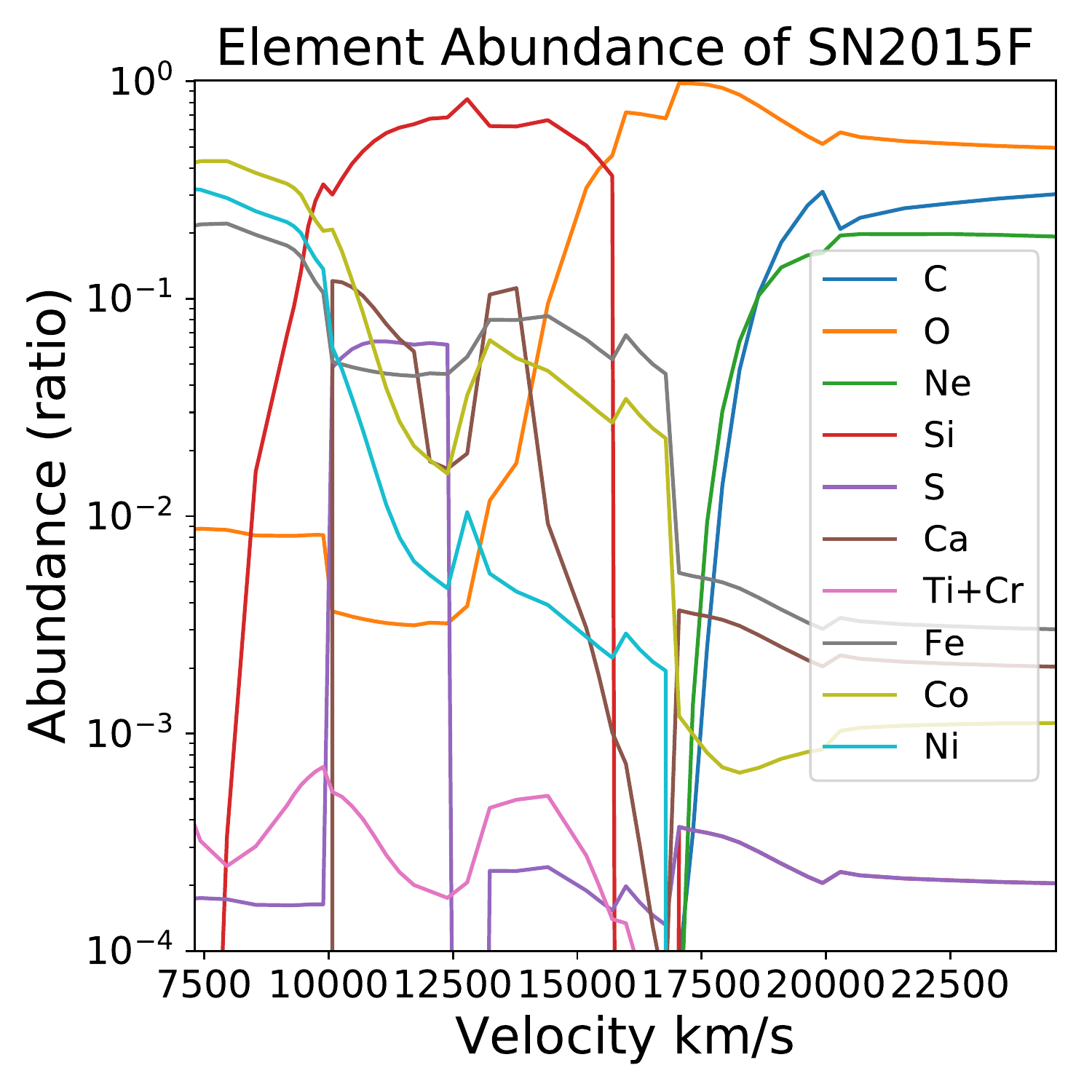}
    \includegraphics[width=0.33\linewidth]{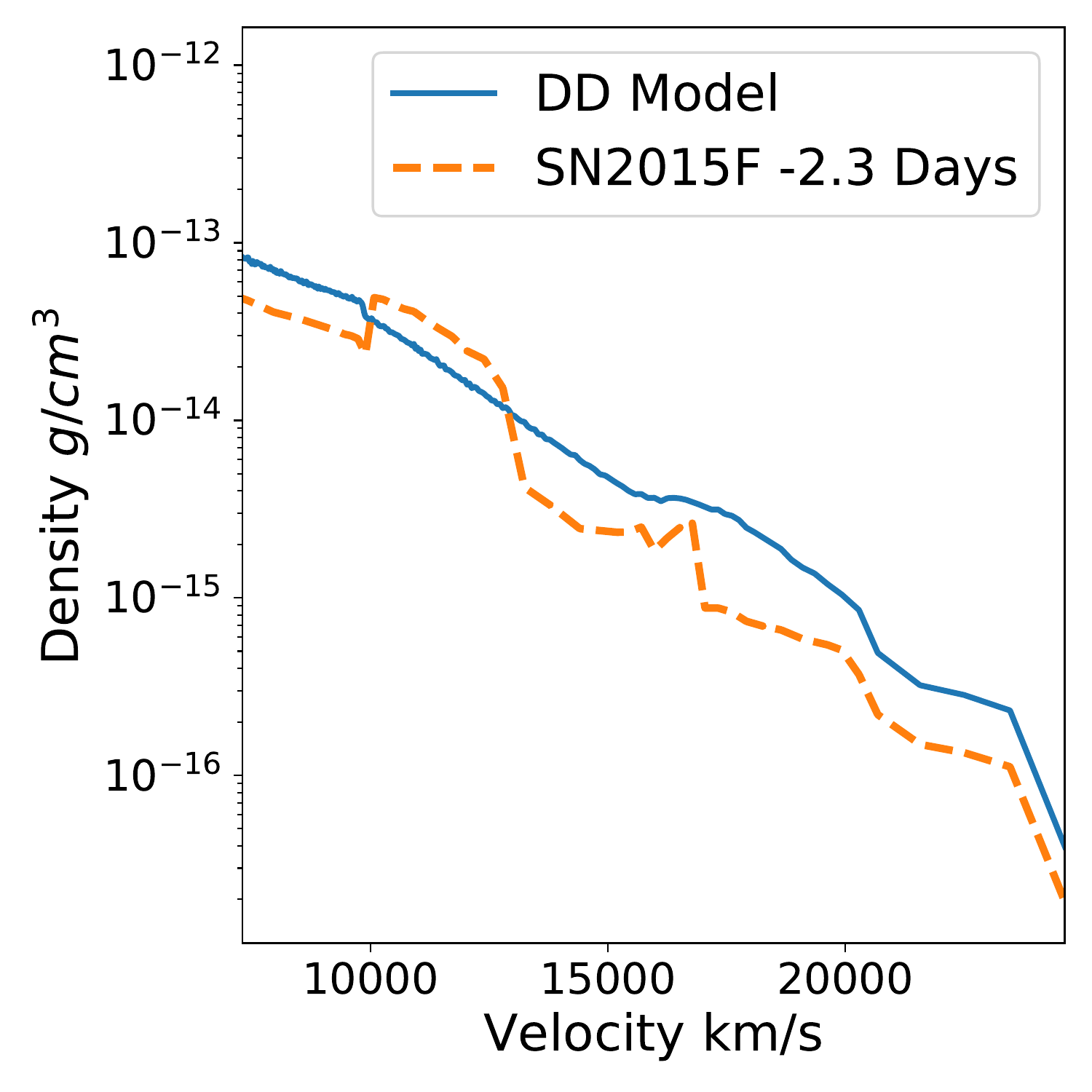}
    \includegraphics[width=0.33\linewidth]{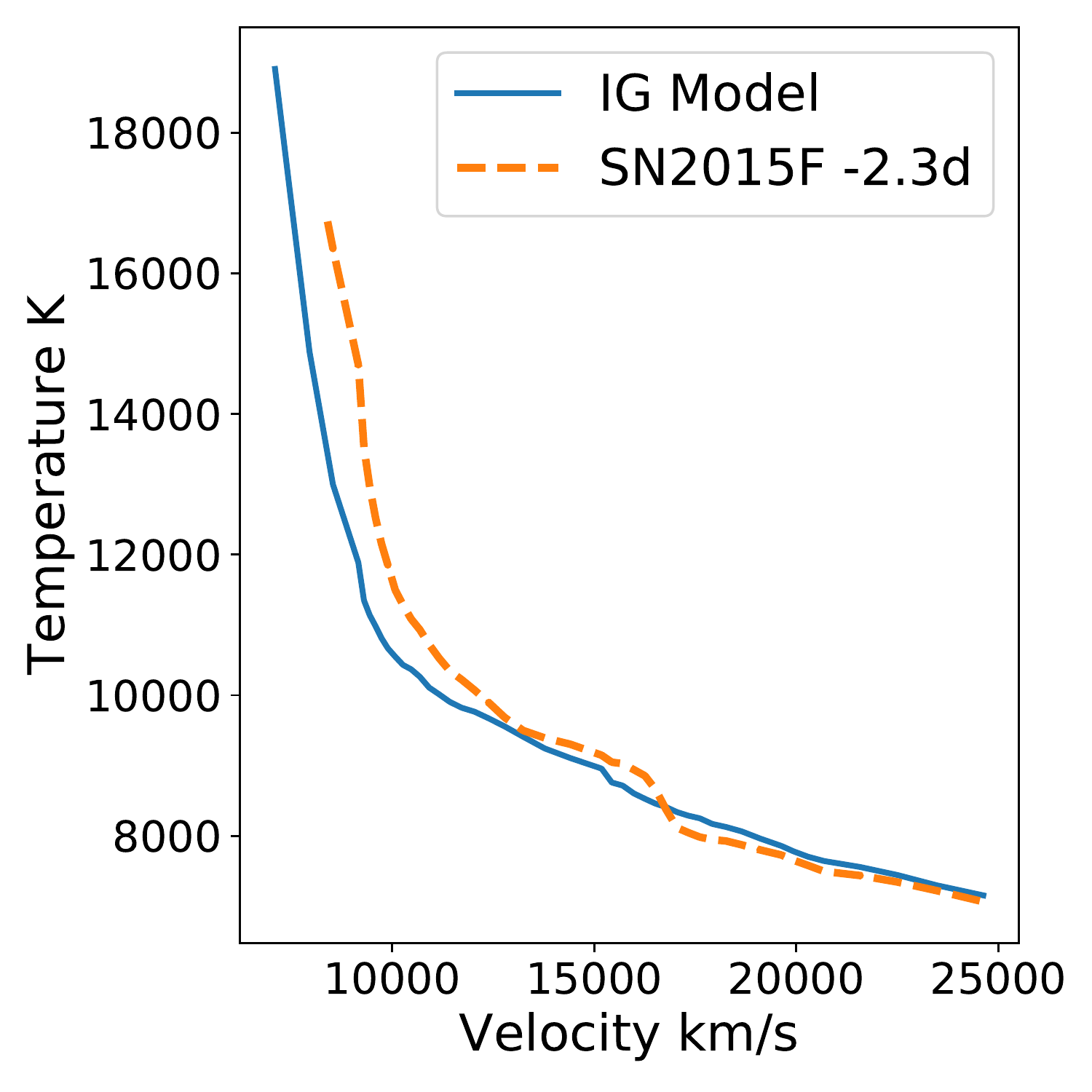}
    \caption{\textbf{a, Upper Panel:} The extinction-corrected observed spectra (blue line) and the TARDIS synthesized spectra (orange line) of SN~2015F at $-$2.6 days. \textbf{b, Lower Left:} The elemental abundances of SN~2015F predicted by MRNN. \textbf{c, Lower Middle:} Density structure of SN~2015F predicted from MRNN (orange line) and the DD model density structure for comparison. \textbf{d, Lower Right:} The temperature structures for SN~2015F spectral model (orange line) and the IGM for comparison (blue line). }\label{fig:SN2015FSpec}
\end{figure}

SN~2015F in NGC~2442 is a slightly sub-luminous supernova with a decline rate $ \Delta M_{B,15}\ =\ 1.35\pm0.03$ \citep{SN2015Fearly}. 
We employed the HST spectrum at $-2.3$ days relative to  $B$-band maximum for elemental abundance predictions and  spectra fitting. 
The host galaxy extinction was corrected using the model from \citet{CCM} with $R_{v}\ =\ 3.1$ and $E(B-V)\ =\ 0.035$ mag.  The Milky Way extinction was corrected with \citet{GCC09} model with $E(B-V)\ =\ 0.175 $ \citep{SN2015F}. The fitting results are shown in Figure~\ref{fig:SN2015FSpec}. Notice the absence of high velocity \ion{Ca}{2} and \ion{O}{1}, and the apparent lower density of the ejecta as is obvious from Figure~~\ref{fig:SN2015FSpec}(c).

\subsubsection{ASASSN-14lp}

ASASSN-14lp is a bright SNIa located in NGC~4666. Its luminosity decline is $\Delta M_{B,15}\ =\ 0.80\pm 0.05$ \citep{asassn14lp}. 
In order to correct the host galaxy extinction, we adopted the \citet{CCM} extinction relation with $R_{v}\ =\ 3.1$ and  $E(B-V)\ =\ 0.33$ mag \citep{asassn14lp}. 
For Mikly Way extinction, we adopted \citet{GCC09} extinction model with $E(B-V)\ =\ 0.021$ mag. 

\begin{figure}[htb!]
    \includegraphics[width=\linewidth]{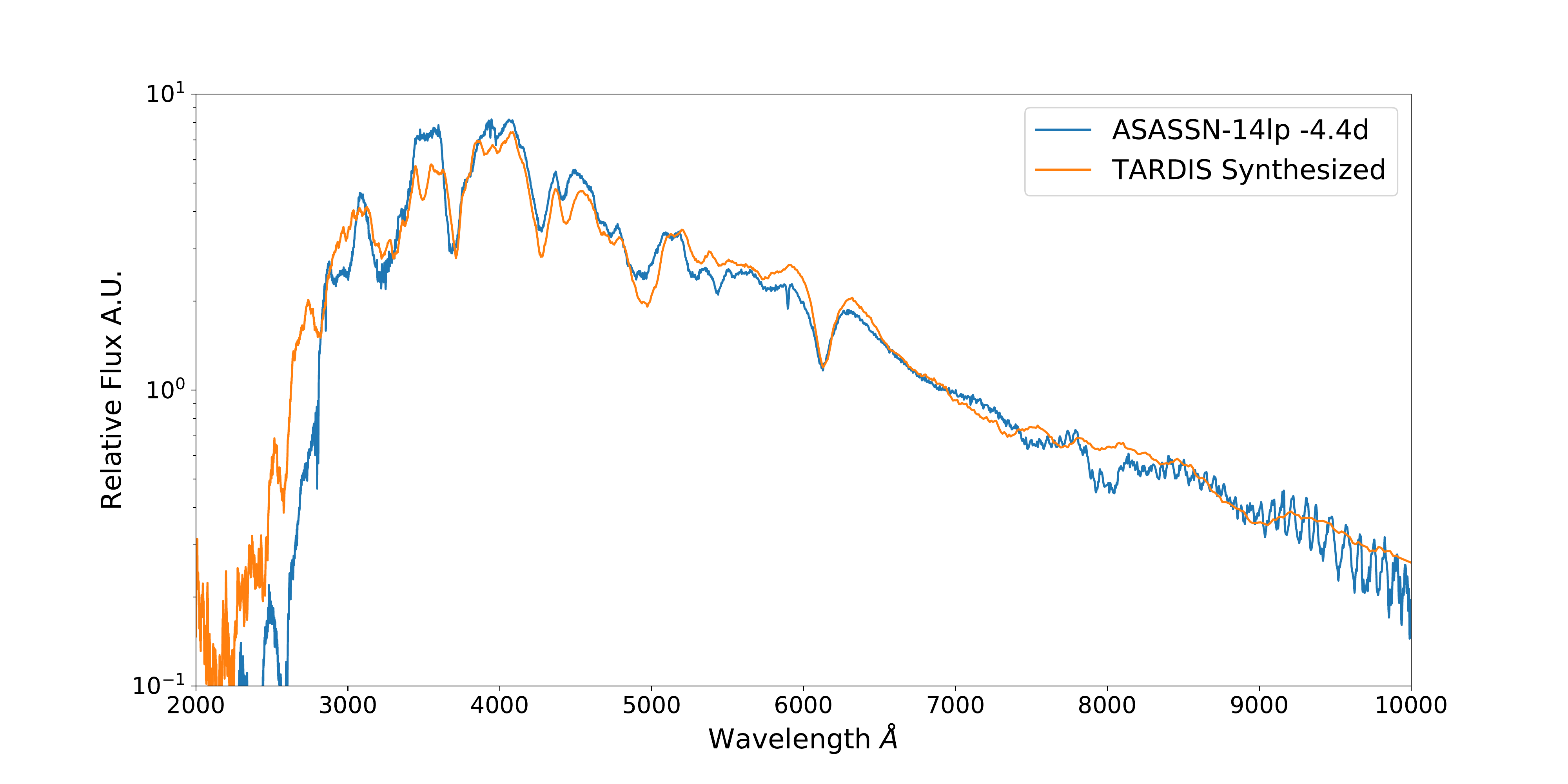}
    \includegraphics[width=0.33\linewidth]{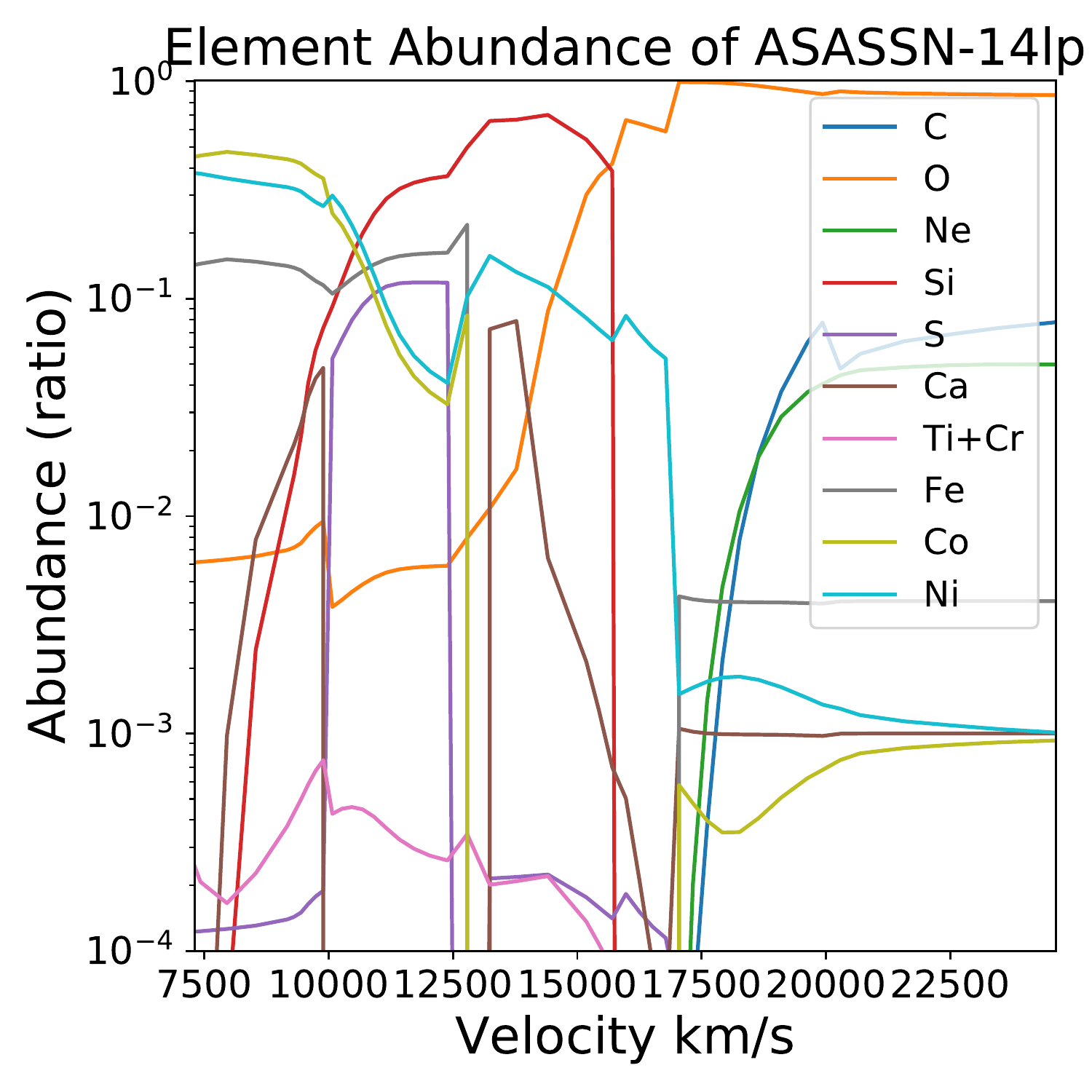}
    \includegraphics[width=0.33\linewidth]{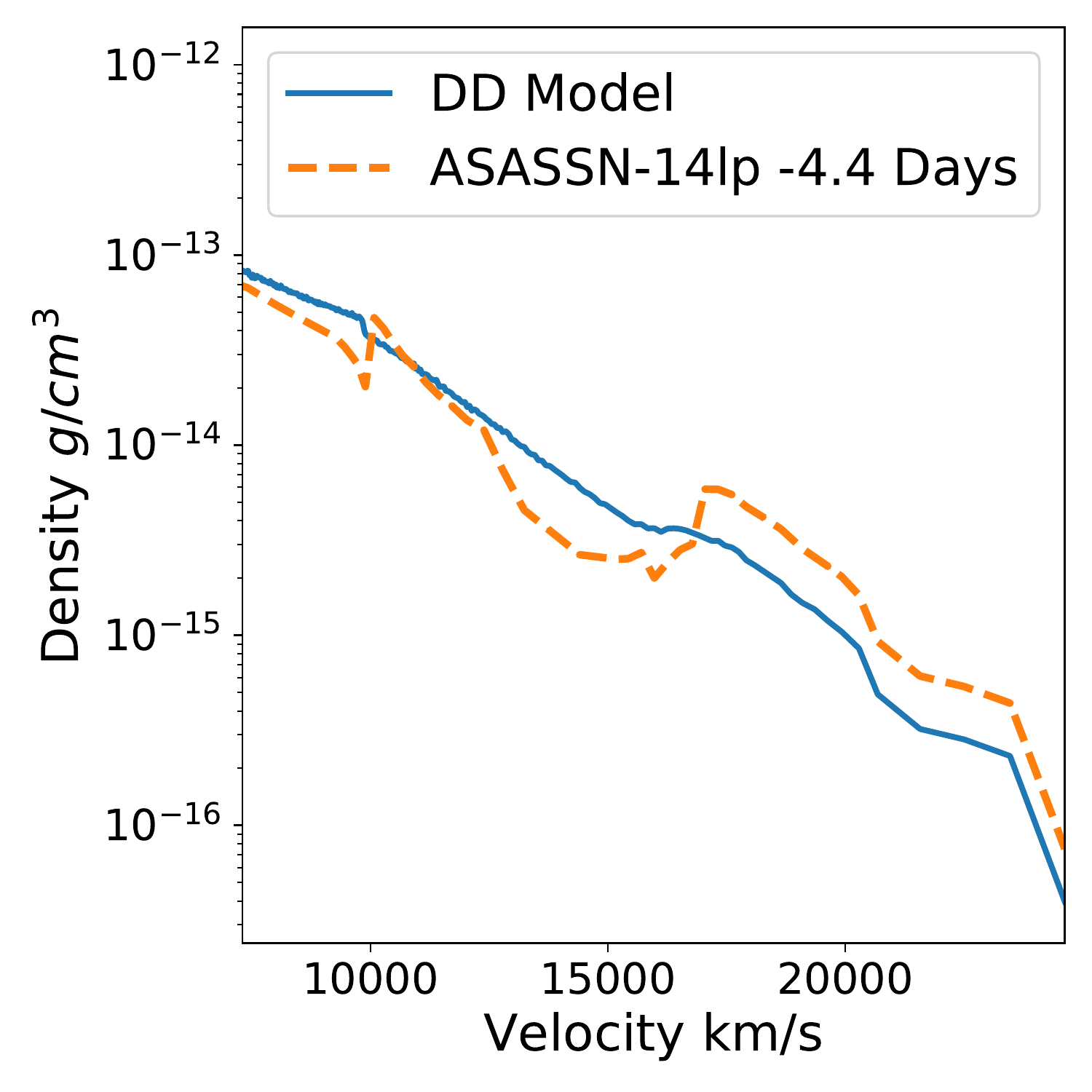}
    \includegraphics[width=0.33\linewidth]{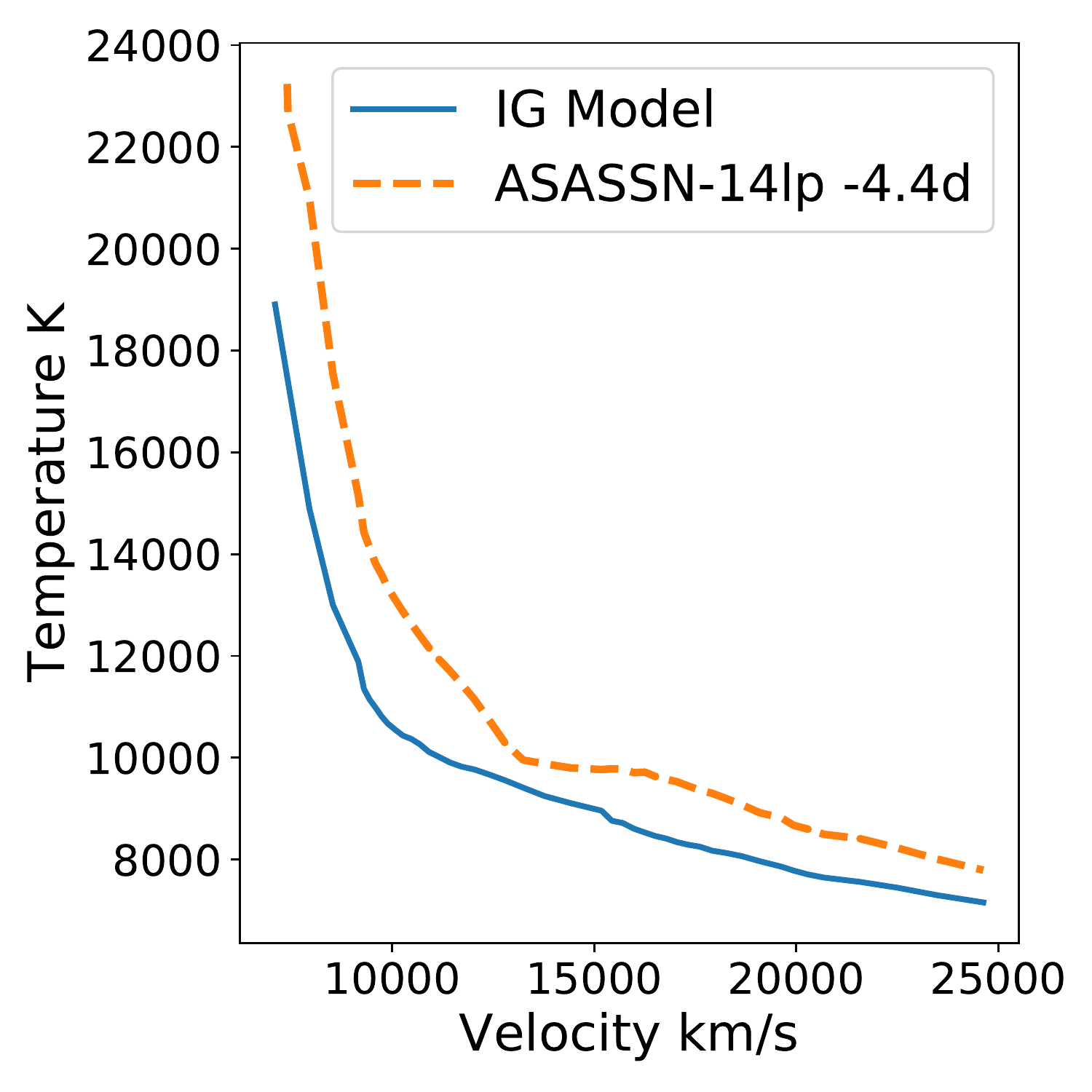}
    \caption{\textbf{(a) Upper Panel:} The extinction-corrected observed spectrum (blue line) and the TARDIS synthesized spectrum (orange line) of ASASSN-14lp at $-$4.4 days. \textbf{(b) Lower Left:} The elemental abundances of ASASSN-14lp predicted by MRNN. \textbf{(c) Lower Middle:} Density structure of ASASSN-14lp predicted by MRNN (orange line) and the DD model density structure for comparison. Both densities are converted to 19 days from $B$ maximum using the $\rho\ \propto\ t^{-3}$ relation. \textbf{(d) Lower Right:} The temperature structure for ASASSN-14lp (orange line) and the IGM temperature structure for comparison (blue line). }\label{fig:ASASSN14lpSpec}
\end{figure}

We used the HST spectrum at $-$4.4 days from  $B$-band maximum  for both the elemental abundance prediction and spectra fittings. 
The results are shown in Figure~\ref{fig:ASASSN14lpSpec}. 

The ejecta show enhanced density at velocity above 17,500 km/sec. The \ion{Ca}{2} H and K, and IR triplet are clearly detected and highly blueshifted. The temperature profile shown in Figure~\ref{fig:ASASSN14lpSpec}(d) is higher than that of the IGM throughout the ejecta. This is consistent with what is typically expected for SNe with slow decline rates. 

\subsubsection{SN~2013dy}

SN~2013dy is located in NGC~7250 with a luminosity decline rate of $\Delta M_{B,15}\ =\ 0.90 \pm 0.03$ \citep{SN2013dy}. Like ASASSN-14lp, SN~2013dy is of the group with slow decline rates.   
We adopted the extinction model of \citet{CCM} with $R_{v}\ =\ 3.1$ and $E(B-V)\ =\ 0.206$ mag to correct the host galaxy extinction, and \citet{GCC09} model with $E(B-V)\ =\ 0.135$ mag for Milky Way reddening correction. 
The extinction parameters $E(B-V)$ were taken from \citet{AbsoluteDistance}. 

\begin{figure}[htb!]
    \includegraphics[width=\linewidth]{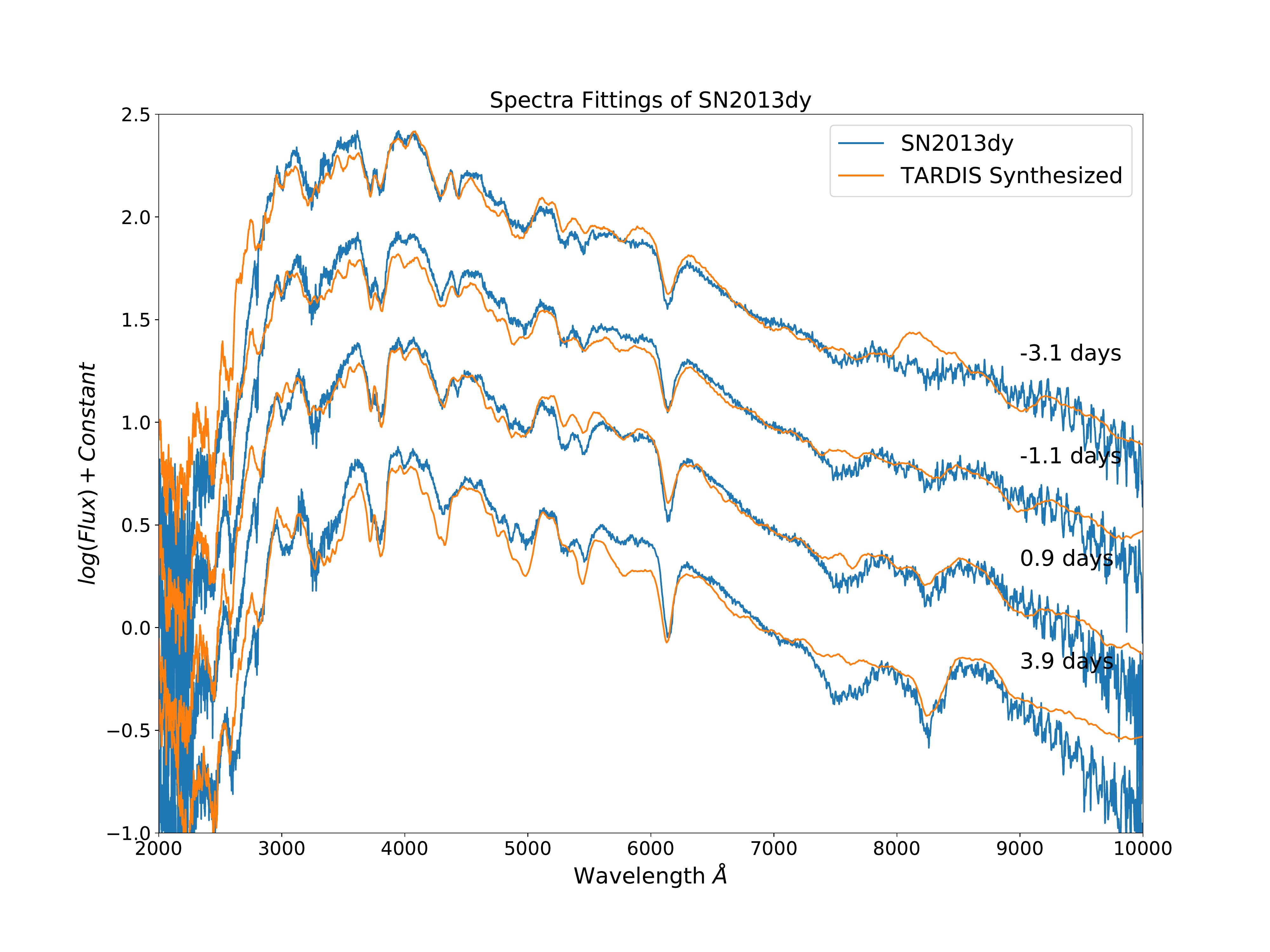}
    \caption{The extinction-corrected observed spectra (blue lines) and the TARDIS synthetic spectra (orange lines) of SN~2013dy at $-3.1$ days, $-1.1$ days, $0.9$ days and $3.9$ days. The fluxes are in logarithmic scale and the spectra of different dates are arbitrarily offset by a constant. }\label{fig:SN2013dySpec}
\end{figure}

The HST spectrum at $-3.1$ days, $-1.1$ days, $0.9$ days and $3.9$ days were used for spectral modeling. The results are shown in Figure~\ref{fig:SN2013dySpec}. 

\begin{figure}
    \includegraphics[width=0.33\linewidth]{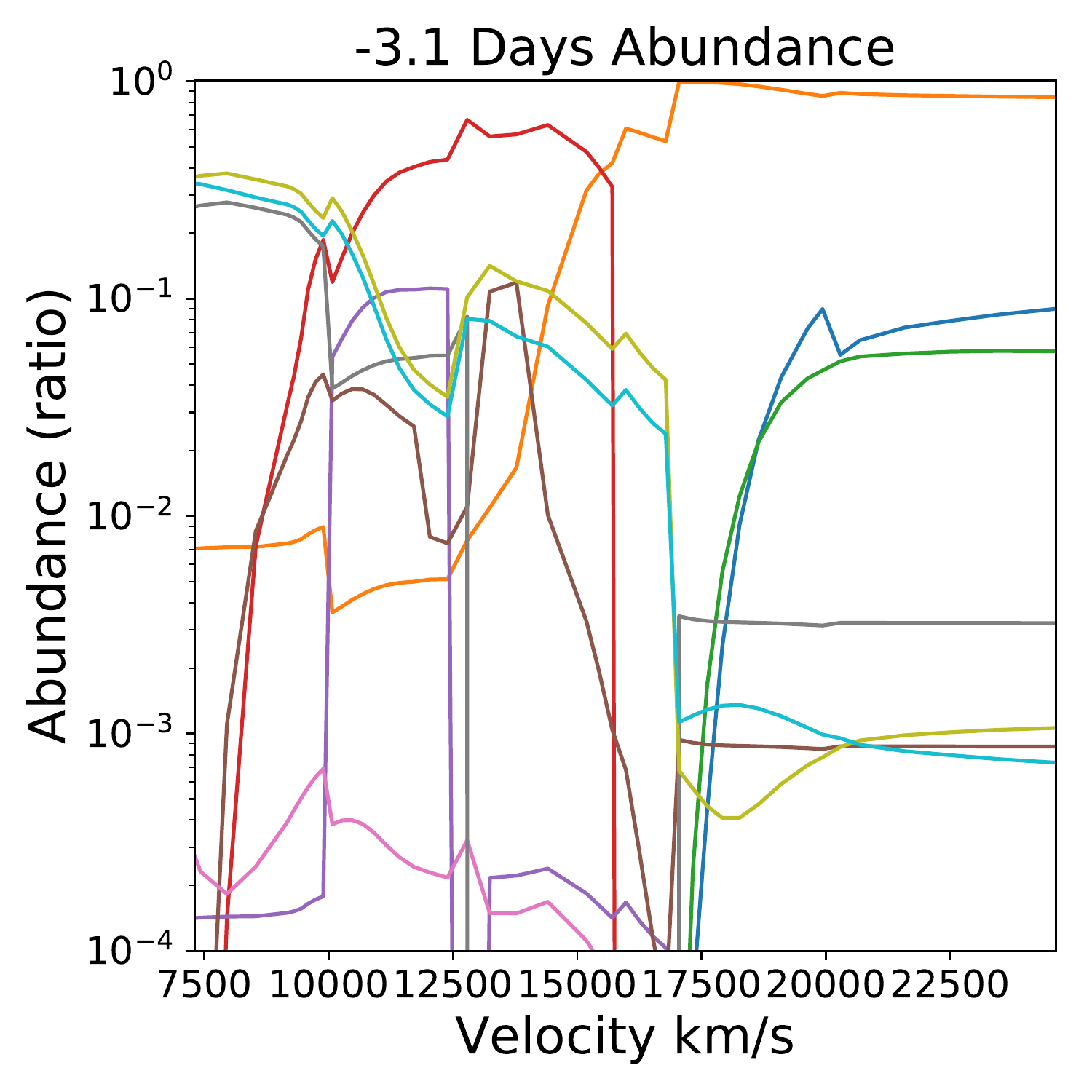}
    \includegraphics[width=0.33\linewidth]{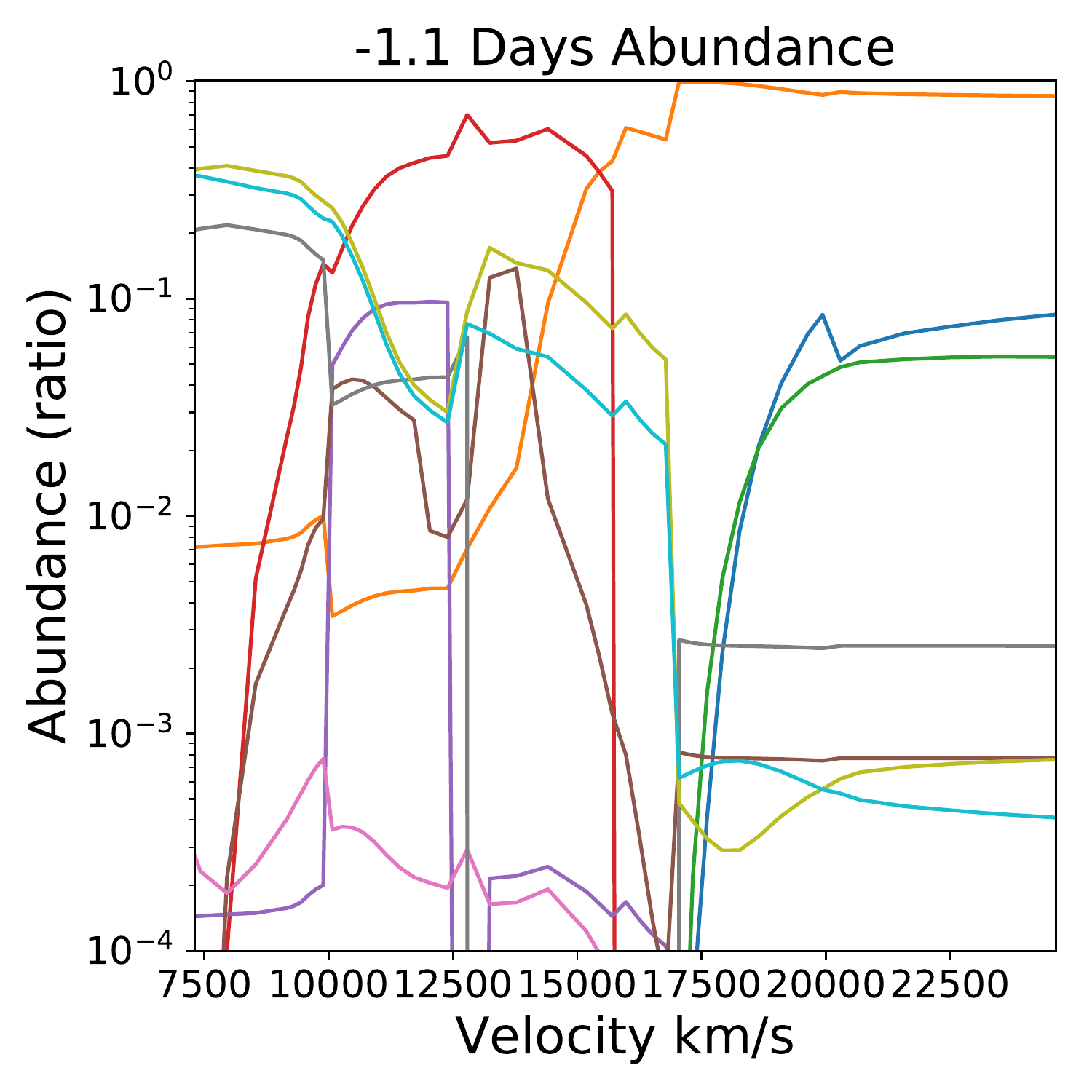}
    \includegraphics[width=0.33\linewidth]{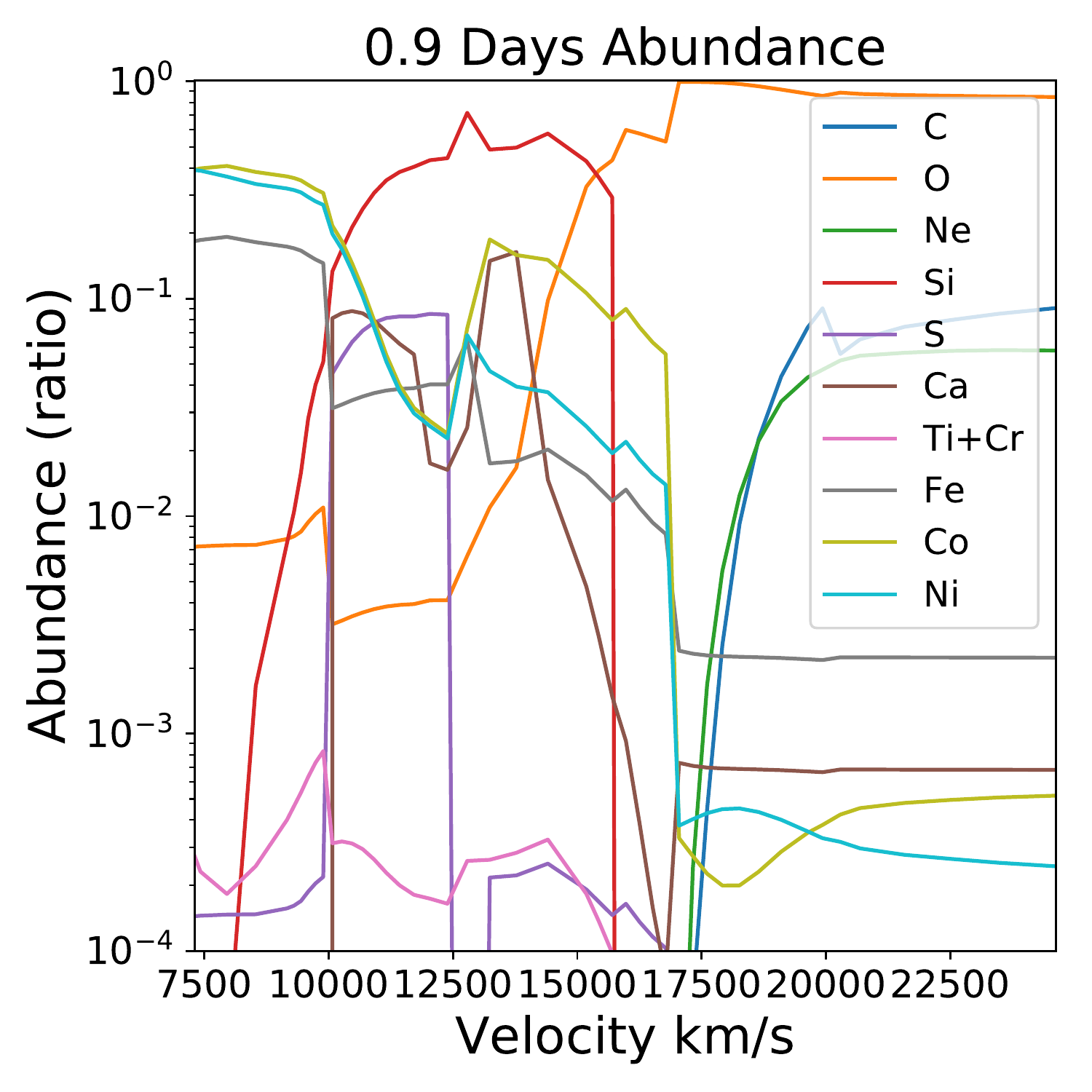}
    \includegraphics[width=0.33\linewidth]{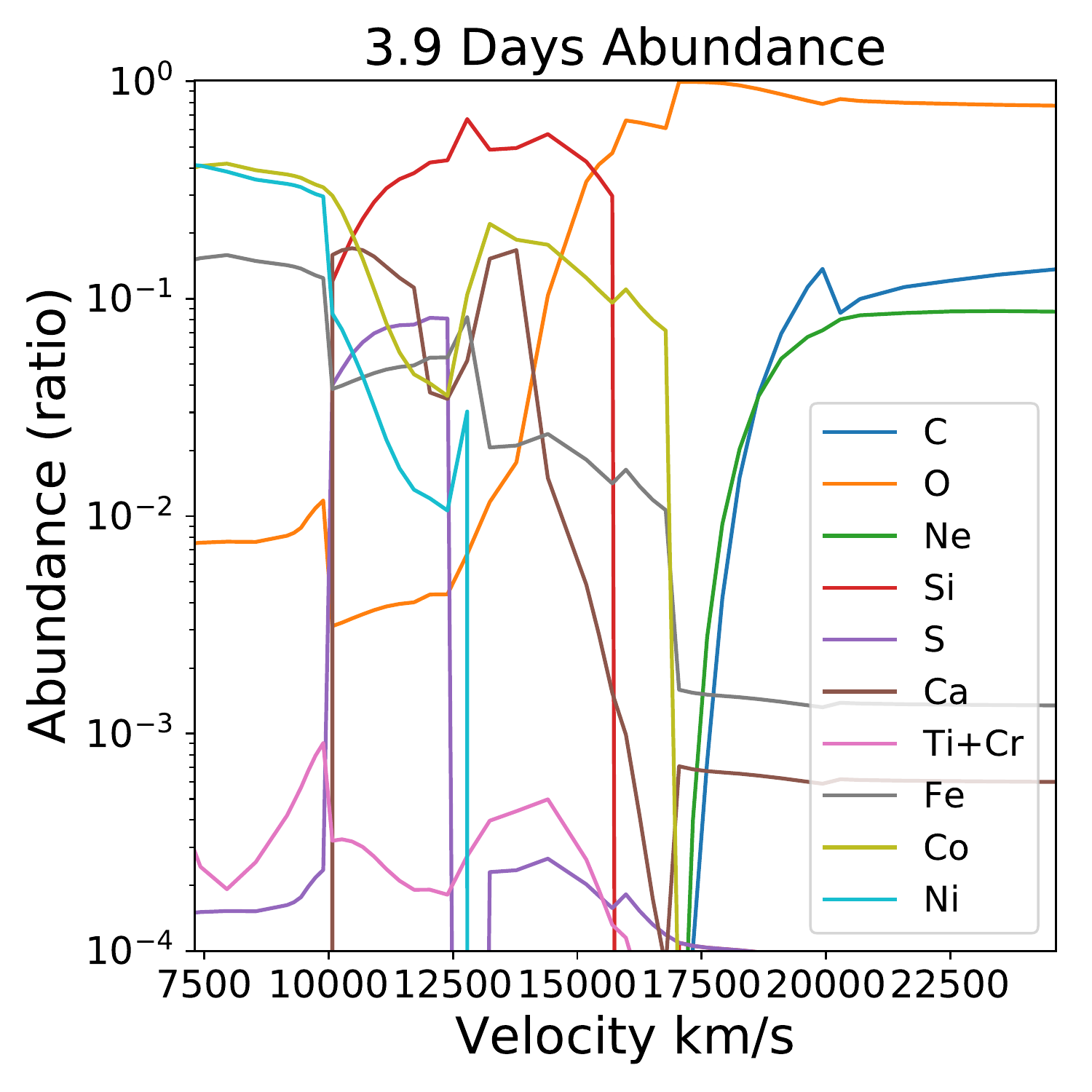}
    \includegraphics[width=0.33\linewidth]{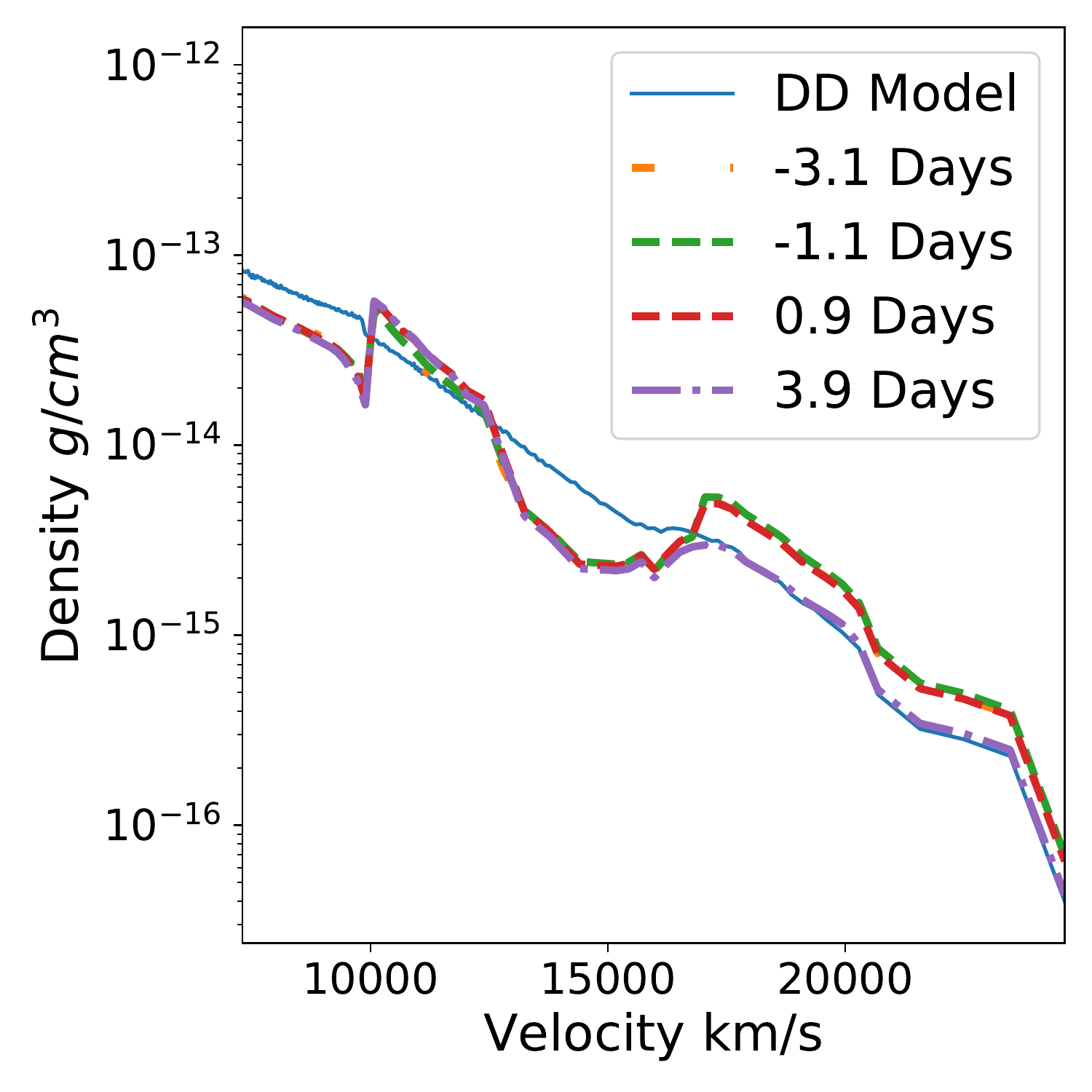}
    \includegraphics[width=0.33\linewidth]{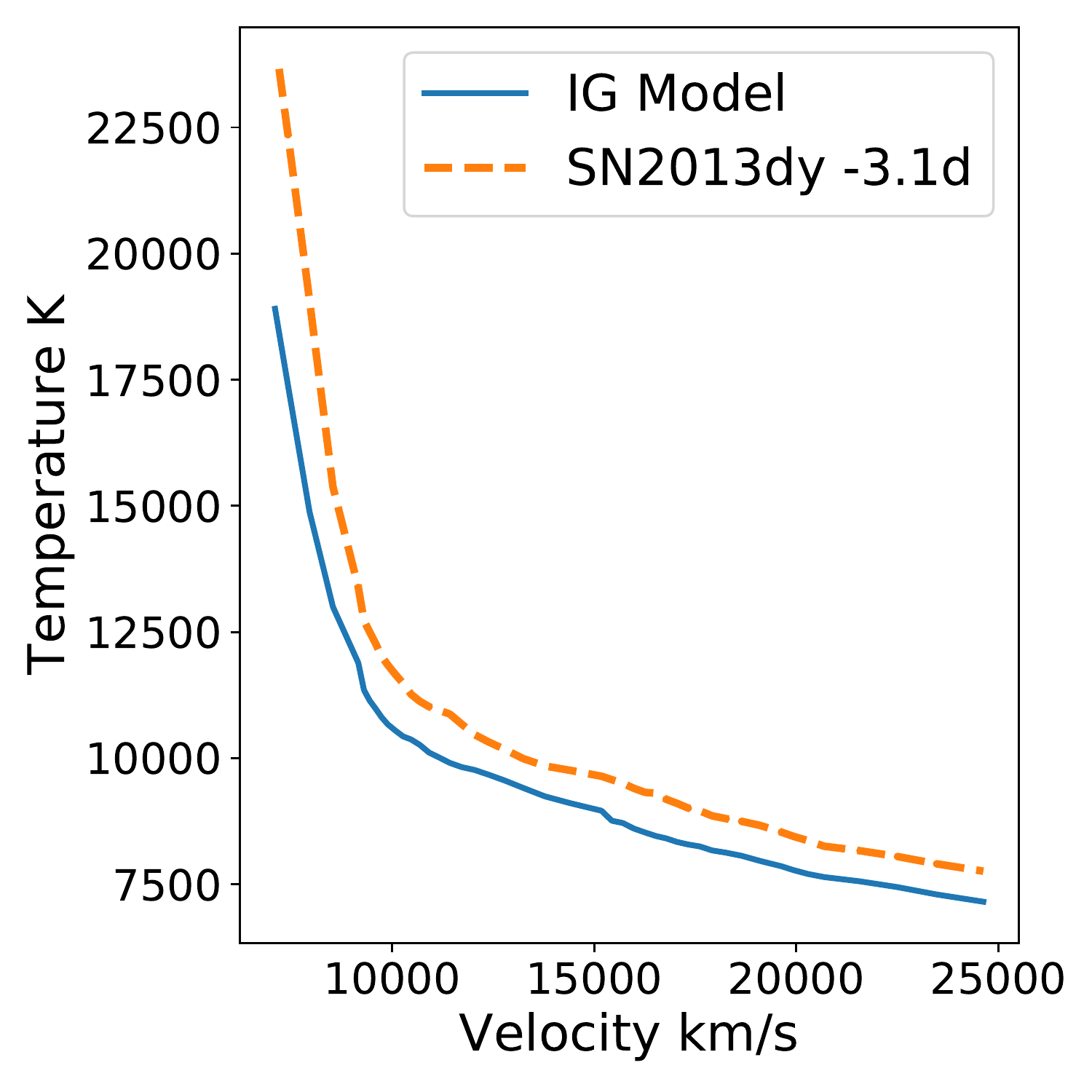}
    \includegraphics[width=0.33\linewidth]{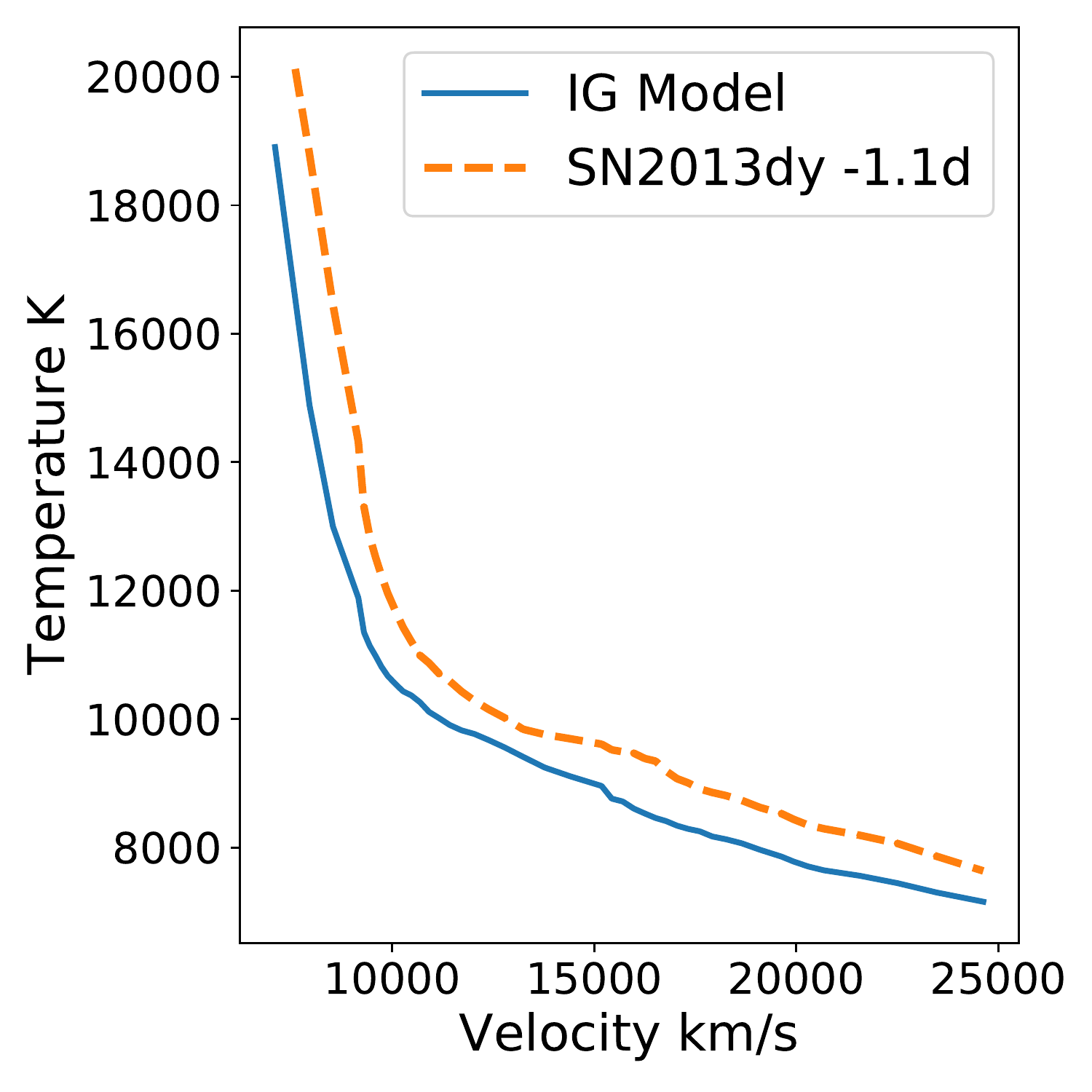}
    \includegraphics[width=0.33\linewidth]{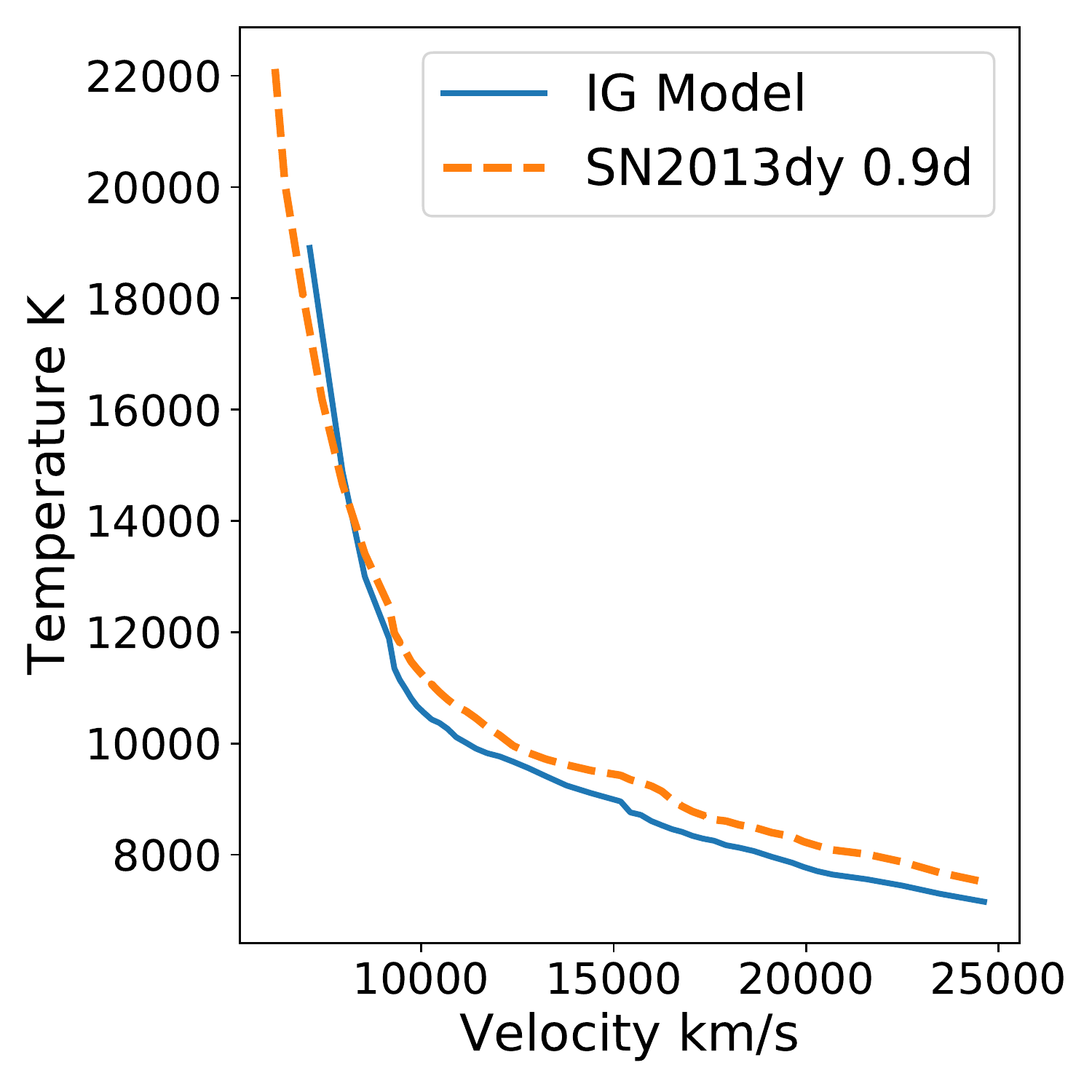}
    \includegraphics[width=0.33\linewidth]{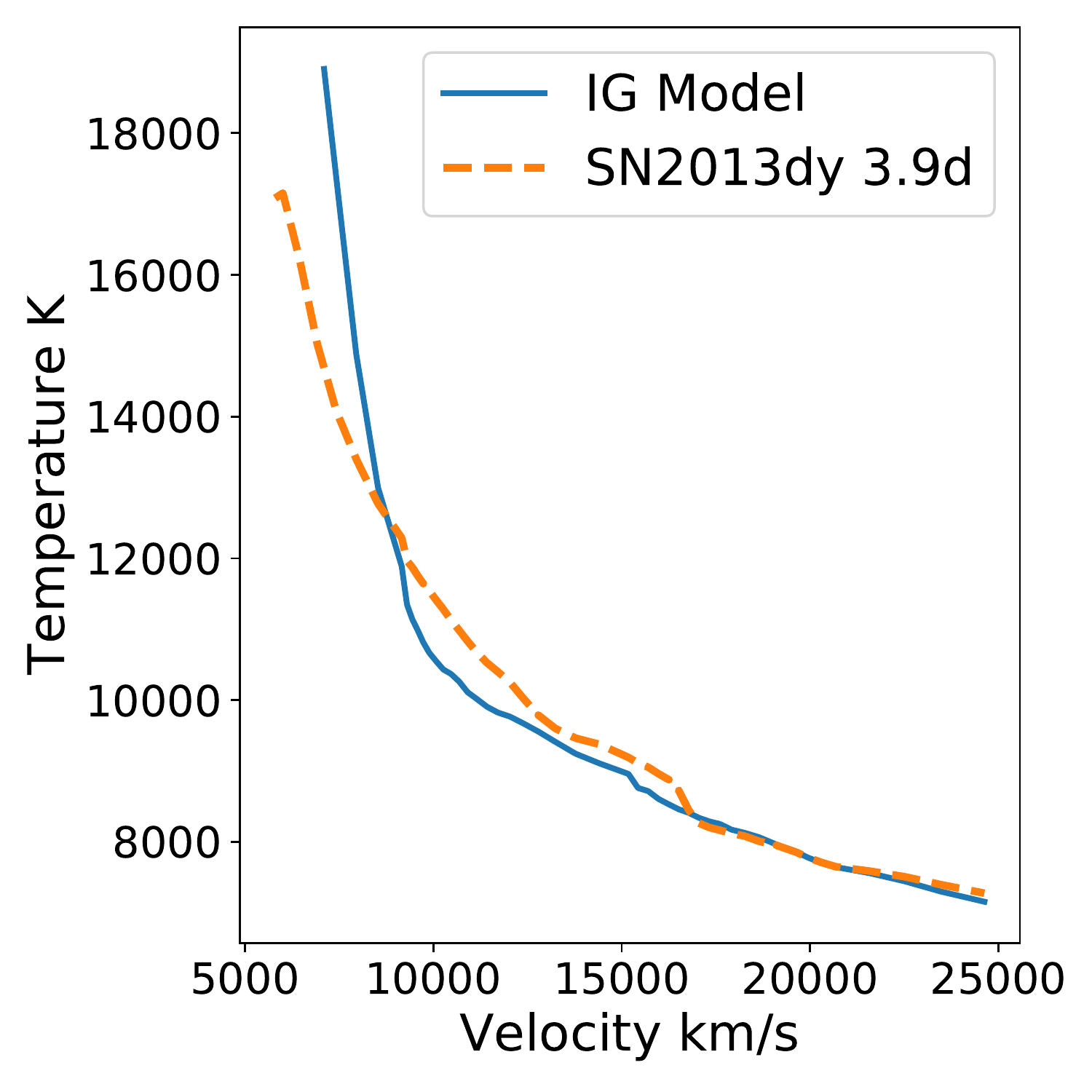}
    \caption{\textbf{(a) Top Left, (b) Top Center, (c) Top Right, and (d) Middle Left:} Elemental abundance structure of SN~2013dy predicted by neural networks from the spectra at $-3.1$, $-1.1$,  $0.9$, and $3.9$ days after $B$-maximum. \textbf{(e) Middle Center} The density structure of SN2013dy predicted from the spectra of $-3.1$, $-1.1$,   $0.9$, and $3.9$ days. The density profiles are converted to the profile of Day 19 using the $\rho\ \propto\ t^{-3}$ relation. \textbf{(f) Middle Right, (g) Lower Left, (h) Lower Center, and (g) Lower Right:}  The temperature structure for  SN2013dy spectrum of day $-3.1$, $-1.1$,  $0.9$, and $3.9$, respectively.  }
\end{figure}

The model to SN~2013dy again show consistently higher density at velocities above 17,500 km/sec and higher temperature before optical maximum than that of the IGM. The \ion{O}{1} feature at 7500 \AA\ is again poorly fit. It implies that the oxygen abundances are too low for the spectra in the spectral library. 

\subsection{Application to SNe with Wavelength Coverage $3000\ - \ 5200$ \AA}

\subsubsection{HST Spectra of 15 PTF Targets}
%In addition, \cite{HSTspec} published 15 more HST UV spectra. The wavelength coverage goes from 3,000 - 5,200 \AA. 

\citet{HSTspec} presented 32 low-redshift ($0.001\ <\ z\ <\ 0.08$) SNIa spectra; the UV spectra were obtained with the HST using the Space Telescope Imaging Spectrograph (STIS). The spectral wavelength coverage of these data is $3,000\ - \ 5,200$ \AA.  Meanwhile, the photometric data are obtained in Palomar Transient Factory \citep[PTF;][]{PTF,PTF2}. There are no published light curves on these targets yet. As our models are built only for data close to optical maximum, we selected only spectra taken between $-$3 and 4 days relative to the date of $B$-band maximum.

\subsubsection{Dust Extinction Correction}\label{sec:Extinction}

To model the observed supernova luminosity, we need to properly treat the dust reddening by the Milky Way and the host galaxies. 
%For the six selected SNe with comprehensive spectral observations from UV to IR, the dust extinction parameter E(B-V) are discussed in previous papers with various calibration methods, using information provided by the interstellar Na I D line, e.g., SN2011fe \citep{SN2011feNature}, and supernova or host-galaxy photometry, e.g., SN2015F \citep{SN2015F}. 
%Accordingly, we adopted the extinction data presented in the literature. 

\begin{figure}[htb!]
    \includegraphics[width=\linewidth]{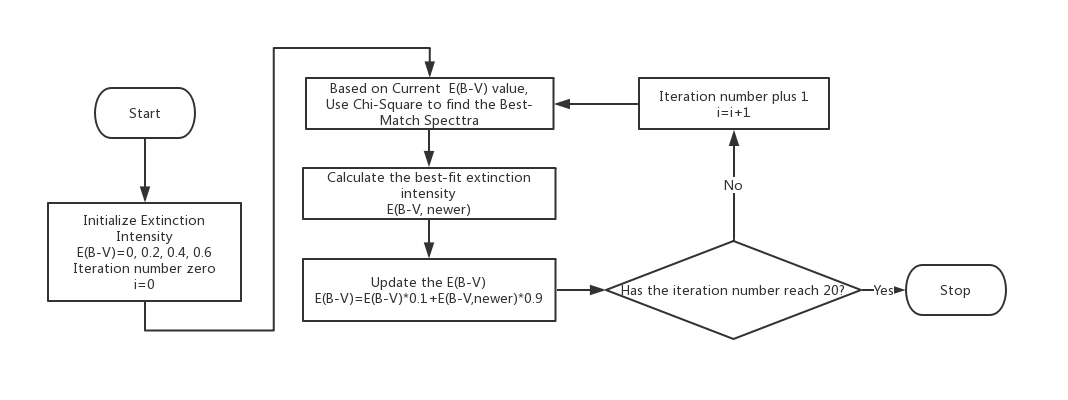}
    \caption{An illustrative flow chart of the algorithm for dust reddening  estimates (see \S\ref{sec:Extinction} for details). }\label{fig:ExtFlow}
\end{figure}

The extinction data of the 15 HST spectra in \citet{HSTspec} are not available. 
In order to correct the reddening effect, we developed an iterative algorithm by searching for the best match of theoretical models and observations with reddening as one of the $\chi^2$ minimization parameters, as is shown in the flow-chart in Figure~\ref{fig:ExtFlow}. The procedure starts with an initial guess of reddening index $E(B-V)$ which is applied to the observed spectrum, it then calculates the RMS of the difference between observed spectrum and all the model spectra in the spectral library to find the best-match spectral model, a new reddening index $ E(B-V)_{newer} $ is calculated assuming the best match spectral model is the true unreddened spectrum, the reddening index is updated using the formula $E(B-V)\ = \ 0.1 \times E(B-V)\ + \ E(B-V)_{newer} \times 0.9$, and the procedure repeats for 20 iterations to ensure convergence of the algorithm. 
%\textcolor{red}{In practice, we find that the estimates to $E(B-V)$ converge to within 0.01 mag in $3\sim 4$ steps.}

%The algorithm first select the spectra from our TARDIS synthesized spectra dataset which matches (how?) the observed spectral features the best \textcolor{blue}{by using Chi-square as the metric}, then use the linear-regression algorithm to calculate the possible extinction intensity denoted as E(B-V,new), which increase the matching fidelity of the selected synthesized spectra and the observed spectra. 
%By combining the estimated E(B-V,new) value and the previous value, we obtain the extinction intensity for the next iteration. 
%In order to acquire a converged result within 20 iterations, we choose the extinction intensity update strategy to be $E(B-V,i+1)=0.1E(B-V,i)+0.9E(B-V,new)$, where i denotes the iteration number and "new" denotes the direct estimated value from linear-regression. 
The algorithm was found to be initial-value sensitive, so we set the initial $E(B-V)$ to be 0, 0.2, 0.4, and 0.6 respectively. With different extinction values and the resulting best-fit spectra, we compare the spectral fitting fidelity and adopt the $E(B-V)$ value and spectral model with the least RMS of the difference between the observed and the model spectra reddened by the $E(B-V)$ value. 

We also tested our extinction correction algorithm on the 11 SN spectra with wavelength coverage of 2000 - 10000 \AA, and compare the derived extinction values with the extinction values in published literature. The results are shown in Table~\ref{tab:ExtinctionTest}. We found our algorithm can reproduce low and intermediate extinctions, while for high extinction case $E(B-V)\ \sim\ 0.3$, the algorithm seems to underestimate the extinction intensity. As it is difficult to assess the precision of the flux calibration of the observed spectra, we consider the values shown in Table~\ref{tab:ExtinctionTest} to be broadly in agreement. 

%So, we only use the algorithm on 3,000 - 5200 \AA spectra, while use the available extinction values for 2,000 - 10,000 \AA spectra. 
\begin{deluxetable}{cccc}[htb!]
    \tablecaption{The Extinction Comparison\label{tab:ExtinctionTest}}
    \tablehead{\colhead{SN name} & \colhead{Phase (days)} & \colhead{$E(B-V)$ From Papers} 
    & \colhead{$E(B-V)$ from Algorithm}   }
    \startdata
    SN2011fe    &  0.4 & 0.000     & 0.021 \\
    SN2011fe    & -2.6 & 0.000     & 0.000    \\
    SN2011fe    &  3.7 & 0.000     & 0.000     \\
    SN2013dy    & -3.1 & 0.341 & 0.222 \\
    SN2013dy    & -1.1 & 0.341 & 0.253 \\
    SN2013dy    &  0.9 & 0.341 & 0.231 \\
    SN2013dy    &  3.9 & 0.341 & 0.253 \\
    SN2011iv    &  0.6 & 0.000    & 0.092 \\
    SN2015F     & -2.3 & 0.210  & 0.129 \\
    ASASSN-14lp & -4.4 & 0.351 & 0.281 \\
    SN2011by    & -0.4 & 0.052 & 0.018 \\
    \enddata
\end{deluxetable}

%Using methods discussed in Section \ref{sec:Extinction}, we attempted to correct the effect of dust extinction by using CCM extinction model \citep{CCM} before inserting these spectra into the neural network. 

In Table \ref{tab:HSText}, we list the extinction values derived from this procedure for the 15 SNe under study. 

\begin{deluxetable}{ccccc}[htb!]
    \tablecaption{The HST observed type Ia SNe from \citet{HSTspec}\label{tab:HSText}}
    \tablehead{\colhead{SN name} & \colhead{Phase (days)} & \colhead{Redshift\tablenotemark{a}} 
    & \colhead{Stretch}  & \colhead{$E(B-V)$} }
    \startdata
    PTF09dlc & 2.8  & 0.068    & $ 1.05\pm 0.03 $  & 0.21 \\
    PTF09dnl & 1.3  & 0.019    & $ 1.05\pm 0.02 $  & 0.04 \\
    PTF09fox & 2.6  & 0.0718   & $ 0.92\pm 0.04 $  & 0.08 \\
    PTF09foz & 2.8  & 0.05     & $ 0.87\pm 0.06 $  & 0.15 \\
    PTF10bjs & 1.9  & 0.0296   & $ 1.08\pm 0.02 $  & 0.02 \\
    PTF10hdv & 3.3  & 0.054    & $ 1.05\pm 0.07 $  & 0.16 \\
    PTF10hmv & 2.5  & 0.032    & $ 1.09\pm 0.01 $  & 0.02 \\
    PTF10icb & 0.8  & 0.086    & $ 0.99\pm 0.03 $  & 0.13 \\
    PTF10mwb & -0.4 & 0.03     & $ 0.94\pm 0.03 $  & 0.02 \\
    PTF10pdf & 2.2  & 0.0757   & $ 1.23\pm 0.03 $  & 0.23 \\
    PTF10qjq & 3.5  & 0.0289   & $ 0.96\pm 0.02 $  & 0.14 \\
    PTF10tce & 3.5  & 0.041    & $ 1.07\pm 0.02 $  & 0.23 \\
    PTF10ufj & 2.7  & 0.07     & $ 0.95\pm 0.02 $  & 0.15 \\
    PTF10xyt & 3.2  & 0.049    & $ 1.07\pm 0.04 $  & 0.24 \\
    SN2009le & 0.3  & 0.017786 & $ 1.08\pm 0.01 $  & 0.17 \\
    \enddata
    \tablenotetext{a}{In \citet{HSTspec}, only cosmological redshifts are given, the collected redshift data here are adopted from WISeREP. }
\end{deluxetable}

The AIAI results, showing the spectral profiles, density structures, chemical structures, and the temperature profiles are shown in Appendix~\ref{sec:hstfitting}.

\section{Discussions and Conclusions}\label{sec:Results}

%\subsection{The Density Profiles}
%Noticeably, several SNe show a density kink at velocities around 13,000 km/sec. This is obviously an important issue as a density kink at such high velocity may be an important discriminator of explosion models of SNIa. The published DD models do not produce such a density kink. It was argued based on physical models that the density profile of delayed-detonation models may be quite robust, and is insensitive the the details of the DD transition \citep{Hoeflich:2017ApJ...846...58H}. The kink can be the results of ejecta-CSM interaction, or alternatively, it may agree with scenarios involving a detonation on the surface before the detonation at the center of the progenitor white dwarf \cite[e.g.,][]{Shen:2018}. However, the total mass of each velocity zone still suffers large errors as some elements such as oxygen and silicon are still poorly trained in the current version of the neural network. It may also be the results of the artificial division of the ejecta into four distinct zones. For the current study, we consider the reality of density kinks to be undetermined, and leave further scrutiny of the density profile to future studies with models using finer velocity zones. 

\subsection{Elemental Abundances and the Light Curve Stretch}

The stretch values are known to be correlated to the SN luminosity \citep{perlmutter1999measurements,Phillips:1993ApJ...413L.105P}. To compare the elemental abundance derived from the best match TARDIS spectra with light curve stretch parameters, we employed the {\tt\string SiFTO} \citep{sifto} program to calculate the stretch factors. The resulting stretch values are listed in Table \ref{tab:Stretch}. 

We may consider the theoretical models as a toolbox to construct empirically parameters to improve the precision of SNIa as standard candles, in a way similar to the light curve shape parameters. This would be particularly interesting for projects based on spectrophotometry, such as has been planned for WFIRST.   
A few more clarifications of the uncertainties of the derived chemical abundances are necessary before carrying out such a study. Based on the testing dataset discussed in Section~\ref{sec:DeepLearning}, we can estimate the $1-\sigma$ limits of the elemental abundances. Not all the predictions from the neural network are reliable due to the sparsity of our parameter space coverage in generating the dataset and the limitations of the sensitivity of the neural network. For example, some of the predictions of Co in Zone 3 are not in the region where the testing dataset has sufficient coverage, as are shown in the upper right panel of Figure~\ref{fig:ResidualSelecter}(b). 
When this situation occurs, we replace the predictions that go above or below the relevant $1-\sigma$ region with an upper limit or lower limit, respectively. Also, for simplicity, we have set the lower limits of the ejecta structure to be 7100 km/s when calculating the abundances of Zone~1 to avoid the effect introduced by a varying photospheric velocity. 

\begin{figure}[htb!]
 %   \minipage{0.7\textwidth}
        \includegraphics[width=0.5\textwidth]{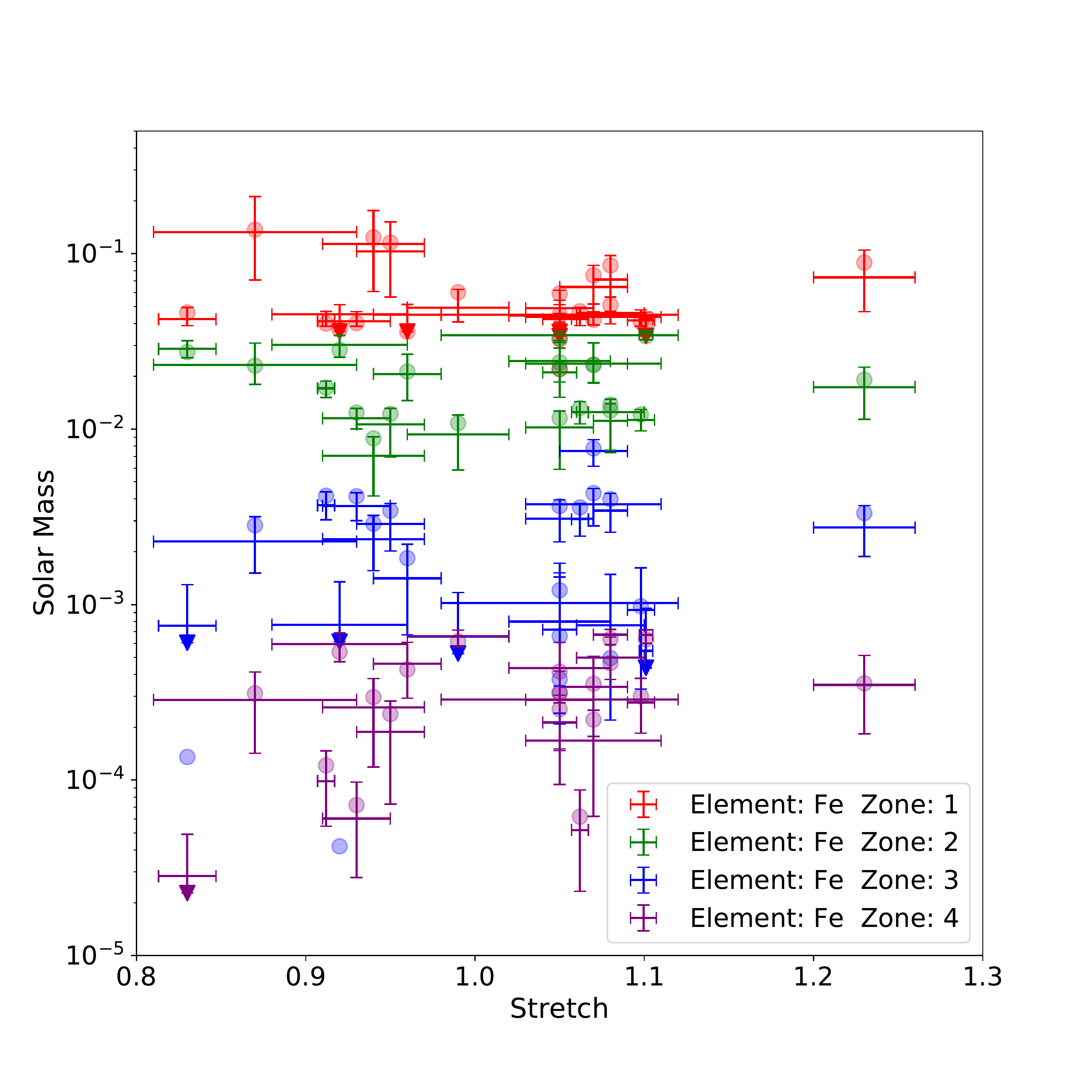}
 %   \endminipage%\hfill
%    \minipage{0.7\textwidth}
        \includegraphics[width=0.5\textwidth]{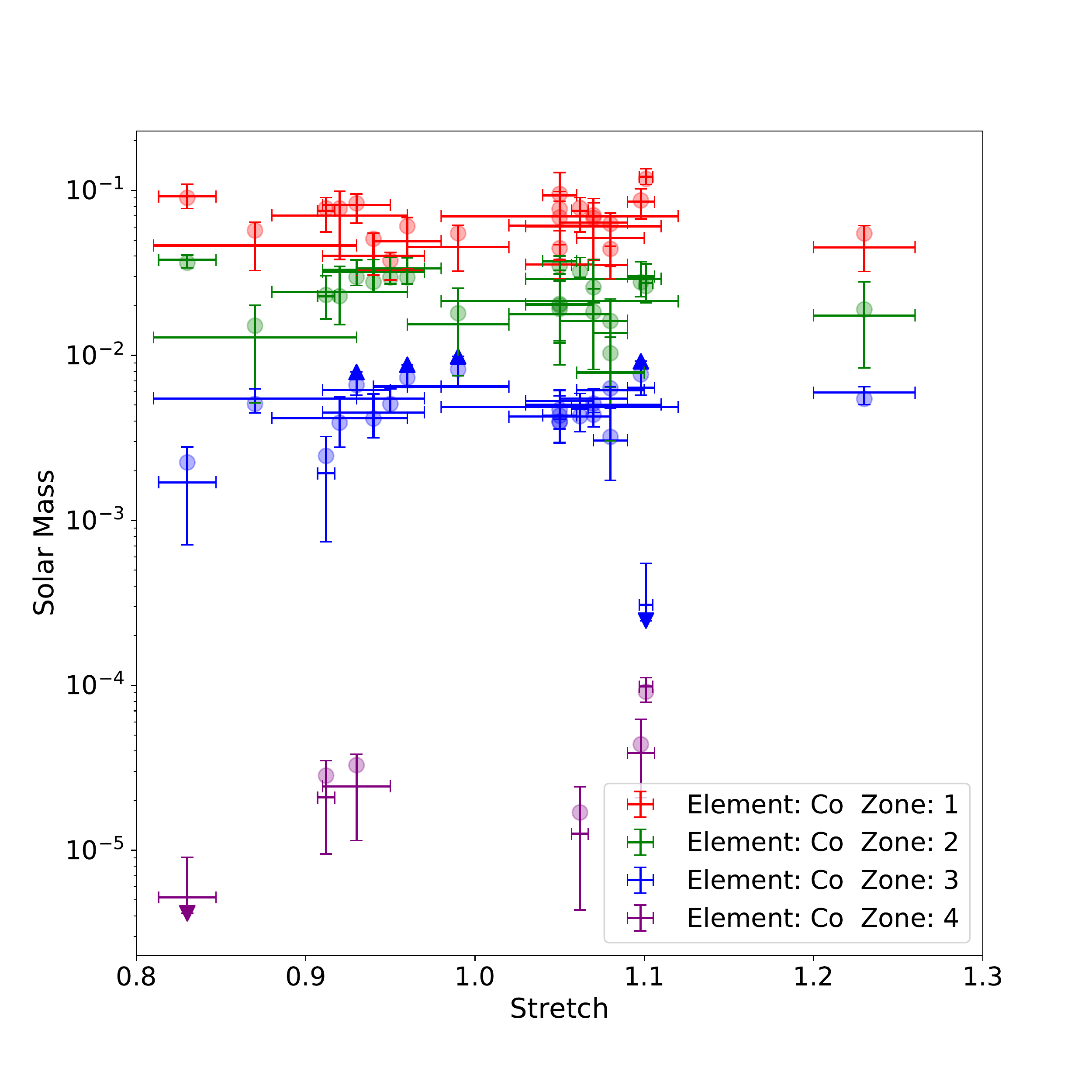}
%    \endminipage%\hfill
 %   \minipage{0.7\textwidth}
        \includegraphics[width=0.5\textwidth]{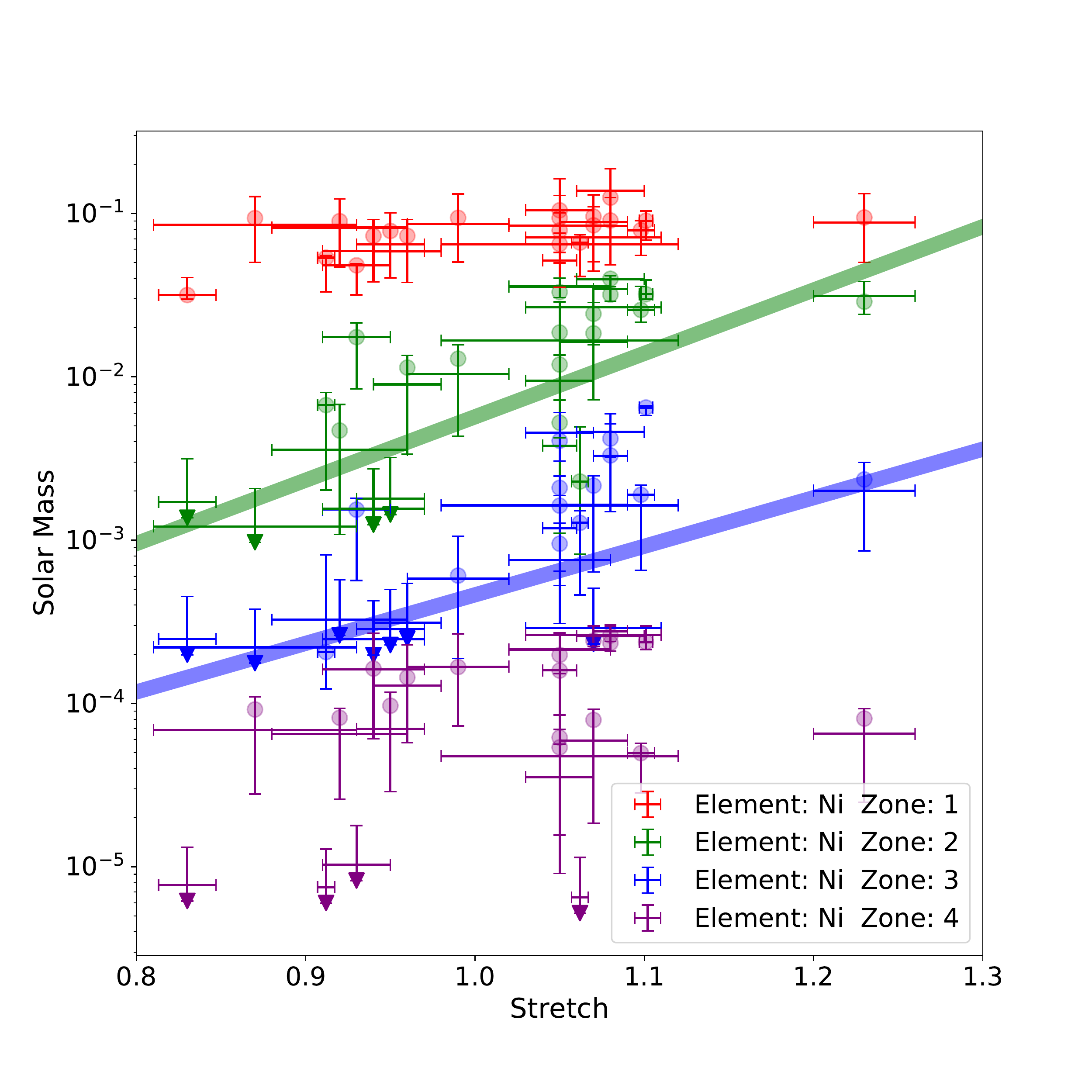}
%    \endminipage%\hfill
    \caption{The correlation between the stretch parameter and the mass of various chemical elements. The lower limits and the upper limits are marked with triangles, the median values are the crosses and the predicted values are transparent circular dots. For simplicity, we choose the lower limit of the ejecta structures to be 7100km/s when calculating the element masses. \textbf{(a) Upper Left:} The correlation between the stretch parameter and Fe mass in Zones 1, 2, 3, and 4. \textbf{(b) Upper Right:} The correlation between the stretch parameter and Co mass in Zones 1, 2, 3, and 4. \textbf{(c)  Lower Left:} The correlation between stretch and Ni mass in Zones 1, 2, 3, and 4. For the neural network WR-Blue (Wavelength Range 3000-5200 \AA), the Co mass in Zone 4 can not be determined so we did not predict the Co mass in zone 4 for the 15 spectra with such wavelength coverage. Notice that the Ni mass are corrected by its radioactive decay, and the predictions are for the $Ni$ mass at the $B-$band maximum rather than the date of observations. We fitted the Ni mass in Zone 2 and Zone 3 with $M_{Ni,2}=7.511\times10^{-5} e^{8.933\times Stretch}M_{\sun}$ and $M_{Ni,3}=4.934\times10^{-7} e^{6.843\times Stretch}M_{\sun}$ respectively. }\label{fig:MD15}
\end{figure}

%The relation between elemental abundances and intrinsic luminosity of the supernovae is among the most interesting ones to explore. 
Taking the elemental masses derived from AIAI at their face values, we show in Figure~\ref{fig:MD15} the correlations between the stretch parameters and Fe, Co, and Ni masses all corrected to the values at 19 days after explosion. 
The correlations for the remaining 22 elements in various zones are presented in Appendix \ref{sec:appendix:MD15}. 

Although we intend to leave a thorough analysis of these correlations to an upcoming paper, we identify immediately from Figure~\ref{fig:MD15} that the Fe, Co and Ni masses in Zone 4  to be more strongly correlated to the stretch parameter than other parameters. Figure~\ref{fig:MD15}~(c) shows also that the mass of Ni in Zones 2, 3, and 4 are all correlated to the stretch parameter. The mass derived from Zone 1 may not be reliable as the photosphere is located in Zone 1. The correlations that can be identified for Fe and Co in Zone 4, and Ni in Zones 2, 3, 4 are suggestive of the presence of radioactive materials at the surface of the supernova ejecta, and are likely critical measures of the luminosities of SNIa. 

It is noteworthy that the Ni mass in Zones 2, 3, and 4 varies by more than one order of magnitude. The masses of Fe and Co do not appear to share such a behavior. No strong correlation is found between the mass of Fe and Co and the stretch factor either. This may be explained if a fraction of Fe and Co are non-radioactive and a dominant  fraction of Ni in these Zones are radioactive. Indeed, 1-D models of thermonuclear explosions predict the existence of a high-density electron capture burning region during the deflagration phase \citep{Hoeflich:1996ApJ...472L..81H,Gerardy_2007} which can lead to the production of a significant amount of non-radioactive Ni and Fe, but little Co. These early deflagration products can be  mixed out to higher velocities layers such as shown in some 3-D models \citep[e.g.,][]{Gamezo:2004,Roepke:2006,GCD,JordanGCD2008ApJ...681.1448J}, they are likely to distort and weaken the correlation between masses of Co and Fe and the light curve shapes of the supernovae. 

\subsection{Time Evolution of Elemental Abundances in SN2011fe and SN2013dy}

Among all six SNe with spectroscopic observations near  $B$-maximum and wavelength coverage over 2000-10000 \AA, there are 3 spectra of SN~2011fe and 4 spectra of SN~2013dy that are compatible with our neural network setups. 
In this section, we investigate the time evolution of the elemental abundances of this two SNe. The time evolution of the masses of radioactive materials can serve as an important check on the fidelity of the elemental abundances we deduce from the neural network whose performance is very difficult to track precisely from first principle mathematical models.

\begin{figure}[htb!]
    \minipage{0.33\textwidth}
        \includegraphics[width=\textwidth]{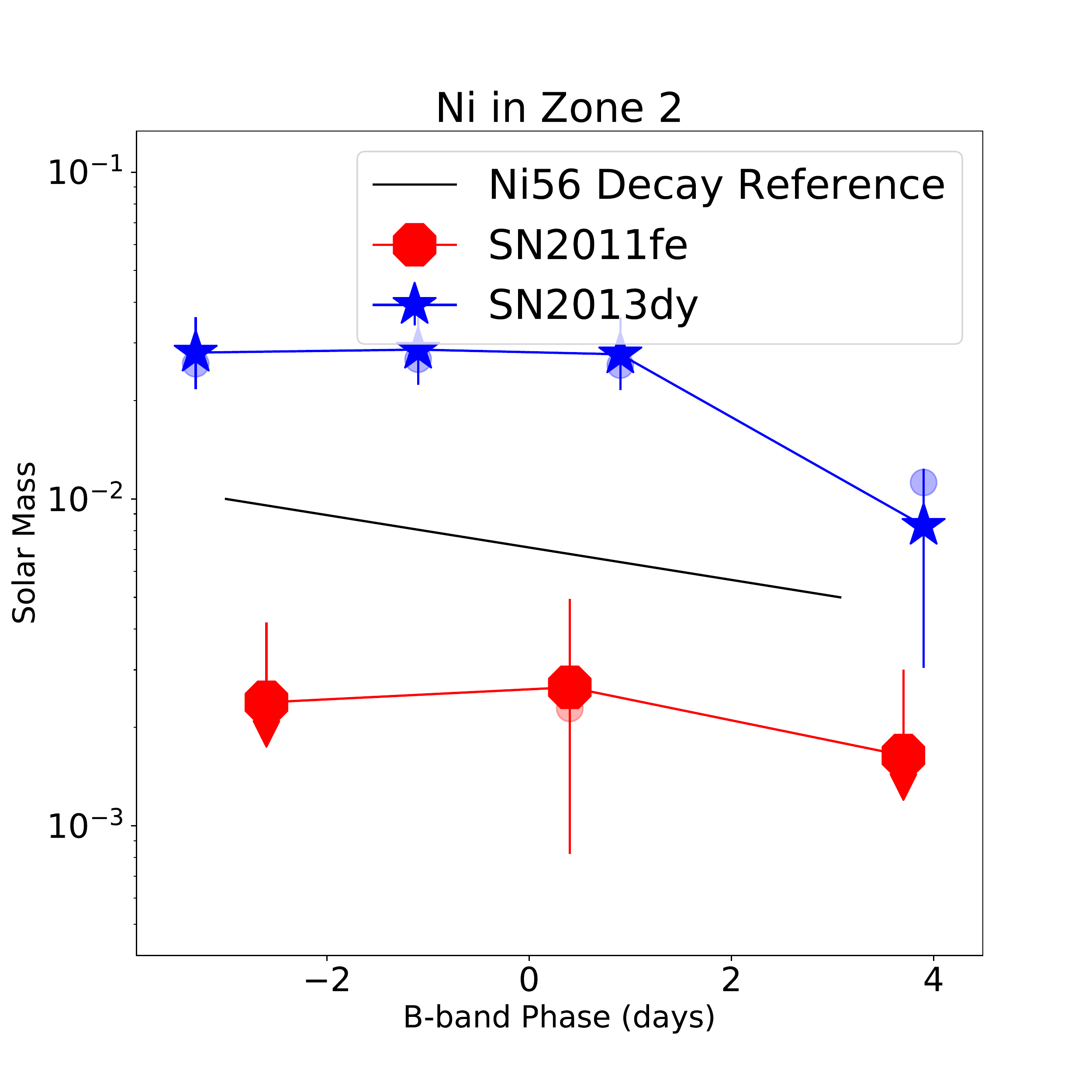}
    \endminipage\hfill
    \minipage{0.33\textwidth}
        \includegraphics[width=\textwidth]{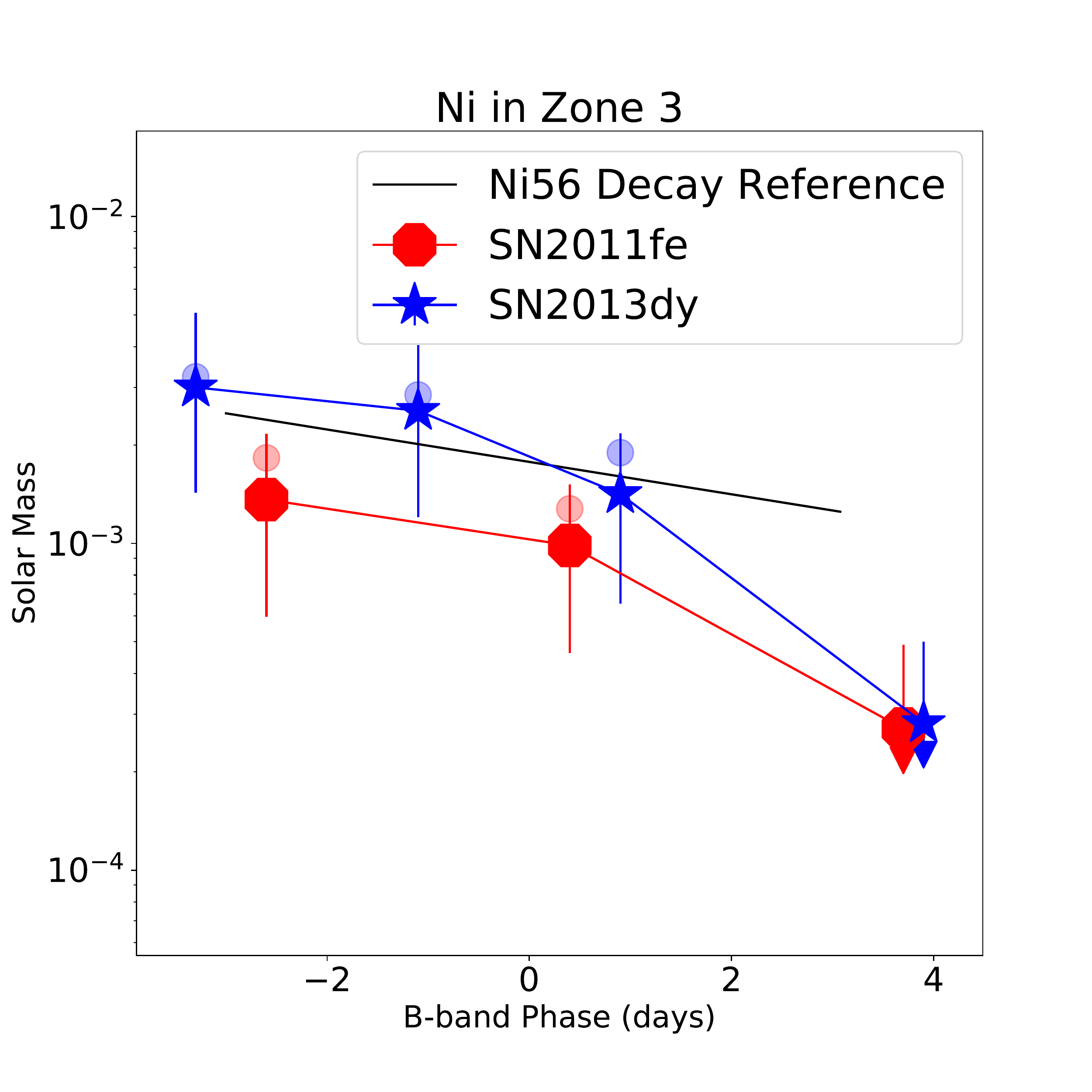}
    \endminipage\hfill
    \minipage{0.33\textwidth}
        \includegraphics[width=\textwidth]{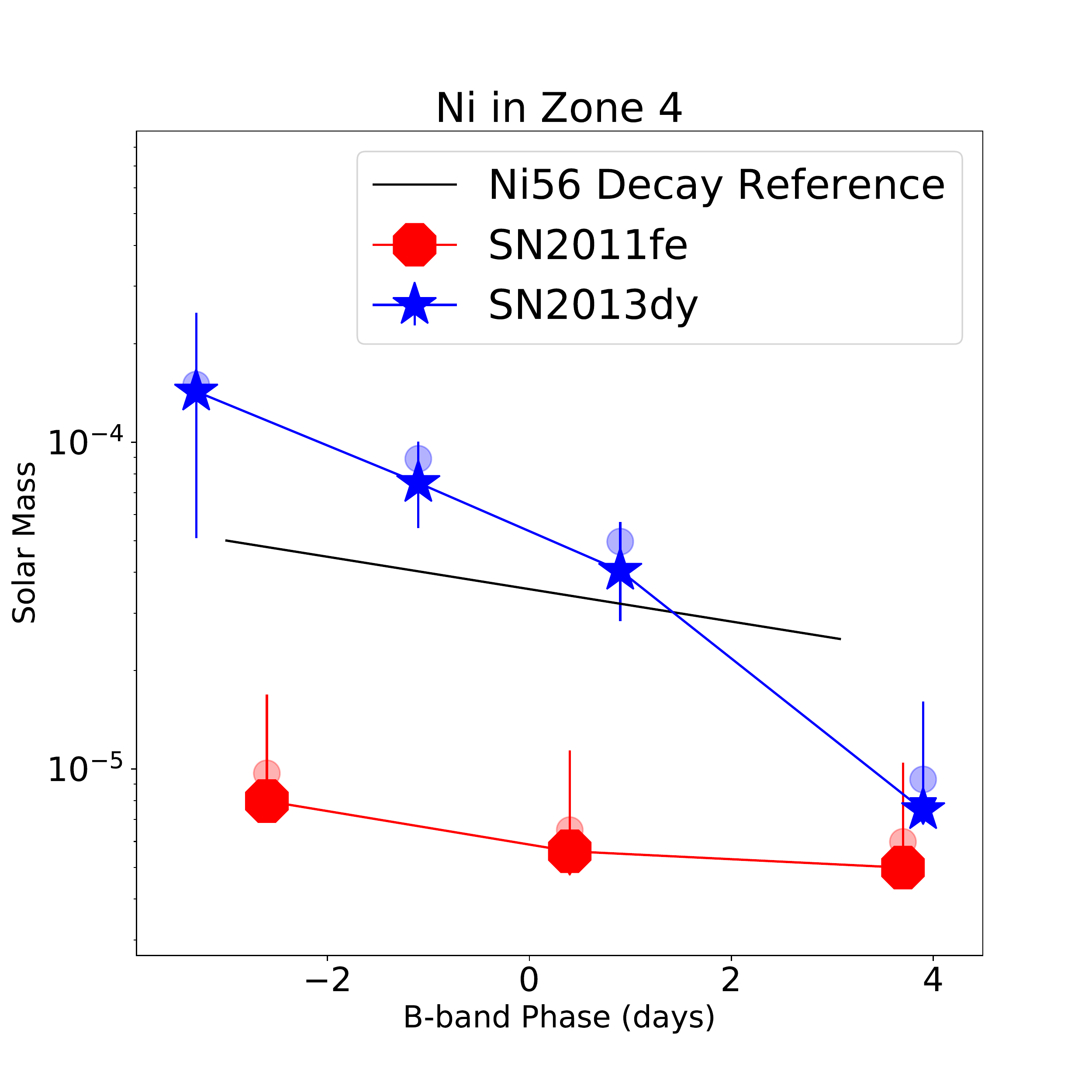}
    \endminipage\hfill
    \minipage{0.33\textwidth}
        \includegraphics[width=\textwidth]{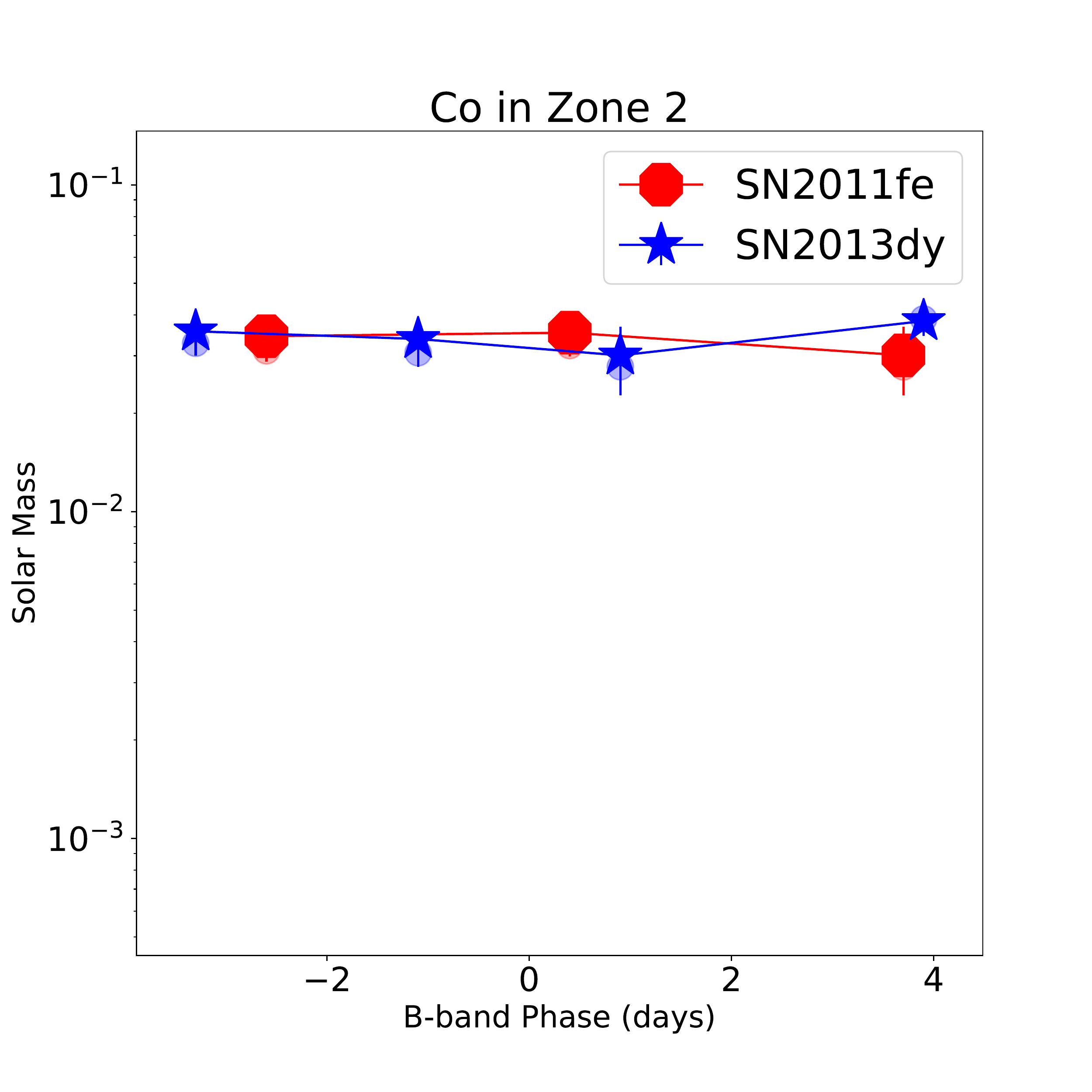}
    \endminipage\hfill
    \minipage{0.33\textwidth}
        \includegraphics[width=\textwidth]{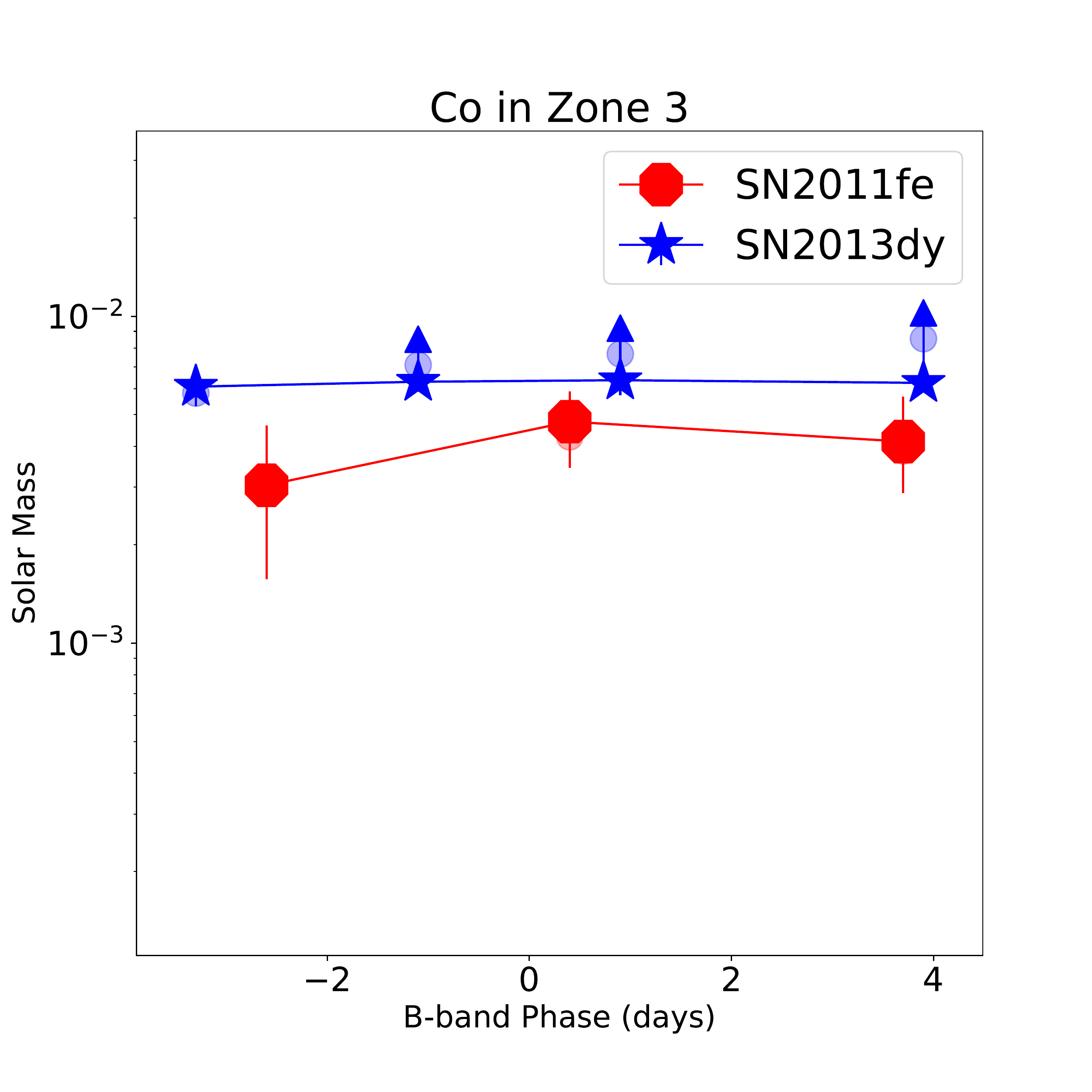}
    \endminipage\hfill
    \minipage{0.33\textwidth}
        \includegraphics[width=\textwidth]{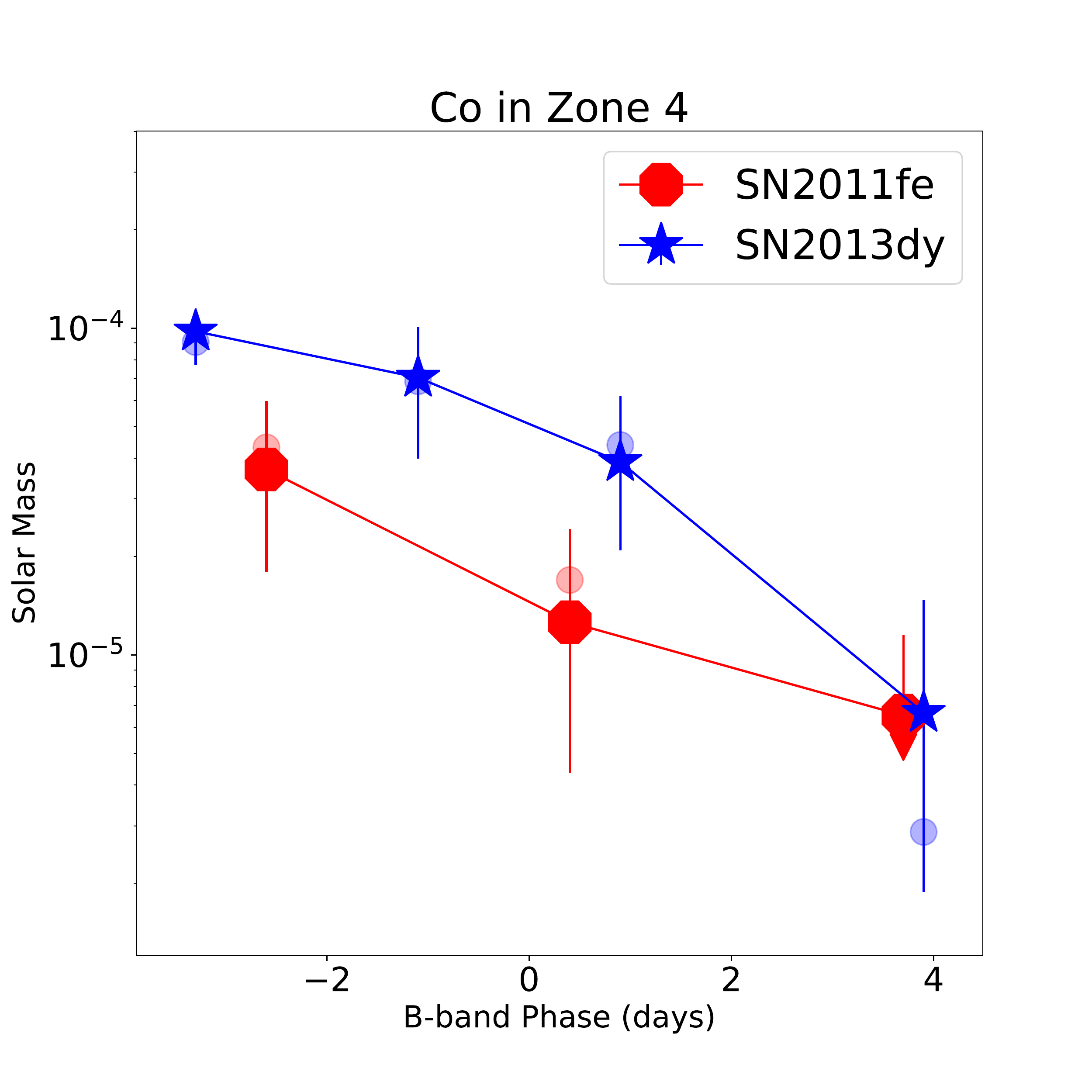}
    \endminipage\hfill
    \minipage{0.33\textwidth}
        \includegraphics[width=\textwidth]{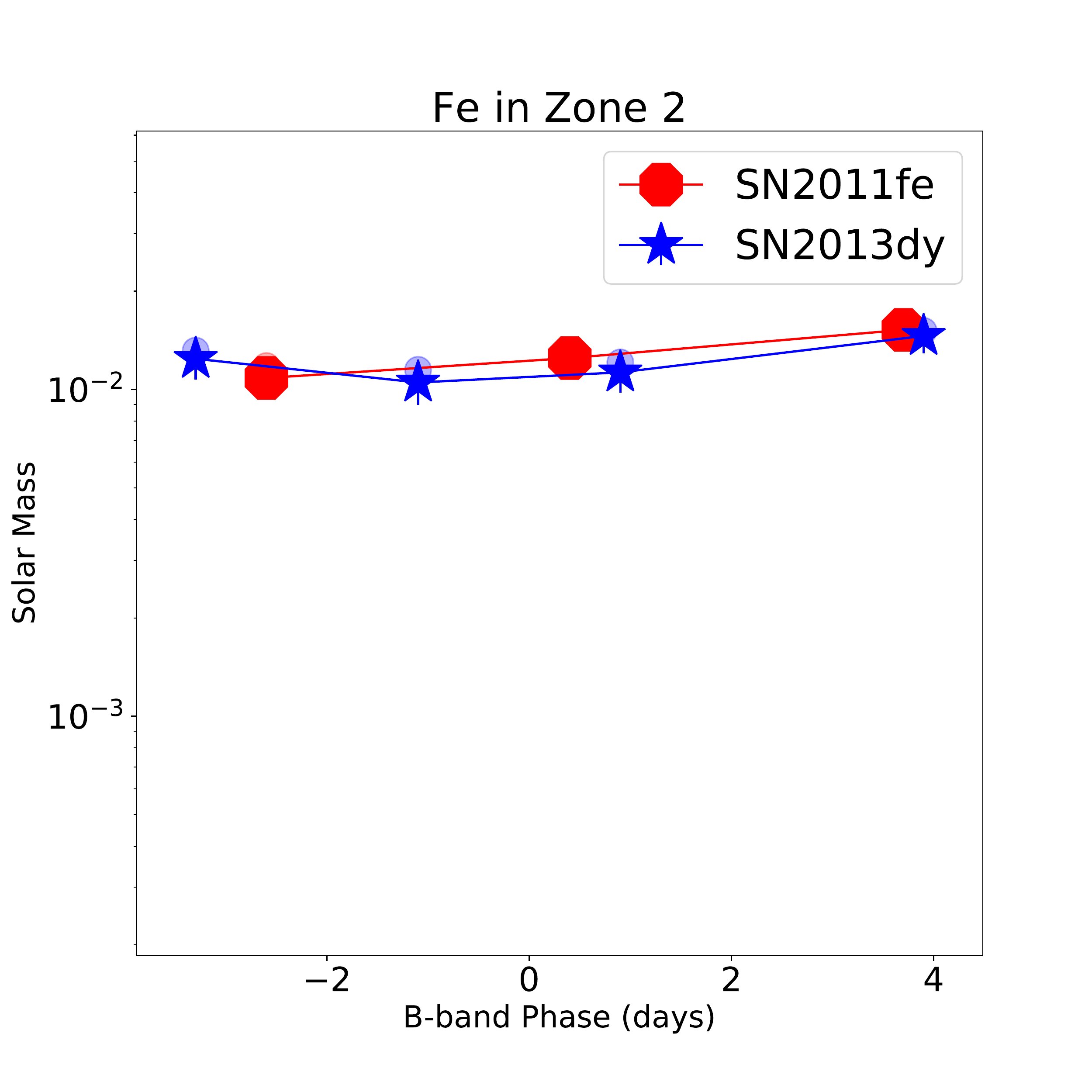}
    \endminipage\hfill
    \minipage{0.33\textwidth}
        \includegraphics[width=\textwidth]{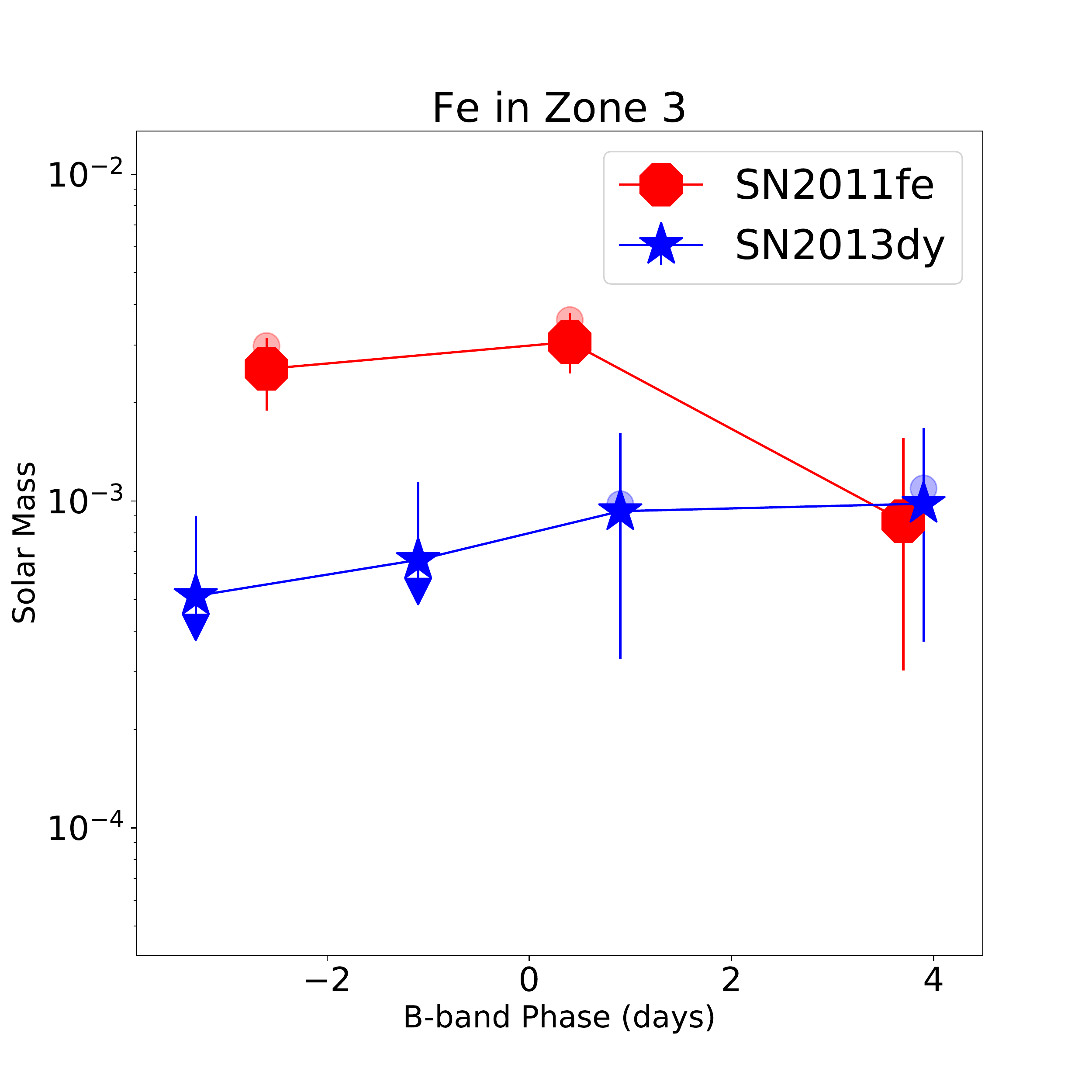}
    \endminipage\hfill
    \minipage{0.33\textwidth}
        \includegraphics[width=\textwidth]{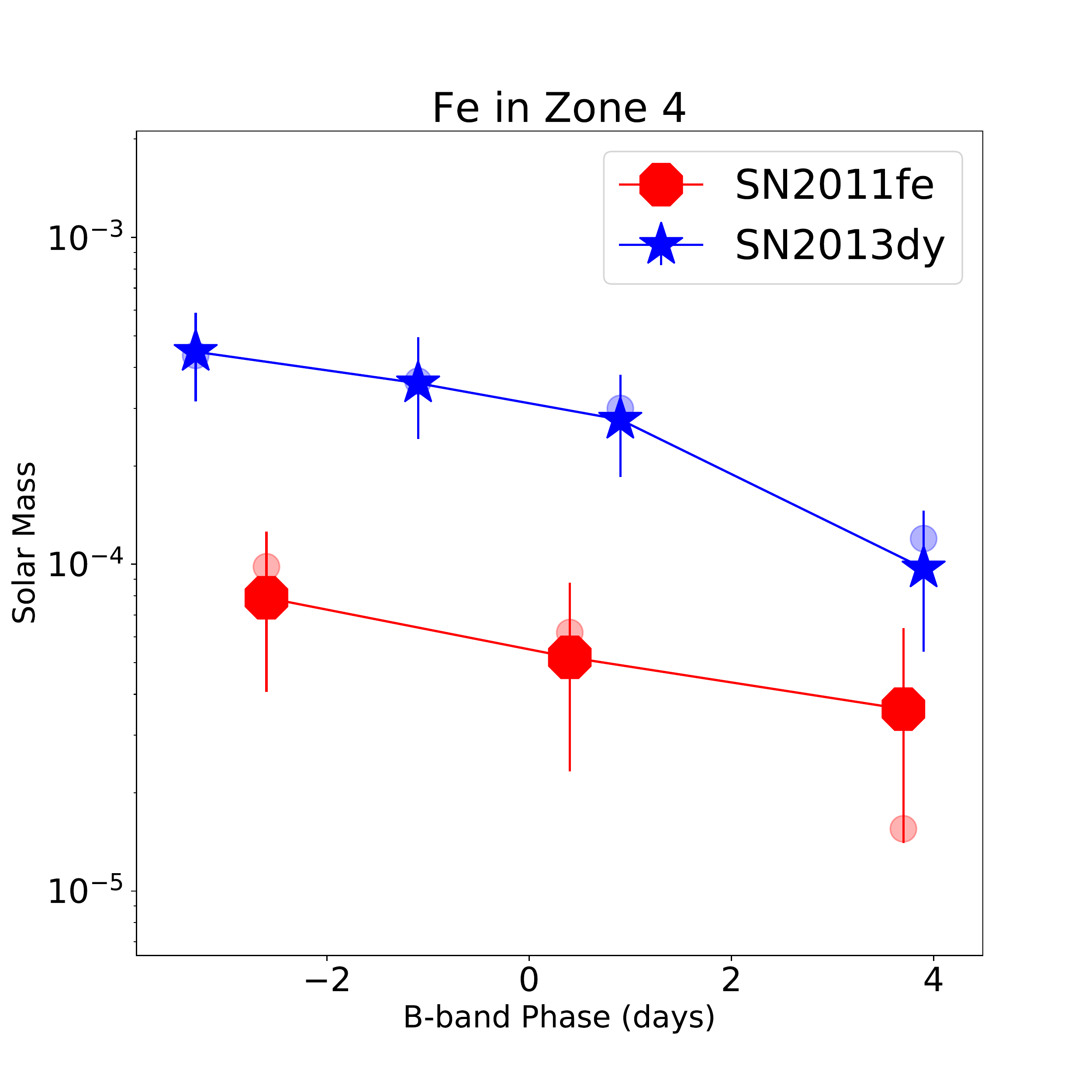}
    \endminipage\hfill
    \caption{The time evolution of Ni mass in different zones of SN2011fe and SN2013dy near $B$-band maximum dates. \textbf{Upper Panel:} From left to right: Ni mass in Zones 2 (a), 3 (b), and 4 (c). The decay rate of Ni$_{56}$ is plotted in black lines for comparison. \textbf{Middle Panel:} From left to right: Co mass in Zones 2 (d), 3 (e), and 4 (f). \textbf{Lower Panel:} From left to right: Fe mass in Zones 2 (g), 3 (h), and 4 (i). }\label{fig:TimeEvol}
\end{figure}

From Figure~\ref{fig:TimeEvol} (a), (b), and (c), we notice a general agreement between the time evolution of the Ni mass in Zone 2, 3, and 4 , respectively, and the radioactive decay rate of $^{56}Ni$. This agreement suggests that most of the $^{56}Ni$ are newly synthesized after the explosion. SN~2013dy shows a significantly larger Nickel mass than SN~2011fe. A significant difference in Ni mass is found for the two SNe in Zones 2, 3, and 4, or, at velocity above 10,000 km/sec. If these difference is true, it may provide an explanation of the difference in the luminosity of the two SNe \cite{SN2013dy}. 

Figure~\ref{fig:TimeEvol} (d), (e), and (f) show the masses of Co in Zone 2, 3, and 4, respectively. Again, SN~2013by is more abundant in Co in Zone 3 and 4 than SN~2011fe. Notice that the mass of Co peaks at around 20 days past explosion if its time variation is related to the radioactive decays of $^{56}Ni$; Figure~\ref{fig:TimeEvol} (d) and (e) are in agreement with this. 
Likewise, more iron is found at the highest velocity (Zone 4) in SN~2013by than in SN~2011fe. Iron seems to be less abundant in SN~2013by than in SN~2011fe in Zone 3.

Moreover, the predicted Ni masses from the spectra 3-4 days past maximum are significantly lower than earlier epochs.  
We surmise such anomaly could due to temperature change which strongly affects the UV spectral features.
When Fe-group element are in low-temperature region, like Zone 3 or 4, the relevant atomic levels are less activated, which result in weaker or the absence of spectral lines in the TARDIS model spectra due to Monte Carlo noise, which may misguide the neural network. 

Curiously enough, the nebular spectra of SN~2013by shows a smaller \ion{Ni}{2}$\lambda$7378\AA/\ion{Fe}{2}$\lambda$7155\AA\ flux ratio than that of SN 2011fe \citep{SN2013dy500}, which may suggest that SN~2013dy produced a lower mass ratio of stable to radioactive iron-group elements than SN~2011fe.
The decay of $^{56}Ni$ leads to an comparatively over abundance of Fe in SN~2013by as compared to that in SN~2011fe.
This is in agreement with what we found in Zones 2, 3, and 4, although the nebular lines measure ejecta at much lower velocities. The primary difference between SN~2011fe and SN~2013by that affects their early UV spectra and likely also their luminosities is the amount of radioactive Ni at velocities above $\sim$~10000 km/sec. 

\citet{SN2013dy} noticed that SN~2013dy is dimmer in the near IR than SN~2011fe. This could also be suggestive of a difference in  chemical structures of the two. It may be related to the excessive amount of radioactive material at the outer layers of SN~2013dy. %which keeps the ejecta at a higher temperature than in SN~2011fe such that its fine structure lines are not excited in the IR. 
The difference in Fe abundance at the highest velocity may also be an indication of a genuine difference of the chemical abundance of the progenitors of the two supernovae if they are primordial to the progenitor. Alternatively, they may also suggest different explosion mechanisms if they are produced during the explosion. Note further that SN~2013dy was discovered to show very strong unburned  C{\sc II} lines before maximum, in contrast to the weak C{\sc II} features of SN~2011fe. It sits on the border of the “normal velocity” SNe Ia and 91T/99aa-like events \citep{SN2013dy} while SN~2011fe is no normal SNIa of normal velocity \citep{WangX:2009,wangX:2013}. Mechanisms such as double detonation may enrich the high velocity ejecta with Fe \citep{WangHan:2012NewAR..56..122W}.

In Appendix \ref{sec:appendix:TimeEvol}, we show the time   evolution of other chemical elements derived from the MRNN. 

\subsection{The Absolute Luminosity}\label{sec:AbsoluteLumi}

%TARDIS allows calculations of absolute fluxes despite physical approximations made in the code that may severely affect the results. The lack of time dependence, a realistic $\gamma$-ray deposition, and the approximation of a sharp photosphere are all sources of concerns. However,  train neural network models with predicting power on the absolute magnitudes based on normalized flux spectra. 
%by integrating the spectra with a Bessel-B band transparency function before all the TARDIS-Synthesized spectra in the training and testing data set are normalized. 
The neural network with the same structure as in Section \ref{sec:DeepLearning} may be constructed to retrieve the luminosity assumed in the synthetic spectral models. 
This opens the possibility of predicting the absolute luminosity of an SNIa based on its spectral shapes. This can be achieved by training the AIAI with predicting power on the absolute magnitude of any filter bands, here chosen to be the Bessel $B$-band.

Unlike elemental abundance predictors, the training on $B-$band luminosities are prone to systematic errors due to TARDIS model limitations. The luminosity is sensitive to both the location of photosphere and the temperature at the photosphere. In reality the location of the photosphere is sensitive to wavelength whereas TARDIS treats the photosphere as wavelength independent. The temperature is sensitive to the UV fluxes and the location of the photosphere is sensitive to the absorption minimum of weak lines. At around optical maximum, the photosphere has receded into the iron rich layers for normal SNIa and the absorption minima of intermediate mass elements are no longer good indicators of the location of the photosphere. Such insensitivity may introduce degeneracy in the dependence of luminosity and spectral profiles: The luminosity of two supernovae may be very different whereas the spectral profiles in the optical may appear to be very similar.

Nonetheless, one may try to study the relations between the luminosities and spectral shapes, in a similar way to what have been done for elemental abundances. However, when the neural networks are trained with multiple iterations, the MSE on the training set decreases while the MSE on the testing set increases. This indicates that the neural network overfits features in the training set, and the results cannot be used for model predictions.
Consequently, we curtailed the performance tolerance to 1 iteration only and adopted a smaller learning rate, which is one-tenth of the value used for training the elemental abundances ($3\times 10^{-6}$ in the first stage and $3\times 10^{-8}$ in the second stage). %\textcolor{red}{what is the number?} 
%After several attempts, we choose the best neural network which has a MSE on the testing data set 0.00043. 

Having done the training and testing, we insert the observed spectra into the trained neural network.
%after the de-redshift, dust-extinction correction and the normalization.
The luminosities of the 11 spectra of the 6 SNe with HST spectra were predicted by the neural network; the results are listed in Table \ref{tab:Bband}. 

We have also estimated the $B$-band maximum luminosities of the  6 SNe using their light curves with SNooPy \citep{SNooPy}. 
They are converted to absolute magnitude as shown in Table~\ref{tab:Bband}. 
%\textcolor{red}{To convert to absolute mag, you need to know the distances, or you need to know the mean magnitude of SNIa if you use a light curve width correction method. Please specify what you did. Also, we need to assign errors to the magnitudes estimated from MRNN.}

%\textcolor{blue}{Referring to the available light curves from the papers in Table \ref{tab:Stretch}, we adopt {\tt\string SNooPy} \citep{SNooPy} to interpolate the B-band observational magnitudes at the spectral observation days. Then, we adopt the host galaxy distances and dust extinction values from the papers in Table \ref{tab:Stretch}, to calculate the absolute B band magnitude listed in Table \ref{tab:Bband}. }

\begin{deluxetable}{ccccccccc}[htb!]
    \tablecaption{B-band Absolute Luminosity Comparisons\label{tab:Bband}}
    \tablehead{\colhead{SN Name} & \colhead{Phase} & \colhead{Obs. Mag} & \colhead{$E(B-V)$} & \colhead{$\mu$}
    & \colhead{MRNN Mag} & \colhead{Abs. Mag} 
    & \colhead{Difference} }
    \startdata
    SN2011fe    & -2.6  & 10.034 & 0     & 29.13 & $-19.96_{-0.12}^{+0.12}$ & -19.10 & 0.86 &  \\
    SN2011fe    &  0.4  &  9.976 & 0     & 29.13 & $-20.00_{-0.11}^{+0.11}$ & -19.15 & 0.85 &  \\
    SN2011fe    &  3.7  & 10.070 & 0     & 29.13 & $-19.96_{-0.12}^{+0.11}$ & -18.96 & 0.90 &  \\
    SN2013dy    & -3.1  & 13.353 & 0.341 & 31.50 & $-20.06_{-0.09}^{+0.08}$ & -19.54 & 0.52 &  \\
    SN2013dy    & -1.1  & 13.294 & 0.341 & 31.50 & $-20.08_{-0.09}^{+0.10}$ & -19.60 & 0.48 &  \\
    SN2013dy    &  0.9  & 13.291 & 0.341 & 31.50 & $-20.02_{-0.09}^{+0.10}$ & -19.61 & 0.41 &  \\
    SN2013dy    &  3.9  & 13.366 & 0.341 & 31.50 & $-19.99_{-0.11}^{+0.11}$ & -19.53 & 0.46 &  \\
    SN2011by    & -0.4  & 12.933 & 0.052 & 31.59 & $-20.03_{-0.09}^{+0.09}$ & -18.87 & 1.16 &  \\
    SN2011iv    &  0.6  & 12.484 & 0     & 31.26 & $-19.87_{-0.09}^{+0.10}$ & -18.82 & 1.05 &  \\
    SN2015F     & -2.3  & 13.590 & 0.210 & 31.89 & $-19.89_{-0.10}^{+0.11}$ & -19.16 & 0.73 &  \\
    ASASSN-14lp & -4.4  & 12.496 & 0.351 & 30.84 & $-20.04_{-0.09}^{+0.09}$ & -19.74 & 0.30 &  \\
    \enddata
    \tablecomments{
        \textbf{Column Names:} SN Name: The name of supernovae. 
        Phase: The days relative to B-band maximum time. 
        Obs. Mag: The observed magnitude, interpolated from the photometry using {\tt\string SNooPy}. 
        $E(B-V)$: The total extinction parameters, including both Milky-Way and host galaxy extinction. 
        $\mu$: The distance modulus, data sources are listed in Table \ref{tab:Stretch}. 
        MRNN Mag: The absolute magnitude predicted by our MRNN, we adopt the median values and the $1-\sigma$ intervals in testing data set. 
        Abs. Mag: The absolute magnitude calculated from the observational magnitude. 
        Difference: The difference between MRNN predicted absolute magnitude and the absolute magnitude deduced from observed light curves. 
    }
\end{deluxetable}

\begin{deluxetable}{cccc}[htb!]
    \tablecaption{B-band Absolute Luminosity Comparisons of 15 SNe from \citet{HSTspec}\label{tab:Bband2}}
    \tablehead{\colhead{Supernova Name} & \colhead{Phase} & \colhead{Stretch} & \colhead{MRNN Luminosity}}
    \startdata
    PTF09dlc & 2.8  & $ 1.05 \pm0.03 $ & $-19.80_{-0.08}^{+0.12}$ \\
    PTF09dnl & 1.3  & $ 1.05 \pm0.02 $ & $-19.91_{-0.1}^{+0.11}$ \\
    PTF09fox & 2.6  & $ 0.92 \pm0.04 $ & $-19.94_{-0.11}^{+0.11}$ \\
    PTF09foz & 2.8  & $ 0.87 \pm0.06 $ & $-19.78_{-0.08}^{+0.13}$ \\
    PTF10bjs & 1.9  & $ 1.08 \pm0.02 $ & $-19.89_{-0.1}^{+0.11}$ \\
    PTF10hdv & 3.3  & $ 1.05 \pm0.07 $ & $-19.85_{-0.09}^{+0.1}$ \\
    PTF10hmv & 2.5  & $ 1.09 \pm0.01 $ & $-19.80_{-0.08}^{+0.12}$ \\
    PTF10icb & 0.8  & $ 0.99 \pm0.03 $ & $-19.93_{-0.11}^{+0.12}$ \\
    PTF10mwb & -0.4 & $ 0.94 \pm0.03 $ & $-19.87_{-0.09}^{+0.11}$ \\
    PTF10pdf & 2.2  & $ 1.23 \pm0.03 $ & $-19.73_{-0.08}^{+0.15}$ \\
    PTF10qjq & 3.5  & $ 0.96 \pm0.02 $ & $-19.92_{-0.1}^{+0.11}$ \\
    PTF10tce & 3.5  & $ 1.07 \pm0.02 $ & $-19.67_{-0.11}^{+0.13}$ \\
    PTF10ufj & 2.7  & $ 0.95 \pm0.02 $ & $-19.75_{-0.08}^{+0.14}$ \\
    PTF10xyt & 3.2  & $ 1.07 \pm0.04 $ & $-19.83_{-0.09}^{+0.11}$ \\
    SN2009le & 0.3  & $ 1.08 \pm0.01 $ & $-19.84_{-0.09}^{+0.1}$ \\
    \enddata
    \tablecomments{The observed luminosity of these SNe are not available due to the lack of published photometric data.}
\end{deluxetable}

The $B-$band magnitudes from the neural network is systematically more luminous than the values deduced from well calibrated light curves. This difference is perhaps the result of the simplified assumption of a sharp photosphere as mentioned above. Despite this limitation, the neural network prediction is largely a measurement of the spectral properties and may be used as an empirical indicator of luminosity, which can still be useful in exploring the diversity of SNIa luminosity based on spectroscopic data. 

\section{Discussion and Summary}\label{sec:Summary}

We have developed a AIAI method using the MRNN for the reconstruction of the chemical and density structures of SNIa models generated using the code TARDIS. 
With this MRNN architecture, we successfully trained and tested the predictive power of the neural network. 
%The neural network contains 16 convolutional and 3 fully connected layers, and is one of the deepest neural network in spectral analysis \textcolor{red}{Is there a paper to cite here?}. It was, half a year ago.  

The relevance of this study to real observations is explored with the limited amount of observational data. With the elemental abundances predicted from the neural network, we successfully derived model fits to the spectra of SN2011fe, SN2011iv, SN2015F, SN2011by, SN2013dy, ASASSN-14lp and other 15 SNIa near their $B$-band maximum. The AIAI allows derivations of chemical structures of these SNe. 

From the AIAI deduced elements, we found that SNIa with higher stretch factor contain larger Ni mass at velocities above 10,000 km/sec (in Zone 2, 3, and 4). We also observed the decline of the mass of $^{56}$Ni due to radioactive decay in some well observed SNe. 

We attempted to predict the $B-$band luminosity from the spectral shapes using the AIAI network. The predicted $B-$band absolute magnitude is systematically higher than the luminosity derived from light curve fits, by $0.47\sim1.26$ magnitudes. 
We surmise the discrepancy is due to approximations of physical processes made in TARDIS. 

Despite these successes, we must stress that the present study is only a preliminary exploration of an exciting approach to SN modeling. The study proves that the combination of deep-learning techniques with physical models of complicated spectroscopic data may yield unique insights to the physical processes in SNIa. However, there are a few caveats that need to be kept in mind, and in a way, these caveats also point to the directions of further improvements:

\begin{itemize}
    \item In TARDIS, some major assumptions need to be kept in mind. The temperature profiles are calculated based on the assumption that the photospheric spectra follow that of black-body radiation. There are radioactive material very close or above to the photosphere so the energy deposition is significantly more complicated than TARDIS assumptions. 
    \item The current implementation of the code only applies to data around optical maximum. 
    \item We sub-divided the ejecta into some artificial grids, this introduces unphysical boundaries that are inconsistent with the physics of nuclear burning in the ejecta. 
   \item The dependence on the adopted baseline model has not been explored yet, models with different density and chemical profiles need to be studied and built into the spectral library. 
   \item The spectral models are drawn from a uniform distribution of parameters around the baseline model, which serves as a plain Bayesian prior for the uncertainty estimation discussed in Section \ref{sec:trainresult}. Different Bayesian priors need to be explored. % This is literally a Bayesian prior. Different Bayesian priors need to be explored.
   \item The highest ejecta velocity explored in this study is 25,000 km/sec, which may be too low for some high velocity supernovae such as SN~2004dt \citep{Wang:2004dt, WangX:2009}. 
   \item The size of the spectral library is still tiny. 
   \item Our MRNN architecture is constructed to be most sensitive to heavily blended spectral features produced by iron group elements. Other neural network architecture such as LSTM \citep{LSTM} may be explored which can improve less blended features. 
\end{itemize}

 On a positive note, the current study are sensitive to spectral variations by construction, and the performance of the AIAI network confirms that. We expect the physical approximations made in TARDIS to have only a weak effect on the derivation of chemical elemental abundances. 

 In summary, we have developed a deep-learning technique to extract physical quantities of supernova spectra. Preliminary application of the methods to a set of observational data proves the method to be powerful. More studies are needed to fully realize the potential of the techniques presented in this study.

%\textcolor{red}{Add citations for software. }
\software{python-keras\citep{keras}, python-dust\_extinction \href{https://github.com/karllark/dust\_extinction}{https://github.com/karllark/dust\_extinction}, TARDIS\citep{TARDIS}, SiFTO\citep{sifto}, SNooPy\citep{SNooPy}}

\acknowledgements

We thank WISeREP \href{https://wiserep.weizmann.ac.il}{https://wiserep.weizmann.ac.il} for the observed SNe spectra. 
We thank Prof. Peter H\"oflich (Florida State University) for the DDT models. 
We thank Prof. Ryan J. Foley (University of California Santa Cruz) for the computer readable version of SN2011by spectra. 
We thank Dr. Kate Maguire (Queen University at Belfast) for the light curves and spectra of the SNe in \citep{HSTspec}. XZ and LW acknowledge supports by an NSF grant AST~1817099.
XZ thank Dr. Wolfgang Kerzendorf (European Southern Observatory) for supportive discussions. 
Portions of this research were conducted with the advanced computing resources provided by Texas A\&M High Performance Research Computing. 
%\textcolor{red}{A new acknowledge format for TARDIS team}
This research made use of \textsc{Tardis}, a community-developed software package for spectral synthesis in supernovae
\citep{TARDIS, kerzendorf_wolfgang_2019_2590539}. 
The development of \textsc{Tardis} received support from the Google Summer of Code initiative and from ESA's Summer of Code in Space program. 
\textsc{Tardis} makes extensive use of Astropy and PyNE. 

%\textcolor{red}{Reference modifications: 1. Remove the URLs. 2. Replace the full journal name with abbreviations, ApJ, ApJS, etc. }

%\bibliography{sneRef}
%\bibliography{hereBibUpDLSY}

\begin{thebibliography}{}
\expandafter\ifx\csname natexlab\endcsname\relax\def\natexlab#1{#1}\fi
\providecommand{\url}[1]{\href{#1}{#1}}

\bibitem[{Abadi {et~al.}(2016)Abadi, Barham, Chen, Chen, Davis, Dean, Devin,
  Ghemawat, Irving, Isard, {et~al.}}]{tensorflow}
Abadi, M., Barham, P., Chen, J., {et~al.} 2016, in 12th {USENIX} Symposium on
  Operating Systems Design and Implementation (OSDI 16), 265--283

\bibitem[{{Abbott} \& {Lucy}(1985)}]{MazzaliMCMC}
{Abbott}, D.~C., \& {Lucy}, L.~B. 1985, \apj, 288, 679

\bibitem[{{Abdi} \& {Nahavandi}(2016)}]{MResNet}
{Abdi}, M., \& {Nahavandi}, S. 2016, arXiv e-prints, arXiv:1609.05672

\bibitem[{Ashall {et~al.}(2018)Ashall, Mazzali, Stritzinger, Hoeflich, Burns,
  Gall, Hsiao, Phillips, Morrell, \& Foley}]{TardisNormal}
Ashall, C., Mazzali, P.~A., Stritzinger, M.~D., {et~al.} 2018, Monthly Notices
  of the Royal Astronomical Society, 477, 153.
\newblock \url{https://doi.org/10.1093/mnras/sty632}

\bibitem[{{Barna} {et~al.}(2018){Barna}, {Szalai}, {Kerzendorf}, {Kromer},
  {Sim}, {Magee}, \& {Leibundgut}}]{IaxFewParameter}
{Barna}, B., {Szalai}, T., {Kerzendorf}, W.~E., {et~al.} 2018, \mnras, 480,
  3609

\bibitem[{{Barna} {et~al.}(2017){Barna}, {Szalai}, {Kromer}, {Kerzendorf},
  {Vink{\'o}}, {Silverman}, {Marion}, \& {Wheeler}}]{AbunTomoSN2011ayTardis}
{Barna}, B., {Szalai}, T., {Kromer}, M., {et~al.} 2017, \mnras, 471, 4865

\bibitem[{{Baron} \& {Hauschildt}(1998)}]{PHOENIX}
{Baron}, E., \& {Hauschildt}, P.~H. 1998, \apj, 495, 370

\bibitem[{{Bialek} {et~al.}(2019){Bialek}, {Fabbro}, {Venn}, {Kumar},
  {O'Briain}, \& {Moo Yi}}]{DLstellar2}
{Bialek}, S., {Fabbro}, S., {Venn}, K.~A., {et~al.} 2019, arXiv e-prints,
  arXiv:1911.02602

\bibitem[{Blondin {et~al.}(2017)Blondin, Dessart, \& Hillier}]{SN1999by}
Blondin, S., Dessart, L., \& Hillier, D.~J. 2017, \mnras, 474, 3931

\bibitem[{{Blondin} {et~al.}(2013){Blondin}, {Dessart}, {Hillier}, \&
  {Khokhlov}}]{DDSequenceSpectra}
{Blondin}, S., {Dessart}, L., {Hillier}, D.~J., \& {Khokhlov}, A.~M. 2013,
  \mnras, 429, 2127

\bibitem[{Branch {et~al.}(2009)Branch, Dang, \& Baron}]{branch2009comparative}
Branch, D., Dang, L.~C., \& Baron, E. 2009, \pasp, 121, 238

\bibitem[{{Bu} {et~al.}(2018){Bu}, {Zeng}, \& {Lei}}]{LAMOST}
{Bu}, Y., {Zeng}, J., \& {Lei}, Z. 2018, arXiv e-prints, arXiv:1805.01617

\bibitem[{Bulla {et~al.}(2015)Bulla, Sim, \& Kromer}]{PolarSynthesis}
Bulla, M., Sim, S.~A., \& Kromer, M. 2015, Monthly Notices of the Royal
  Astronomical Society, 450, 967.
\newblock \url{https://doi.org/10.1093/mnras/stv657}

\bibitem[{Burns {et~al.}(2010)Burns, Stritzinger, Phillips, Kattner, Persson,
  Madore, Freedman, Boldt, Campillay, Contreras, \& et~al.}]{SNooPy}
Burns, C.~R., Stritzinger, M., Phillips, M.~M., {et~al.} 2010, The Astronomical
  Journal, 141, 19.
\newblock \url{http://dx.doi.org/10.1088/0004-6256/141/1/19}

\bibitem[{{Cardelli} {et~al.}(1989){Cardelli}, {Clayton}, \& {Mathis}}]{CCM}
{Cardelli}, J.~A., {Clayton}, G.~C., \& {Mathis}, J.~S. 1989, \apj, 345, 245

\bibitem[{Cartier {et~al.}(2016)Cartier, Sullivan, Firth, Pignata, Mazzali,
  Maguire, Childress, Arcavi, Ashall, Bassett, Crawford, Frohmaier, Galbany,
  Gal-Yam, Hosseinzadeh, Howell, Inserra, Johansson, Kasai, McCully, Prajs,
  Prentice, Schulze, Smartt, Smith, Smith, Valenti, \& Young}]{SN2015Fearly}
Cartier, R., Sullivan, M., Firth, R.~E., {et~al.} 2016, \mnras, 464, 4476

\bibitem[{Chollet {et~al.}(2015)}]{keras}
Chollet, F., {et~al.} 2015, Keras, \url{https://keras.io}, ,

\bibitem[{{Chomiuk} {et~al.}(2012){Chomiuk}, {Soderberg}, {Moe}, {Chevalier},
  {Rupen}, {Badenes}, {Margutti}, {Fransson}, {Fong}, \&
  {Dittmann}}]{SN2011feEVLA}
{Chomiuk}, L., {Soderberg}, A.~M., {Moe}, M., {et~al.} 2012, \apj, 750, 164

\bibitem[{{Cikota} {et~al.}(2019){Cikota}, {Patat}, {Wang}, {Wheeler}, {Bulla},
  {Baade}, {H{\"o}flich}, {Cikota}, {Clocchiatti}, {Maund}, {Stevance}, \&
  {Yang}}]{Cikota:2019MNRAS.tmp.2014C}
{Cikota}, A., {Patat}, F., {Wang}, L., {et~al.} 2019, \mnras, 2014

\bibitem[{{Conley} {et~al.}(2008){Conley}, {Sullivan}, {Hsiao}, {Guy},
  {Astier}, {Balam}, {Balland}, {Basa}, {Carlberg}, {Fouchez}, {Hardin},
  {Howell}, {Hook}, {Pain}, {Perrett}, {Pritchet}, \& {Regnault}}]{sifto}
{Conley}, A., {Sullivan}, M., {Hsiao}, E.~Y., {et~al.} 2008, \apj, 681, 482

\bibitem[{Fabbro {et~al.}(2017)Fabbro, Bialek, O'Briain, Kielty, Jahandar,
  Monty, \& Venn}]{DLstellar}
Fabbro, S., Bialek, S., O'Briain, T., {et~al.} 2017, \mnras, 475, 2978

\bibitem[{{Fitzpatrick}(1999)}]{F99}
{Fitzpatrick}, E.~L. 1999, \pasp, 111, 63

\bibitem[{{Foley} {et~al.}(2018){Foley}, {Hoffmann}, {Macri}, {Riess}, {Brown},
  {Filippenko}, {Graham}, \& {Milne}}]{SN2011by2}
{Foley}, R.~J., {Hoffmann}, S.~L., {Macri}, L.~M., {et~al.} 2018, arXiv
  e-prints, arXiv:1806.08359

\bibitem[{{Foley} {et~al.}(2012){Foley}, {Kromer}, {Howie Marion}, {Pignata},
  {Stritzinger}, {Taubenberger}, {Challis}, {Filippenko}, {Folatelli},
  {Hillebrandt}, {Hsiao}, {Kirshner}, {Li}, {Morrell}, {R{\"o}pke},
  {Ciaraldi-Schoolmann}, {Seitenzahl}, {Silverman}, {Simcoe}, {Berta},
  {Ivarsen}, {Newton}, {Nysewander}, \& {Reichart}}]{SN2011ivHST}
{Foley}, R.~J., {Kromer}, M., {Howie Marion}, G., {et~al.} 2012, \apjl, 753, L5

\bibitem[{{Foreman-Mackey} {et~al.}(2013){Foreman-Mackey}, {Hogg}, {Lang}, \&
  {Goodman}}]{emcee}
{Foreman-Mackey}, D., {Hogg}, D.~W., {Lang}, D., \& {Goodman}, J. 2013, \pasp,
  125, 306

\bibitem[{{Gall} {et~al.}(2018){Gall}, {Stritzinger}, {Ashall}, {Baron},
  {Burns}, {Hoeflich}, {Hsiao}, {Mazzali}, {Phillips}, {Filippenko},
  {Anderson}, {Benetti}, {Brown}, {Campillay}, {Challis}, {Contreras}, {Elias
  de la Rosa}, {Folatelli}, {Foley}, {Fraser}, {Holmbo}, {Marion}, {Morrell},
  {Pan}, {Pignata}, {Suntzeff}, {Taddia}, {Torres Robledo}, \&
  {Valenti}}]{SN2011iv}
{Gall}, C., {Stritzinger}, M.~D., {Ashall}, C., {et~al.} 2018, \aap, 611, A58

\bibitem[{{Gamezo} {et~al.}(2004){Gamezo}, {Khokhlov}, \& {Oran}}]{Gamezo:2004}
{Gamezo}, V.~N., {Khokhlov}, A.~M., \& {Oran}, E.~S. 2004, \prl, 92, 211102

\bibitem[{{Gamezo} {et~al.}(2003){Gamezo}, {Khokhlov}, {Oran}, {Chtchelkanova},
  \& {Rosenberg}}]{3dimDDTScience}
{Gamezo}, V.~N., {Khokhlov}, A.~M., {Oran}, E.~S., {Chtchelkanova}, A.~Y., \&
  {Rosenberg}, R.~O. 2003, Science, 299, 77

\bibitem[{{Garavini} {et~al.}(2007){Garavini}, {Nobili}, {Taubenberger},
  {Pastorello}, {Elias-Rosa}, {Stanishev}, {Blanc}, {Benetti}, {Goobar},
  {Mazzali}, {Sanchez}, {Salvo}, {Schmidt}, \& {Hillebrandt}}]{SN2005cf}
{Garavini}, G., {Nobili}, S., {Taubenberger}, S., {et~al.} 2007, \aap, 471, 527

\bibitem[{Gerardy {et~al.}(2007)Gerardy, Meikle, Kotak, Hoflich, Farrah,
  Filippenko, Foley, Lundqvist, Mattila, Pozzo, Sollerman, Dyk, \&
  Wheeler}]{Gerardy_2007}
Gerardy, C.~L., Meikle, W. P.~S., Kotak, R., {et~al.} 2007, The Astrophysical
  Journal, 661, 995.
\newblock \url{https://doi.org/10.1086%2F516728}

\bibitem[{{Gordon} {et~al.}(2009){Gordon}, {Cartledge}, \& {Clayton}}]{GCC09}
{Gordon}, K.~D., {Cartledge}, S., \& {Clayton}, G.~C. 2009, \apj, 705, 1320

\bibitem[{{Graham} {et~al.}(2015){Graham}, {Foley}, {Zheng}, {Kelly},
  {Shivvers}, {Silverman}, {Filippenko}, {Clubb}, \&
  {Ganeshalingam}}]{SN2011by}
{Graham}, M.~L., {Foley}, R.~J., {Zheng}, W., {et~al.} 2015, \mnras, 446, 2073

\bibitem[{{Graur} {et~al.}(2018{\natexlab{a}}){Graur}, {Zurek}, {Cara}, {Rest},
  {Seitenzahl}, {Shappee}, {Shara}, \& {Riess}}]{asassn14lplate}
{Graur}, O., {Zurek}, D.~R., {Cara}, M., {et~al.} 2018{\natexlab{a}}, \apj,
  866, 10

\bibitem[{{Graur} {et~al.}(2018{\natexlab{b}}){Graur}, {Zurek}, {Rest},
  {Seitenzahl}, {Shappee}, {Fisher}, {Guillochon}, {Shara}, \&
  {Riess}}]{SN2015F}
{Graur}, O., {Zurek}, D.~R., {Rest}, A., {et~al.} 2018{\natexlab{b}}, \apj,
  859, 79

\bibitem[{{Hachinger} {et~al.}(2009){Hachinger}, {Mazzali}, {Taubenberger},
  {Pakmor}, \& {Hillebrandt}}]{SN2005bl}
{Hachinger}, S., {Mazzali}, P.~A., {Taubenberger}, S., {Pakmor}, R., \&
  {Hillebrandt}, W. 2009, \mnras, 399, 1238

\bibitem[{{He} {et~al.}(2015){He}, {Zhang}, {Ren}, \& {Sun}}]{ResNet}
{He}, K., {Zhang}, X., {Ren}, S., \& {Sun}, J. 2015, arXiv e-prints,
  arXiv:1512.03385

\bibitem[{{Hillier} \& {Miller}(1998)}]{CMFGEN}
{Hillier}, D.~J., \& {Miller}, D.~L. 1998, \apj, 496, 407

\bibitem[{Hochreiter \& Schmidhuber(1997)}]{LSTM}
Hochreiter, S., \& Schmidhuber, J. 1997, Neural Computation, 9, 1735.
\newblock \url{https://doi.org/10.1162/neco.1997.9.8.1735}

\bibitem[{{Hoeflich} {et~al.}(1996{\natexlab{a}}){Hoeflich}, {Khokhlov},
  {Wheeler}, {Phillips}, {Suntzeff}, \& {Hamuy}}]{Hoeflich:1996ApJ...472L..81H}
{Hoeflich}, P., {Khokhlov}, A., {Wheeler}, J.~C., {et~al.} 1996{\natexlab{a}},
  \apj, 472, L81

\bibitem[{{Hoeflich} {et~al.}(1996{\natexlab{b}}){Hoeflich}, {Wheeler},
  {Hines}, \& {Trammell}}]{HoeflichPol1996ApJ...459..307H}
{Hoeflich}, P., {Wheeler}, J.~C., {Hines}, D.~C., \& {Trammell}, S.~R.
  1996{\natexlab{b}}, \apj, 459, 307

\bibitem[{{Hoeflich} {et~al.}(2017){Hoeflich}, {Hsiao}, {Ashall}, {Burns},
  {Diamond}, {Phillips}, {Sand}, {Stritzinger}, {Suntzeff}, {Contreras},
  {Krisciunas}, {Morrell}, \& {Wang}}]{Hoeflich:2017ApJ...846...58H}
{Hoeflich}, P., {Hsiao}, E.~Y., {Ashall}, C., {et~al.} 2017, \apj, 846, 58

\bibitem[{{Huang} {et~al.}(2016){Huang}, {Liu}, {van der Maaten}, \&
  {Weinberger}}]{DNN}
{Huang}, G., {Liu}, Z., {van der Maaten}, L., \& {Weinberger}, K.~Q. 2016,
  arXiv e-prints, arXiv:1608.06993

\bibitem[{{Ioffe} \& {Szegedy}(2015)}]{BatchNorm}
{Ioffe}, S., \& {Szegedy}, C. 2015, arXiv e-prints, arXiv:1502.03167

\bibitem[{{Iwamoto} {et~al.}(1999){Iwamoto}, {Brachwitz}, {Nomoto},
  {Kishimoto}, {Umeda}, {Hix}, \& {Thielemann}}]{NucleosynthesisReview}
{Iwamoto}, K., {Brachwitz}, F., {Nomoto}, K., {et~al.} 1999, \apjs, 125, 439

\bibitem[{{Jordan} {et~al.}(2008){Jordan}, {Fisher}, {Townsley}, {Calder},
  {Graziani}, {Asida}, {Lamb}, \& {Truran}}]{JordanGCD2008ApJ...681.1448J}
{Jordan}, G.~C., I., {Fisher}, R.~T., {Townsley}, D.~M., {et~al.} 2008, \apj,
  681, 1448

\bibitem[{{Kasen} {et~al.}(2002){Kasen}, {Branch}, {Baron}, \&
  {Jeffery}}]{DanielKasonLineAnalysis}
{Kasen}, D., {Branch}, D., {Baron}, E., \& {Jeffery}, D. 2002, \apj, 565, 380

\bibitem[{{Kasen} \& {Plewa}(2005)}]{GCD2}
{Kasen}, D., \& {Plewa}, T. 2005, \apjl, 622, L41

\bibitem[{{Kasen} {et~al.}(2006){Kasen}, {Thomas}, \& {Nugent}}]{SEDONA}
{Kasen}, D., {Thomas}, R.~C., \& {Nugent}, P. 2006, \apj, 651, 366

\bibitem[{Kerzendorf {et~al.}(2019)Kerzendorf, Nöbauer, Sim, Lietzau,
  Jančauskas, Vogl, Mishin, Tsamis, Boyle, Gupta, Desai, Klauser, Beaujean,
  Suban-Loewen, Heringer, Shingles, Barna, Gautam, Patel, Barbosa, Varanasi,
  Reinecke, Bylund, Bentil, Rajagopalan, Jain, Singh, Talegaonkar, Sofiatti,
  Patel, Yap, Wahi, \& Gupta}]{kerzendorf_wolfgang_2019_2590539}
Kerzendorf, W., Nöbauer, U., Sim, S., {et~al.} 2019, tardis-sn/tardis: TARDIS
  v3.0 alpha2, , , doi:10.5281/zenodo.2590539

\bibitem[{Kerzendorf \& Sim(2014)}]{TARDIS}
Kerzendorf, W.~E., \& Sim, S.~A. 2014, \mnras, 440, 387

\bibitem[{{Khokhlov}(1991{\natexlab{a}})}]{KhokhlovDDT}
{Khokhlov}, A.~M. 1991{\natexlab{a}}, \aap, 246, 383

\bibitem[{{Khokhlov}(1991{\natexlab{b}})}]{Khokhlov:1991A&A...245..114K}
---. 1991{\natexlab{b}}, \aap, 245, 114

\bibitem[{Kingma \& Ba(2014)}]{adam}
Kingma, D.~P., \& Ba, J. 2014, CoRR, abs/1412.6980, arXiv:1412.6980

\bibitem[{Krizhevsky(2012)}]{cifardata}
Krizhevsky, A. 2012, University of Toronto

\bibitem[{Krizhevsky {et~al.}(2012)Krizhevsky, Sutskever, \& Hinton}]{AlexNet}
Krizhevsky, A., Sutskever, I., \& Hinton, G.~E. 2012, in Advances in Neural
  Information Processing Systems 25, ed. F.~Pereira, C.~J.~C. Burges,
  L.~Bottou, \& K.~Q. Weinberger (Curran Associates, Inc.), 1097--1105

\bibitem[{Kromer \& Sim(2009)}]{ARTIS}
Kromer, M., \& Sim, S.~A. 2009, Monthly Notices of the Royal Astronomical
  Society, 398, 1809.
\newblock \url{https://doi.org/10.1111/j.1365-2966.2009.15256.x}

\bibitem[{{Law} {et~al.}(2009){Law}, {Kulkarni}, {Dekany}, {Ofek}, {Quimby},
  {Nugent}, {Surace}, {Grillmair}, {Bloom}, {Kasliwal}, {Bildsten}, {Brown},
  {Cenko}, {Ciardi}, {Croner}, {Djorgovski}, {van Eyken}, {Filippenko}, {Fox},
  {Gal-Yam}, {Hale}, {Hamam}, {Helou}, {Henning}, {Howell}, {Jacobsen},
  {Laher}, {Mattingly}, {McKenna}, {Pickles}, {Poznanski}, {Rahmer}, {Rau},
  {Rosing}, {Shara}, {Smith}, {Starr}, {Sullivan}, {Velur}, {Walters}, \&
  {Zolkower}}]{PTF2}
{Law}, N.~M., {Kulkarni}, S.~R., {Dekany}, R.~G., {et~al.} 2009, \pasp, 121,
  1395

\bibitem[{LeCun \& Hinton(2015)}]{DeepLearning}
LeCun, Y.~Bengio, Y., \& Hinton, G. 2015, Nature, 521, 436

\bibitem[{{Lentz} {et~al.}(2001){Lentz}, {Baron}, {Branch}, \&
  {Hauschildt}}]{SN1984A}
{Lentz}, E.~J., {Baron}, E., {Branch}, D., \& {Hauschildt}, P.~H. 2001, \apj,
  547, 402

\bibitem[{Liu {et~al.}(2019)Liu, Gibson, Mills, \& Osadchy}]{Dynamic}
Liu, J., Gibson, S.~J., Mills, J., \& Osadchy, M. 2019, Chemometrics and
  Intelligent Laboratory Systems, 184, 175

\bibitem[{{Lucy}(1971)}]{Lucy:1971ApJ...163...95L}
{Lucy}, L.~B. 1971, \apj, 163, 95

\bibitem[{{Lucy}(1999)}]{MazzilMCMC3}
---. 1999, \aap, 345, 211

\bibitem[{{Lucy}(2002)}]{macroatom}
---. 2002, \aap, 384, 725

\bibitem[{{Maguire} {et~al.}(2012){Maguire}, {Sullivan}, {Ellis}, {Nugent},
  {Howell}, {Gal-Yam}, {Cooke}, {Mazzali}, {Pan}, {Dilday}, {Thomas}, {Arcavi},
  {Ben-Ami}, {Bersier}, {Bianco}, {Fulton}, {Hook}, {Horesh}, {Hsiao}, {James},
  {Podsiadlowski}, {Walker}, {Yaron}, {Kasliwal}, {Laher}, {Law}, {Ofek},
  {Poznanski}, \& {Surace}}]{HSTspec}
{Maguire}, K., {Sullivan}, M., {Ellis}, R.~S., {et~al.} 2012, \mnras, 426, 2359

\bibitem[{{Mazzali}(2000)}]{MazzilMCMC4}
{Mazzali}, P.~A. 2000, \aap, 363, 705

\bibitem[{{Mazzali} \& {Lucy}(1993)}]{Mazzali:1993A&A...279..447M}
{Mazzali}, P.~A., \& {Lucy}, L.~B. 1993, \aap, 279, 447

\bibitem[{{Mazzali} {et~al.}(1993){Mazzali}, {Lucy}, {Danziger}, {Gouiffes},
  {Cappellaro}, \& {Turatto}}]{MazzilMCMC2}
{Mazzali}, P.~A., {Lucy}, L.~B., {Danziger}, I.~J., {et~al.} 1993, \aap, 269,
  423

\bibitem[{{Mazzali} {et~al.}(2014){Mazzali}, {Sullivan}, {Hachinger}, {Ellis},
  {Nugent}, {Howell}, {Gal-Yam}, {Maguire}, {Cooke}, {Thomas}, {Nomoto}, \&
  {Walker}}]{rho11fe}
{Mazzali}, P.~A., {Sullivan}, M., {Hachinger}, S., {et~al.} 2014, \mnras, 439,
  1959

\bibitem[{{Mont{\'u}far} {et~al.}(2014){Mont{\'u}far}, {Pascanu}, {Cho}, \&
  {Bengio}}]{NetworkLayers}
{Mont{\'u}far}, G., {Pascanu}, R., {Cho}, K., \& {Bengio}, Y. 2014, arXiv
  e-prints, arXiv:1402.1869

\bibitem[{{Munari} {et~al.}(2013){Munari}, {Henden}, {Belligoli}, {Castellani},
  {Cherini}, {Righetti}, \& {Vagnozzi}}]{2011feLC}
{Munari}, U., {Henden}, A., {Belligoli}, R., {et~al.} 2013, \na, 20, 30

\bibitem[{Nidever {et~al.}(2015)Nidever, Holtzman, Prieto, Beland, Bender,
  Bizyaev, Burton, Desphande, Fleming, P{\'{e}}rez, Hearty, Majewski,
  M{\'{e}}sz{\'{a}}ros, Muna, Nguyen, Schiavon, Shetrone, Skrutskie, Sobeck, \&
  Wilson}]{Nidever:2015}
Nidever, D.~L., Holtzman, J.~A., Prieto, C.~A., {et~al.} 2015, \aj, 150, 173

\bibitem[{{Nomoto} {et~al.}(1984){Nomoto}, {Thielemann}, \& {Yokoi}}]{W7}
{Nomoto}, K., {Thielemann}, F.-K., \& {Yokoi}, K. 1984, \apj, 286, 644

\bibitem[{{Nugent} {et~al.}(2011){Nugent}, {Sullivan}, {Cenko}, {Thomas},
  {Kasen}, {Howell}, {Bersier}, {Bloom}, {Kulkarni}, {Kandrashoff},
  {Filippenko}, {Silverman}, {Marcy}, {Howard}, {Isaacson}, {Maguire},
  {Suzuki}, {Tarlton}, {Pan}, {Bildsten}, {Fulton}, {Parrent}, {Sand},
  {Podsiadlowski}, {Bianco}, {Dilday}, {Graham}, {Lyman}, {James}, {Kasliwal},
  {Law}, {Quimby}, {Hook}, {Walker}, {Mazzali}, {Pian}, {Ofek}, {Gal-Yam}, \&
  {Poznanski}}]{SN2011feNature}
{Nugent}, P.~E., {Sullivan}, M., {Cenko}, S.~B., {et~al.} 2011, \nat, 480, 344

\bibitem[{Pan {et~al.}(2015)Pan, Foley, Kromer, Fox, Zheng, Challis, Clubb,
  Filippenko, Folatelli, Graham, Hillebrandt, Kirshner, Lee, Pakmor, Patat,
  Phillips, Pignata, Röpke, Seitenzahl, Silverman, Simon, Sternberg,
  Stritzinger, Taubenberger, Vinko, \& Wheeler}]{SN2013dy500}
Pan, Y.-C., Foley, R.~J., Kromer, M., {et~al.} 2015, \mnras, 452, 4307

\bibitem[{{Parrent} {et~al.}(2010){Parrent}, {Branch}, \&
  {Jeffery}}]{Parrent:2010}
{Parrent}, J., {Branch}, D., \& {Jeffery}, D. 2010, {SYNOW: A Highly
  Parameterized Spectrum Synthesis Code for Direct Analysis of SN Spectra}, , ,
  ascl:1010.055

\bibitem[{{Patat} {et~al.}(2013){Patat}, {Cordiner}, {Cox}, {Anderson},
  {Harutyunyan}, {Kotak}, {Palaversa}, {Stanishev}, {Tomasella}, {Benetti},
  {Goobar}, {Pastorello}, \& {Sollerman}}]{SN2011fePatat}
{Patat}, F., {Cordiner}, M.~A., {Cox}, N.~L.~J., {et~al.} 2013, \aap, 549, A62

\bibitem[{Perlmutter {et~al.}(1999)Perlmutter, Aldering, Goldhaber, Knop,
  Nugent, Castro, Deustua, Fabbro, Goobar, Groom,
  {et~al.}}]{perlmutter1999measurements}
Perlmutter, S., Aldering, G., Goldhaber, G., {et~al.} 1999, \apj, 517, 565

\bibitem[{{Phillips}(1993)}]{Phillips:1993ApJ...413L.105P}
{Phillips}, M.~M. 1993, \apjl, 413, L105

\bibitem[{{Plewa} {et~al.}(2004){Plewa}, {Calder}, \& {Lamb}}]{GCD}
{Plewa}, T., {Calder}, A.~C., \& {Lamb}, D.~Q. 2004, \apjl, 612, L37

\bibitem[{Poludnenko {et~al.}(2019)Poludnenko, Chambers, Ahmed, Gamezo, \&
  Taylor}]{Poludnenkoeaau7365}
Poludnenko, A.~Y., Chambers, J., Ahmed, K., Gamezo, V.~N., \& Taylor, B.~D.
  2019, Science, 366

\bibitem[{Poludnenko {et~al.}(2011)Poludnenko, Gardiner, \&
  Oran}]{Poludnenko:2011}
Poludnenko, A.~Y., Gardiner, T.~A., \& Oran, E.~S. 2011, Phys. Rev. Lett., 107,
  054501

\bibitem[{{Pskovskii}(1977)}]{Pskovskii:1977}
{Pskovskii}, I.~P. 1977, Sov. Ast., 21, 675

\bibitem[{{Rau} {et~al.}(2009){Rau}, {Kulkarni}, {Law}, {Bloom}, {Ciardi},
  {Djorgovski}, {Fox}, {Gal-Yam}, {Grillmair}, {Kasliwal}, {Nugent}, {Ofek},
  {Quimby}, {Reach}, {Shara}, {Bildsten}, {Cenko}, {Drake}, {Filippenko},
  {Helfand}, {Helou}, {Howell}, {Poznanski}, \& {Sullivan}}]{PTF}
{Rau}, A., {Kulkarni}, S.~R., {Law}, N.~M., {et~al.} 2009, \pasp, 121, 1334

\bibitem[{Riess {et~al.}(1996)Riess, Press, \& Kirshner}]{riess1996precise}
Riess, A.~G., Press, W.~H., \& Kirshner, R.~P. 1996, \apj, 473, 88

\bibitem[{Riess {et~al.}(1998)Riess, Filippenko, Challis, Clocchiatti, Diercks,
  Garnavich, Gilliland, Hogan, Jha, Kirshner,
  {et~al.}}]{riess1998observational}
Riess, A.~G., Filippenko, A.~V., Challis, P., {et~al.} 1998, \aj, 116, 1009

\bibitem[{{R{\"o}pke} {et~al.}(2006){R{\"o}pke}, {Gieseler}, {Reinecke},
  {Travaglio}, \& {Hillebrandt}}]{Roepke:2006}
{R{\"o}pke}, F.~K., {Gieseler}, M., {Reinecke}, M., {Travaglio}, C., \&
  {Hillebrandt}, W. 2006, \aap, 453, 203

\bibitem[{{Rubin} {et~al.}(2013){Rubin}, {Knop}, {Rykoff}, {Aldering},
  {Amanullah}, {Barbary}, {Burns}, {Conley}, {Connolly}, {Deustua}, {Fadeyev},
  {Fakhouri}, {Fruchter}, {Gibbons}, {Goldhaber}, {Goobar}, {Hsiao}, {Huang},
  {Kowalski}, {Lidman}, {Meyers}, {Nordin}, {Perlmutter}, {Saunders},
  {Spadafora}, {Stanishev}, {Suzuki}, {Wang}, \& {Supernova Cosmology
  Project}}]{Rubin:2013}
{Rubin}, D., {Knop}, R.~A., {Rykoff}, E., {et~al.} 2013, \apj, 763, 35

\bibitem[{Savitzky \& Golay(1964)}]{SGfilter}
Savitzky, A., \& Golay, M. J.~E. 1964, Analytical Chemistry, 36, 1627

\bibitem[{{Shappee} {et~al.}(2016){Shappee}, {Piro}, {Holoien}, {Prieto},
  {Contreras}, {Itagaki}, {Burns}, {Kochanek}, {Stanek}, {Alper}, {Basu},
  {Beacom}, {Bersier}, {Brimacombe}, {Conseil}, {Danilet}, {Dong}, {Falco},
  {Grupe}, {Hsiao}, {Kiyota}, {Morrell}, {Nicolas}, {Phillips}, {Pojmanski},
  {Simonian}, {Stritzinger}, {Szczygie{\l}}, {Taddia}, {Thompson},
  {Thorstensen}, {Wagner}, \& {Wo{\'z}niak}}]{asassn14lp}
{Shappee}, B.~J., {Piro}, A.~L., {Holoien}, T.~W.-S., {et~al.} 2016, \apj, 826,
  144

\bibitem[{{Shen} {et~al.}(2018){Shen}, {Boubert}, {G{\"a}nsicke}, {Jha},
  {Andrews}, {Chomiuk}, {Foley}, {Fraser}, {Gromadzki}, {Guillochon}, {Kotze},
  {Maguire}, {Siebert}, {Smith}, {Strader}, {Badenes}, {Kerzendorf}, {Koester},
  {Kromer}, {Miles}, {Pakmor}, {Schwab}, {Toloza}, {Toonen}, {Townsley}, \&
  {Williams}}]{Shen:2018}
{Shen}, K.~J., {Boubert}, D., {G{\"a}nsicke}, B.~T., {et~al.} 2018, \apj, 865,
  15

\bibitem[{Silverman {et~al.}(2013)Silverman, Ganeshalingam, \&
  Filippenko}]{BSNIPV}
Silverman, J.~M., Ganeshalingam, M., \& Filippenko, A.~V. 2013, Monthly Notices
  of the Royal Astronomical Society, 430, 1030.
\newblock \url{https://doi.org/10.1093/mnras/sts674}

\bibitem[{{Simonyan} \& {Zisserman}(2014)}]{VGG16}
{Simonyan}, K., \& {Zisserman}, A. 2014, arXiv e-prints, arXiv:1409.1556

\bibitem[{{Smartt} {et~al.}(2017){Smartt}, {Chen}, {Jerkstrand}, {Coughlin},
  {Kankare}, {Sim}, {Fraser}, {Inserra}, {Maguire}, {Chambers}, {Huber},
  {Kr{\"u}hler}, {Leloudas}, {Magee}, {Shingles}, {Smith}, {Young}, {Tonry},
  {Kotak}, {Gal-Yam}, {Lyman}, {Homan}, {Agliozzo}, {Anderson}, {Angus},
  {Ashall}, {Barbarino}, {Bauer}, {Berton}, {Botticella}, {Bulla}, {Bulger},
  {Cannizzaro}, {Cano}, {Cartier}, {Cikota}, {Clark}, {De Cia}, {Della Valle},
  {Denneau}, {Dennefeld}, {Dessart}, {Dimitriadis}, {Elias-Rosa}, {Firth},
  {Flewelling}, {Fl{\"o}rs}, {Franckowiak}, {Frohmaier}, {Galbany},
  {Gonz{\'a}lez-Gait{\'a}n}, {Greiner}, {Gromadzki}, {Guelbenzu},
  {Guti{\'e}rrez}, {Hamanowicz}, {Hanlon}, {Harmanen}, {Heintz}, {Heinze},
  {Hernandez}, {Hodgkin}, {Hook}, {Izzo}, {James}, {Jonker}, {Kerzendorf},
  {Klose}, {Kostrzewa-Rutkowska}, {Kowalski}, {Kromer}, {Kuncarayakti},
  {Lawrence}, {Lowe}, {Magnier}, {Manulis}, {Martin-Carrillo}, {Mattila},
  {McBrien}, {M{\"u}ller}, {Nordin}, {O'Neill}, {Onori}, {Palmerio},
  {Pastorello}, {Patat}, {Pignata}, {Podsiadlowski}, {Pumo}, {Prentice}, {Rau},
  {Razza}, {Rest}, {Reynolds}, {Roy}, {Ruiter}, {Rybicki}, {Salmon}, {Schady},
  {Schultz}, {Schweyer}, {Seitenzahl}, {Smith}, {Sollerman}, {Stalder},
  {Stubbs}, {Sullivan}, {Szegedi}, {Taddia}, {Taubenberger}, {Terreran}, {van
  Soelen}, {Vos}, {Wainscoat}, {Walton}, {Waters}, {Weiland}, {Willman},
  {Wiseman}, {Wright}, {Wyrzykowski}, \& {Yaron}}]{TardisKilonova}
{Smartt}, S.~J., {Chen}, T.-W., {Jerkstrand}, A., {et~al.} 2017, \nat, 551, 75

\bibitem[{{Stehle} {et~al.}(2005){Stehle}, {Mazzali}, {Benetti}, \&
  {Hillebrandt}}]{AbundanceTomography}
{Stehle}, M., {Mazzali}, P.~A., {Benetti}, S., \& {Hillebrandt}, W. 2005,
  \mnras, 360, 1231

\bibitem[{{Thomas}(2013)}]{Thomas:2013}
{Thomas}, R.~C. 2013, {SYN++: Standalone SN spectrum synthesis}, , ,
  ascl:1308.008

\bibitem[{Thomas {et~al.}(2011)Thomas, Nugent, \& Meza}]{SYNAPPS}
Thomas, R.~C., Nugent, P.~E., \& Meza, J. 2011, \pasp, 123, 237

\bibitem[{{Timmes} {et~al.}(2003){Timmes}, {Brown}, \& {Truran}}]{Timmes:2003}
{Timmes}, F.~X., {Brown}, E.~F., \& {Truran}, J.~W. 2003, \apjl, 590, L83

\bibitem[{{Vink{\'o}} {et~al.}(2018){Vink{\'o}}, {Ordasi}, {Szalai},
  {S{\'a}rneczky}, {B{\'a}nyai}, {B{\'{\i}}r{\'o}}, {Borkovits}, {Heged{\"u}s},
  {Hodos{\'a}n}, {Kelemen}, {Klagyivik}, {Kriskovics}, {Kun}, {Marion},
  {Marschalk{\'o}}, {Moln{\'a}r}, {Nagy}, {P{\'a}l}, {Silverman},
  {Szak{\'a}ts}, {Szegedi-Elek}, {Sz{\'e}kely}, {Szing}, {Vida}, \&
  {Wheeler}}]{AbsoluteDistance}
{Vink{\'o}}, J., {Ordasi}, A., {Szalai}, T., {et~al.} 2018, \pasp, 130, 064101

\bibitem[{{Vogl} {et~al.}(2019){Vogl}, {Sim}, {Noebauer}, {Kerzendorf}, \&
  {Hillebrandt}}]{TardisII}
{Vogl}, C., {Sim}, S.~A., {Noebauer}, U.~M., {Kerzendorf}, W.~E., \&
  {Hillebrandt}, W. 2019, \aap, 621, A29

\bibitem[{{Wang} \& {Han}(2012)}]{WangHan:2012NewAR..56..122W}
{Wang}, B., \& {Han}, Z. 2012, \nar, 56, 122

\bibitem[{Wang {et~al.}(2006)Wang, Baade, Hoflich, Wheeler, Kawabata, Khokhlov,
  Nomoto, \& Patat}]{Wang:2004dt}
Wang, L., Baade, D., Hoflich, P., {et~al.} 2006, \apj, 653, 490

\bibitem[{{Wang} {et~al.}(2003){Wang}, {Goldhaber}, {Aldering}, \&
  {Perlmutter}}]{WangCMAGIC2003}
{Wang}, L., {Goldhaber}, G., {Aldering}, G., \& {Perlmutter}, S. 2003, \apj,
  590, 944

\bibitem[{{Wang} {et~al.}(2006){Wang}, {Strovink}, {Conley}, {Goldhaber},
  {Kowalski}, {Perlmutter}, \& {Siegrist}}]{WangStrovink}
{Wang}, L., {Strovink}, M., {Conley}, A., {et~al.} 2006, \apj, 641, 50

\bibitem[{Wang \& Wheeler(2008)}]{Wang:2008ARAA}
Wang, L., \& Wheeler, J.~C. 2008, \araa, 46, 433

\bibitem[{{Wang} {et~al.}(1996){Wang}, {Wheeler}, {Li}, \&
  {Clocchiatti}}]{Wang:1996ApJ...467..435W}
{Wang}, L., {Wheeler}, J.~C., {Li}, Z., \& {Clocchiatti}, A. 1996, \apj, 467,
  435

\bibitem[{{Wang} {et~al.}(2013){Wang}, {Wang}, {Filippenko}, {Zhang}, \&
  {Zhao}}]{wangX:2013}
{Wang}, X., {Wang}, L., {Filippenko}, A.~V., {Zhang}, T., \& {Zhao}, X. 2013,
  Science, 340, 170

\bibitem[{{Wang} {et~al.}(2009){Wang}, {Filippenko}, {Ganeshalingam}, {Li},
  {Silverman}, {Wang}, {Chornock}, {Foley}, {Gates}, {Macomber}, {Serduke},
  {Steele}, \& {Wong}}]{WangX:2009}
{Wang}, X., {Filippenko}, A.~V., {Ganeshalingam}, M., {et~al.} 2009, \apjl,
  699, L139

\bibitem[{{Yang} {et~al.}(2019){Yang}, {Hoeflich}, {Baade}, {Maund}, {Wang},
  {Brown}, {Stevance}, {Arcavi}, {Burke}, {Cikota}, {Clocchiatti}, {Gal-Yam},
  {Graham}, {Hiramatsu}, {Hosseinzadeh}, {Howell}, {Jha}, {McCully}, {Patat},
  {Sand}, {Schulze}, {Spyromilio}, {Valenti}, {Vinko}, {Wang}, {Wheeler},
  {Yaron}, \& {Zhang}}]{Yang:2019arXiv190310820Y}
{Yang}, Y., {Hoeflich}, P.~A., {Baade}, D., {et~al.} 2019, arXiv e-prints,
  arXiv:1903.10820

\bibitem[{{Zhai} {et~al.}(2016){Zhai}, {Zhang}, {Wang}, {Zhang}, {Liu},
  {Brown}, {Huang}, {Zhao}, {Chang}, {Yi}, {Wang}, {Xin}, {Wang}, {Lun},
  {Zhang}, {Fan}, {Zheng}, \& {Bai}}]{SN2013dy}
{Zhai}, Q., {Zhang}, J.-J., {Wang}, X.-F., {et~al.} 2016, \aj, 151, 125

\end{thebibliography}

\clearpage

\appendix

\section{The MSE for elements and zones}\label{sec:appendix:mse}

In this section, we present two sets of the mean-squared-error. 
Table \ref{tab:Residual} is for the neural networks on 2000-10000 \AA spectra, which contains 34 trainable element-zone combinations. 
The other Table \ref{tab:Residual2} is for the neural networks on 3000-5200 \AA spectra, which contains 32 trainable element-zone combinations. 

\begin{deluxetable}{c|cccc}[htb!]
    \tablecaption{MSE for elements 6 (C) to 28 (Ni) in 4 Zones}
    \tablehead{\colhead{Element} & \colhead{Zone 1} & \colhead{Zone 2} 
    & \colhead{Zone 3} & \colhead{Zone 4}}
    \startdata
    C  & 0.1144 & 0.1155 & 0.1171 & 0.1106 \\
    N  & 0.1159 & 0.1137 & 0.1165 & 0.1147 \\
    O  & 0.1147 & 0.1146 & 0.1155 & 0.0822 \\
    F  & 0.1128 & 0.1166 & 0.1159 & 0.1158 \\
    Ne & 0.1147 & 0.1131 & 0.1174 & 0.1154 \\
    Na & 0.1134 & 0.1127 & 0.1125 & 0.1121 \\
    Mg & 0.1120 & 0.0135 & 0.0886 & 0.0509 \\
    Al & 0.1143 & 0.1146 & 0.1149 & 0.1156 \\
    Si & 0.0686 & 0.0177 & 0.0265 & 0.1173 \\
    P  & 0.1161 & 0.1142 & 0.1147 & 0.1145 \\
    S  & 0.1162 & 0.0678 & 0.1174 & 0.1134 \\
    Cl & 0.1109 & 0.1136 & 0.1159 & 0.1153 \\
    Ar & 0.1147 & 0.0964 & 0.1121 & 0.1127 \\
    K  & 0.1114 & 0.1122 & 0.1153 & 0.1155 \\
    Ca & 0.0672 & 0.0285 & 0.0615 & 0.0248 \\
    Sc & 0.1152 & 0.1145 & 0.1135 & 0.0670 \\
    Ti & 0.0974 & 0.0673 & 0.0587 & 0.0546 \\
    V  & 0.0624 & 0.0286 & 0.1096 & 0.1149 \\
    Cr & 0.1150 & 0.1119 & 0.1164 & 0.1164 \\
    Mn & 0.0416 & 0.0241 & 0.1105 & 0.1081 \\
    Fe & 0.0052 & 0.0078 & 0.0090 & 0.0347 \\
    Co & 0.0136 & 0.0409 & 0.0567 & 0.0931 \\
    Ni & 0.0143 & 0.0434 & 0.0585 & 0.0813 \\
    \enddata
    \tablecomments{For all $23\times 4$ neural networks involved in this table, they are trained on 10000 spectra. The MSEs are tested on 1829 spectra testing data set. Wavelength between 2000 and 10000 \AA are used as input. }
\end{deluxetable}

\begin{deluxetable}{c|cccc}[htb!]
    \tablecaption{MSE for elements 6 to 28 in 4 Zones}
    \tablehead{\colhead{Element} & \colhead{Zone 1} & \colhead{Zone 2} 
    & \colhead{Zone 3} & \colhead{Zone 4}}
    \startdata
    C   &  0.1158  &   0.1149  &   0.1160  &   0.1130  \\ 
    N   &  0.1156  &   0.1162  &   0.1146  &   0.1129  \\ 
    O   &  0.1158  &   0.1150  &   0.1147  &   0.1022  \\ 
    F   &  0.1137  &   0.1137  &   0.1153  &   0.1145  \\ 
    Ne  &  0.1135  &   0.1164  &   0.1154  &   0.1132  \\ 
    Na  &  0.1146  &   0.1144  &   0.1124  &   0.1160  \\ 
    Mg  &  0.1148  &   0.0733  &   0.0989  &   0.0602  \\ 
    Al  &  0.1148  &   0.1152  &   0.1147  &   0.1134  \\ 
    Si  &  0.0901  &   0.0652  &   0.0593  &   0.1140  \\ 
    P   &  0.1148  &   0.1151  &   0.1138  &   0.1152  \\ 
    S   &  0.1166  &   0.0750  &   0.1144  &   0.1146  \\ 
    Cl  &  0.1150  &   0.1145  &   0.1144  &   0.1164  \\ 
    Ar  &  0.1146  &   0.0992  &   0.1143  &   0.1152  \\ 
    K   &  0.1150  &   0.1157  &   0.1149  &   0.1136  \\ 
    Ca  &  0.0959  &   0.0775  &   0.0788  &   0.0600  \\ 
    Sc  &  0.1149  &   0.1151  &   0.1157  &   0.0677  \\ 
    Ti  &  0.1046  &   0.0738  &   0.0676  &   0.0521  \\ 
    V   &  0.0925  &   0.0524  &   0.1145  &   0.1150  \\ 
    Cr  &  0.1137  &   0.1154  &   0.1144  &   0.1123  \\ 
    Mn  &  0.0757  &   0.0644  &   0.1089  &   0.1153  \\ 
    Fe  &  0.0473  &   0.0311  &   0.0136  &   0.0465  \\ 
    Co  &  0.0372  &   0.0507  &   0.0568  &   0.1055  \\ 
    Ni  &  0.0462  &   0.0657  &   0.0583  &   0.0876  \\
    \enddata
    \tablecomments{For all $23\times 4$ neural networks involved in this table, they are trained on 10000 spectra. The MSEs are tested on 1829 spectra testing data set. Wavelength between 3000 and 5200 \AA are used as input. }
\end{deluxetable}

\clearpage

\section{The Relations between Stretch and Elemental Abundance}\label{sec:appendix:MD15}
The relations between the stretch parameter and the chemical abundance are shown in this appendix for intermediate mass elements. No obvious correlation is identified.

\begin{figure}[htb!]
    \minipage{0.33\textwidth}
        \includegraphics[width=\textwidth]{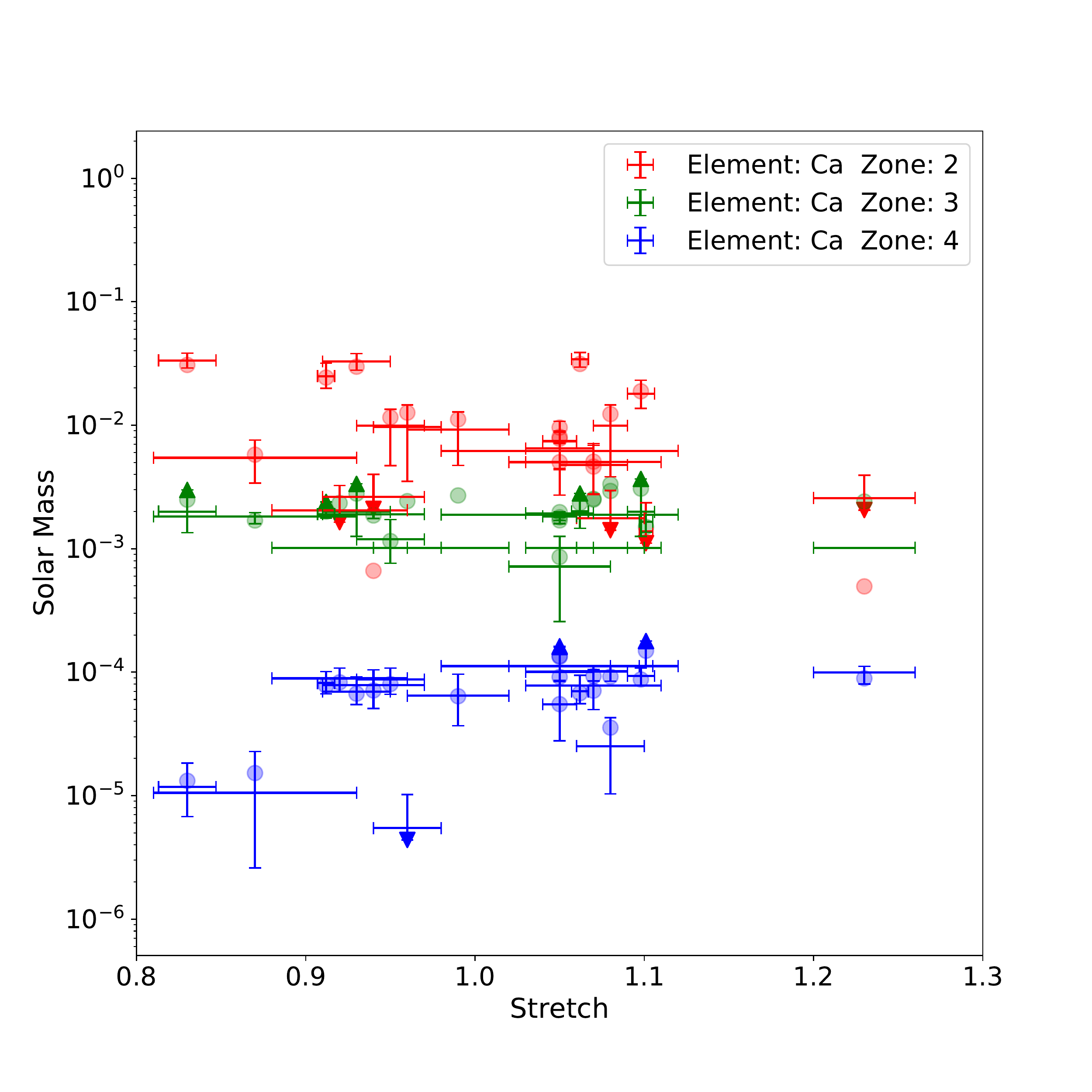}
    \endminipage\hfill
    \minipage{0.33\textwidth}
        \includegraphics[width=\textwidth]{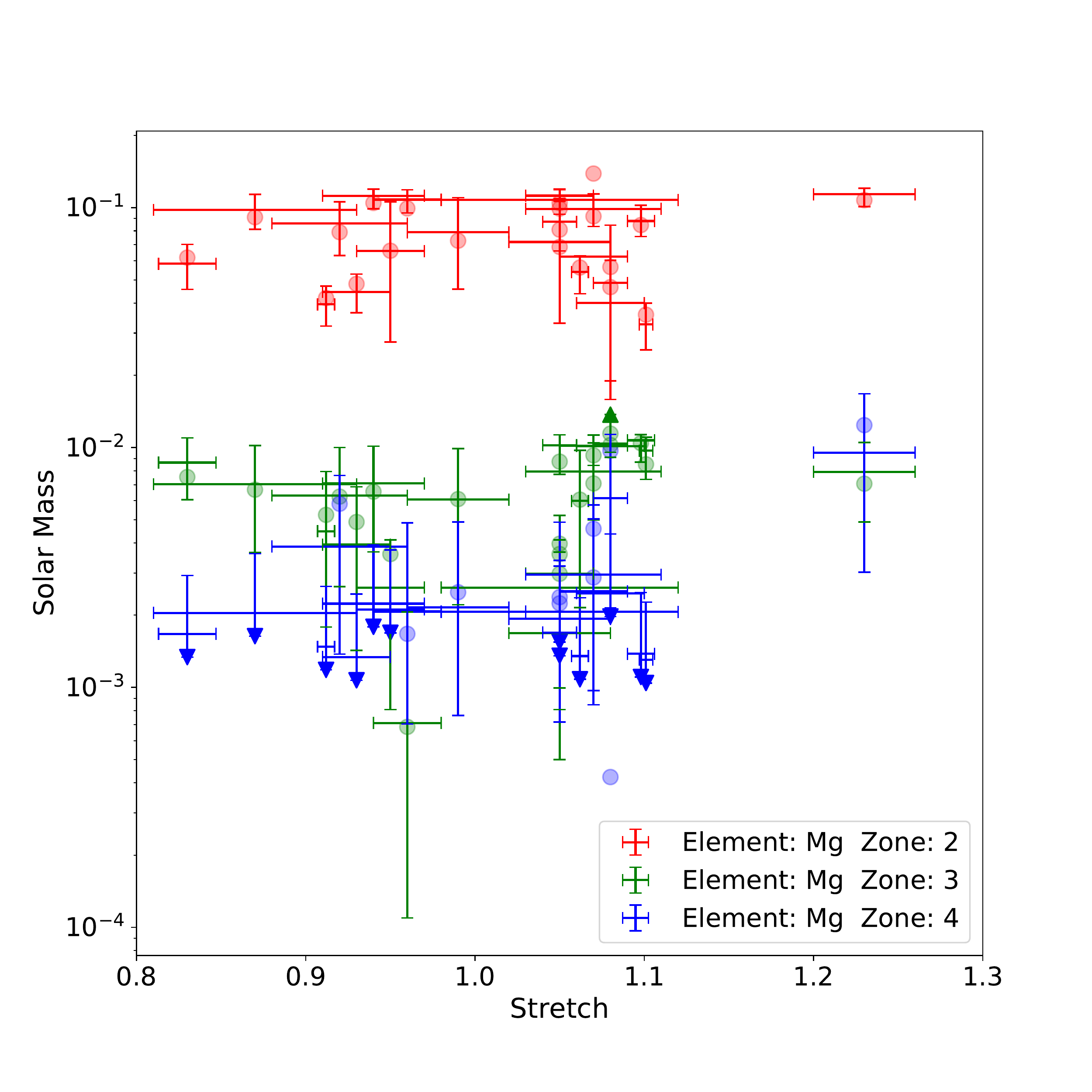}
    \endminipage\hfill
    \minipage{0.33\textwidth}
        \includegraphics[width=\textwidth]{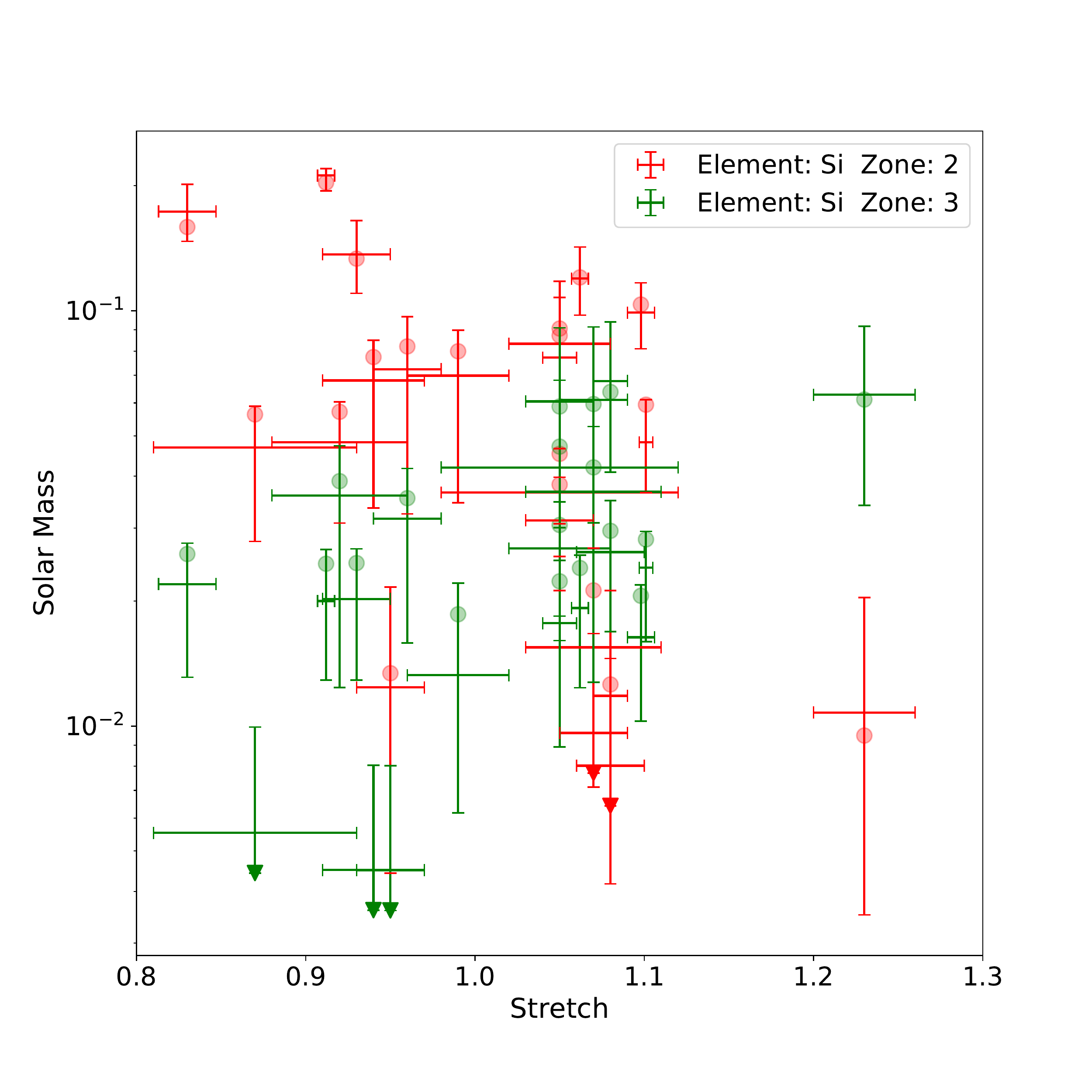}
    \endminipage\hfill
    \minipage{0.33\textwidth}
        \includegraphics[width=\textwidth]{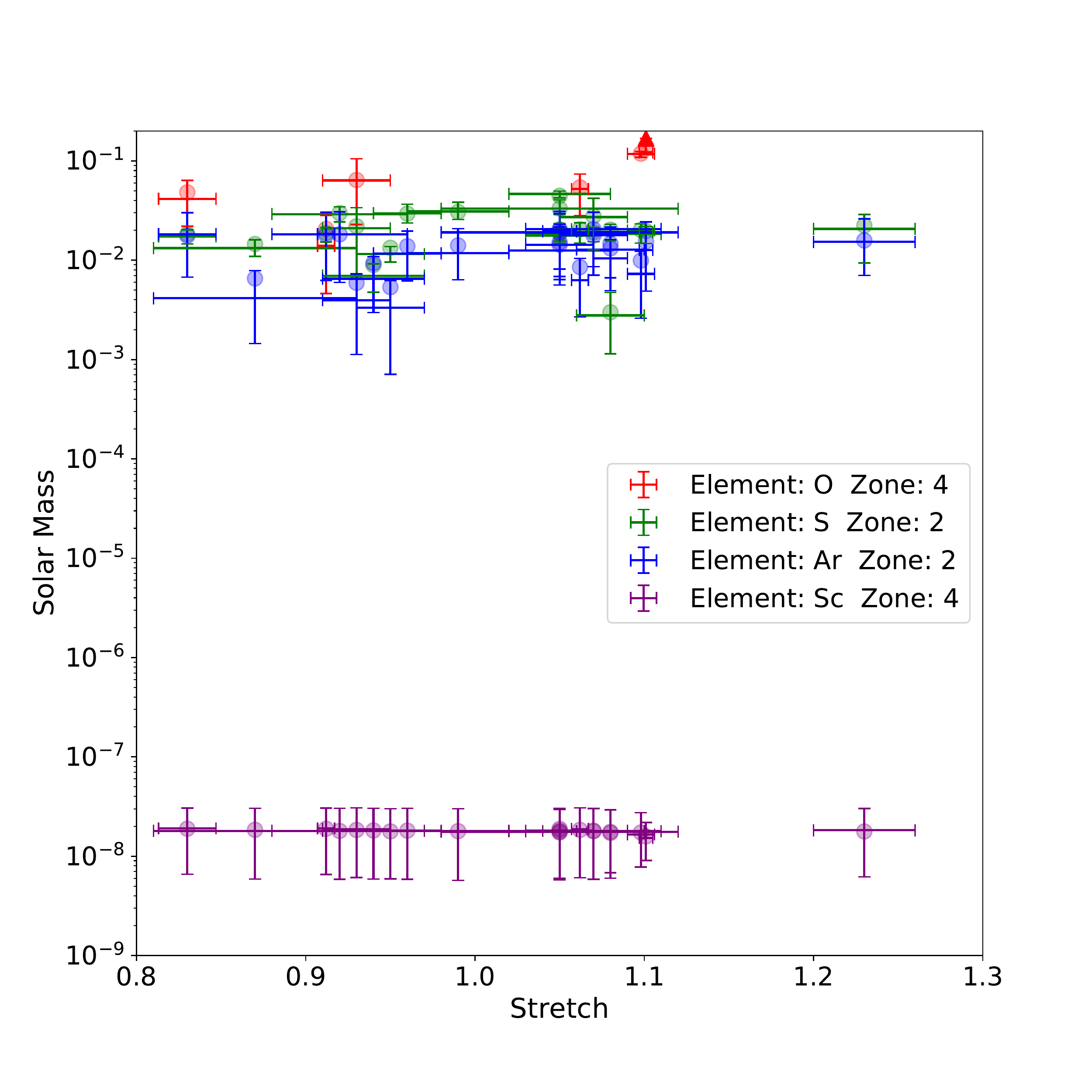}
    \endminipage\hfill
    \minipage{0.33\textwidth}
        \includegraphics[width=\textwidth]{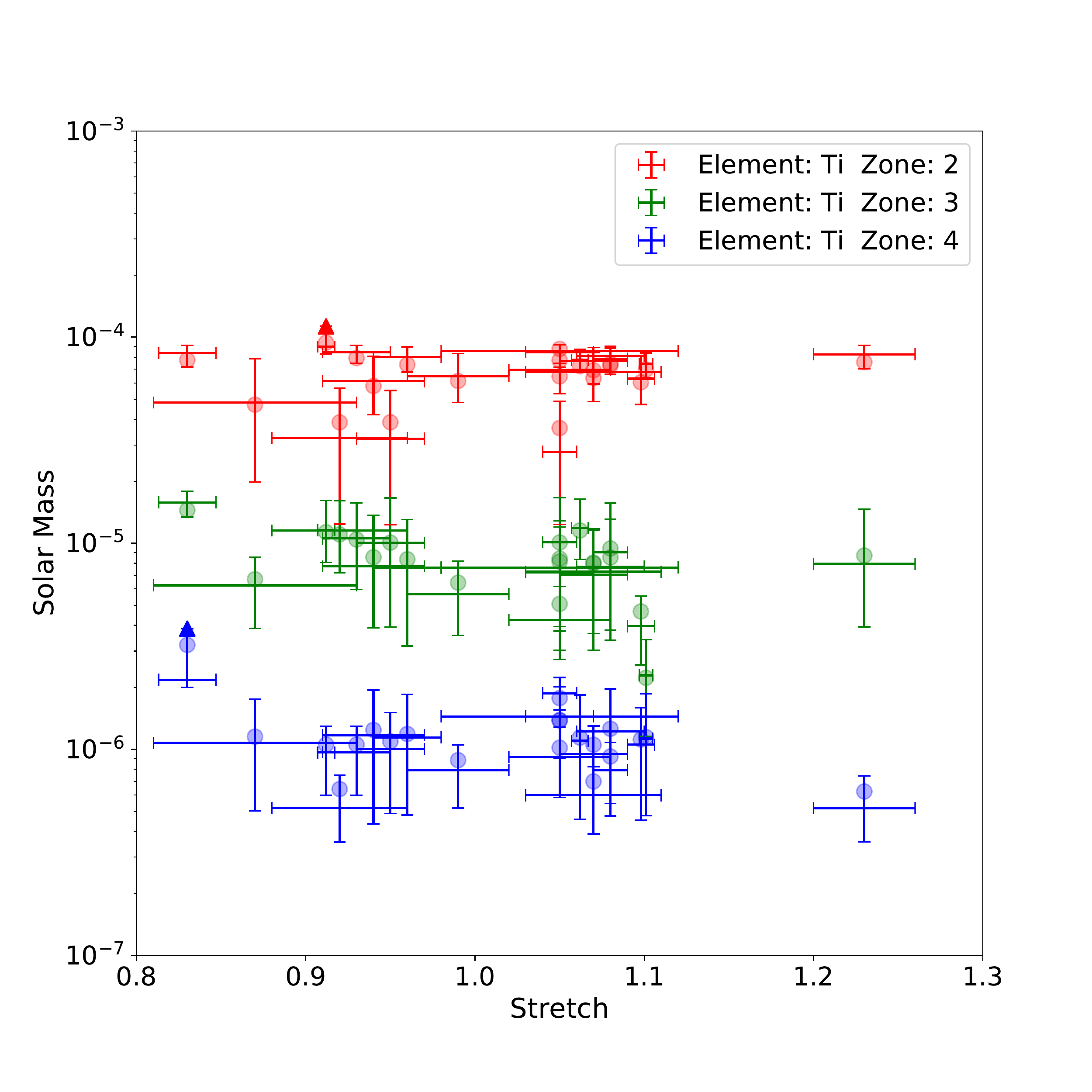}
    \endminipage\hfill
    \minipage{0.33\textwidth}
        \includegraphics[width=\textwidth]{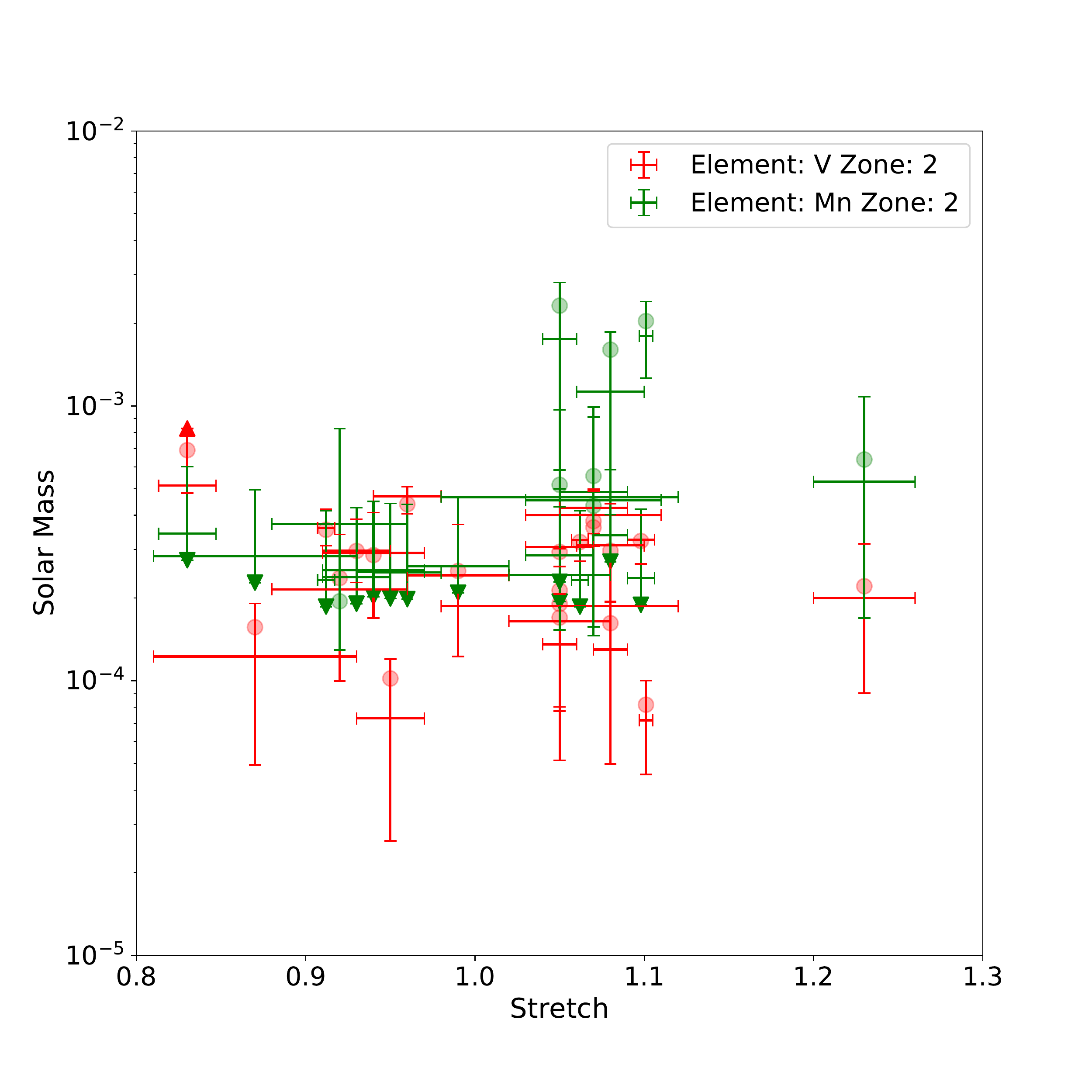}
    \endminipage\hfill
    \caption{The masses of intermediate mass elements are compared with the stretch factors. No obvious correlation is found.}
    \label{fig:appendix:MD15}
\end{figure}

\clearpage

\section{15 HST Spectra Fitting Results}\label{sec:hstfitting}

\begin{figure}[htb!]
    \includegraphics[width=\linewidth]{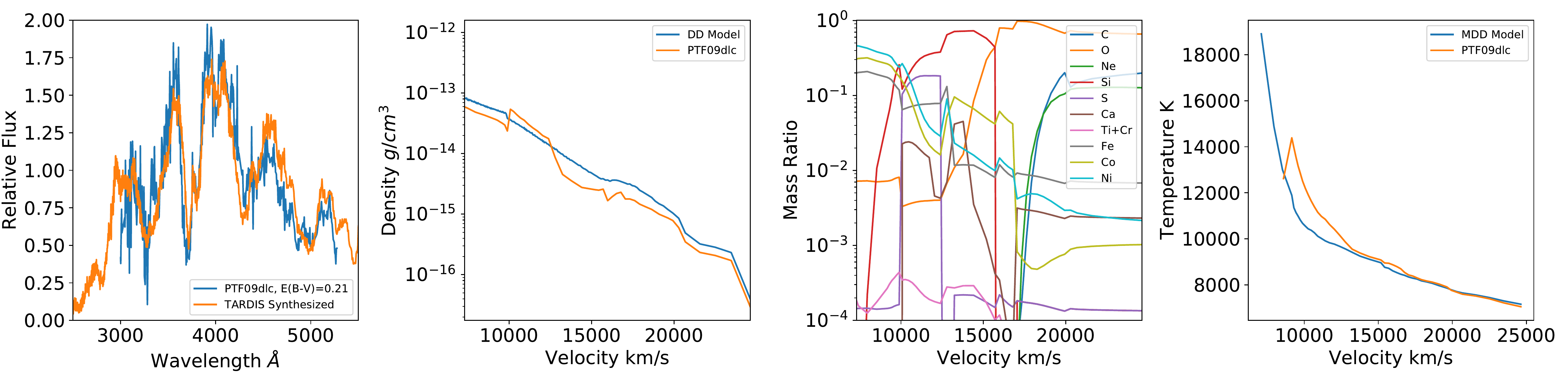}
    \includegraphics[width=\linewidth]{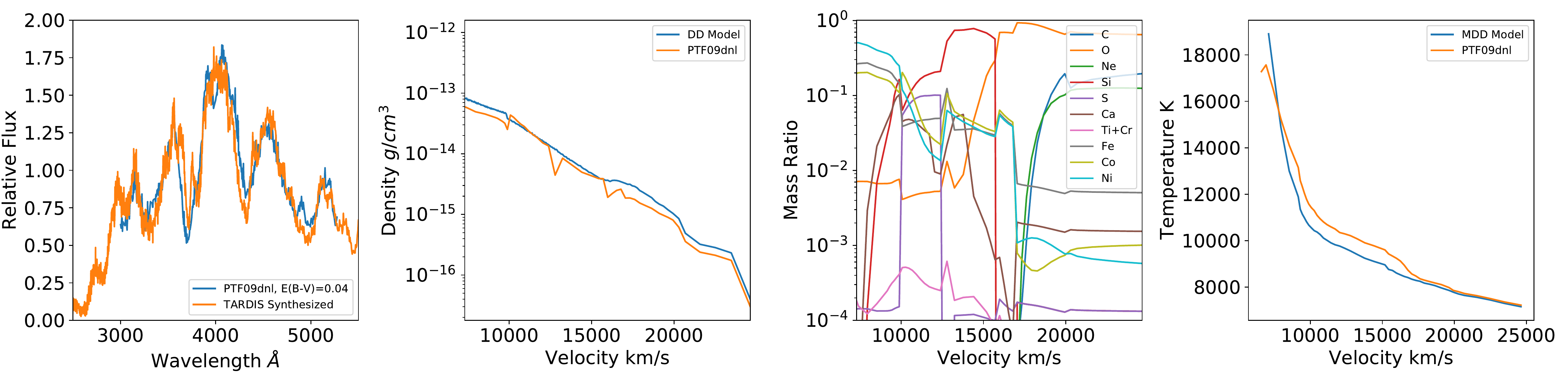}
    \includegraphics[width=\linewidth]{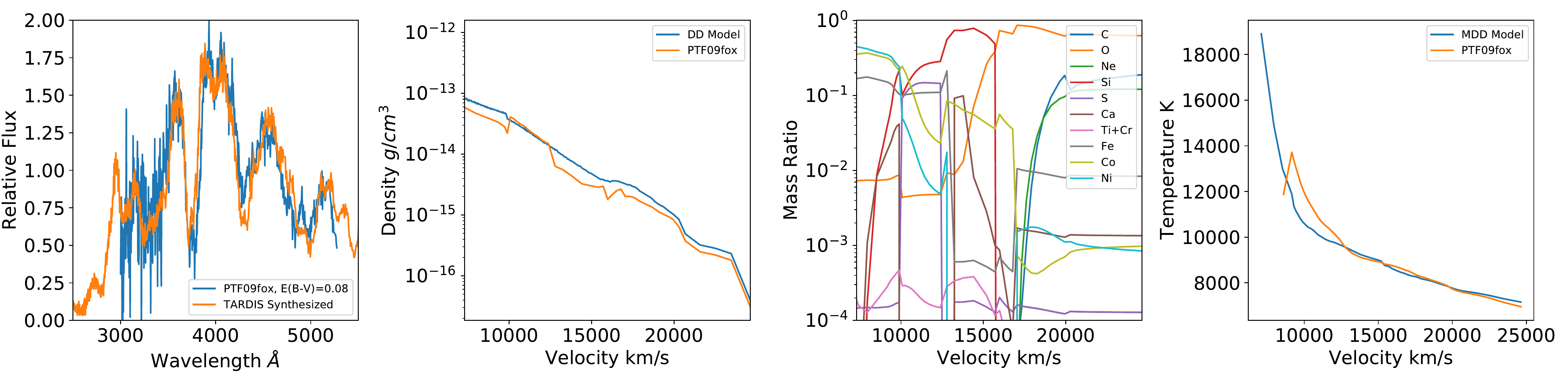}
    \includegraphics[width=\linewidth]{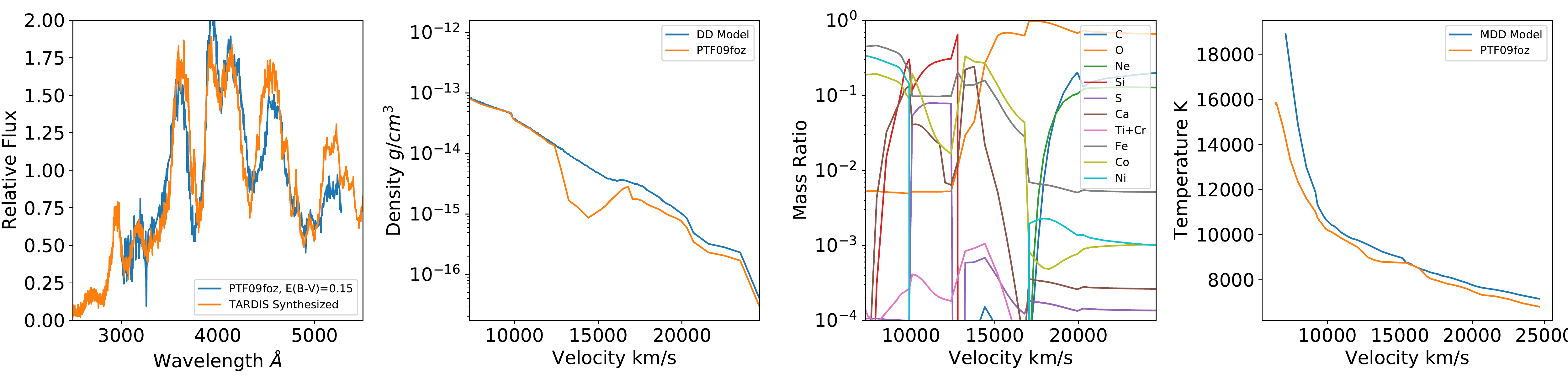}
    \caption{From upper to lower rows: models for PTF09dlc, PTF09dnl, PTF09fox, PTF09foz. From left to right columns: 1. The observed spectra (blue line) after dust extinction correction using the CCM model, $E(B-V)$ values given in the legend. The TARDIS synthesized spectra (orange line) calculated with the MRNN derived chemical structure. 2. The density of the ejecta derived from the MRNN (orange line) and the IG model density as a comparison (blue line), all densities are converted to 19 days after explosion. 3. Elemental abundances derived from MRNN, and is used for TARDIS spectra calculation. 4. Temperature structure for the synthesized spectra (orange line) and the temperature for IG model as comparisons (blue line). }
    \label{fig:hstfitting1}
\end{figure}

\begin{figure}[htb!]
    \includegraphics[width=\linewidth]{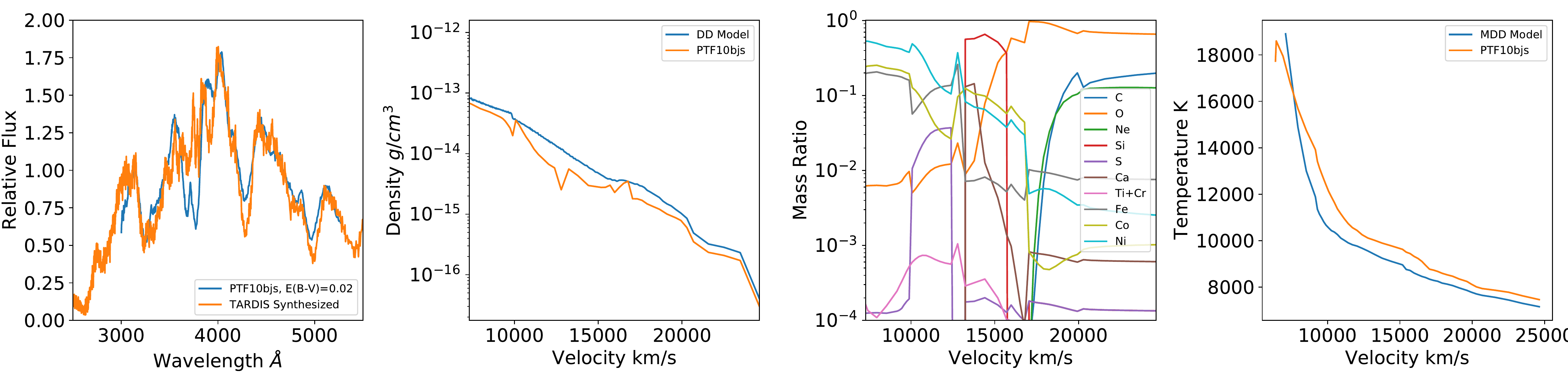}
    \includegraphics[width=\linewidth]{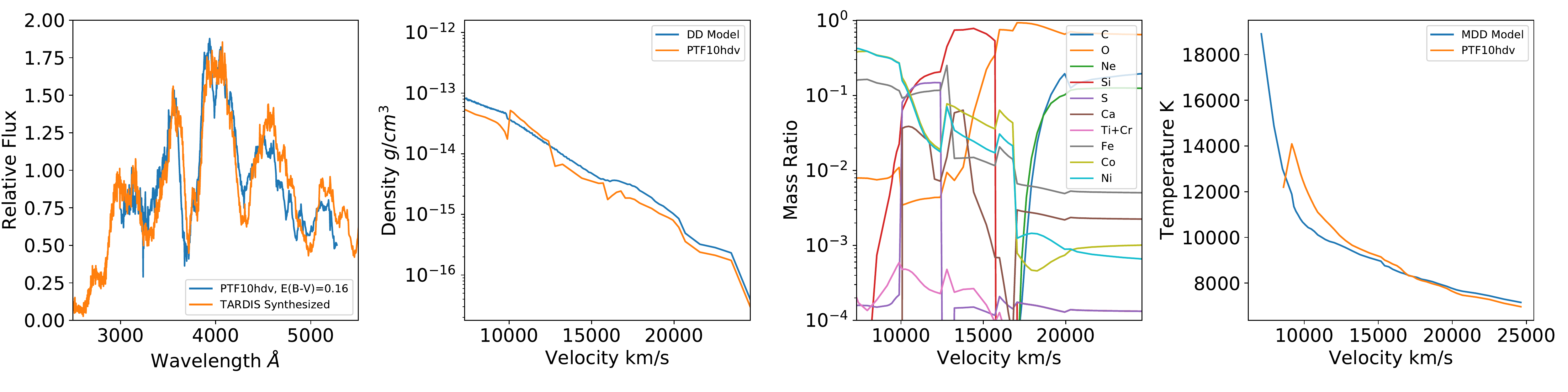}
    \includegraphics[width=\linewidth]{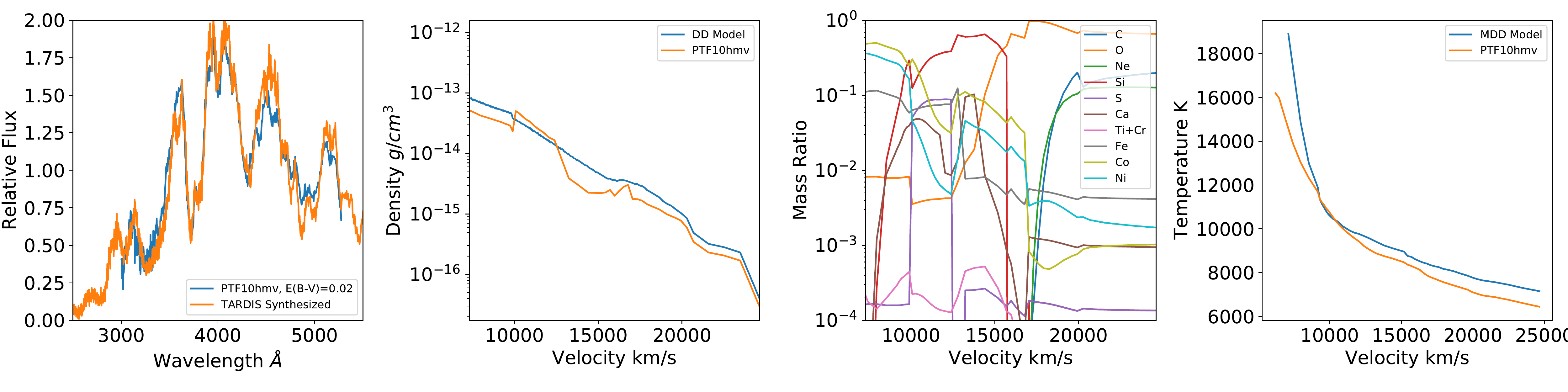}
    \includegraphics[width=\linewidth]{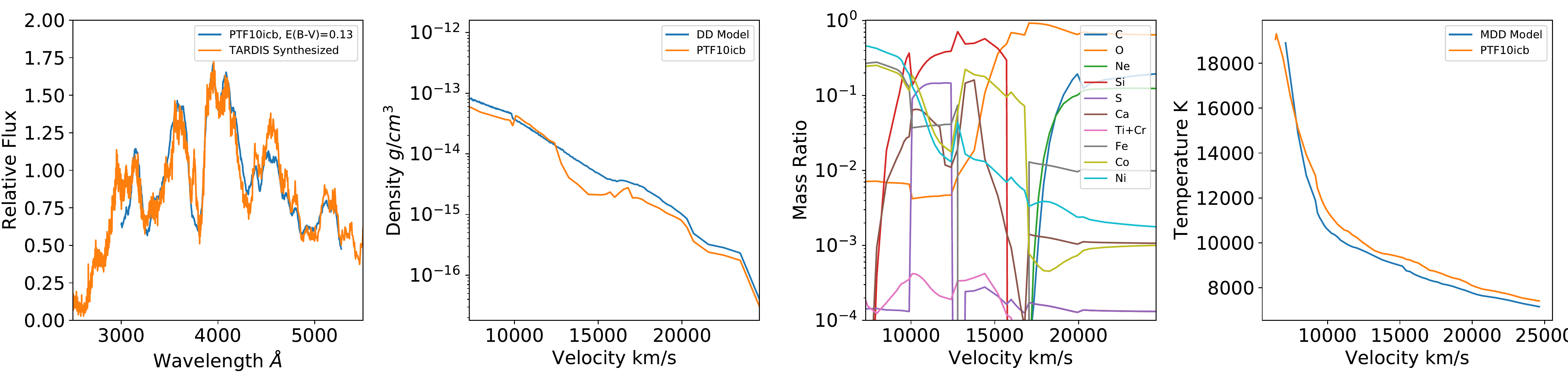}
    \caption{The same as Figure~\ref{fig:hstfitting1}, but for PTF10bjs, PTF10hdv, PTF10hmv, and PTF10icb. 
     %   From Left to Right: 1. The observed spectra (blue line) after dust extinction correction using CCM model, E(B-V) values given in the legend. The TARDIS synthesized spectra (orange line). 2. The density of the ejecta structure for TARDIS spectra calculation (orange line) and the IG model density as a comparison (blue line), all densities are converted to 19 days after explosion. 3. Element abundances of each shells, which is used for TARDIS spectra calculation. 4. Temperature structure for the synthesized spectra (orange line) and the temperature for IG model as a comparison (blue line). 
 }
        \label{fig:hstfitting2}
\end{figure}

\begin{figure}[htb!]
    \includegraphics[width=\linewidth]{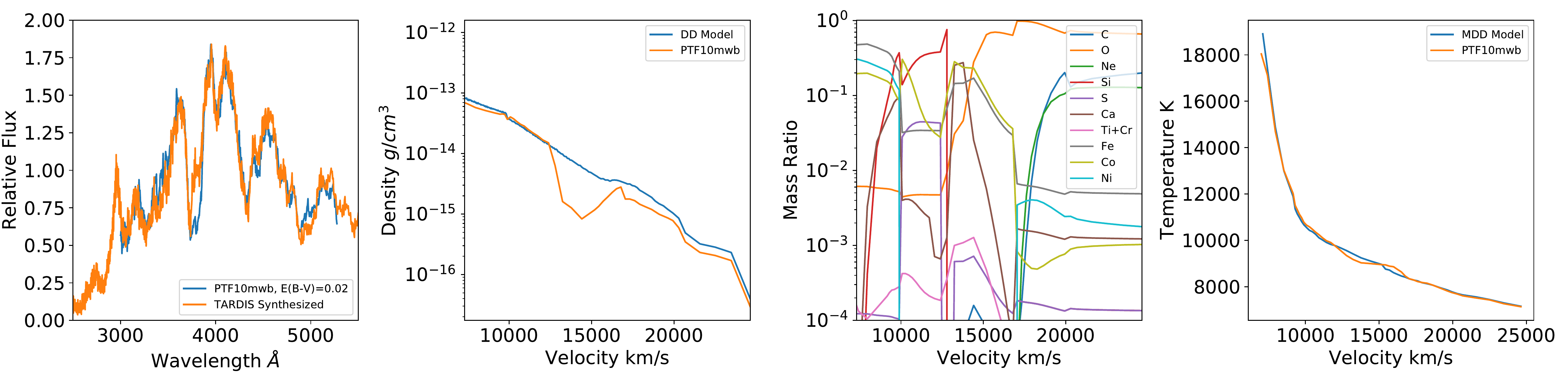}
    \includegraphics[width=\linewidth]{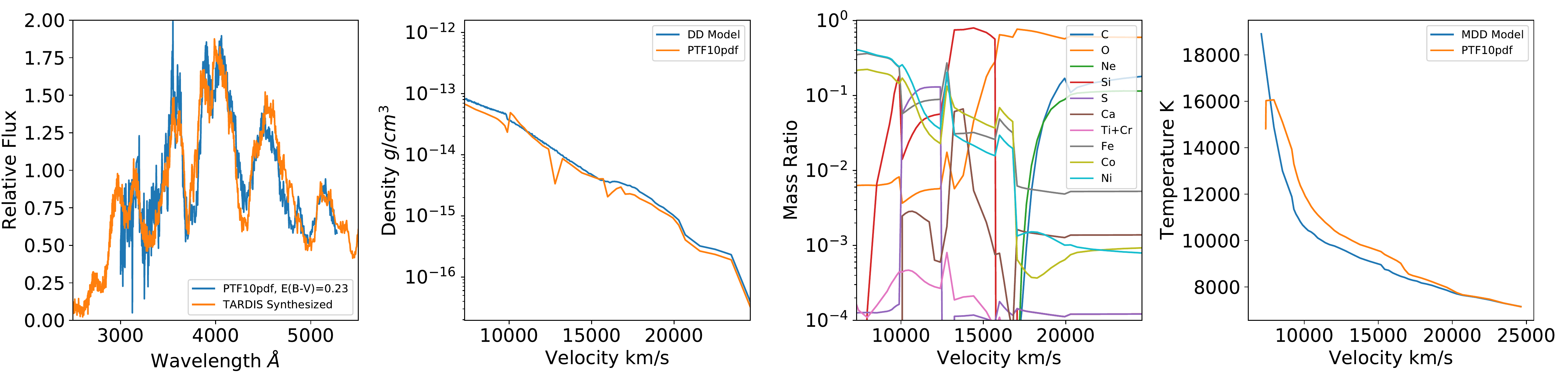}
    \includegraphics[width=\linewidth]{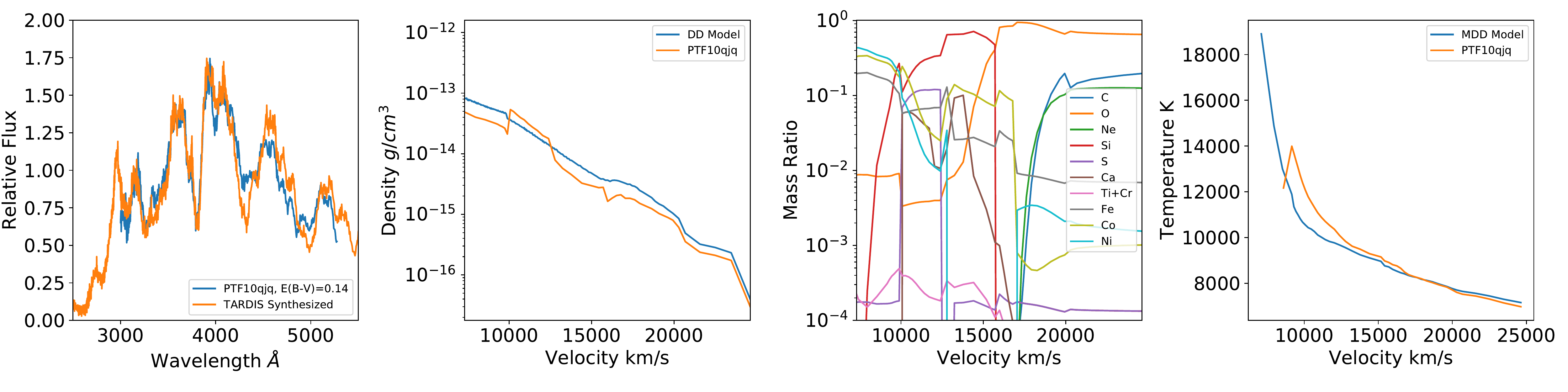}
    \includegraphics[width=\linewidth]{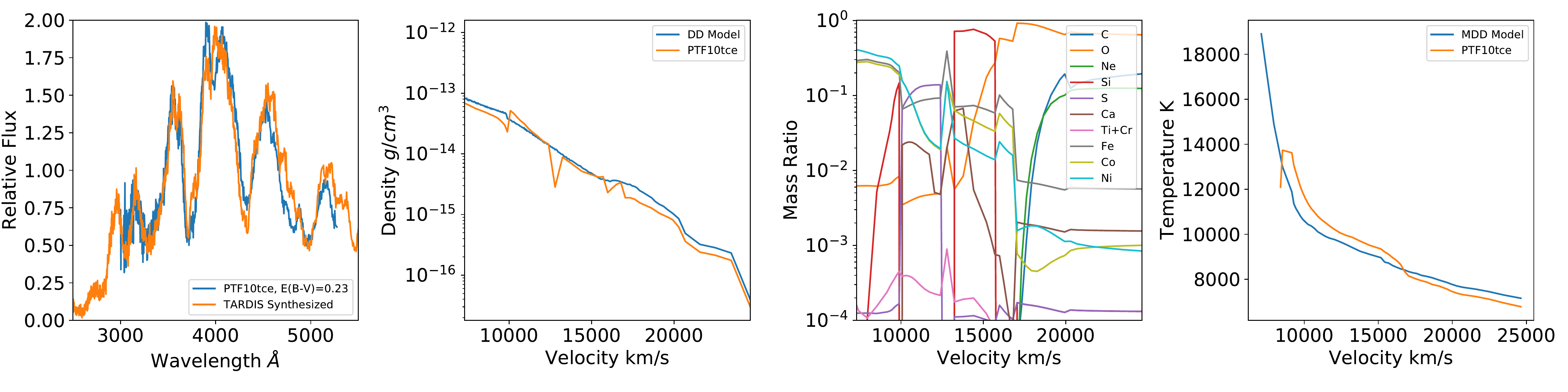}
    \caption{The same as Figure~\ref{fig:hstfitting1}, but for PTF10mwb, PTF10pdf, and PTF10qjq.}
    %From Upper to Lower: PTF10mwb, PTF10pdf, PTF10qjq, PTF10tce. From Left to Right: 1. The observed spectra (blue line) after dust extinction correction using CCM model, E(B-V) values given in the legend. The TARDIS synthesized spectra (orange line). 2. The density of the ejecta structure for TARDIS spectra calculation (orange line) and the IG model density as a comparison (blue line), all densities are converted to 19 days after explosion. 3. Element abundances of each shells, which is used for TARDIS spectra calculation. 4. Temperature structure for the synthesized spectra (orange line) and the temperature for IG model as a comparison (blue line). }
        \label{fig:hstfitting3}
\end{figure}

\begin{figure}
    \includegraphics[width=\linewidth]{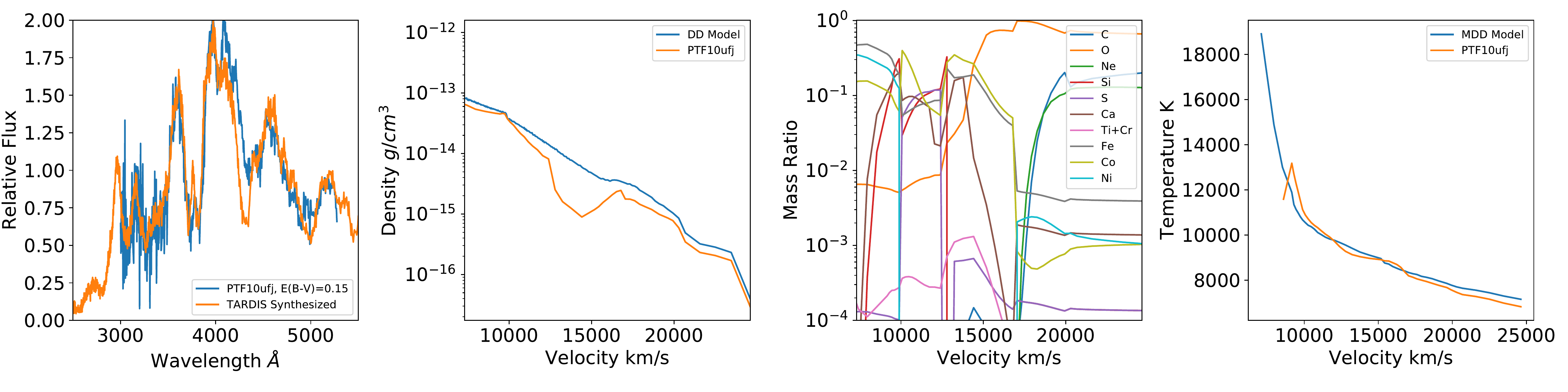}
    \includegraphics[width=\linewidth]{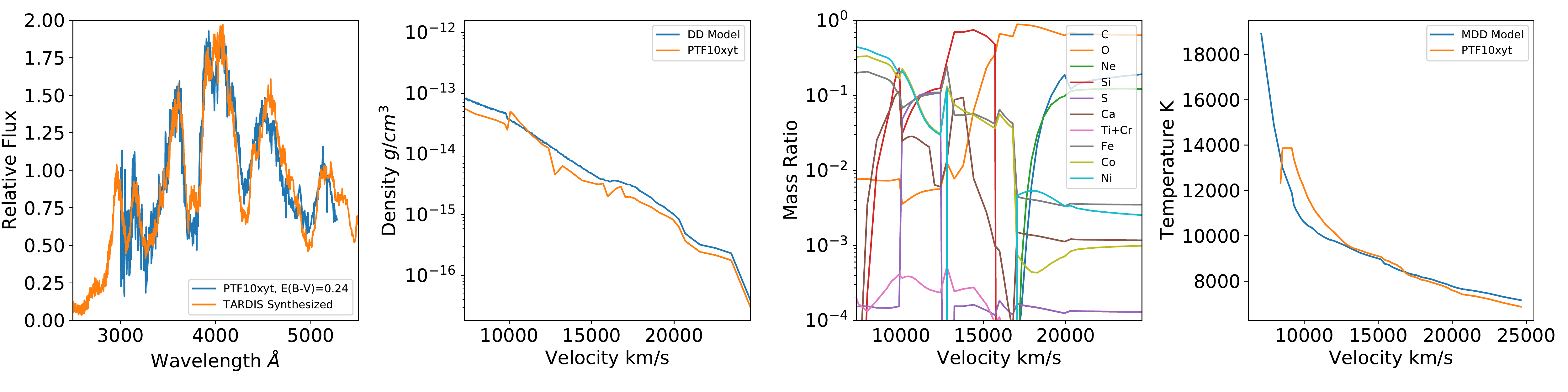}
    \includegraphics[width=\linewidth]{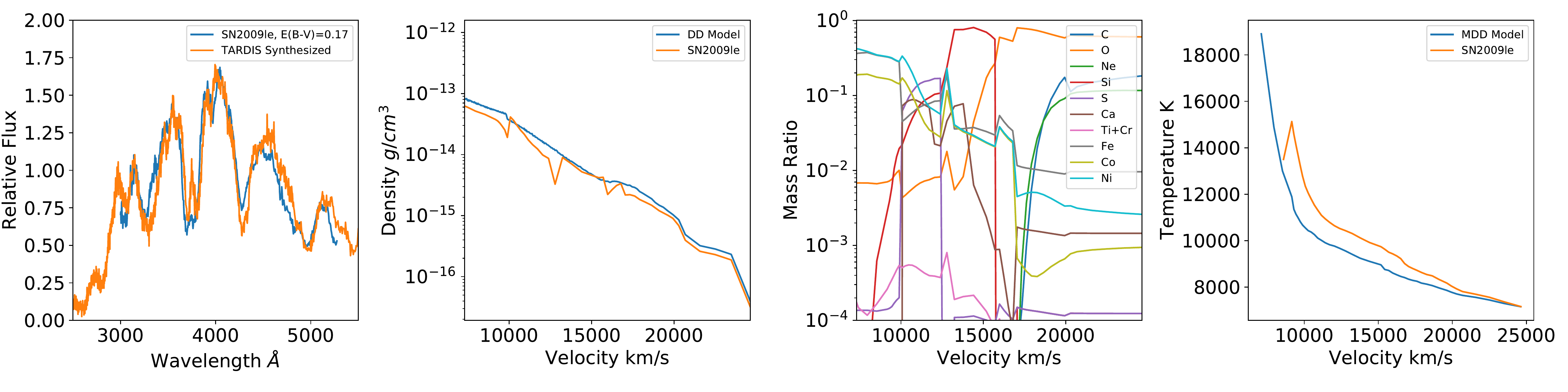}
    \caption{The same as Figure~\ref{fig:hstfitting1}, but for 
    PTF10ufj, PTF10xyt, and SN2009le.}
    %From Upper to Lower: PTF10ufj, PTF10xyt, SN2009le. From Left to Right: 1. The observed spectra (blue line) after dust extinction correction using CCM model, E(B-V) values given in the legend. The TARDIS synthesized spectra (orange line). 2. The density of the ejecta structure for TARDIS spectra calculation (orange line) and the IG model density as a comparison (blue line), all densities are converted to 19 days after explosion. 3. Element abundances of each shells, which is used for TARDIS spectra calculation. 4. Temperature structure for the synthesized spectra (orange line) and the temperature for IG model as a comparison (blue line). }
        \label{fig:hstfitting4}
\end{figure}

In this Section, we present the spectral fitting results of the 15 HST UV spectra, including the synthetic spectra, elemental abundances, density and the temperature profiles (Figures~\ref{fig:hstfitting1}, Figures~\ref{fig:hstfitting2}, Figures~\ref{fig:hstfitting3}, and Figures~\ref{fig:hstfitting4}). 

\clearpage

\section{The Time Evolution of Elements}\label{sec:appendix:TimeEvol}

\begin{figure}[htb!]
    \minipage{0.33\textwidth}
        \includegraphics[width=\textwidth]{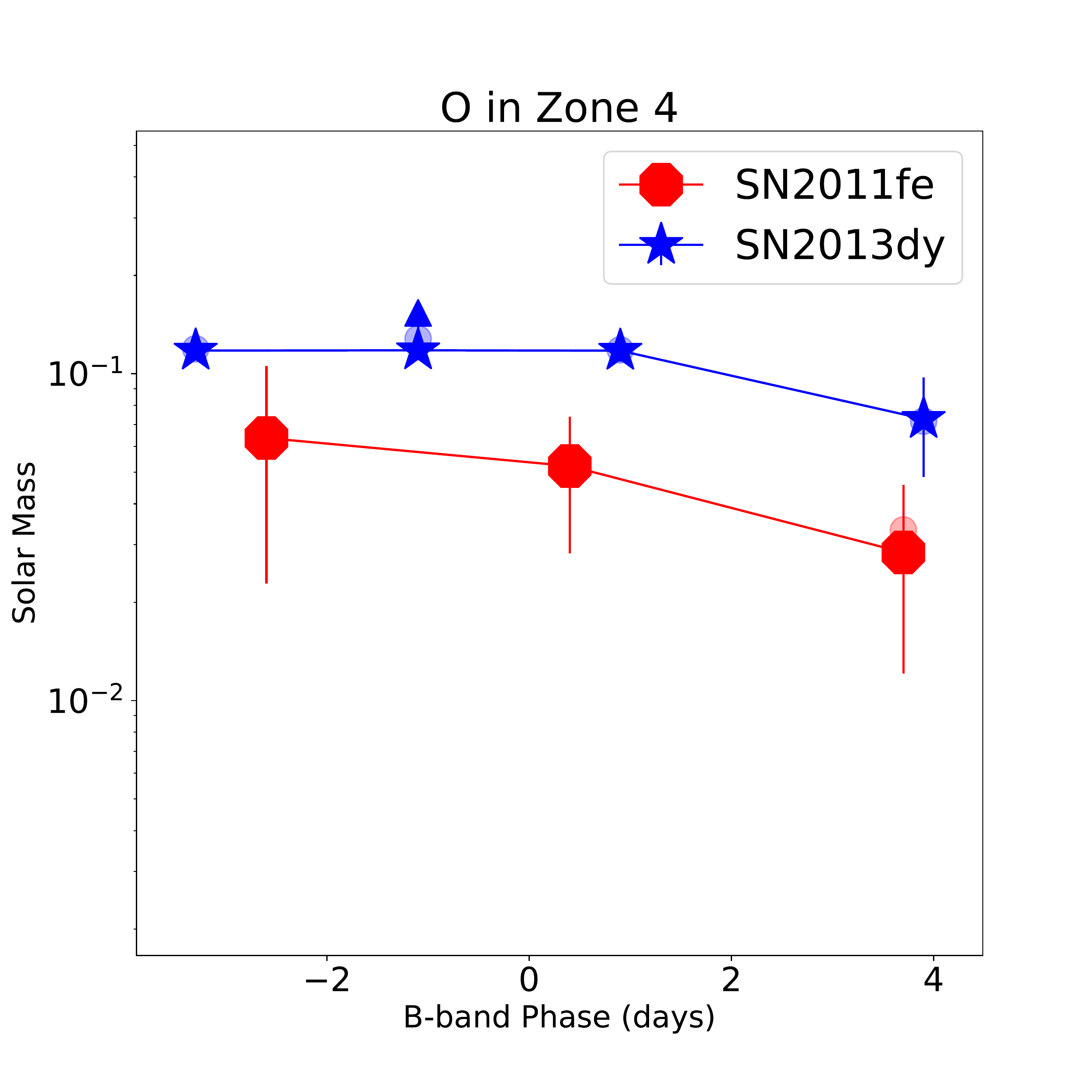}
    \endminipage\hfill
    \minipage{0.33\textwidth}
        \includegraphics[width=\textwidth]{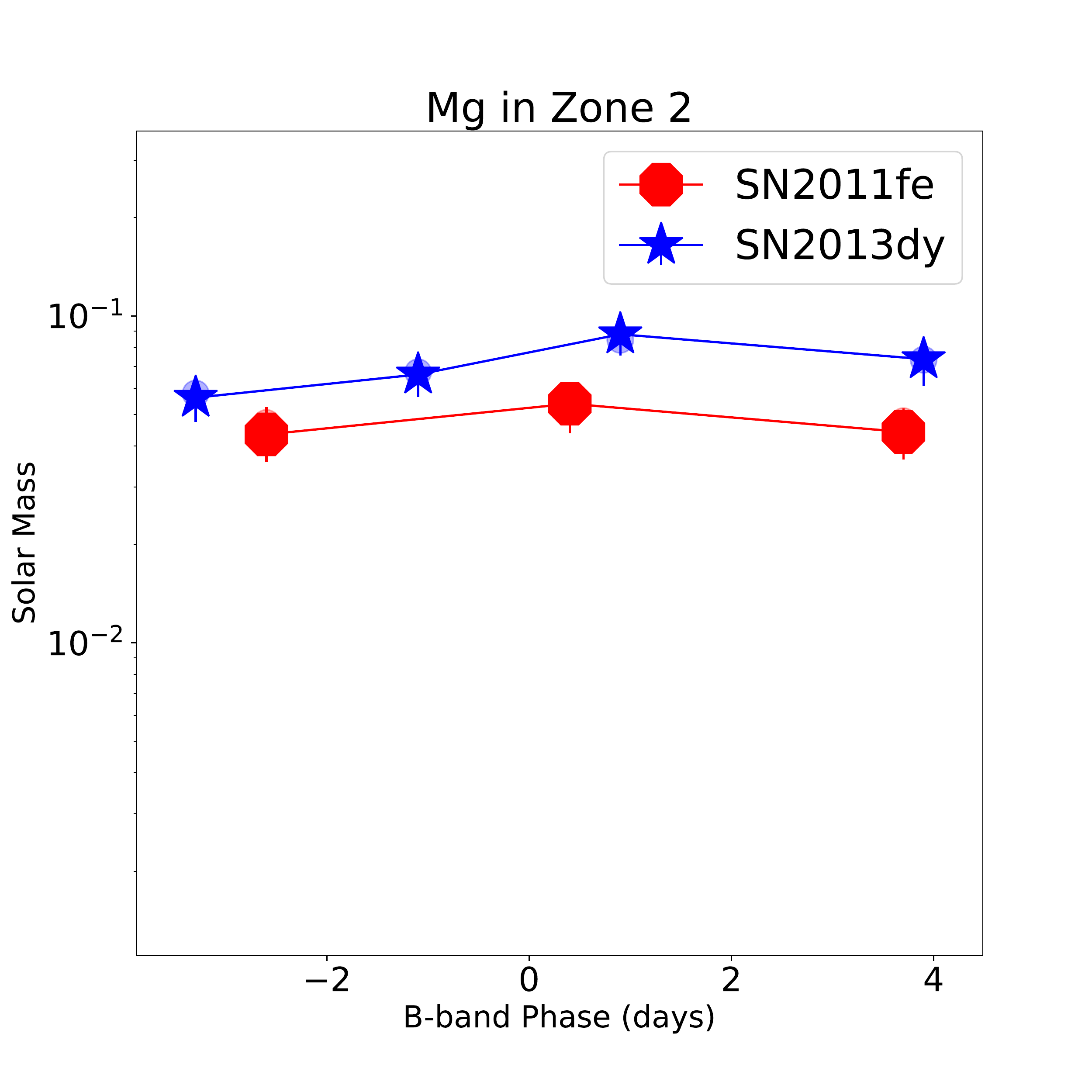}
    \endminipage\hfill
    \minipage{0.33\textwidth}
        \includegraphics[width=\textwidth]{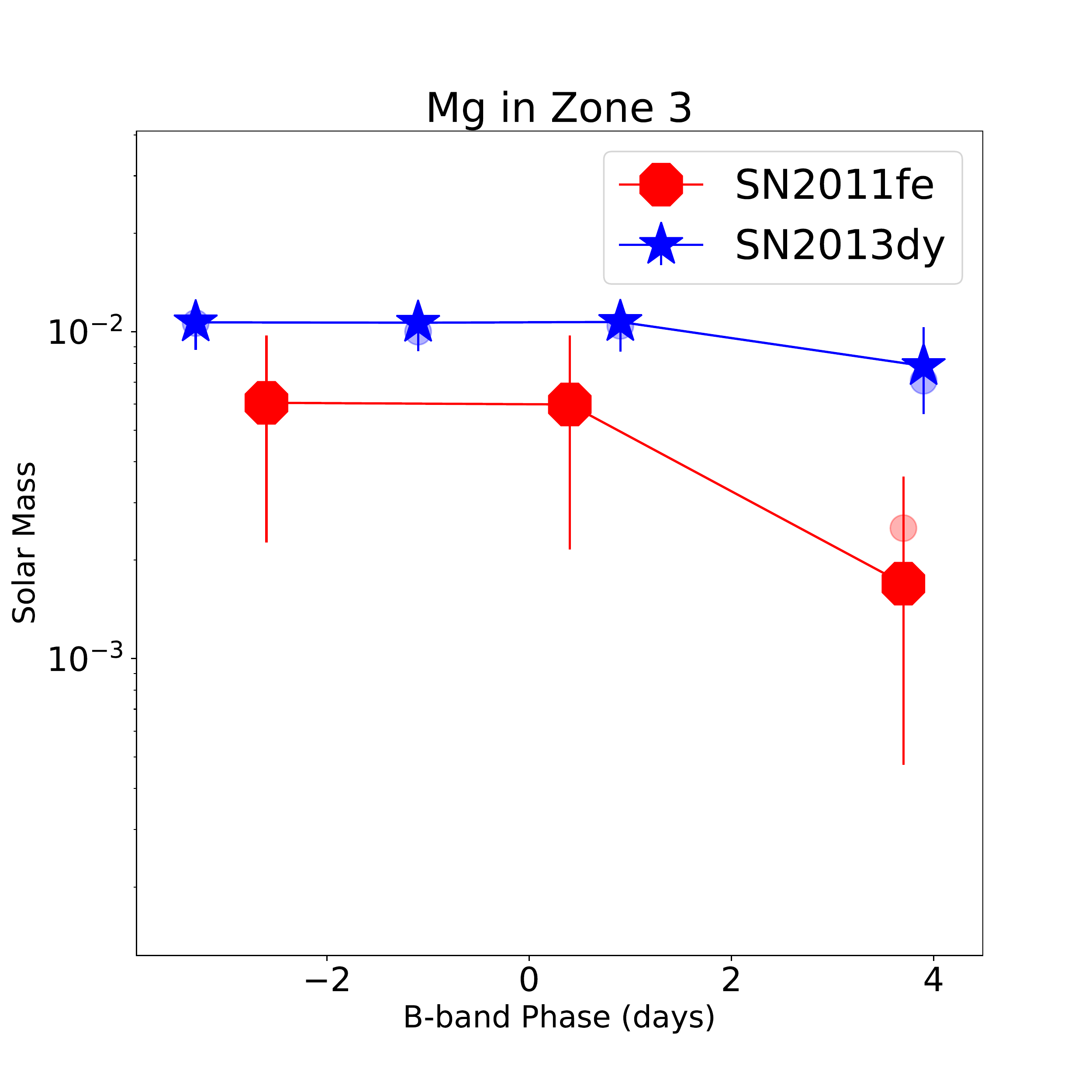}
    \endminipage\hfill
    \minipage{0.33\textwidth}
        \includegraphics[width=\textwidth]{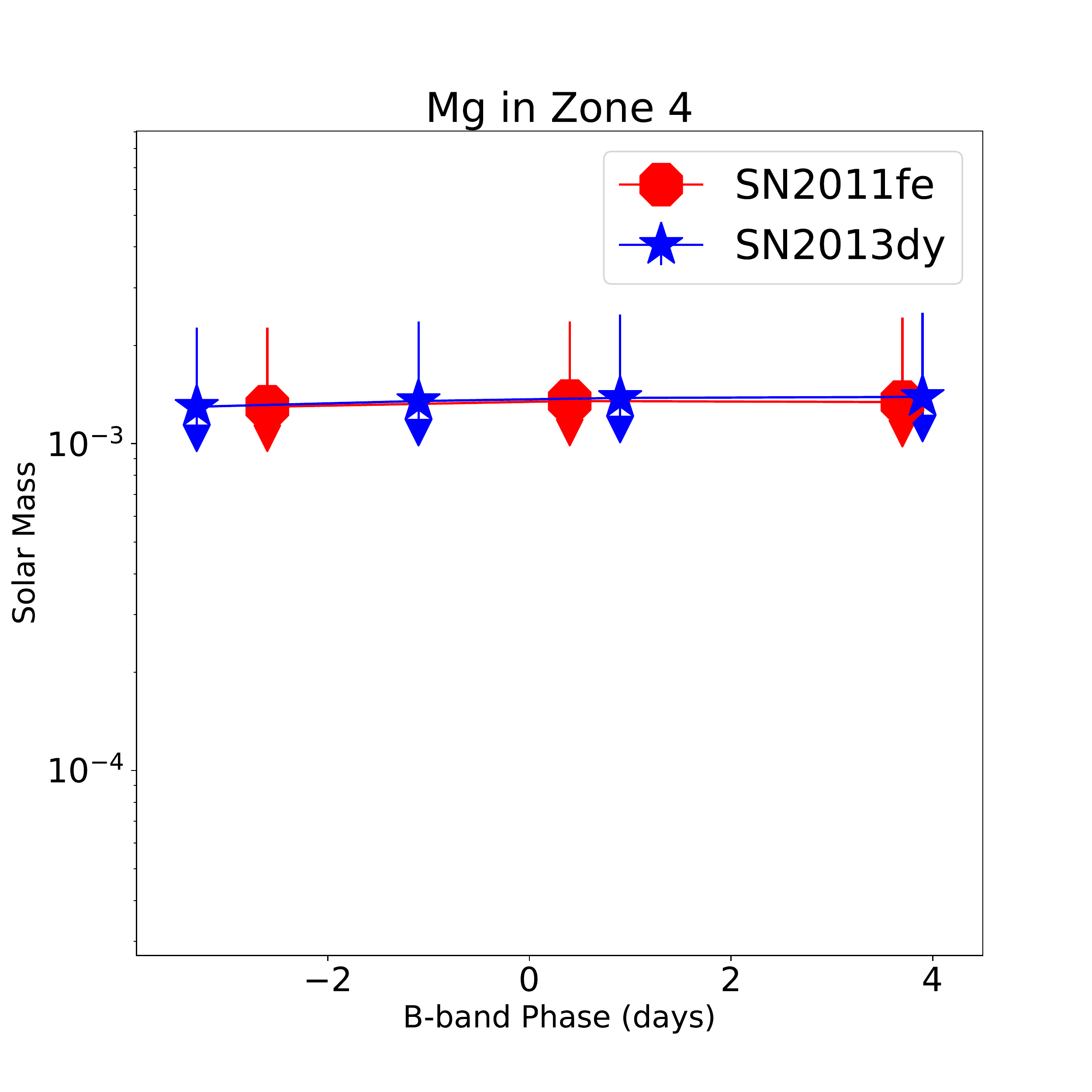}
    \endminipage\hfill
    \minipage{0.33\textwidth}
        \includegraphics[width=\textwidth]{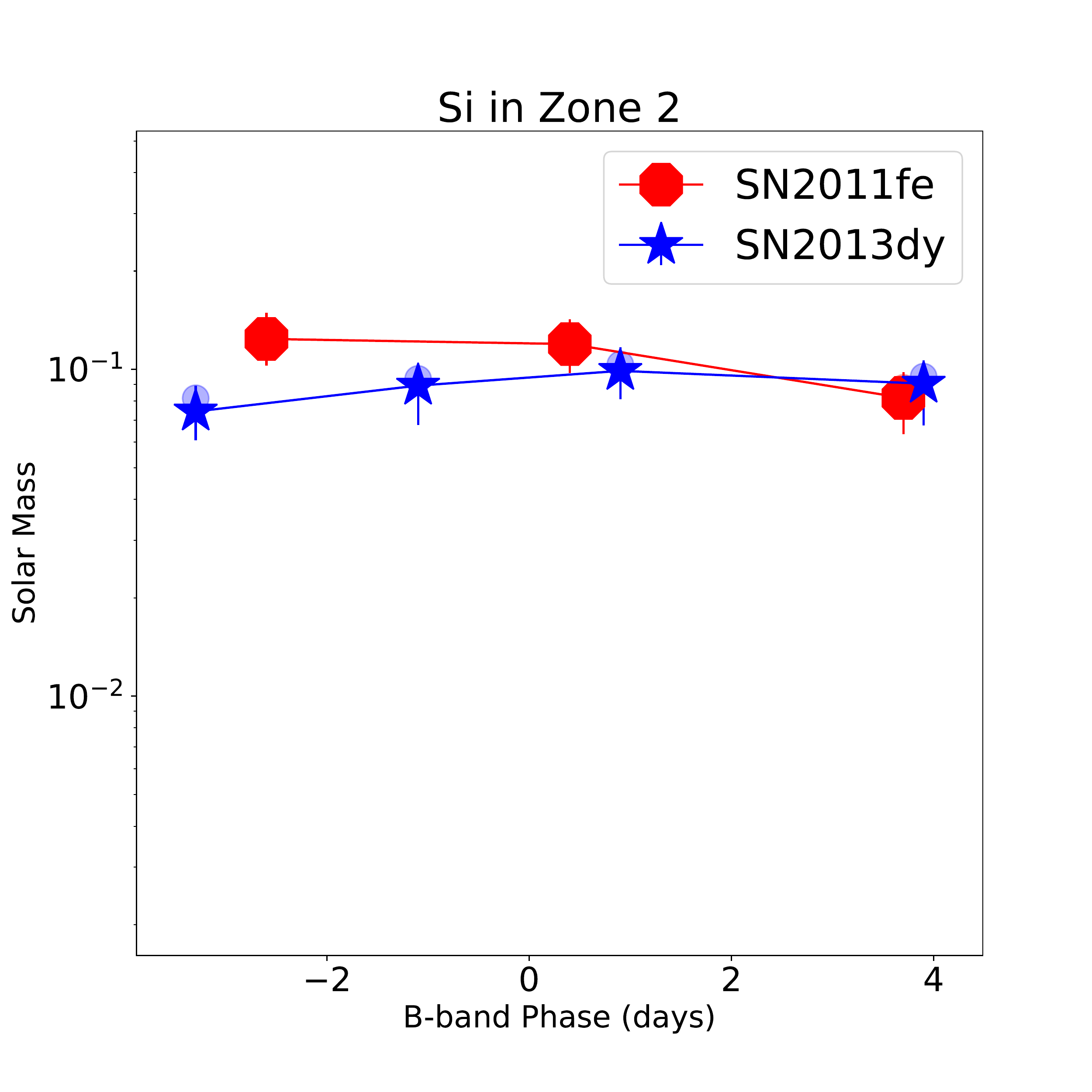}
    \endminipage\hfill
    \minipage{0.33\textwidth}
        \includegraphics[width=\textwidth]{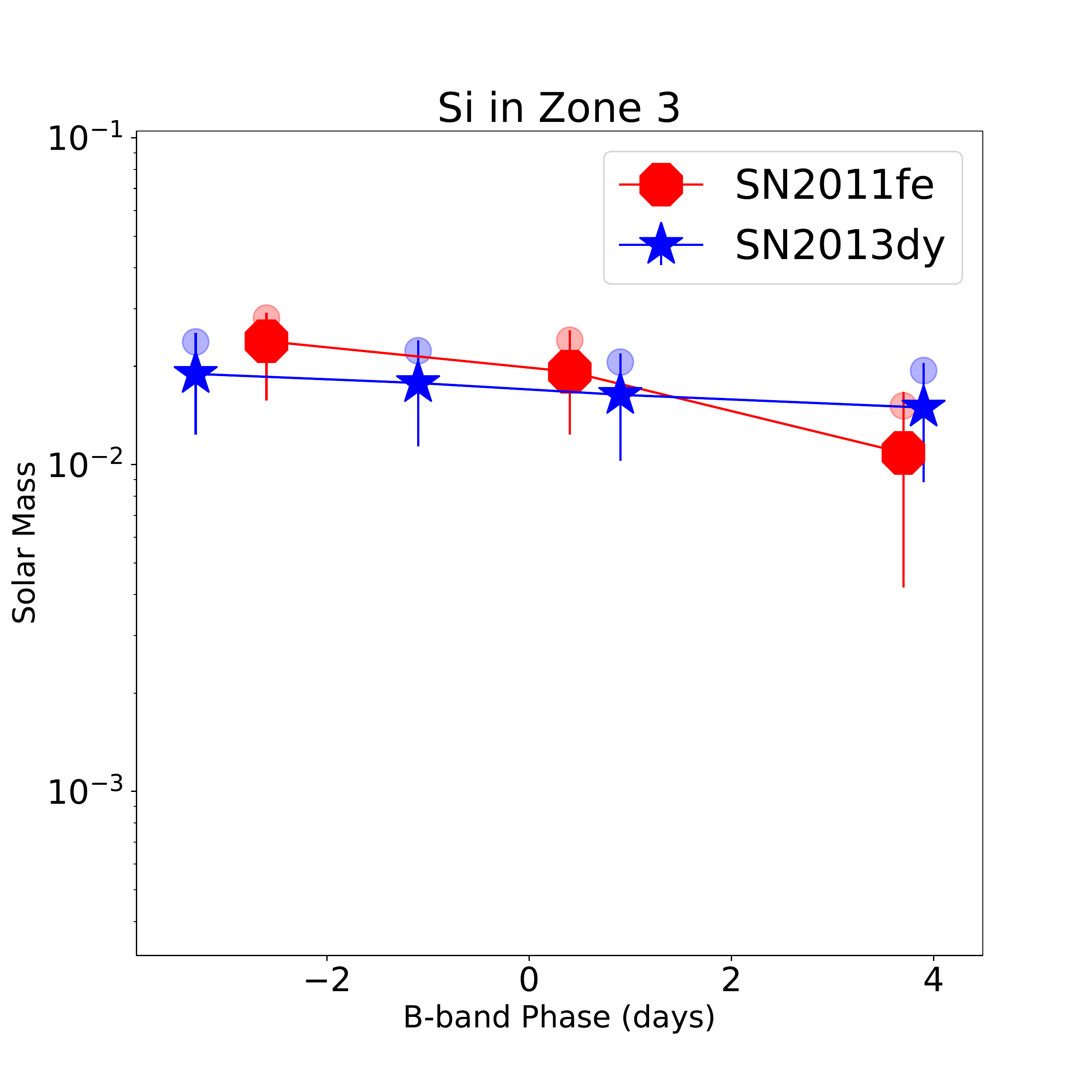}
    \endminipage\hfill
    \minipage{0.33\textwidth}
        \includegraphics[width=\textwidth]{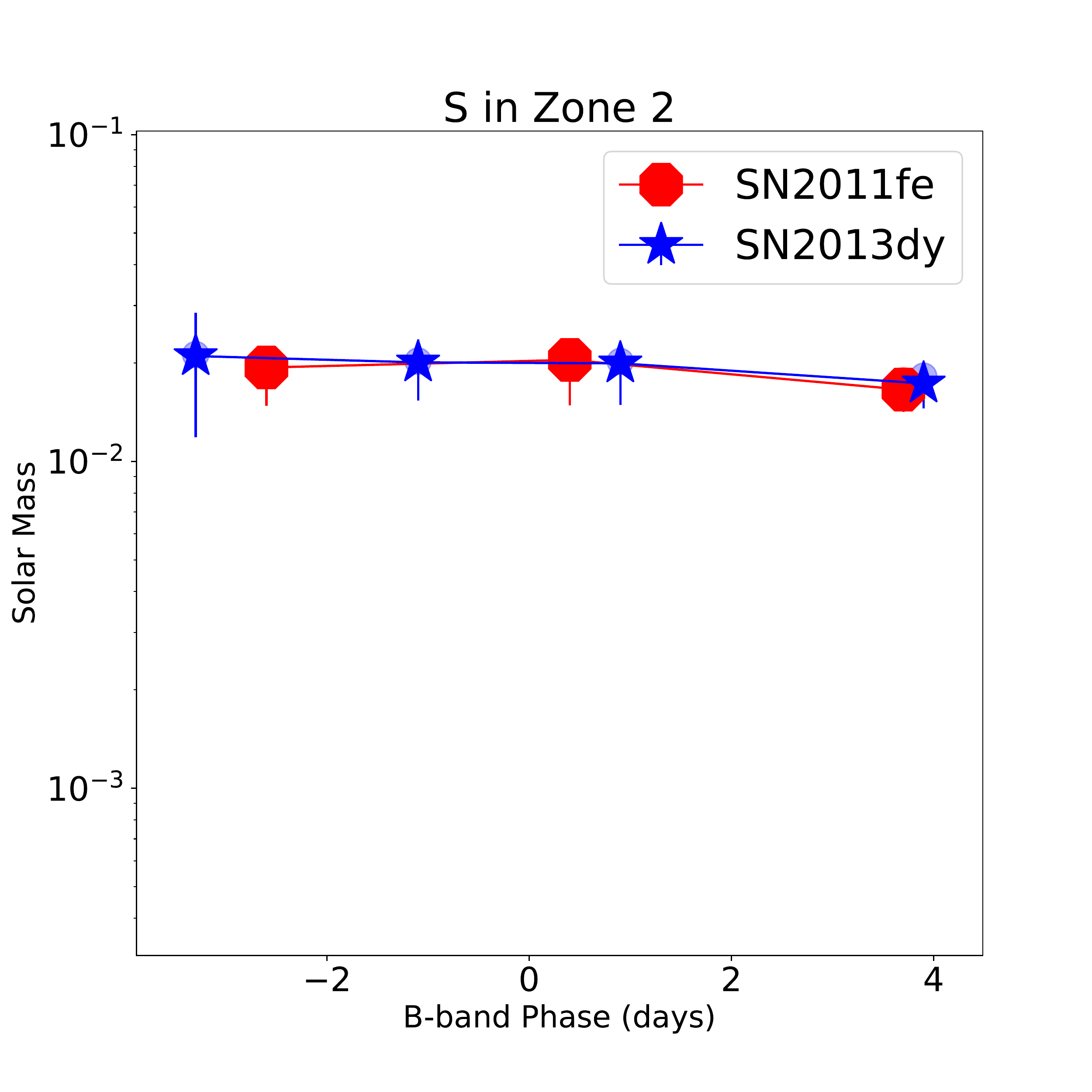}
    \endminipage\hfill
    \minipage{0.33\textwidth}
        \includegraphics[width=\textwidth]{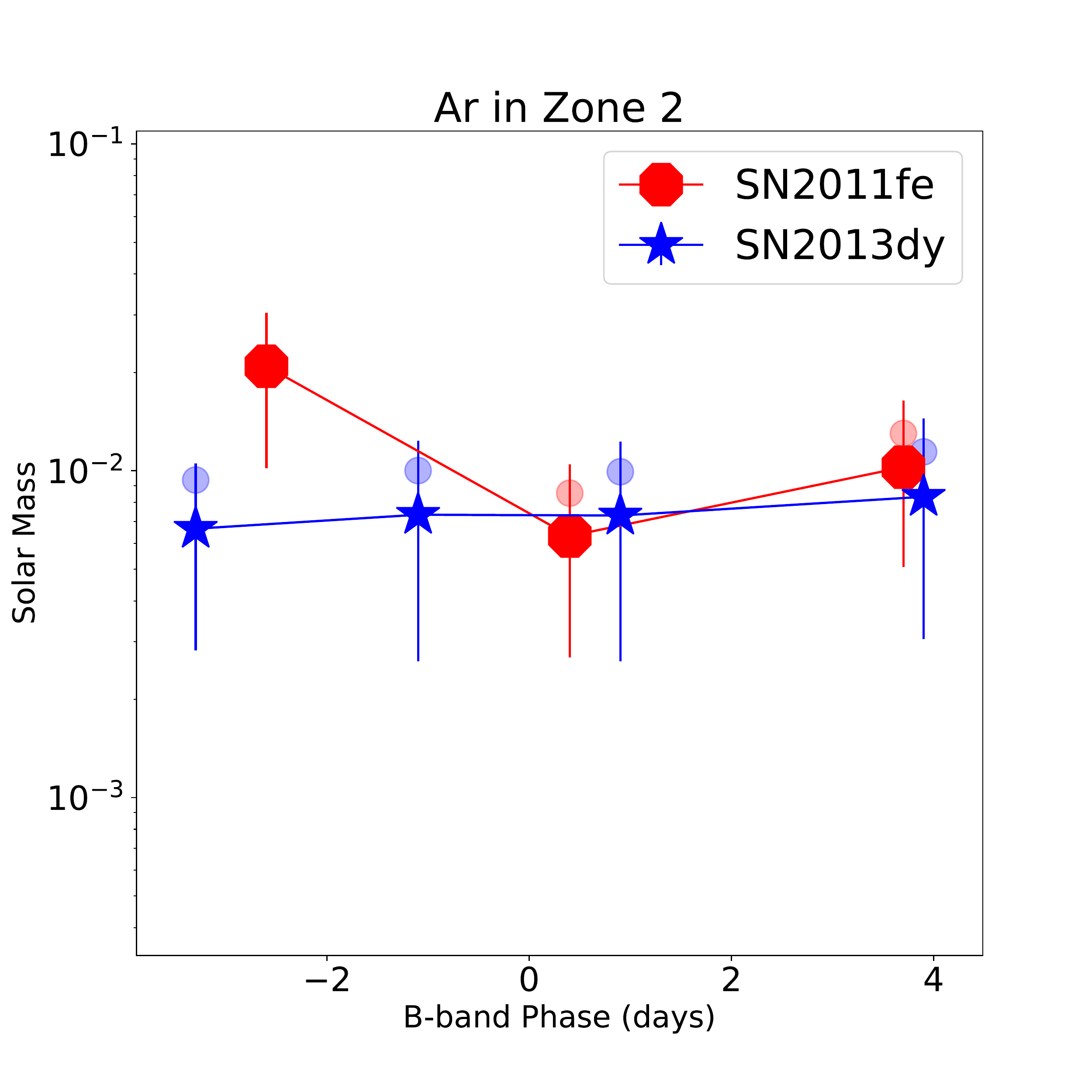}
    \endminipage\hfill
    \minipage{0.33\textwidth}
        \includegraphics[width=\textwidth]{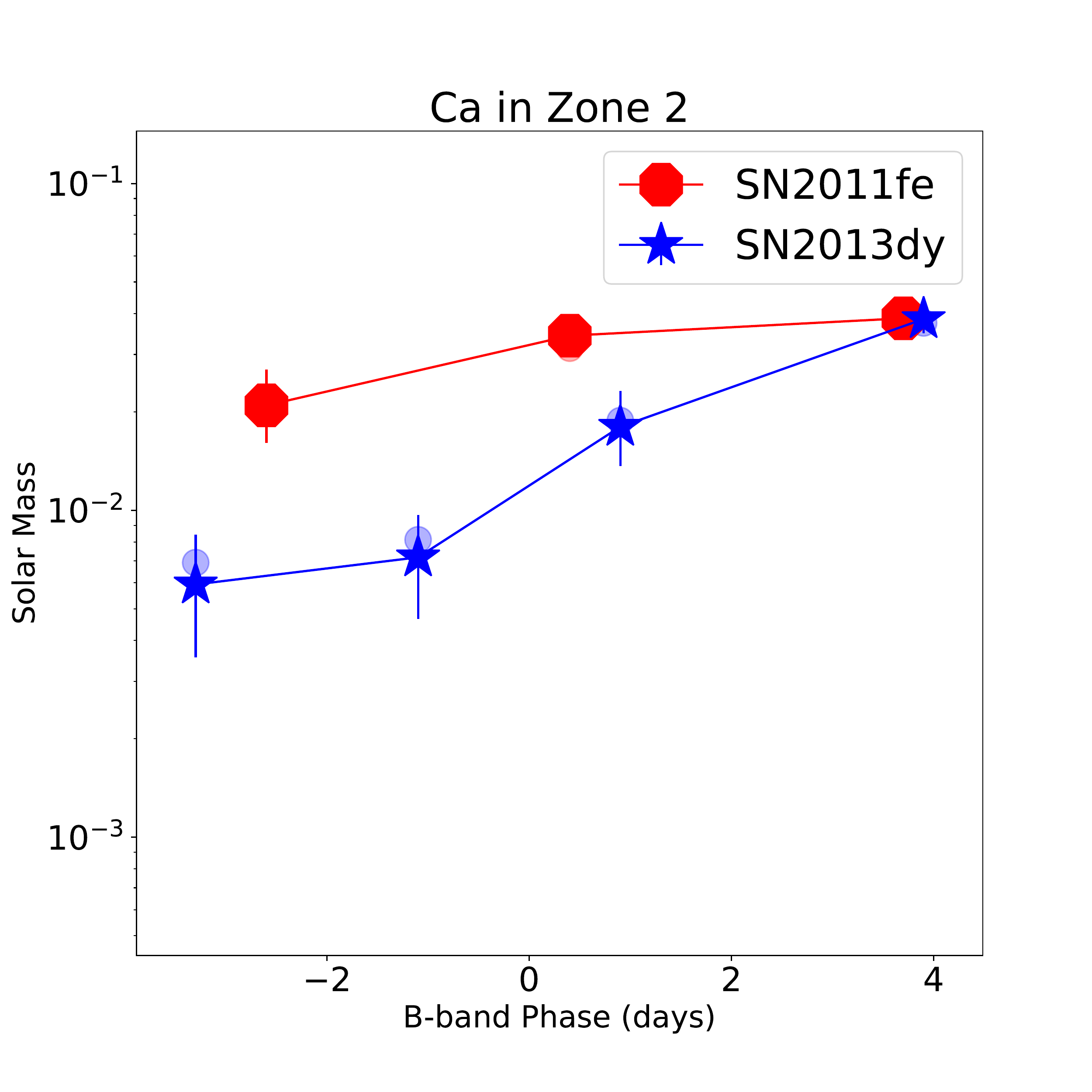}
    \endminipage\hfill
    \caption{The masses of elements in different zones predicted by the neural network, and their  evolution  with the  time after explosion. Red lines are for SN~2011fe, blue lines are for  SN~2013dy. Transparent circular dots are the 
    predictions from neural network, the crosses with error bars are the median value from the testing dataset, error bars indicate the $1-\sigma$ limits, and upper and lower limits are marked with triangles. The elements and zones are labeled on the titles of every panels. \label{fig:RestElementEvolution1}}
\end{figure}

\begin{figure}[htb!]
    \minipage{0.33\textwidth}
        \includegraphics[width=\textwidth]{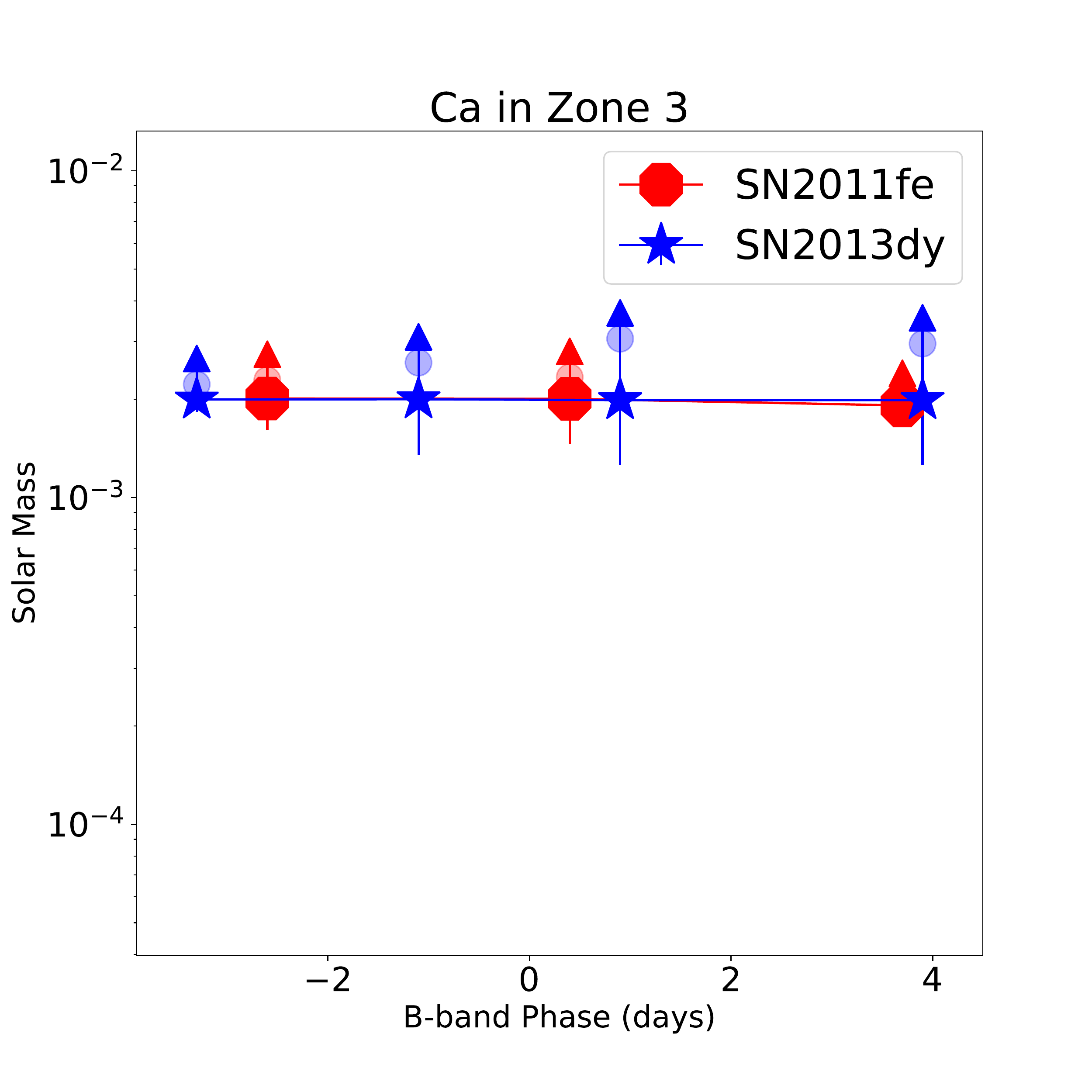}
    \endminipage\hfill
    \minipage{0.33\textwidth}
        \includegraphics[width=\textwidth]{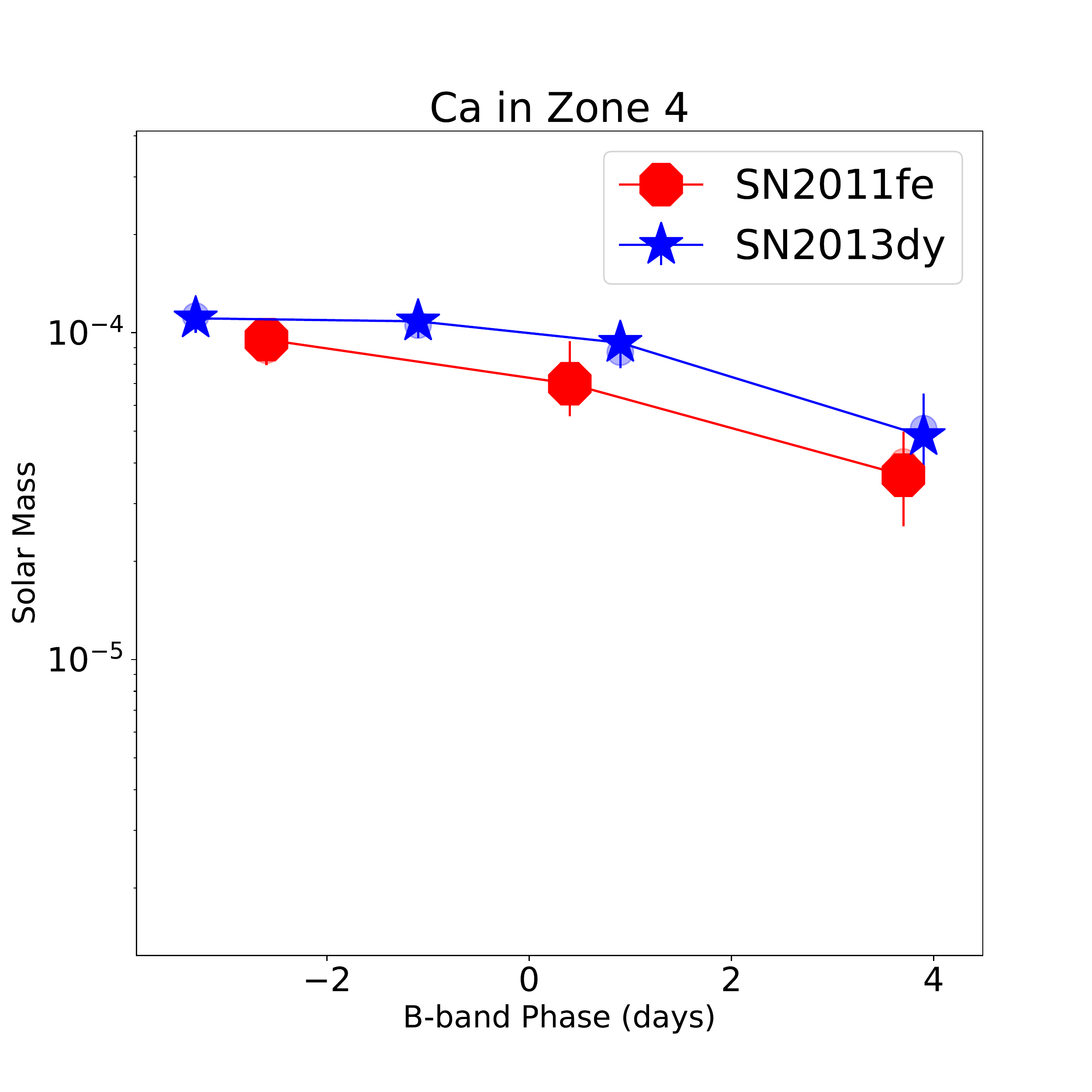}
    \endminipage\hfill
    \minipage{0.33\textwidth}
        \includegraphics[width=\textwidth]{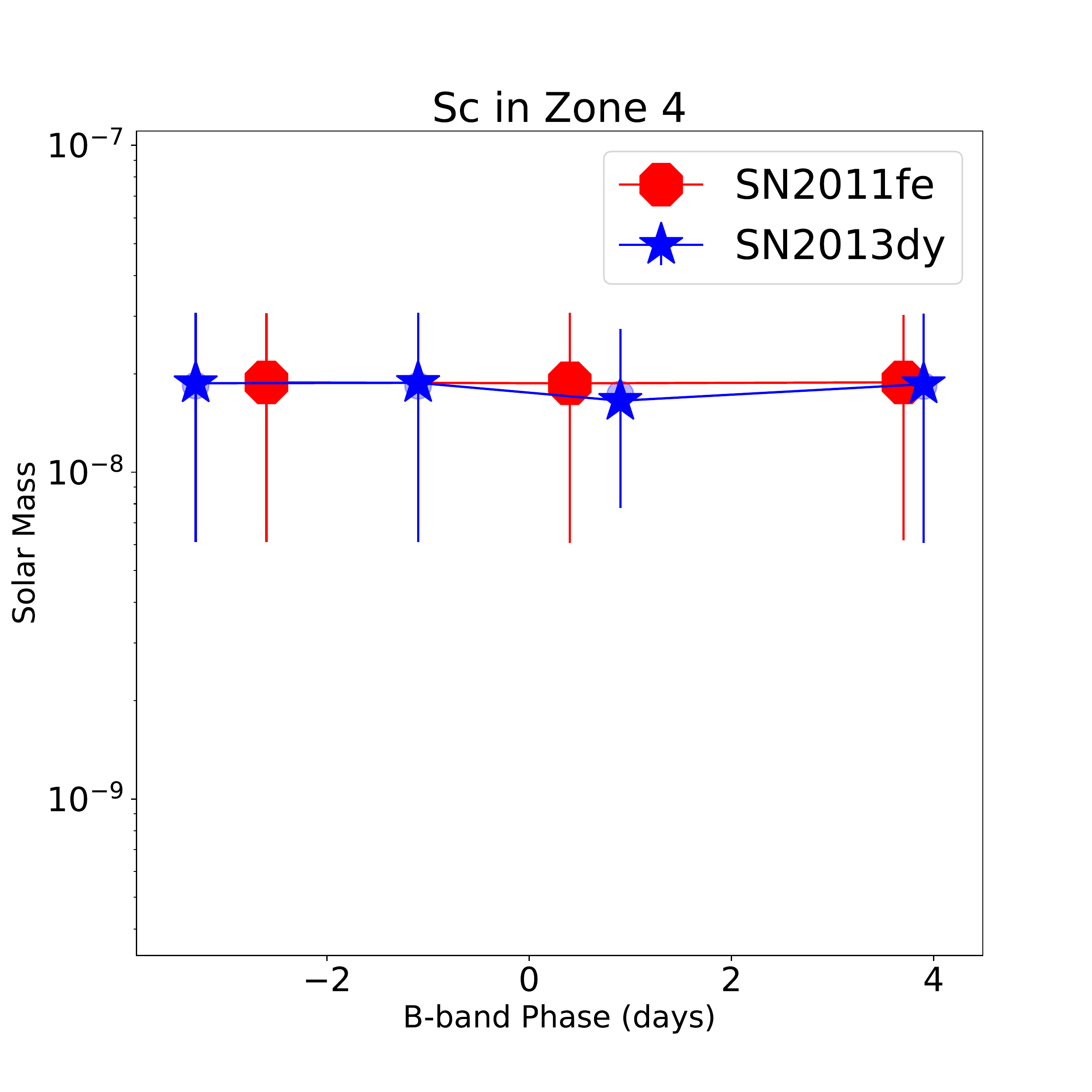}
    \endminipage\hfill
    \minipage{0.33\textwidth}
        \includegraphics[width=\textwidth]{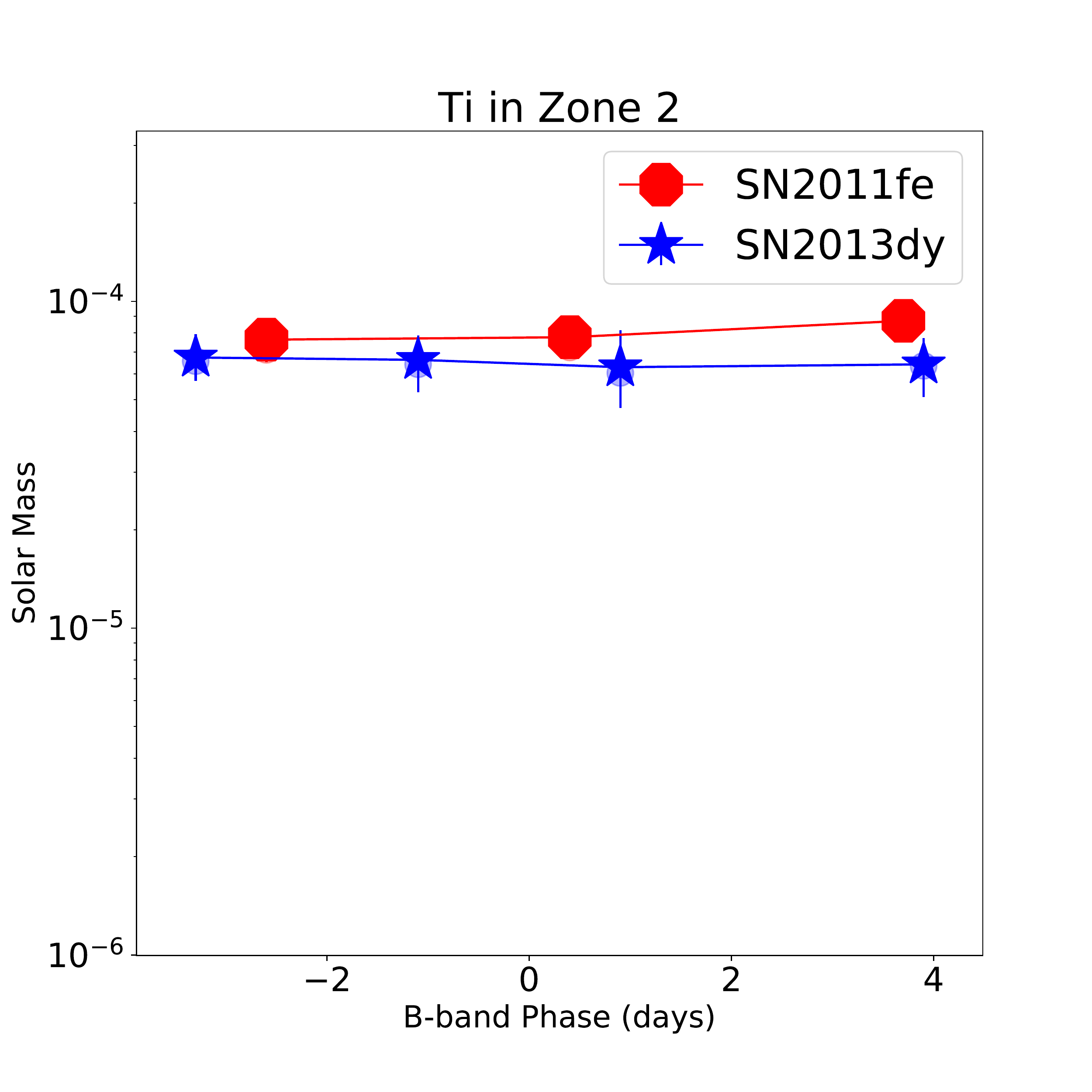}
    \endminipage\hfill
    \minipage{0.33\textwidth}
        \includegraphics[width=\textwidth]{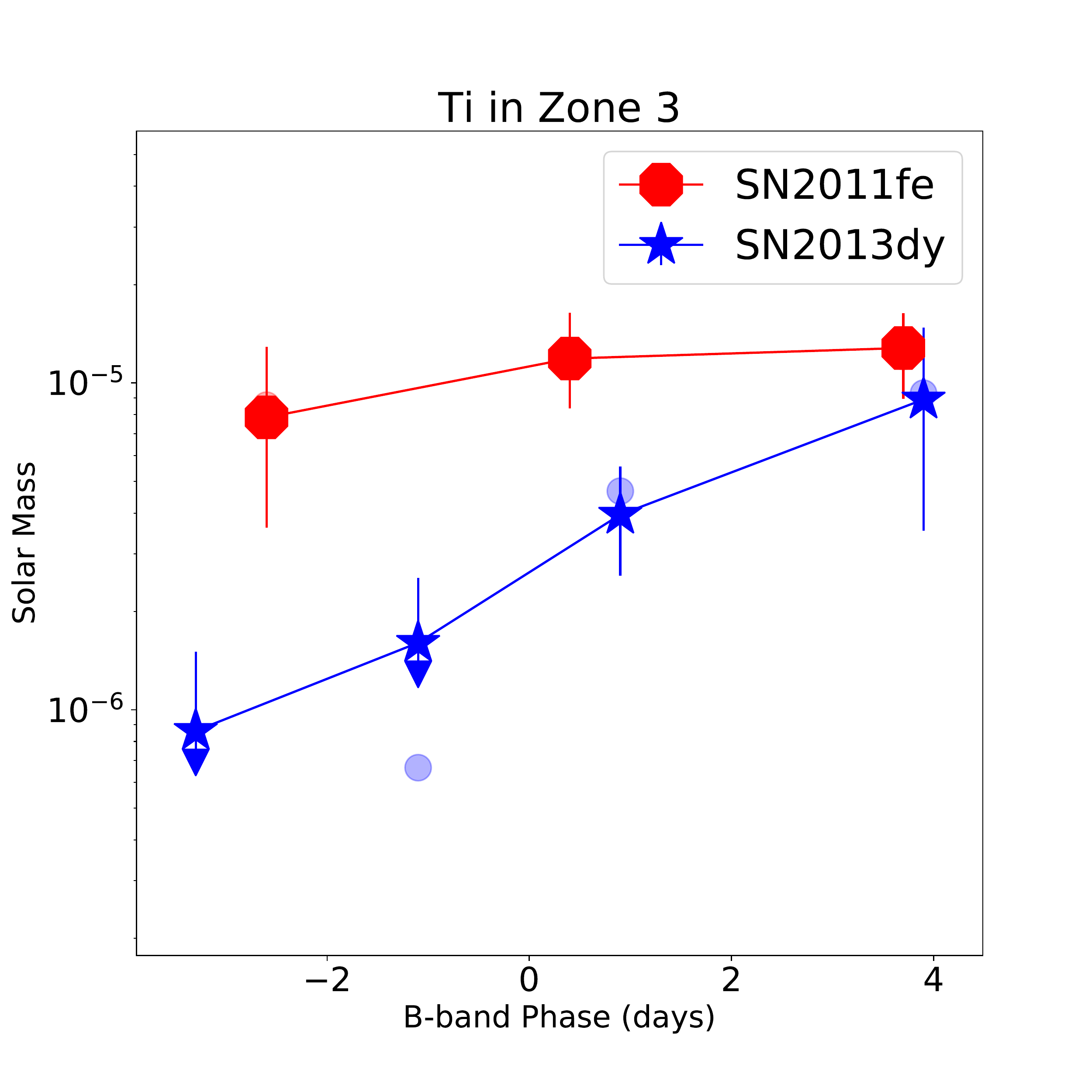}
    \endminipage\hfill
    \minipage{0.33\textwidth}
        \includegraphics[width=\textwidth]{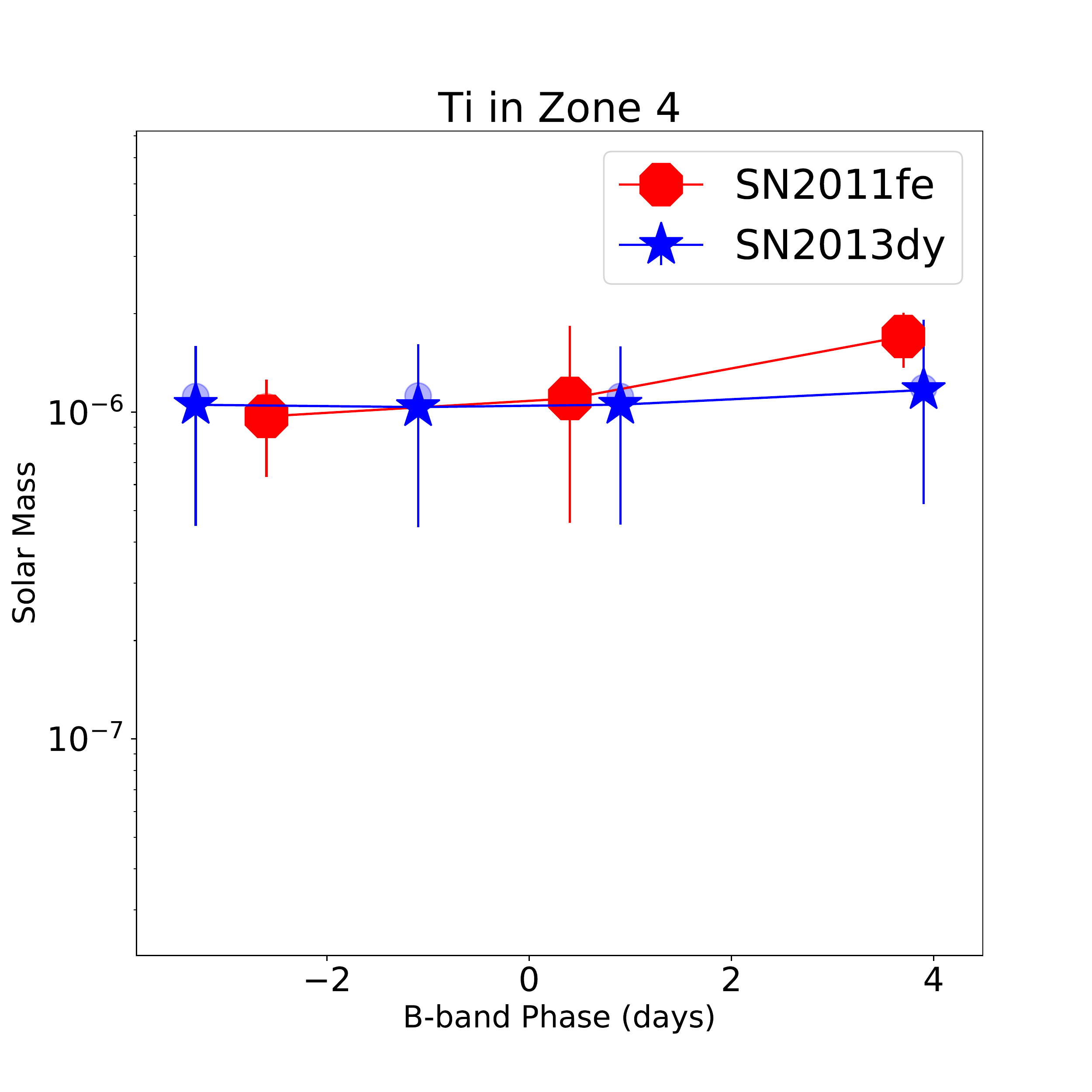}
    \endminipage\hfill
    \minipage{0.33\textwidth}
        \includegraphics[width=\textwidth]{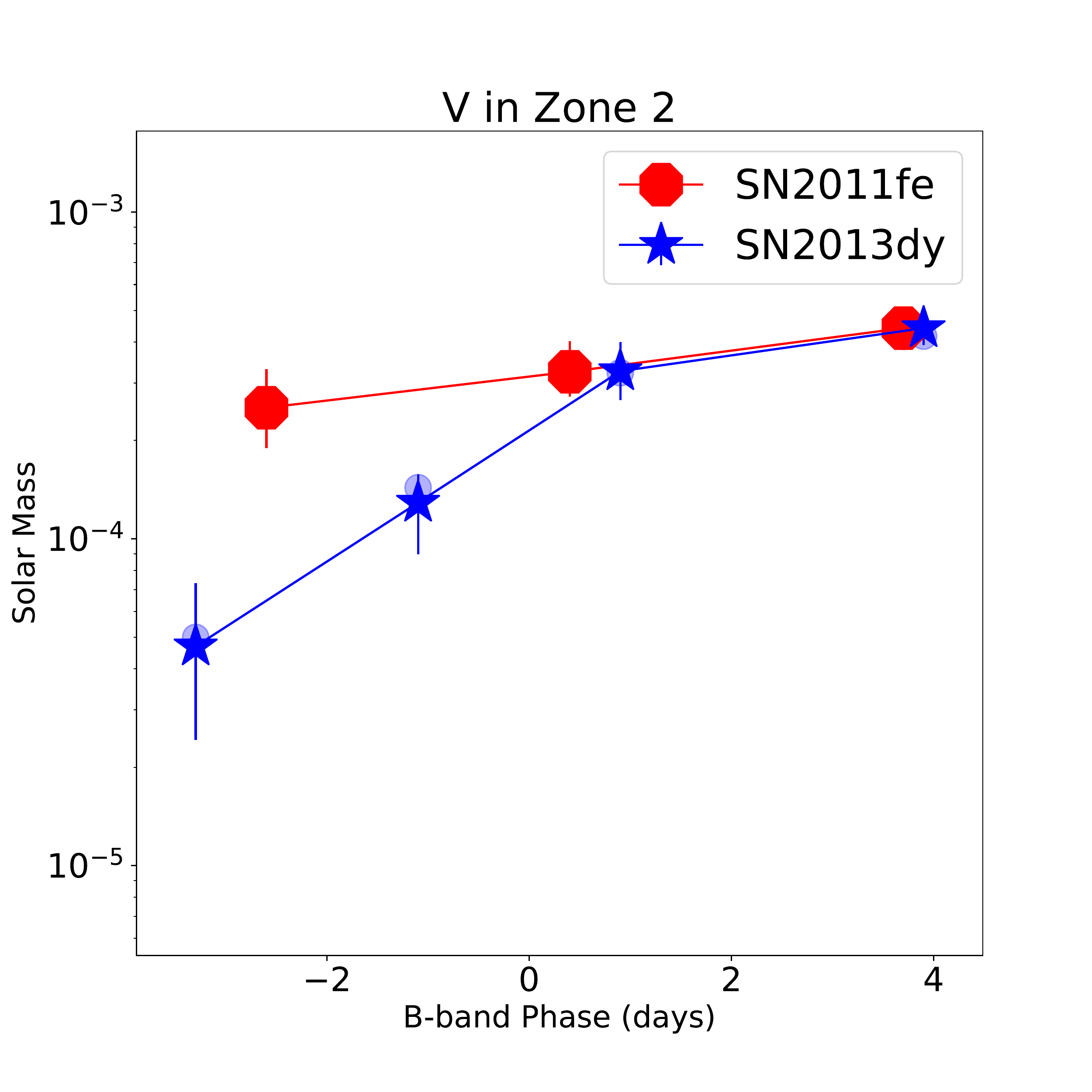}
    \endminipage\hfill
    \minipage{0.33\textwidth}
        \includegraphics[width=\textwidth]{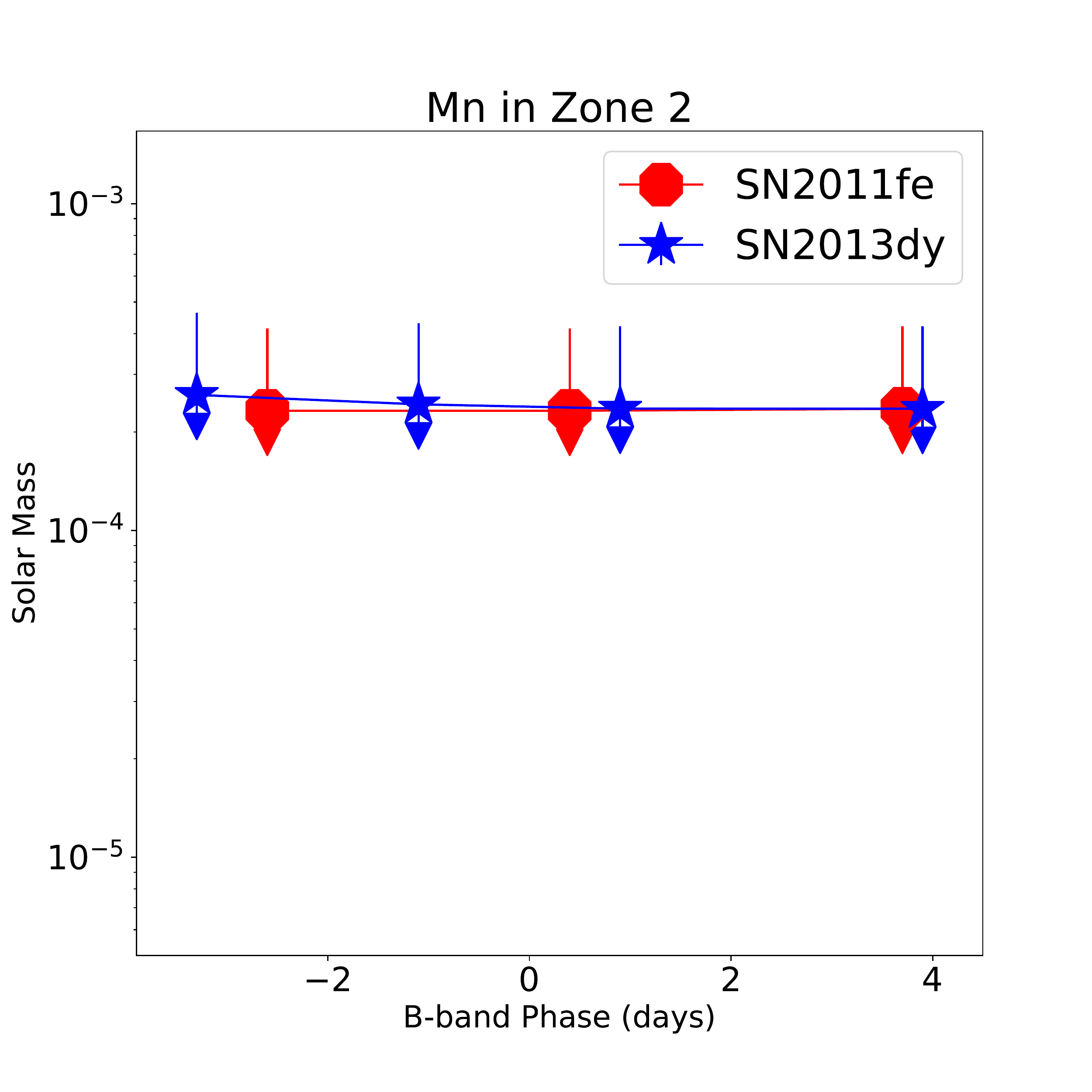}
    \endminipage\hfill
    \minipage{0.33\textwidth}
        \includegraphics[width=\textwidth]{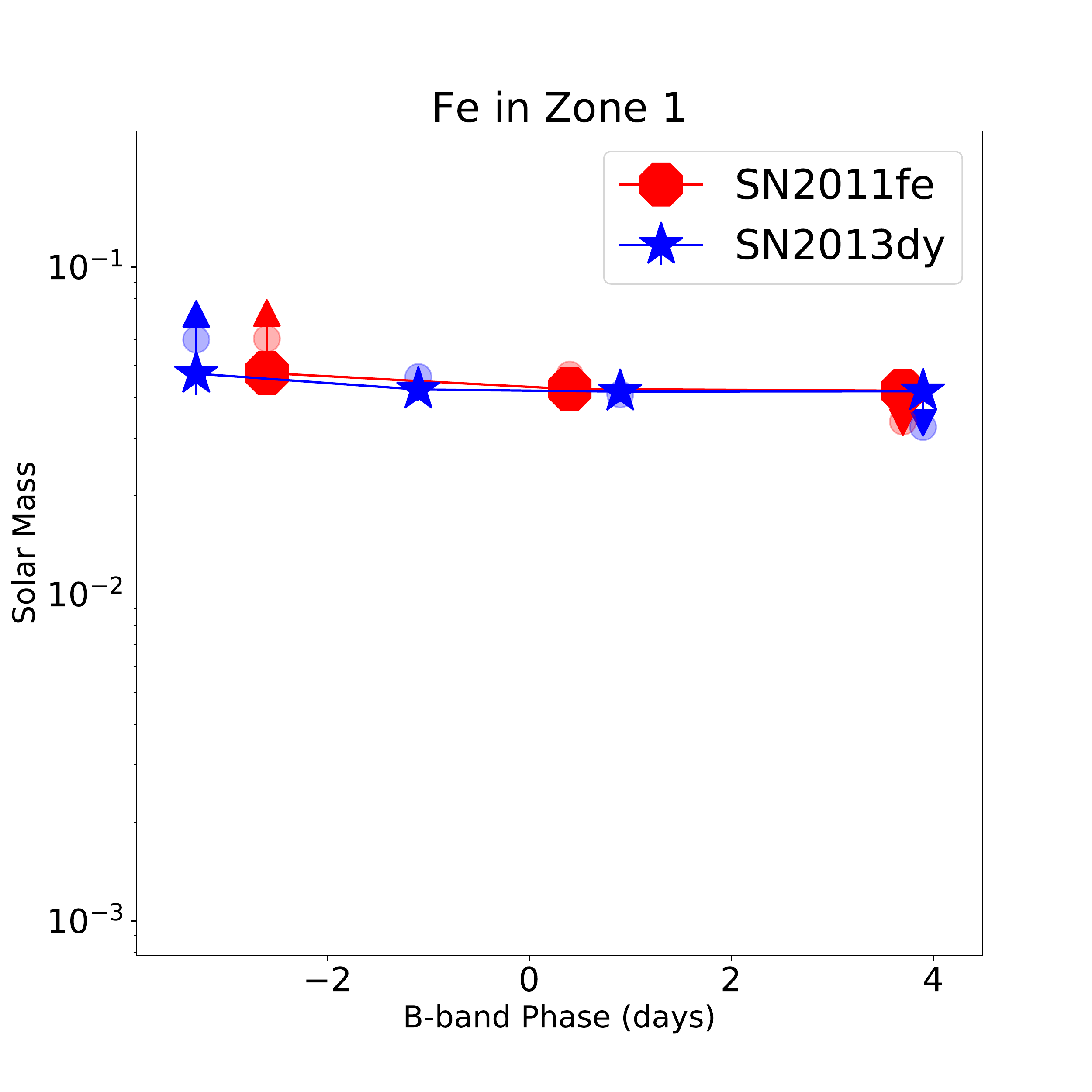}
    \endminipage\hfill
    
    \caption{Same as Figure \ref{fig:RestElementEvolution1}, but for different elements and zones \label{fig:RestElementEvolution2}}
\end{figure}

\begin{figure}[htb!]
    \minipage{0.33\textwidth}
        \includegraphics[width=\textwidth]{TimeEvol/26_2.pdf}
    \endminipage\hfill
    \minipage{0.33\textwidth}
        \includegraphics[width=\textwidth]{TimeEvol/26_3.pdf}
    \endminipage\hfill
    \minipage{0.33\textwidth}
        \includegraphics[width=\textwidth]{TimeEvol/26_4.pdf}
    \endminipage\hfill
    \minipage{0.33\textwidth}
        \includegraphics[width=\textwidth]{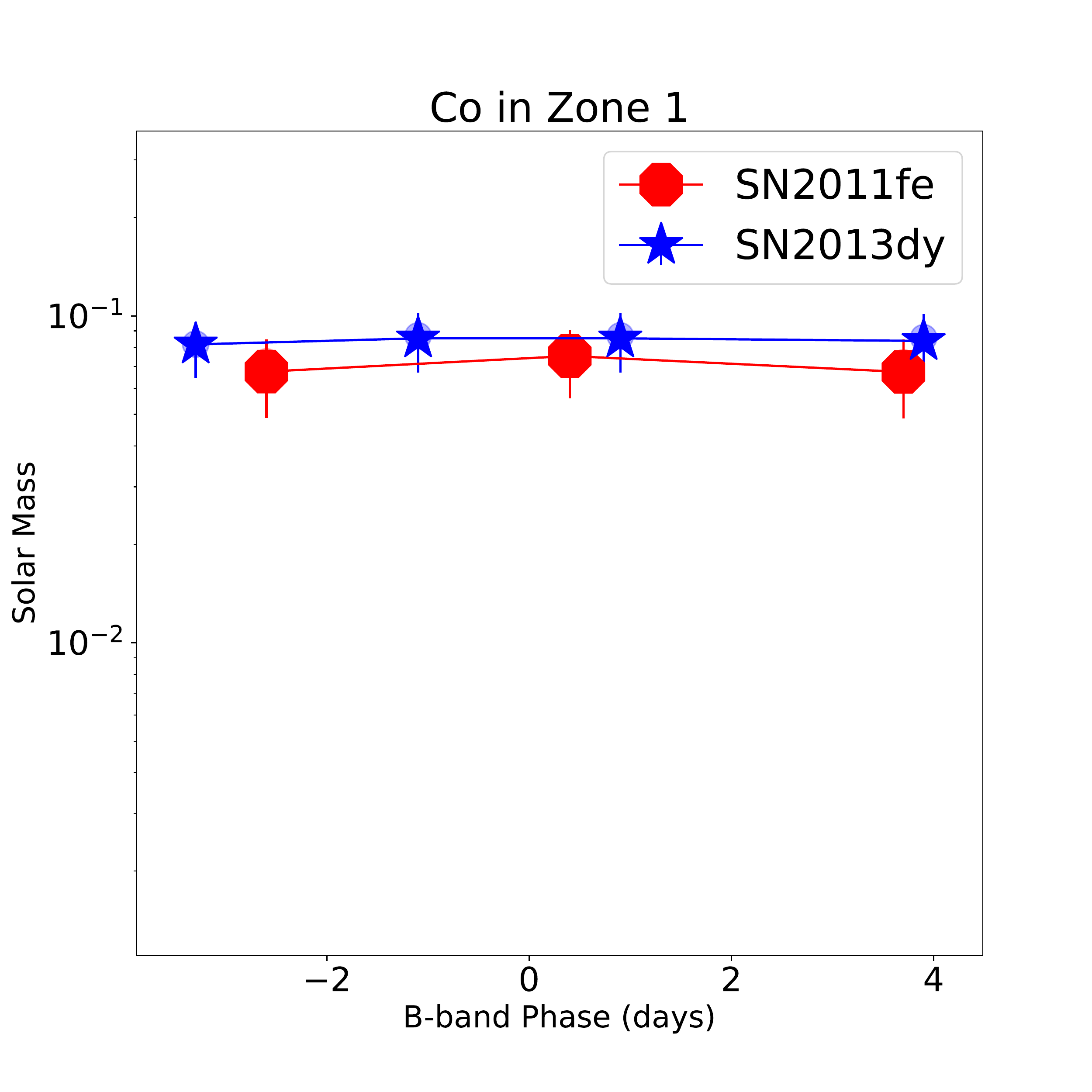}
    \endminipage\hfill
    \minipage{0.33\textwidth}
        \includegraphics[width=\textwidth]{TimeEvol/27_2.pdf}
    \endminipage\hfill
    \minipage{0.33\textwidth}
        \includegraphics[width=\textwidth]{TimeEvol/27_3.pdf}
    \endminipage\hfill
    \minipage{0.33\textwidth}
        \includegraphics[width=\textwidth]{TimeEvol/27_4.pdf}
    \endminipage\hfill
    \minipage{0.33\textwidth}
        \includegraphics[width=\textwidth]{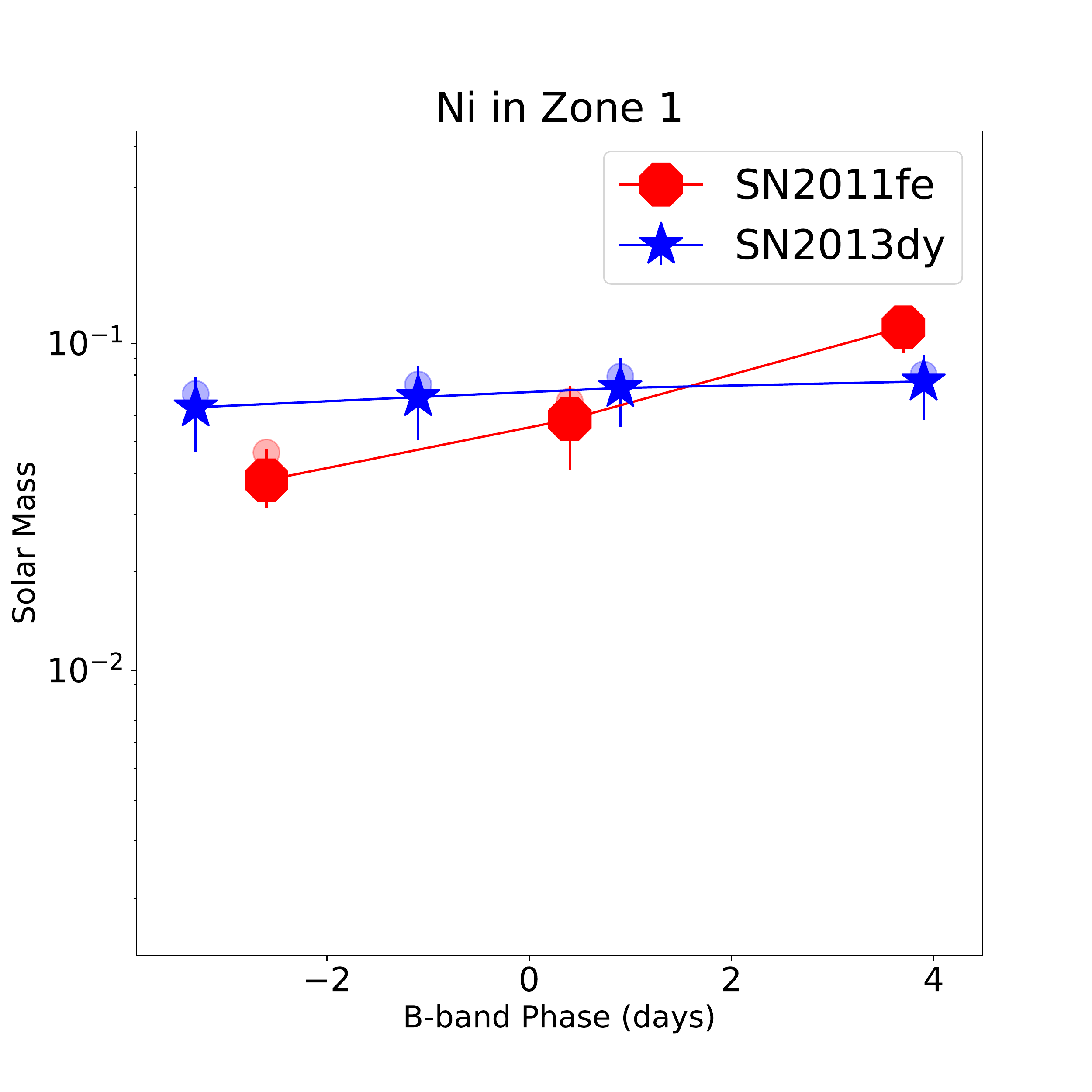}
    \endminipage\hfill
    \minipage{0.33\textwidth}
        \includegraphics[width=\textwidth]{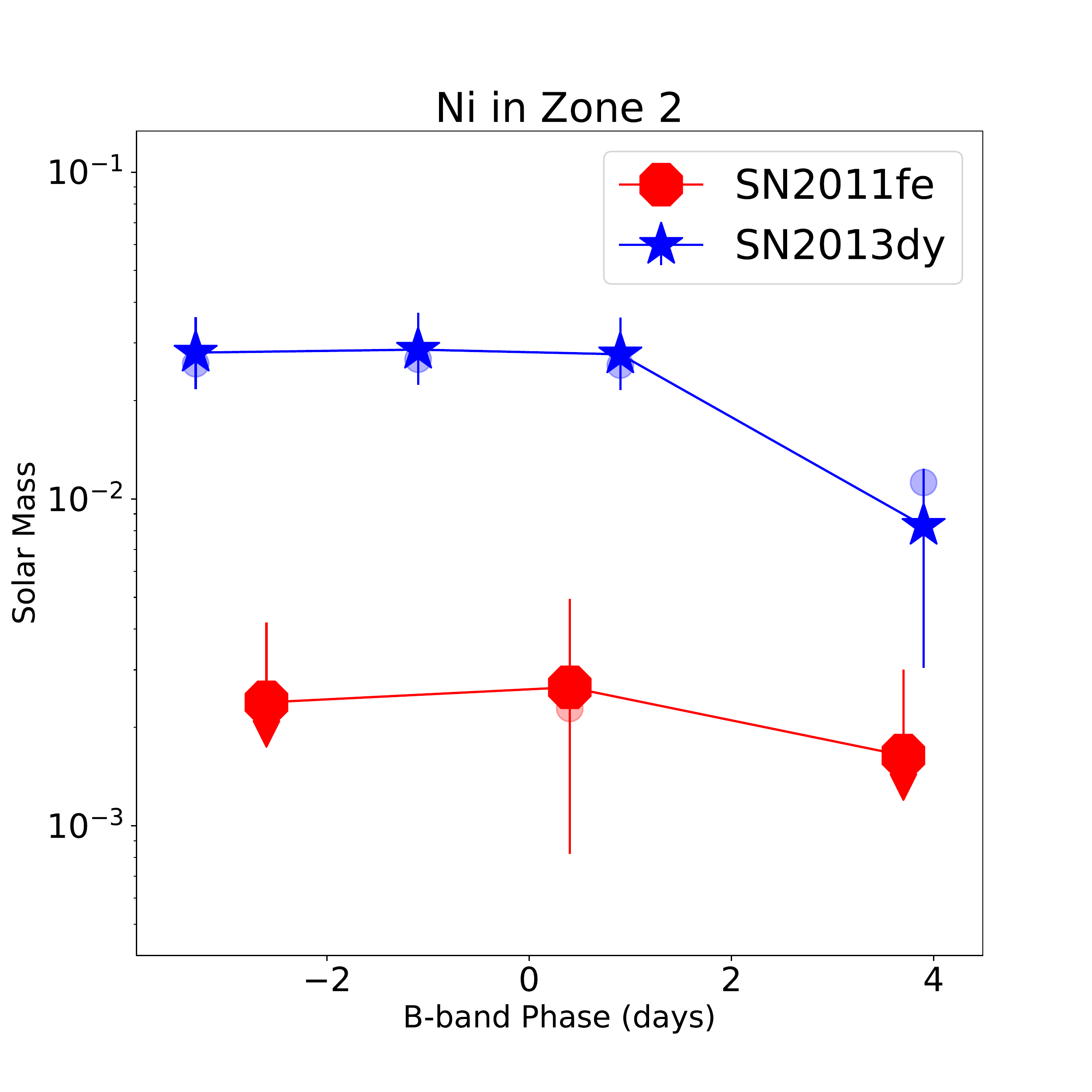}
    \endminipage\hfill
    
    \caption{Same as Figure \ref{fig:RestElementEvolution1}, but for different elements and zones \label{fig:RestElementEvolution3}}
\end{figure}

\begin{figure}[htb!]
    \minipage{0.33\textwidth}
        \includegraphics[width=\textwidth]{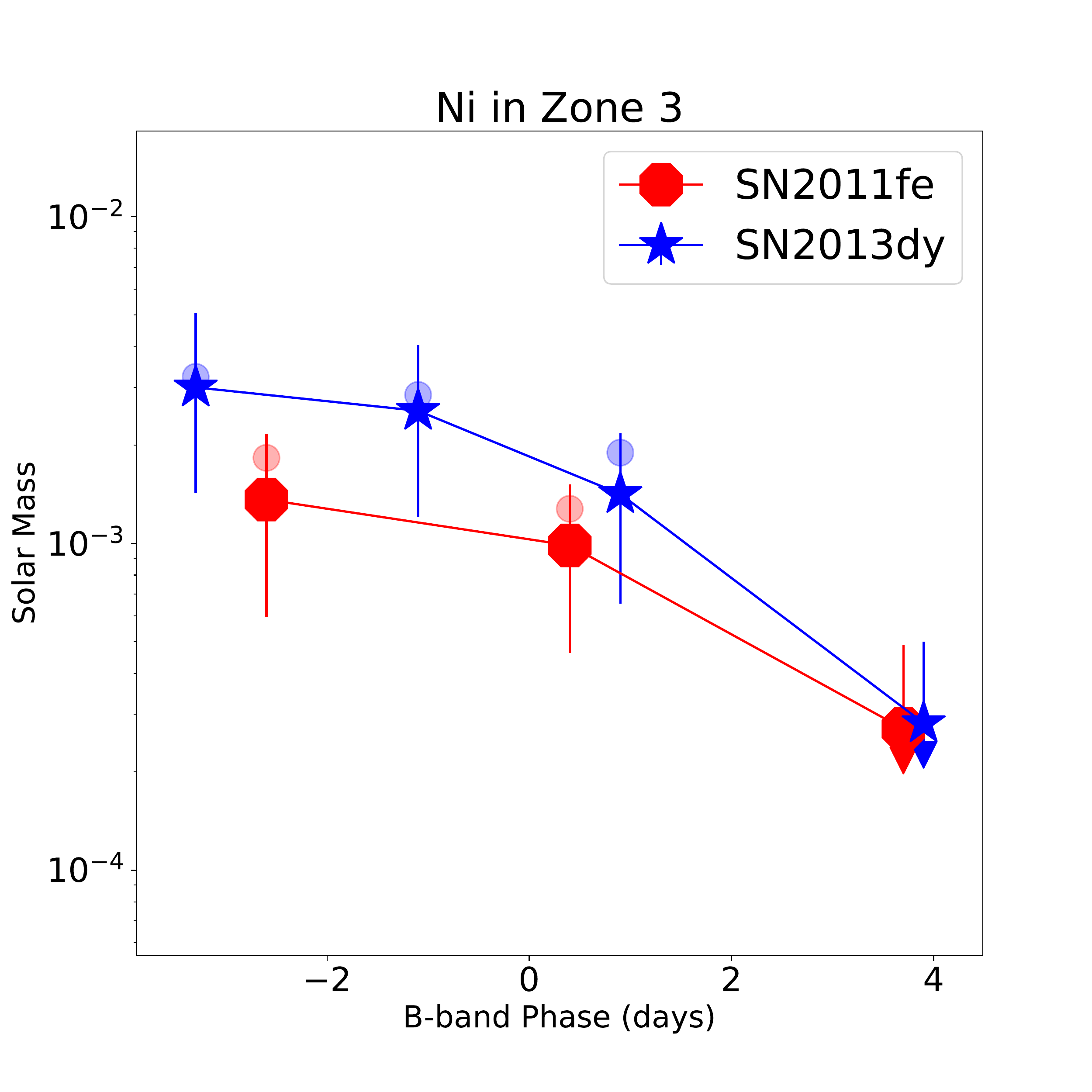}
    \endminipage\hfill
    \minipage{0.33\textwidth}
        \includegraphics[width=\textwidth]{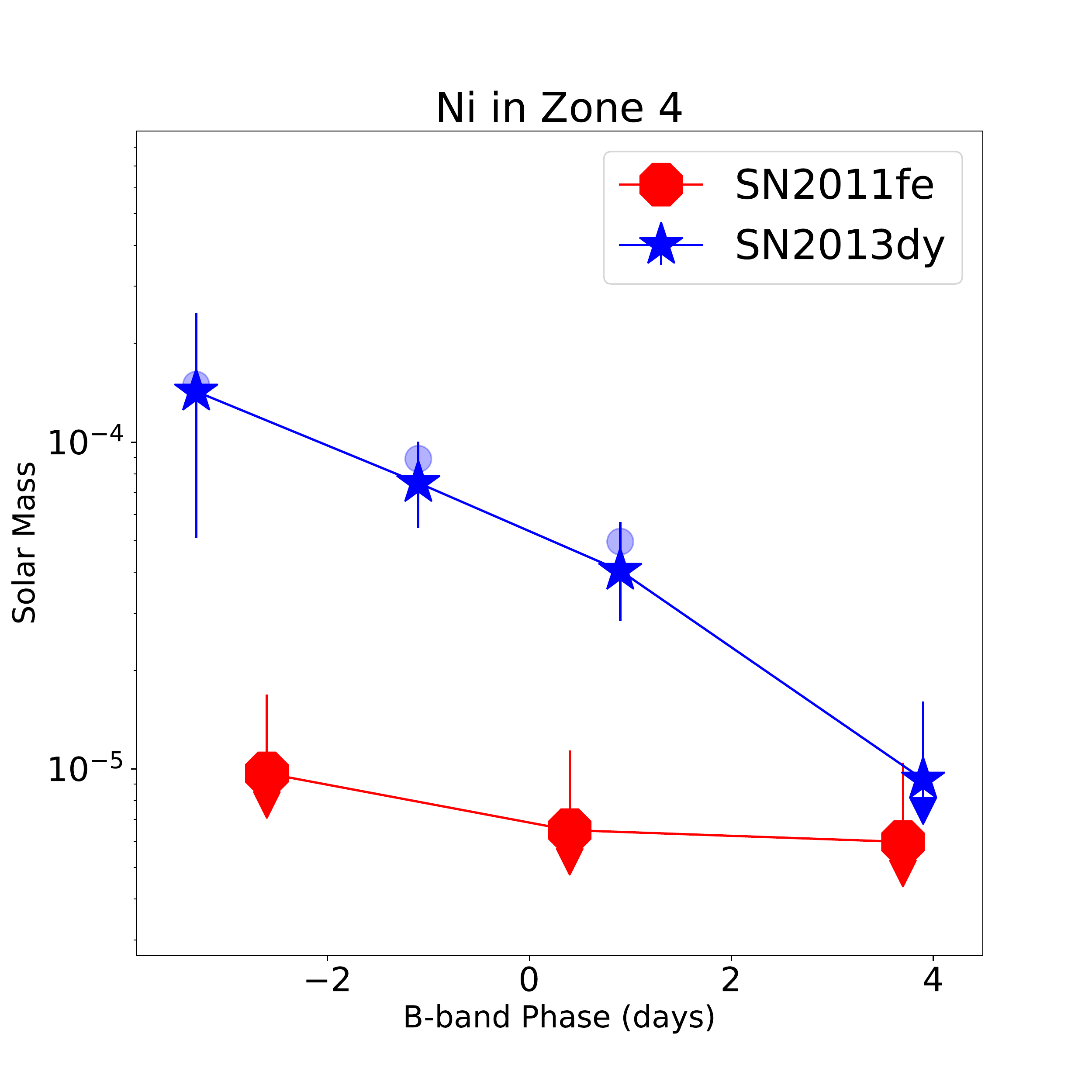}
    \endminipage\hfill
    \caption{Same as Figure \ref{fig:RestElementEvolution1}, but for different elements and zones \label{fig:RestElementEvolution4}}
\end{figure}

We show here (Figures~\ref{fig:RestElementEvolution1}, \ref{fig:RestElementEvolution2}, \ref{fig:RestElementEvolution3}, and \ref{fig:RestElementEvolution4}) the time evolution of the masses of intermediate mass elements. %The data are consistent with no time evolution with time.

\clearpage

\clearpage

\clearpage

\section{The fidelity of 1-$\sigma$ from Testing Set}

There are two caveats for our neural network. First, the MRNN using MSE as loss function is not designed to mimic the input parameter distributions. Second, the parameter space used in Section~\ref{Sec:SpecLib} is \textit{a priori} and not necessarily similar to the real SN elemental mass - zone distribution. 
Therefore, the results from median value on the testing set may be biased, and we directly adopt the predictions from neural network to calculate the spectra in Section~\ref{sec:fitresults}. 
However, the $1-\sigma$ errors from the testing set can be indicative of the sensitivity of MRNNs on different elements and zones, and allow us to assess the fitting fidelity. 

\begin{figure}[htb!]
\includegraphics[width=\linewidth]{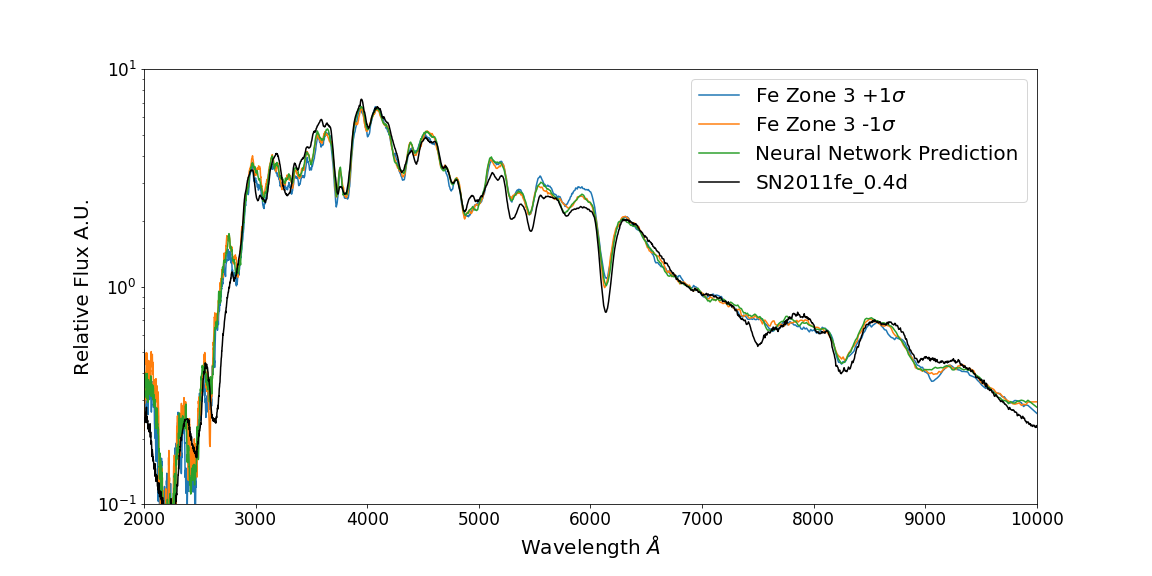}
\includegraphics[width=\linewidth]{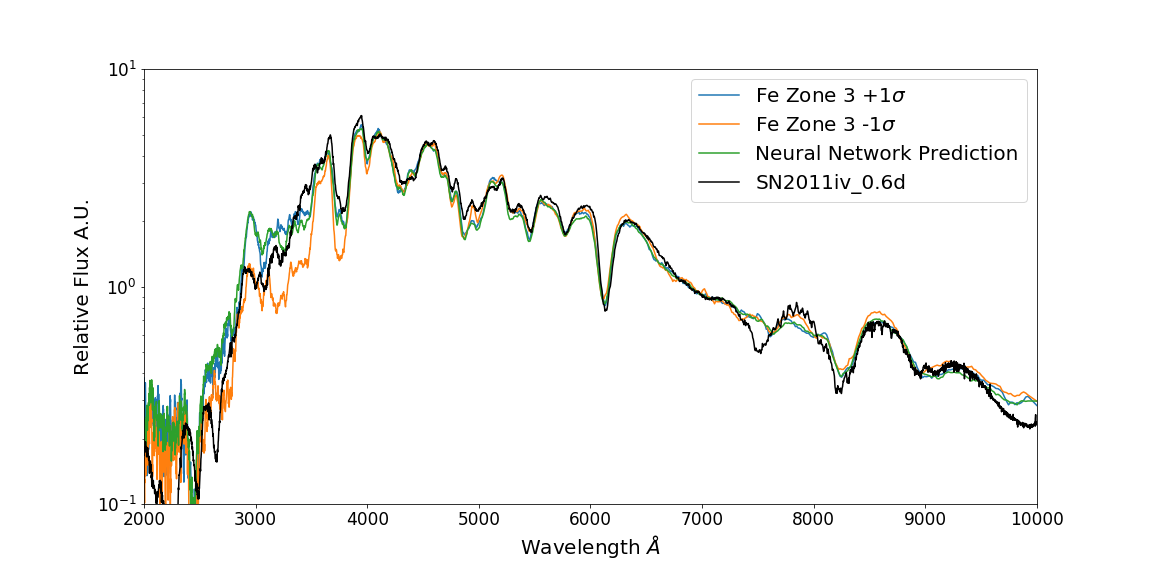}
\caption{Top: SN~2011fe on day 0.4: The observed spectrum (black line) is compared with the TARIDS spectra calculated using the MRNN estimated chemical structure (green line), with the Fe abundance in Zone 3 enhanced by 1 $\sigma$ (blue line), and reduced by 1 $\sigma$ (orange line). Bottom: the same as the Top, but for SN2011iv at 0.6 days. }\label{fig:OneSigma}
\end{figure}

\begin{figure}[htb!]
\includegraphics[width=\linewidth]{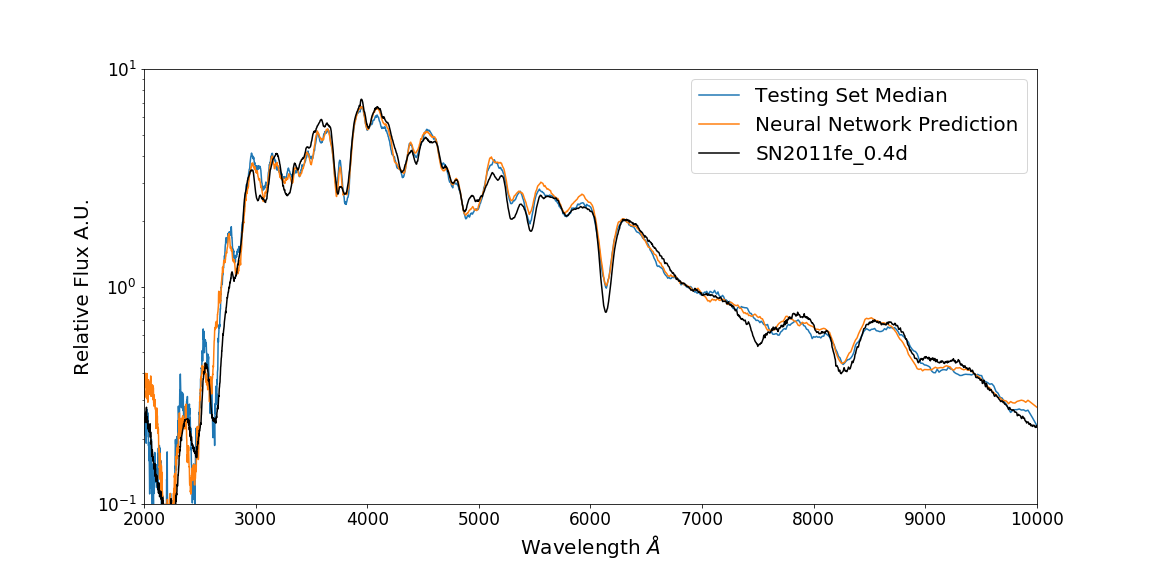}
\includegraphics[width=\linewidth]{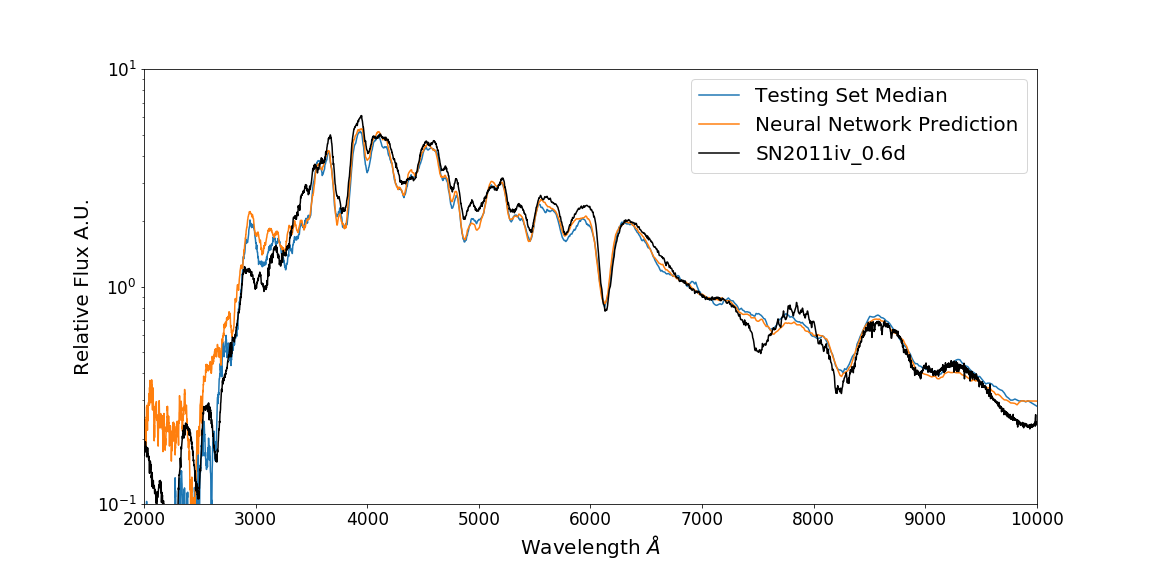}
\caption{Top: The observed spectrum of SN~2011fe at day 0.4 (black), compared with the TARDIS spectrum calculated using MRNN predicted ejecta structure (orange), and the TARDIS spectrum calculated using the median values as the estimates of the ejecta structure (blue). This demonstrates that the predicted spectrum is robust to the methods of ejecta structure estimation. Bottom: The same as Top panel, but for SN~2011iv at 0.6 days. 
}\label{fig:Median}
\end{figure}

\begin{figure}[htb!]
\includegraphics[width=\linewidth]{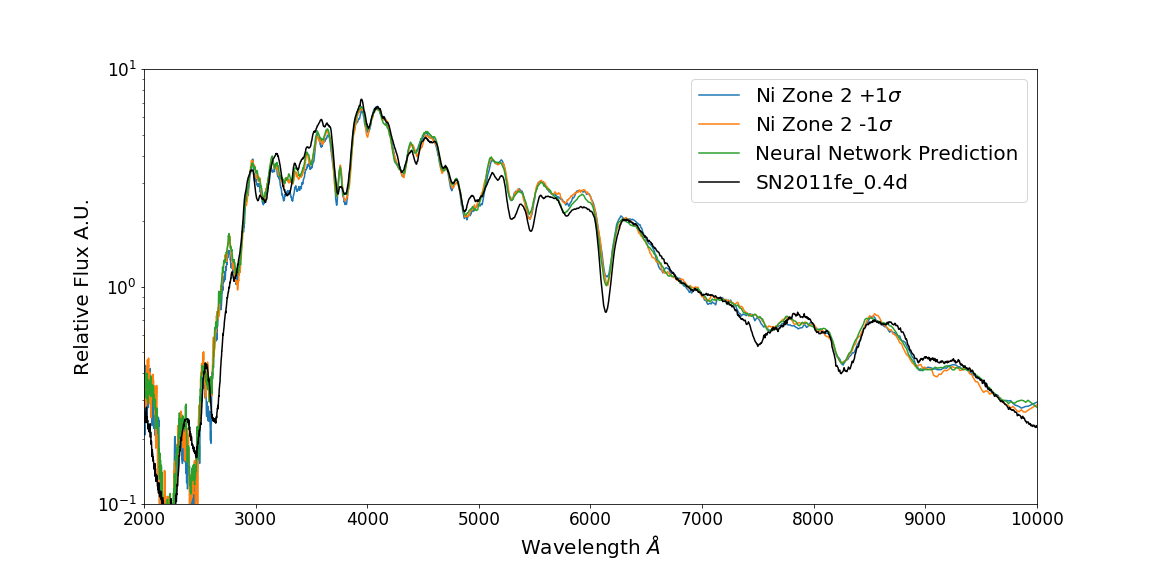}
\includegraphics[width=\linewidth]{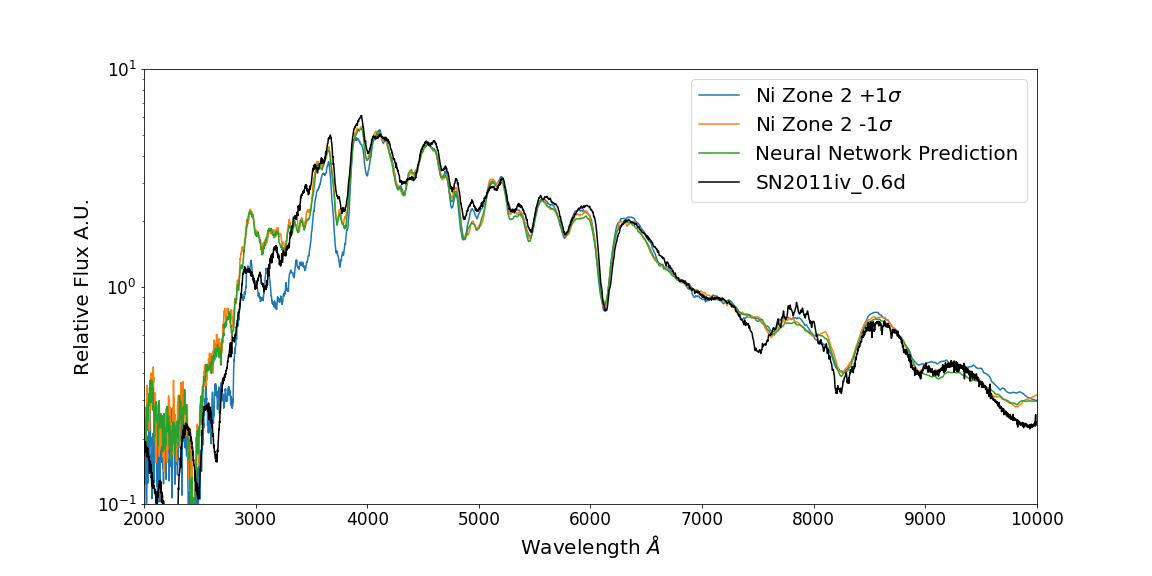}
\caption{The same as Figure~\ref{fig:OneSigma}, but now with Ni in Zone 2 enhanced or reduced by 1 $\sigma$.}\label{fig:OneSigmaNi2}
\end{figure}

\begin{figure}[htb!]
\includegraphics[width=\linewidth]{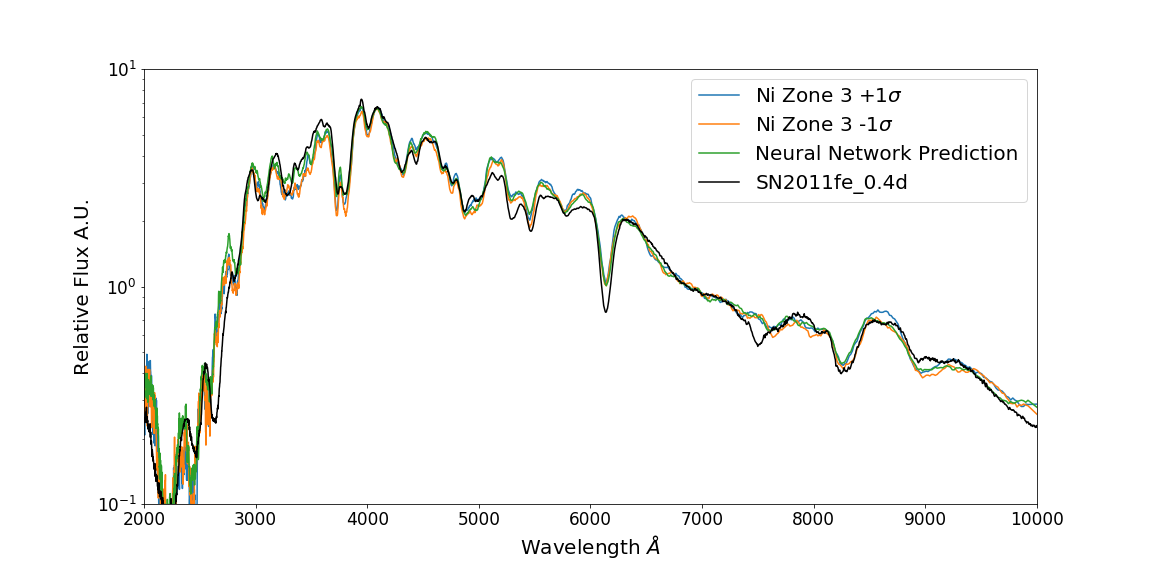}
\includegraphics[width=\linewidth]{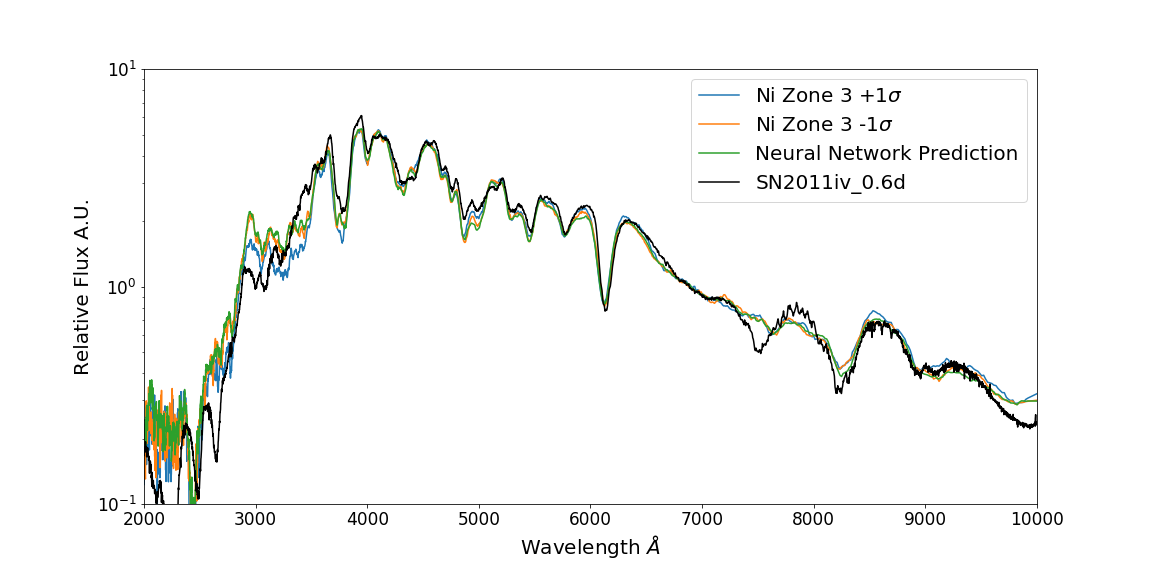}
\caption{The same as Figure~\ref{fig:OneSigma}, but now with Ni in Zone 3 enhanced or reduced by 1 $\sigma$.}\label{fig:OneSigmaNi3}
\end{figure}

We calculated the spectra using the median values as the mass estimates, as shown in Figure \ref{fig:Median}. 
Moreover, we modified the abundances of Fe in Zone 2, Ni in Zone 2 and Ni in Zone 3 mass by $\pm 1\sigma$ based on neural network predictions of the 11 spectra with a  wavelength range of 2000-10000 \AA, to evaluate their effect on the spectral profiles, these are shown in Figure~\ref{fig:OneSigma}, Figure~\ref{fig:OneSigmaNi2} and 
Figure~\ref{fig:OneSigmaNi3}, respectively. These calculations prove that the results from both the median estimates and the direct estimates of TARDIS model parameters reproduce the observations well.

\clearpage

\clearpage

\clearpage

\end{document}